\documentclass[english,twocolumn,rmp,aps]{revtex4}

\newcommand{\vq}{{\vec q}}

\newcommand{\vf}{v_{\rm F}}

\newcommand{\kf}{k_{\rm F}}

\newcommand{\tp}{t_\perp}

\newcommand{\Hca}{\mathcal{H}}

\newcommand{\vk}{{\bf k}}

\usepackage[T1]{fontenc}
\usepackage[latin1]{inputenc}
\usepackage{amsmath}
\usepackage{color}
\usepackage{graphicx}
\usepackage{amssymb}

\makeatletter

\usepackage{bm}
\usepackage{babel}
\makeatother
\begin{document}

\title{The electronic properties of graphene}

\author{ A. H. Castro Neto$^1$, F. Guinea$^2$, N. M. R. Peres$^3$,
K. S. Novoselov$^4$, and A. K. Geim$^4$}

\affiliation{$^1$Department of Physics, Boston University, 590
  Commonwealth Avenue, Boston, MA 02215,USA}

\affiliation{$^2$Instituto de Ciencia de Materiales de
  Madrid. CSIC. Cantoblanco. E-28049 Madrid, Spain}

\affiliation{$^3$Center of Physics and Department of
  Physics, Universidade do Minho, P-4710-057, Braga, Portugal}

\affiliation{$^4$Department of Physics and Astronomy, University of Manchester,
Manchester, M13 9PL, UK}

\date{\today{}}

\begin{abstract}
{\bf
This article reviews the basic theoretical aspects of graphene, a one atom
thick allotrope of carbon, with unusual two-dimensional Dirac-like electronic
excitations. The Dirac electrons can be controlled by application of external
electric and magnetic fields, or by altering sample geometry and/or
topology. We show that the Dirac electrons behave in unusual ways in
tunneling, confinement, and integer quantum Hall effect. We discuss
the electronic properties of graphene stacks and show that they vary with
stacking order and number of layers. Edge (surface) states in graphene are
strongly dependent on the edge termination (zigzag or armchair) and affect
the physical properties of nanoribbons. We also discuss how different types
of disorder modify the Dirac equation leading to unusual spectroscopic and
transport properties. The effects of electron-electron and electron-phonon
interactions in single layer and multilayer graphene are also presented.
}
\end{abstract}


\maketitle

\maketitle
\tableofcontents{}

%------------------------------------------------------------------------------
\section{Introduction}
\label{intro}
%------------------------------------------------------------------------------

\begin{figure}
\begin{center}
\includegraphics*[width=9cm]{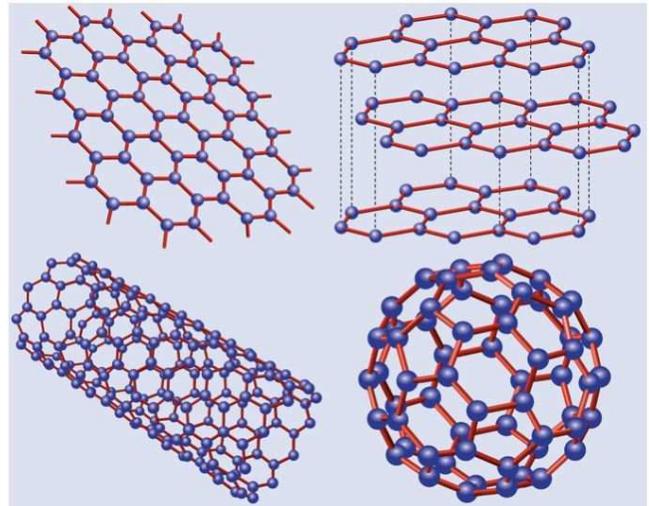}
\end{center}
\caption{\label{family}(Color online) Graphene (top left) is a honeycomb lattice of carbon
atoms. Graphite (top right) can be viewed a stack of graphene layers. Carbon nanotubes are rolled-up cylinders
of graphene (bottom left). Fullerenes (C$_{60}$) are molecules consisting of wrapped graphene
by the introduction of pentagons on the hexagonal lattice \cite{pw}.}
\end{figure}

Carbon is the {\it materia prima} for life on the planet and the basis of all organic chemistry.
Because of the flexibility of its bonding, carbon-based systems show an unlimited number of
different structures with an equally large variety of physical properties. These physical properties are,
in great part, the result of the dimensionality of these structures. Among systems with only
carbon atoms, graphene - a two-dimensional (2D)  allotrope of carbon - plays an important role since
it is the basis for the understanding of the electronic properties in other allotropes. Graphene
is made out of carbon atoms arranged on a honeycomb structure made out of hexagons (see Fig.~\ref{family}),
and can be thought as composed of benzene rings stripped out from their hydrogen atoms \cite{P72}.
Fullerenes \cite{fullerenes} are molecules where carbon atoms are arranged spherically, and hence, from the
physical point of view, are zero-dimensional objects with discrete energy states.
Fullerenes can be obtained from graphene with the introduction of pentagons (that create
positive curvature defects), and hence, fullerenes can be thought as wrapped up graphene.
Carbon nanotubes \cite{nanotubes,nanotube_review} are obtained by rolling graphene along a given direction
and reconnecting the
carbon bonds. Hence, carbon nanotubes have only hexagons and can be thought as one-dimensional
(1D) objects. Graphite, a three dimensional (3D) allotrope of carbon, became widely known to
mankind after the invention of the pencil in 1564 \cite{pencil} and its usefulness as
an instrument for writing comes
from the fact that graphite is made out of stacks of graphene layers that are weakly coupled
by van der Waals forces. Hence, when one presses a pencil against a sheet of paper one is
actually producing graphene stacks and, somewhere among them, there could be individual graphene layers.
Although graphene is the mother for all these different allotropes and has been presumably produced every time
someone writes with a pencil, it was only isolated 440 years after its invention \cite{Netal04}.
The reason is that, first, no one actually expected graphene to exist in the free state and, second,
even with the benefit of hindsight, no experimental tools existed to search for one-atom-thick-flakes among the pencil debris covering macroscopic areas \cite{GMac07}. Graphene was eventually spotted due to
the subtle
optical effect it creates on top of a cleverly chosen SiO$_2$ substrate \cite{Netal04} that
allows its observation with an ordinary optical microscope \cite{blake07,ARF07,CHLQHGNF07}.
Hence, graphene is relatively straightforward to make, but not so easy to find.

The structural flexibility of graphene is reflected in its electronic properties. The sp$^2$
hybridization between one s-orbital and two p-orbitals leads to a trigonal planar structure
with a formation of a $\sigma$-bond between carbon atoms which are separated by $1.42$ \AA.
The $\sigma$-band is responsible for the robustness of the lattice structure in all allotropes.
Due to the Pauli principle these bands have a filled shell and hence, form a deep
valence band. The unaffected p-orbital, which is perpendicular to the planar structure, can bind
covalently with neighboring carbon atoms leading to the formation of a $\pi$-band. Since
each p-orbital has one extra electron, the $\pi$-band is half-filled.

Half-filled bands in transition elements have played an important role in the physics of strongly
correlated systems since, due to its strong tight binding character, the Coulomb energies are very
large, leading to strong collective effects, magnetism, and insulating behavior due to correlation
gaps or Mottness \cite{mottness}. In fact, Linus Pauling proposed in the 1950's that, on the basis
of the electronic properties of benzene, graphene should be a resonant valence bond structure
(RVB) \cite{P72}. RVB states have become very popular in the literature of transition metal oxides,
and particularly in studies of cuprate oxides superconductors \cite{hightc}. This point of view,
should be contrasted with contemporaneous band structure studies of graphene \cite{W47}
that found it to be a semimetal with unusual linearly dispersing electronic excitations called
Dirac electrons. While most of the current experimental data in graphene supports the band
structure point of view, the role of the electron-electron interactions in graphene is a subject
of intense research.

It was P. R. Wallace who in 1946 wrote the first papers on the band structure of graphene
and showed the unusual semimetallic behavior in this material \cite{W47}. At that point
in time, the thought of a purely 2D structure was a mere fantasy and Wallace's studies of
graphene served him as a starting point to study graphite, a very important material
for nuclear reactors in the post-World War II era. During the following years, the study of
graphite culminated with the Slonczewski-Weiss-McClure (SWM) band structure of graphite which provided
a detailed description of the electronic properties in this material \cite{SW58,M57}
and was very successful in describing the experimental data
\cite{M58,SMS64,WFD65,BN58,SS60,M64,DSM77}. Interestingly enough, from 1957 to 1968, the assignment
of the electron and hole states within the SWM model
were the opposite to what is accepted today. In 1968, Schroeder {\it et al.}
\cite{SDJ68} established the currently accepted location of electron and hole pockets \cite{mcclure_review}.
The SWM model has been revisited in recent years because of its inability to describe the
van der Waals-like interactions between graphene planes, a problem that requires the understanding
of many-body effects that go beyond the band structure description \cite{langreth}. These issues,
however, do not arise in the context of a single graphene crystal but they show up with great
importance when graphene layers are stacked on top of each other, as in the case, for instance, of
the bilayer graphene. Stacking can change the electronic properties considerably and
the layering structure can be used in order to control the electronic properties.

One of the most interesting aspects of the graphene problem is that its low energy excitations
are massless, chiral, Dirac fermions. In neutral graphene the chemical potential crosses
exactly the Dirac point. This particular dispersion, that is only valid at low energies, mimics
the physics of quantum electrodynamics (QED) for massless fermions except by the fact
that in graphene the Dirac fermions move with a speed $v_F$ which is 300 times smaller than the speed
of light, $c$. Hence, many of the unusual properties of QED
can show up in graphene but at much smaller speeds \cite{pw,KNG06,KN0703}. Dirac fermions
behave in very unusual
ways when compared to ordinary electrons if subjected to magnetic fields,
leading to new physical phenomena \cite{PGC06,GS05} such as the anomalous integer
quantum Hall effect (IQHE) measured experimentally \cite{Netal05,Zetal05b}. Besides being
qualitatively different from the IQHE observed in Si and GaAlAs (heterostructures) devices \cite{iqhe_si}, the
IQHE in graphene can be observed at room temperature because of the large cyclotron energies
for ``relativistic'' electrons \cite{Netal07}. In fact, the anomalous
IQHE is the trademark of Dirac fermion behavior.

Another particularly interesting feature of Dirac fermions is their insensitivity to external
electrostatic potentials due to the so-called Klein paradox, that is, the fact that Dirac fermions
can be transmitted with probability one through a classically forbidden
region \cite{zuber06,calogeracos99}.  In fact, Dirac fermions behave in a
very unusual way in the presence of confining potentials leading to the phenomenon
of {\it zitterbewegung}, or jittery motion of the wavefunction
\cite{zuber06}. In graphene these electrostatic
potentials can be easily generated by disorder. Since disorder is
unavoidable in any material, there has been great interest in trying to understand how disorder
affects the physics of electrons in graphene and its transport properties.
In fact, under certain conditions, Dirac fermions
are immune to localization effects observed in ordinary electrons \cite{localization} and it
has been established experimentally that electrons can propagate without scattering over large
distances of the order of micrometers in graphene \cite{Netal04}. The sources
of disorder in graphene are many and can vary from ordinary effects commonly found in semiconductors,
such as ionized impurities in the Si substrate, to adatoms and various molecules adsorbed in the graphene
surface, to more unusual defects such as ripples associated with the soft structure of
graphene \cite{meyer07}. In fact, graphene is unique in the sense that it shares properties of
soft membranes \cite{nelson} and at the same time it behaves in a metallic way, so that the Dirac
fermions propagate on a locally curved space. Here, analogies with problems of quantum gravity
become apparent \cite{qgravity}. The softness of graphene is related with the fact that it
has out-of-plane vibrational modes (phonons)
that cannot be found in 3D solids. These flexural modes, responsible for the bending properties of
graphene, also account for the lack of long range structural order in soft membranes leading
the phenomenon of crumpling \cite{nelson}. Nevertheless, the presence of a substrate or scaffolds that
hold graphene in place, can stabilize a certain degree of order in graphene but leaves behind
the so-called ripples (which can be viewed as frozen flexural modes).

It was realized very early on that graphene should also present unusual mesoscopic effects \cite{nuno_cond,Kats0703}.
These effects have their origin in  the boundary conditions required for the wavefunctions in mesoscopic samples with various types of edges graphene can have \cite{Nakada96,WFA+99,PGC06,AkBe07}. The most studied edges, zigzag
and armchair, have drastically different electronic properties. Zigzag edges can sustain edge (surface)
states and resonances that are not present in the armchair case. Moreover, when coupled to conducting
leads, the boundary conditions for a graphene ribbon strongly affects its
conductance and
the chiral Dirac nature of the fermions in graphene can be exploited for applications where one can
control the valley flavor of the electrons besides its charge, the so-called valleytronics \cite{valleytronics}.
Furthermore, when superconducting contacts are attached to graphene, they lead to the development
of supercurrent flow and Andreev processes characteristic of superconducting proximity effect \cite{Hetal06}.
The fact that Cooper pairs can propagate so well in graphene attests for the robust electronic coherence in this
material. In fact, quantum interference phenomena such as weak localization, universal conductance fluctuations \cite{Metal06}, and
the Aharonov-Bohm
effect in graphene rings have already been observed experimentally \cite{ROWHSVM07,RTBBM07}.
The ballistic electronic propagation in graphene can be used for field effect
devices such as p-n \cite{CF06,TSAB07,HSSTYG07,LEBK07,WDM07,CFA07,ZF07,FGNS07}
and p-n-p \cite{OTB07} junctions, and
as ``neutrino'' billiards \cite{BM87,MWCZL07}. It has also been suggested that Coulomb interactions are considerably
enhanced in smaller geometries, such as graphene quantum dots \cite{MPVP07},
leading to unusual Coulomb blockade effects \cite{geim_review} and perhaps to magnetic phenomena such as the Kondo
effect. The amazing transport properties of graphene allow for their use in a
plethora of applications ranging from single molecule detection \cite{Setal07,WNMVKGL07}
to spin injection \cite{HGNSB07,OSNNSS07,CCF07,TJPJW07}.

Because of its unusual structural and electronic flexibility, graphene can be tailored
chemically and/or structurally in many different ways: deposition of metal atoms
\cite{ULCN07,CM07} or molecules \cite{Setal07,WNMVKGL07,LPP07} on top;
intercalation (as it is done in graphite intercalated compounds \cite{gic,DDFM83,TK85});
incorporation of nitrogen and/and boron in its structure \cite{nit_bor,MMSF07}
(in analogy with what has been done in nanotubes \cite{nano_boron});
using different substrates that modify the electronic structure \cite{ZGFFHLGCNL07,Vetal07b,GKBKB07,FNPMDBH07,CBMLB07,DCS07}.
The control of graphene properties can be extended in new directions allowing for creation of graphene-based
systems with magnetic and superconducting properties \cite{UCN07} that are unique in their 2D properties.
Although the graphene field is still in its infancy, the scientific and technological possibilities
of this new material seem to be unlimited. The understanding and control of the properties of this material
can open doors for a new frontier in electronics. As the current status of the experiment and potential
applications have recently been reviewed \cite{geim_review}, in this article we mostly concentrate on
the theory and more technical aspects of electronic properties of this exciting new material.

%------------------------------------------------------------------------------
\section{Elementary electronic properties of graphene}
\label{elementary}
%------------------------------------------------------------------------------

\subsection{Single layer: tight-binding approach}
\label{single}

\begin{figure}
\begin{center}
\includegraphics*[width=8cm]{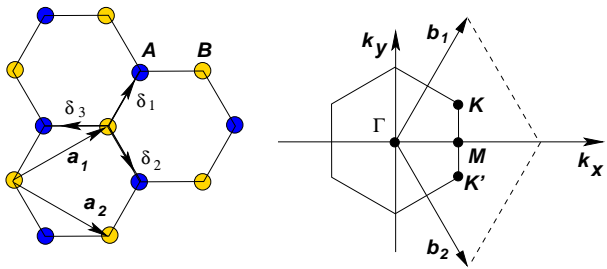}
\end{center}
\caption{\label{cap:bz}(Color online)
Left: Lattice structure of graphene, made out of two interpenetrating
triangular lattices ($\bm a_1$ and $\bm a_2$ are the lattice unit vectors,
and $\bm \delta_i$, $i=1,2,3$ are the nearest neighbor vectors);
Right: corresponding Brillouin zone. The Dirac cones are located at the K and K' points.}
\end{figure}

Graphene is made out of carbon atoms arranged in hexagonal structure as shown in
Fig. \ref{cap:bz}. The structure is not a Bravais lattice but can be seen as a triangular
lattice with a basis of two atoms per unit cell. The lattice vectors can be written as:
\begin{equation}
{\bm{a_1}}= \frac a 2 (3,\sqrt 3)\,,
\hspace{1cm}
{\bm{a_2}} = \frac a 2 (3,-\sqrt 3)\,,
\end{equation}
where $a \approx 1.42$ \AA \, is the carbon-carbon distance. The reciprocal lattice vectors are given by:
\begin{equation}
{\bm{b_1}} = \frac {2\pi}{3a} (1,\sqrt 3)\,,
\hspace{1cm}
{\bm{b_2}} = \frac {2\pi}{3a} (1,-\sqrt 3)\,.
\end{equation}
Of particular importance for the physics of graphene are the two points $K$ and $K'$ at the corners of the
graphene Brillouin zone (BZ). These are named Dirac points for reasons that will become clear later.
Their positions in momentum space are given by:
\begin{equation}
\label{kpoint}
{\bm K}=\left(\frac {2\pi}{3a},\frac {2\pi}{3\sqrt 3 a}
\right)\,, \hspace{1cm}
{\bm K'}=\left(\frac {2\pi}{3a},-\frac {2\pi}{3\sqrt 3 a}
\right)\,.
\end{equation}
The three nearest neighbors vectors in real space are given by:
\begin{equation}
{\bm{\delta_1}} =  \frac a 2 (1,\sqrt 3)
\hspace{0.85cm}
{\bm{\delta_2}} = \frac a 2 (1,-\sqrt 3)
\hspace{0.85cm}
{\bm{\delta_3}} = - a(1,0)
\end{equation}
while the six second-nearest neighbors
are located at: ${\bm \delta'_1}=\pm {\bm a_1},
{\bm \delta'_2}=\pm {\bm a_2}, {\bm \delta'_3}=\pm ({\bm a_2}-{\bm a_1})$.

\begin{figure}[!ht]
\begin{center}
\includegraphics*[width=8cm]{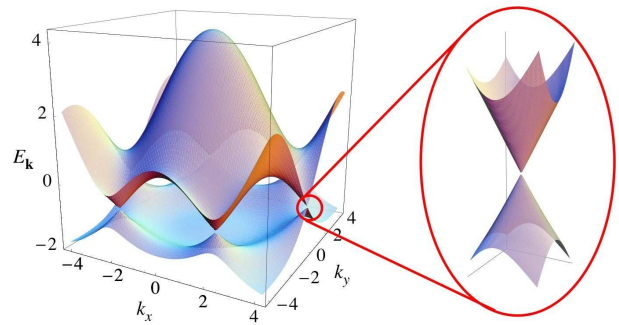}
\end{center}
\caption{(Color online) Left: Energy spectrum (in units of $t$)
for finite values of $t$ and $t'$, with
$t=$2.7 eV  and $t'=0.2t$. Right: zoom-in
of the energy bands close to one of the  Dirac points.
\label{cap:bands}}
\end{figure}

The tight-binding Hamiltonian for electrons in graphene considering that
electrons can hop both to nearest and
next nearest neighbor atoms has the form (we use units such that $\hbar=1$):
\begin{eqnarray}
\label{H1}
H=&-&t \sum_{\langle i,j \rangle,\sigma}
\left(a_{\sigma,i}^{\dag} b_{\sigma,j}
+ {\rm h.c.} \right)
\nonumber
\\
  &-& t'
\sum_{\langle \langle i,j \rangle \rangle,\sigma}
\left(a_{\sigma,i}^{\dag}a_{\sigma,j} +
   b_{\sigma,i}^{\dag}b_{\sigma,j} + {\rm h.c.} \right) \, ,
\end{eqnarray}
where $a_{i,\sigma}$ ($a^{\dag}_{i,\sigma}$) annihilates (creates) an electron
with spin $\sigma$ ($\sigma = \uparrow,\downarrow$) on site ${\bf R}_i$ on sublattice A (an equivalent definition is used for sublattice B), $t$ ($\approx 2.8$ eV) is the nearest neighbor
hopping energy (hopping between different sublattices), $t'$
\footnote[1]{The value of $t'$ is not
well known but {\it ab initio} calculations \cite{Retal02} find $0.02 t \lesssim t' \lesssim 0.2t$
depending on the tight-binding parameterization. These calculations
also include the effect of a third nearest neighbors
hopping, which has a value of around $0.07$ eV. A tight binding
fit to cyclotron resonance experiments \cite{DCNNG07} finds $t'\approx 0.1$ eV.}is the next nearest neighbor hopping energy (hopping in the same sublattice).
 The energy bands derived from this Hamiltonian
have the form \cite{W47}:
\begin{equation}
\label{E1}
\begin{array}{l}
\displaystyle{E_{\pm}(\mathbf k)=\pm t \sqrt{3 + f(\mathbf k)}
- t' f(\mathbf k)} \, , \\ \\
\displaystyle{f(\mathbf k)=2\cos\left(\sqrt{3}k_ya\right)+
4\cos\left(\frac{\sqrt{3}}{2}k_ya\right)\cos\left(\frac{3}{2}k_xa\right)} \, ,
\end{array}
\end{equation}
where the plus sign applies to the upper ($\pi$) and the minus
sign the lower ($\pi^*$) band.
It is clear from (\ref{E1}) that the spectrum is symmetric
around zero energy if $t'=0$. For finite values of $t'$ the electron-hole
symmetry is broken and the $\pi$ and $\pi^*$ bands become asymmetric.
In Fig.~\ref{cap:bands} we show the full band structure of graphene
with both $t$ and $t'$. In the same figure we also show
a zoom in of the band structure close to one of the Dirac points (at the K or K' point in the BZ).
This dispersion can be obtained by expanding the full band structure, eq.(\ref{E1}), close
to the ${\bf K}$ (or ${\bf K'}$) vector, eq.(\ref{kpoint}), as: ${\bf k} = {\bf K} + {\bf q}$,
with $|{\bf q}| \ll |{\bf K}|$ \cite{W47}:
\begin{equation}
E_{\pm}(\mathbf q) \approx \pm v_F\vert \mathbf q\vert + {\cal O}((q/K)^2) \, ,
\label{eq:conical}
\end{equation}
where $\bm q$ is the momentum measured relatively to the Dirac points
and  $v_F$ represents the Fermi velocity, given by
$v_F=3ta/2$, with a value $v_F\simeq 1\times 10^6$ m/s.
This result was first obtained by Wallace \cite{W47}.

The most striking
difference between this result and the usual case, $\epsilon({\bf q}) = q^2/(2 m)$
where $m$ is the electron mass, is that the Fermi velocity in (\ref{eq:conical})
does not depend on the energy or momentum: in the usual case we have $v = k/m =
\sqrt{2 E/m}$ and hence the velocity changes substantially with energy.
The expansion of the spectrum around the Dirac point including $t'$ up to
second order in $q/K$ is given by:
\begin{equation}
E_{\pm}(\mathbf q)\simeq 3 t' \pm v_{\scriptscriptstyle{F}}  \lvert \mathbf q \rvert
- \left(\frac{9 t' a^2}{4} \pm \frac{3 t a^2}{8} \sin(3 \theta_{{\bf q}})\right) |{\bf q}|^2 \,,
\label{Efull}
\end{equation}
where
\begin{eqnarray}
\theta_{{\bf q}} = \arctan\left(\frac{q_x}{q_y}\right) \, ,
\label{angle}
\end{eqnarray}
is the angle in momentum space.
Hence, the presence of $t'$ shifts in energy the position of the Dirac point and
breaks electron-hole symmetry. Notice that up to order $(q/K)^2$ the dispersion depends
on the direction in momentum space and has a three fold symmetry. This is the so-called
trigonal warping of the electronic spectrum \cite{gic,ANS98}.

\subsubsection{Cyclotron mass}
\label{cyclotron}

\begin{figure}[!ht]
\begin{center}
\includegraphics*[width=10cm]{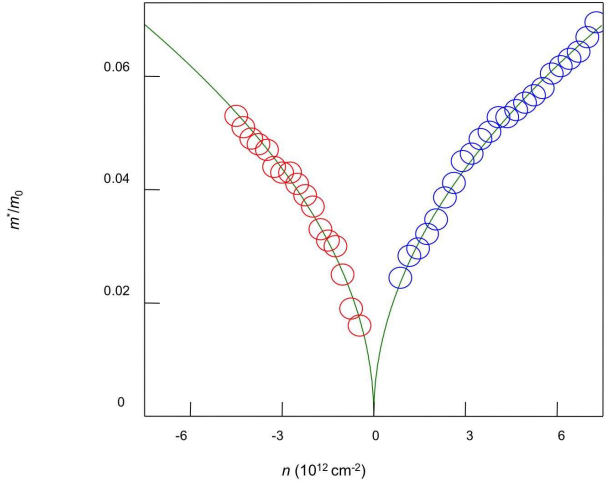}
\end{center}
\caption{(Color online) Cyclotron mass of charge carriers in graphene as a function of their
concentration $n$. Positive and negative $n$ correspond to electrons and holes, respectively.
Symbols are the experimental data extracted from temperature dependence of the SdH oscillations;
solid curves is the best fit to Eq. (\ref{eq:msrootn}). $m_0$ is the free electron mass. Adapted from \onlinecite{Netal05}.
\label{mcexp}}
\end{figure}

The energy dispersion (\ref{eq:conical}) resembles the energy
of ultra-relativistic particles; these particles are
quantum mechanically described by
the massless Dirac equation (see  section \ref{sec:continuous}
for more on this analogy).
An immediate consequence of this
massless Dirac-like dispersion
is a cyclotron mass that depends on the electronic density as its
square root \cite{Netal05,Zetal05b}.
The cyclotron mass is defined, within the semiclassical
approximation \cite{Ashcroft}, as
\begin{equation}
m^\ast = \frac {1}{2\pi}\left[\frac {\partial A(E)}{\partial E}\right]_{E=E_F}\,,
\label{cmass}
\end{equation}
with $A(E)$ the area in $k-$space enclosed by the orbit and given by:
\begin{equation}
A(E) = \pi q(E)^2=\pi \frac {E^2}{v_F^2}\,.
\label{AE}
\end{equation}
Using (\ref{AE}) in (\ref{cmass}) one obtains:
\begin{equation}
m^\ast = \frac {E_F}{v^2_F}=\frac {k_F}{v_F}\,.
\label{mc2}
\end{equation}
The electronic density, $n$, is related to the Fermi momentum, $k_F$, as
$k^2_F/\pi=n$ (with contributions from the two Dirac
points ${\bm K}$ and  ${\bm K'}$ and spin included) which leads to:
\begin{equation}
\label{eq:msrootn}
m^\ast =\frac {\sqrt{\pi}}{v_F}\sqrt n\,.
\end{equation}
Fitting (\ref{eq:msrootn}) to the experimental data (see Fig.\ref{mcexp}) provides an estimation
for the Fermi velocity and the hopping parameter as
$v_F \approx 10^{6} \, \textrm{ms$^{-1}$}$ and $t \approx 3 \, \textrm{eV}$, respectively.
The experimental observation of the $\sqrt n$ dependence of the
cyclotron mass provides evidence for the existence of massless Dirac
quasiparticles in graphene \cite{Jetal07,Netal05,Zetal05b,DCNNG07} ~-~ the usual parabolic (Schr\"odinger)
dispersion implies a constant cyclotron mass.

\subsubsection{Density of states}
\label{dos}

\begin{figure}[!ht]
\begin{center}
\includegraphics*[width=8cm]{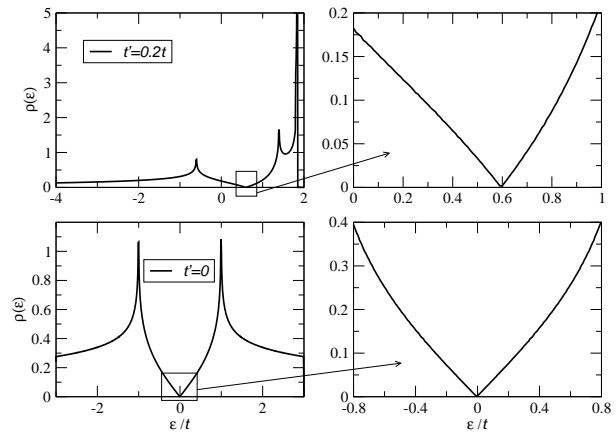}
\end{center}
\caption{(Color online) Density of states per unit cell as a function of energy (in units
of $t$) computed
from the energy dispersion (\ref{E1}),
$t' =0.2 t$ (top) and for $t'=0$ (bottom). Also shown is a zoom in of the
density of states close to the neutrality point of one electron
per site. For the case $t'=0$ the electron-hole nature of the spectrum
is apparent and the density of states close to the neutrality point can
be approximated by $\rho(\epsilon)\propto |\epsilon$|.
\label{Fig:dos}}
\end{figure}

The density of states per unit cell, derived from (\ref{E1}), is given in Fig. \ref{Fig:dos}
for both $t'=0$ and $t'\neq 0$, showing in both cases a semimetallic behavior \cite{W47,BK05}. For $t'=0$ it is possible
to derive an analytical expression for the density of states per unit cell, which has the form \cite{HN53}:
\begin{equation}\label{Dos13}
\begin{array}{l}

\rho(E)=\displaystyle{\frac{4}{\pi^2} \frac{\lvert E \rvert}{t^2} \frac{1}{\sqrt{Z_0}}
\mathbf{F}\left(\frac{\pi}{2}, \sqrt{\frac{Z_1}{Z_0}}\right)}\\

Z_0\!=\!\left\{\begin{array}{ll}
\displaystyle{\left(1\!+\!\Big\lvert
\frac{E}{t}\Big\rvert\right)^2\!-\!\frac{\left(\left(\frac{E}{t}\right)^2\!-\!1\right)^2}{4}} \quad;\quad
-t\!\leq\! E \!\leq\! t \\ \\

\displaystyle{4\Big\lvert \frac{E}{t}\Big\rvert} \quad;\quad -3t \leq E \leq
-t \lor t\leq E \leq 3t\end{array} \right. \\ \\

Z_1\!=\!\left\{\begin{array}{ll}
\displaystyle{4\Big\lvert \frac{E}{t}\Big\rvert} \quad;\quad -t \leq E \leq t \\ \\

\displaystyle{\!\!\left(\!1\!+\!\Big\lvert
\frac{E}{t}\Big\rvert\!\right)^2\!\!-\!\!\frac{\left(\!\left(\frac{E}{t}\right)^2\!-\!\!1\!\right)^2}{4}}
;-3t \!\leq\! E \!\leq\! -t \lor t\leq E \!\leq\! 3t \end{array} \right.
\end{array}
\end{equation}
where $\mathbf{F}(\pi/2,x)$ is the complete elliptic integral
of the first kind. Close to the Dirac point the dispersion
is approximated by (\ref{eq:conical}) and the expression for the
density of states per unit cell is given by
(with a degeneracy of four included):
\begin{equation}
  \rho(E) = \frac {2A_c} {\pi}\frac {\vert E\vert}{v^2_F}\,
\label{rho0}
\end{equation}
where $A_c$ is the unit cell area given by
$A_c=3\sqrt 3 a^2/2$.
It is worth noting that the density
of states for graphene is very different from the density
of states of carbon nanotubes \cite{SFDD092a,SFDD092b}.
The latter show $1/\sqrt E$ singularities due to the 1D nature
of their electronic spectrum, which comes about
due to the quantization of the momentum in the direction
perpendicular to the tube axis. From this perspective,
graphene nanoribbons, which also have momentum quantization perpendicular
to the ribbon  length, have properties very similar to carbon nanotubes.

%------------------------------------------------------------------------------
\subsection{Dirac fermions}
\label{sec:continuous}
%------------------------------------------------------------------------------

Let us consider Hamiltonian (\ref{H1}) with $t'=0$ and consider
the Fourier transform of the electron operators:
\begin{equation}
a_n =\frac 1{\sqrt{N_c}}\sum_{\bm k}e^{-i\bm k\cdot\bm R_n}
a(\bm k),,
\end{equation}
where $N_c$ is the number of unit cells. Using this transformation,
let us write  the field $a_n$ as a sum of two terms,
coming from expanding the Fourier sum around   ${\bm K'}$ and  ${\bm K}$.
This  produces an approximation for the representation of the
field  $a_n$  as a sum of two new fields, written as
\begin{eqnarray}
a_n &\simeq &e^{-i\bm K\cdot\bm R_n}a_{1,n}+
e^{-i\bm K'\cdot\bm R_n}a_{2,n}\,,
\nonumber
\\
b_n &\simeq &e^{-i\bm K\cdot\bm R_n}b_{1,n}+
e^{-i\bm K'\cdot\bm R_n}b_{2,n}\,,
\label{projection}
\end{eqnarray}
where the index $i=1$ ($i=2$) refers to the K (K') point.
These new fields, $a_{i,n}$ and $b_{i,n}$
are assumed to vary slowly over the unit cell. The
procedure for deriving
a theory that is valid close to the Dirac point
consists in using this representation in the tight-binding Hamiltonian and
expanding the operators up to a linear order in $\bm \delta$. In the derivation
one uses the fact that
$
 \sum_{\bm \delta}e^{\pm i\bm K\cdot \bm\delta}=
\sum_{\bm \delta}e^{\pm i\bm K'\cdot \bm\delta}=0.
$
After some straightforward algebra we arrive at \cite{semenoff84}:
\begin{widetext}
\begin{eqnarray}
H&\simeq&-t\int  dxdy
\hat\Psi^\dag_1(\bm r)
\left[
\left(
\begin{array}{cc}
0 & 3a(1-i\sqrt 3)/4\\
-3a(1+i\sqrt 3)/4&0
\end{array}
\right)\partial_x
\right.
+
\left.
\left(
\begin{array}{cc}
0 & 3a(-i-\sqrt 3)/4\\
-3a(i-\sqrt 3)/4&0
\end{array}
\right)\partial_y
\right]
\hat
\Psi_1(\bm r)\nonumber\\
&+&\hspace{1.7cm}\hat\Psi^\dag_2(\bm r)
\left[
\left(
\begin{array}{cc}
0 & 3a(1+i\sqrt 3)/4\\
-3a(1-i\sqrt 3)/4&0
\end{array}
\right)\partial_x +
\left(
\begin{array}{cc}
0 & 3a(i-\sqrt 3)/4\\
-3a(-i-\sqrt 3)/4&0
\end{array}
\right)\partial_y
\right]
\hat
\Psi_2(\bm r)\nonumber\\
&=&-i v_F \int dxdy\left(\hat\Psi^\dag_1(\bm r)\bm \sigma\cdot
\nabla \hat\Psi_1(\bm r)+
\hat \Psi^\dag_2(\bm r) \bm \sigma^*\cdot \nabla
\hat\Psi_2(\bm r)\right)\,,
\label{H2}
\end{eqnarray}
\end{widetext}
with Pauli matrices $\bm\sigma=(\sigma_x,\sigma_y)$,
$\bm\sigma^*=(\sigma_x,-\sigma_y)$, and
$\hat\Psi^\dag_i=(a^\dag_i,b^\dag_i)$ ($i=1,2$). It is clear that the
effective
Hamiltonian (\ref{H2})  is made of two copies of the massless Dirac-like
Hamiltonian, one holding for $\bm p$  around $\bm  K$  and other
for  $\bm p$  around $\bm K'$. Notice that, in first quantized language,
the two-component electron wavefunction, $\psi({\bf r})$, close to the K point,
obeys the  2D Dirac equation:
\begin{eqnarray}
- i v_F \bm \sigma \cdot \nabla \psi({\bf r}) = E \psi({\bf r}) \, .
\label{diraceq}
\end{eqnarray}

The wavefunction, in momentum space, for the momentum around $\bm K$ has the form:
\begin{equation}
\psi_{\pm,{\bf K}}({\bf k}) = \frac 1 {\sqrt 2}
\left(
\begin{array}{c}
e^{-i\theta_{{\bf k}}/2}\\
\pm e^{i\theta_{{\bf k}}/2}
\end{array}
\right) \, ,
\label{psiK}
\end{equation}
for $H_K = v_F \bm\sigma \cdot {\bf k}$,
where the $\pm$ signs correspond to the eigenenergies
$E=\pm v_F k$, that is, for the $\pi$ and $\pi^*$ band, respectively,
and $\theta_{{\bf k}}$ is given by (\ref{angle}).
The wavefunction for the momentum around
$\bm K'$ has the form:
\begin{equation}
\psi_{\pm,{\bf K'}}({\bf k}) = \frac 1 {\sqrt 2}
\left(
\begin{array}{c}
e^{i\theta_{{\bf k}}/2}\\
\pm e^{- i \theta_{{\bf k}}/2}
\end{array}
\right) \, ,
\label{psiKp}
\end{equation}
for $H_{K'} = v_F \bm\sigma^* \cdot {\bf k}$.
Notice that the wavefunctions at ${\bf K}$ and ${\bf K'}$ are related by
time reversal symmetry: if we set the origin of coordinates in momentum space in the
M-point of the BZ (see Fig.\ref{cap:bz}), time reversal becomes equivalent
to a reflection along the $k_x$ axis, that is, $(k_x,k_y) \to (k_x,-k_y)$.
Also note that if the phase $\theta$ is rotated by $2\pi$ the wavefunction
changes sign indicating a phase of $\pi$ (in the literature this is commonly called a Berry's phase). This change of phase by $\pi$ under
rotation is characteristic of spinors. In fact, the wavefunction is a two
component spinor.

A relevant quantity used to
characterize the eigenfunctions is their helicity defined
as the projection of the momentum operator along the (pseudo)spin
direction. The quantum mechanical operator for the helicity has the
form:
\begin{equation}
  \hat h = \frac {1} 2 {\bm \sigma}\cdot\frac {\bm p}{\vert \bm p\vert}\,.
\end{equation}
It is clear from the definition of $\hat h$ that the
states $\psi_{\bm K}(\bm r)$
and $\psi_{\bm K'}(\bm r)$ are also eigenstates of $\hat h$:
\begin{equation}
\hat h\psi_{\bm K}(\bm r)=\pm\frac{1}{2}\psi_{\bm K}(\bm r),
\label{eq:helicity}
\end{equation}
and an equivalent equation for $\psi_{\bm K'}(\bm r)$ with inverted sign.
Therefore electrons (holes) have a positive (negative)
helicity. Equation (\ref{eq:helicity}) implies that $\bm{\sigma}$
 has its two eigenvalues either in the direction of
($\Uparrow$)
or against ($\Downarrow$) the  momentum ${\bm p}$.
This property says that the states of
the system close to the Dirac point have well defined {\it chirality} or
helicity. Notice that chirality is not defined in regards to the
real spin of the electron (that has not yet appeared in the problem) but to
a pseudo-spin variable associated with the two components of the wavefunction.
The helicity values are good quantum numbers as long as the Hamiltonian (\ref{H2})
is valid. Therefore the existence of  helicity quantum numbers
holds only as an asymptotic property, which is well defined
close to the Dirac points $\bm K$ and $\bm K'$. Either at larger
energies or due to the presence of a finite $t'$ the helicity stops
being a good quantum number.

%------------------------------------------------------------------------------
\subsubsection{Chiral Tunneling and Klein paradox}
\label{klein}
%------------------------------------------------------------------------------

In this section we want to address the
scattering of chiral electrons in two dimensions
by a square barrier \cite{KNG06,KMT}. The one dimensional scattering
of chiral electrons was discussed earlier in the context of carbon
nanotubes \cite{ANS98,Louie99}

\begin{figure}
\begin{center}
\includegraphics*[width=7cm]{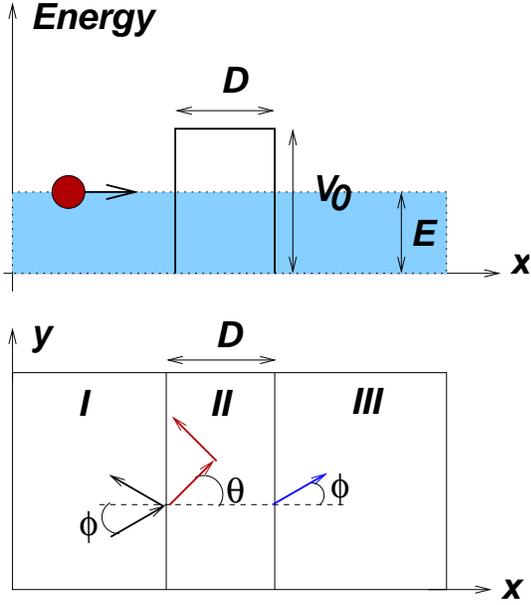}
\end{center}
\caption{\label{Fig:Klein}(Color online) Top: Schematic picture
of the scattering of Dirac electrons by a square potential.
Bottom: definition of the angles $\phi$ and $\theta$ used
in the scattering formalism in the three
regions I, II, and III.}
\end{figure}

We start by noticing that by a gauge transformation the wavefunction (\ref{psiK}) can
be written as:
\begin{equation}
\psi_{\bm K}(\bm k) = \frac 1 {\sqrt 2}
\left(
\begin{array}{c}
1\\
\pm e^{i\theta_k}
\end{array}
\right) \,.
\label{psiKb}
\end{equation}
We further assume that the scattering does not mix the momenta
around $\bm K$ and $\bm K'$ points.
In Fig.~\ref{Fig:Klein} we depict the scattering process
due to the square barrier of width $D$.

The wavefunction in the different regions can be written in
terms of incident and reflected waves. In region I we have:
\begin{eqnarray}
\psi_I(\bm r) &=& \frac 1 {\sqrt 2}
\left(
\begin{array}{c}
1\\
 se^{i\phi}
\end{array}
\right)e^{i (k_xx+k_yy)}
\nonumber\\
&+&
\frac r {\sqrt 2}
\left(
\begin{array}{c}
1\\
 se^{i(\pi-\phi)}
\end{array}
\right)e^{i (-k_xx+k_yy)}\,,
\label{psiI}
\end{eqnarray}
with $\phi = \arctan(k_y/k_x)$, $k_x=k_F\cos\phi$,
$k_y=k_F\sin\phi$ and $k_F$ the Fermi momentum. In region II we have:
\begin{eqnarray}
\psi_{II}(\bm r) &=& \frac a {\sqrt 2}
\left(
\begin{array}{c}
1\\
 s'e^{i\theta}
\end{array}
\right)e^{i (q_xx+k_yy)}
\nonumber\\
&+&
 \frac b {\sqrt 2}
\left(
\begin{array}{c}
1\\
s' e^{i(\pi-\theta)}
\end{array}
\right)e^{i (-q_xx+k_yy)}\,,
\label{psiII}
\end{eqnarray}
with $\theta=\arctan(k_y/q_x)$ and
\begin{equation}
q_x=\sqrt{(V_0-E)^2/(v_F^2)-k_y^2},
\end{equation}
and finally in region III we have a transmitted wave only:
\begin{equation}
\psi_{III}(\bm r) = \frac t {\sqrt 2}
\left(
\begin{array}{c}
1\\
 se^{i\phi}
\end{array}
\right)e^{i (k_xx+k_yy)}\,,
\end{equation}
with $s={\rm sgn}(E)$ and $s'={\rm sgn}(E-V_0)$.
The coefficients $r$, $a$, $b$ and $t$ are determined from the continuity
of the wavefunction, which implies that the wavefunction
has to obey the conditions
$\psi_I(x=0,y)=\psi_{II}(x=0,y)$ and $\psi_{II}(x=D,y)=\psi_{III}(x=D,y)$.
Unlike the Sch\"odinger equation we only need to match the wavefunction
but not its derivative.
The transmission through the barrier is obtained from
$T(\phi)=tt^\ast$ and has the form:
\begin{equation}
T(\!\phi\!) \!=\! \frac {\cos^2\theta\cos^2\phi}
{[\cos(Dq_x)\cos\phi\cos\theta]^2\!+\!\sin^2(\!Dq_x\!)(\!1\!\!-\!\!ss'\!\sin\phi\sin\theta\!)^2}\,.
\label{Trans}
\end{equation}
This expression does not take into account a contribution from evanescent
waves in region II, which is usually negligible, unless the chemical potential in region II is at the Dirac energy (see section \ref{transdirac}).

\begin{figure}
\begin{center}
\includegraphics*[width=8cm]{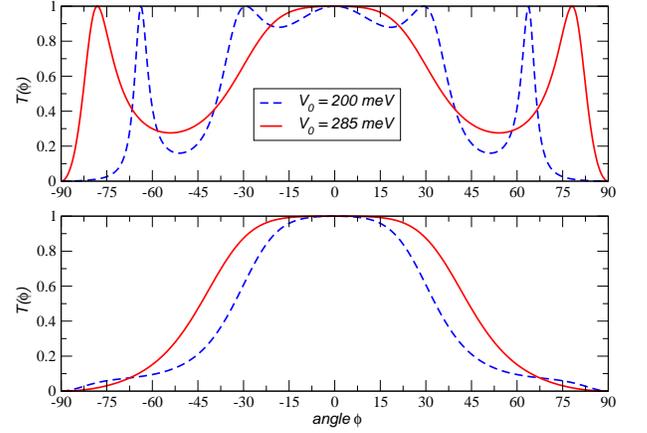}
\end{center}
\caption{\label{Fig:Klein_angle}(Color online)
Angular behavior of $T(\phi)$  for two different
values of $V_0$: $V_0=200$  meV  dashed line,
$V_0=285$ meV solid line. The remaining parameters
are $D=110$ nm (top), $D=50$ nm (bottom) $E=80$ meV, $k_F=2\pi/\lambda$,
$\lambda=50$ nm.}
\end{figure}

Notice that $T(\phi)=T(-\phi)$ and
for values of $Dq_x$ satisfying the relation
$Dq_x=n\pi$, with $n$ an integer, the barrier becomes
completely transparent since $T(\phi)=1$, independently
of the value of $\phi$. Also, for normal incidence ($\phi\rightarrow 0$
and $\theta\rightarrow 0$) and for any value
of $Dq_x$ one obtains $T(0)=1$, and the barrier
is again totally transparent. This result is a manifestation of the
Klein paradox \cite{zuber06,calogeracos99}
and does not occur for non-relativistic electrons. In this latter case and for normal incidence, the
transmission is always smaller than one. In the limit
$\vert V_0\vert\gg\vert E \vert$, eq.~(\ref{Trans}) has the following
asymptotic form
\begin{equation}
  T(\phi)\simeq \frac {\cos^2\phi}{1-\cos^2(Dq_x)\sin^2\phi}\,.
\end{equation}

In Fig.~\ref{Fig:Klein_angle}
we show the angular dependence of $T(\phi)$ for two different
values of the potential $V_0$; it is clear  that there
are several directions for which the transmission is one.
Similar calculations were done for a graphene bilayer \cite{KNG06}
with its most distinctive behavior being the absence of tunneling
in the forward ($k_y=0$) direction.

The simplest example of a potential barrier is a square potential
discussed previously.
When intervalley scattering and the lack of symmetry between sublattices
are neglected, a potential barrier shows no reflection for electrons
incident in the normal direction \cite{KNG06}.
Even when the barrier separates regions where the
Fermi surface is electron like on one side and hole like on the
other, a normally incident electron continues propagating as a hole
with 100\% efficiency. This phenomenon is another manifestation of
the chirality of the Dirac electrons within each valley, which
prevents backscattering in general. The transmission and reflection
probabilities of electrons at different angles depend on the
potential profile along the barrier. A slowly varying barrier is
more efficient in reflecting electrons at non-zero incident
angles \cite{CF06}.

Electrons moving through a barrier separating p- and n-doped
graphene, a p-n junction, are transmitted as holes. The relation
between the velocity and the momentum for a hole is the inverse of
that for an electron. This implies that, if the momentum parallel to
the barrier is conserved, the velocity of the quasiparticle is
inverted. When the incident electrons emerge from a source, the
transmitting holes are focused into an image of the source. This
behavior is the same as that of photons moving in a medium with
negative reflection index \cite{CFA07}. Similar effects can occur in graphene
quantum dots, where the inner and outer regions contain electrons
and holes, respectively \cite{CPP07}. Note that the fact that barriers
do not impede the transmission of normally incident electrons does
not preclude the existence of sharp resonances, due to the
confinement of electrons with a finite parallel momentum. This leads
to the possibility of fabricating quantum dots with potential
barriers \cite{SE07}. Finally, at half-filling, due to disorder
graphene can be
divided in electron and hole charge puddles \cite{KNG06,MAULSKY07}.
Transport is determined
by the transmission across the p-n junctions between these
puddles \cite{CFAA07,Shk07}. There is a rapid progress in the measurement of
transport
properties of graphene ribbons with additional top gates that play the role of
tunable potential barriers \cite{HOZK07,WDM07,HSSTYG07,OJEALK07,LEBK07}.

A magnetic field and potential fluctuations break both
inversion symmetry of the lattice and time reversal symmetry.
The combination of these effects break also the symmetry between the
two valleys. The transmission coefficient becomes
valley dependent, and, in general, electrons from different valleys propagate
along different paths. This opens the possibility of manipulating
the valley index \cite{TSAB07} (valleytronics) in a way similar to
the control of the spin in mesoscopic devices (spintronics). For
large magnetic fields, a p-n junction separates regions with
different quantized Hall conductivities. At the junction, chiral
currents can flow at both edges \cite{AL0704}, inducing backscattering
between the Hall currents at the edges of the sample.

The scattering of electrons near the Dirac point by
graphene-superconductor junctions differs from Andreev scattering
process in normal metals \cite{TB06}. When the distance between the Fermi energy
and the Dirac energy is smaller than the superconducting gap, the
superconducting interaction hybridizes quasiparticles from one band
with quasiholes in the other. As in the case of scattering at a p-n
junction, the trajectories of the incoming electron and reflected
hole (note that hole here is meant as in the BCS theory of
superconductivity) are different from those in similar
processes in metals with only one type of carrier \cite{BS06,MS07}.

\subsubsection{Confinement and zitterbewegung}
\label{zitter}

Zitterbewegung, or jittery motion of the wavefunction of the Dirac problem,
occurs when one tries to confine the Dirac electrons \cite{zuber06}. Localization
of a wavepacket leads, due to the Heisenberg principle, to uncertainty in the
momentum. For a Dirac particle with zero rest mass, uncertainty in the momentum
translates into uncertainty in the energy of the particle as well (this should
be contrasted with the non-relativistic case where the position-momentum
uncertainty relation is independent of the energy-time uncertainty relation).
Thus, for a ultra-relativistic particle, a particle-like state can have hole-like
states in its time evolution. Consider,
for instance, if one tries to construct a wave packet at some time $t=0$,
and let us assume, for simplicity, that this packet has a Gaussian shape
of width $w$ with momentum close to ${\bf K}$:
\begin{eqnarray}
\psi_0({\bf r}) = \frac{e^{-r^2/(2 w^2)}}{\sqrt{\pi} w} \, e^{i {\bf K} \cdot {\bf r}} \phi \, ,
\label{wf0}
\end{eqnarray}
where $\phi$ is spinor composed of positive energy states (associated with $\psi_{+,{\bf K}}$ of
(\ref{psiK})). The eigenfunction of the Dirac
equation can be written in terms of the solution (\ref{psiK}) as:
\begin{eqnarray}
\psi({\bf r},t) = \int \frac{d^2 k}{(2 \pi)^2} \sum_{a=\pm 1} \alpha_{a,{\bf k}} \psi_{a,{\bf K}}({\bf k})
e^{-i a ({\bf k} \cdot {\bf r}+v_F k t)}  \, ,
\label{wft}
\end{eqnarray}
where $\alpha_{\pm,{\bf k}}$ are Fourier coefficients.
We can rewrite (\ref{wf0}) in terms of (\ref{wft}) by inverse Fourier transform
and find that:
\begin{eqnarray}
\alpha_{\pm,{\bf k}} = \sqrt{\pi} w e^{-k^2 w^2/2} \psi^{\dag}_{\pm,{\bf K}}({\bf k}) \phi \, .
\end{eqnarray}
Notice that the relative weight of positive energy states with respect to negative energy states,
$|\alpha_+/\alpha_-|$, given by (\ref{psiK}) is one, that is, there are as many positive energy states as
negative energy states in a wavepacket. Hence, these will cause the wavefunction to
be delocalized at any time $t \neq 0$. Thus, a wave packet of electron-like states has hole-like components, a result that puzzled many researchers in the early days of QED \cite{zuber06}.

\begin{figure}[ht]
\begin{center}
\includegraphics*[width=6.cm,angle=0]{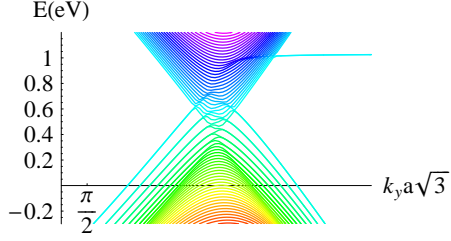}
\end{center}
\caption{(Color on line) Energy spectrum (in units of $t$) for a graphene ribbon $600 a$ wide, as a function of the momentum $k$ along the ribbon (in units of $1/(\sqrt{3} a)$), in the presence of confining potential with $V_0=1$ eV, $\lambda=180 a$.}
\label{hifield}
\end{figure}

Consider the tight-binding description \cite{nuno_confinement,CAC07} of Sec.~\ref{single} when
a potential $V_i$ on site ${\bf R}_i$ is added to the problem:
\begin{eqnarray}
H_e = \sum_i V_i n_i \, ,
\end{eqnarray}
where $n_i$ is the local electronic density. For simplicity, we assume that the confining potential
is 1D, that is, that  $V_i$ vanishes in the bulk but becomes large at the edge of
the sample. Let us assume a potential
that decays exponentially away from the edges into the bulk with a penetration depth, $\lambda$.
In Fig.~\ref{hifield} we show  the electronic spectrum for a graphene ribbon of width
$L=600 a$, in the presence of a confining potential,
\begin{eqnarray}
V(x) = V_0 \left[ e^{ -( x-L/2 ) / \lambda} + e^{ -( L/2-x ) / \lambda} \right] \, ,
\end{eqnarray}
where $x$ is the direction of confinement and  $V_0$ the strength of the potential.
One can clearly see that in the presence of the confining potential the
electron-hole symmetry is broken and, for $V_0>0$, the hole part of the spectrum
is strongly distorted. In particular, for $k$ close to the Dirac point, we see
that the hole dispersion is given by:
$E_{n,\sigma=-1}(k) \approx -\gamma_n k^2 - \zeta_n k^4$ where
$n$ is a positive integer, and $\gamma_n <0$ ($\gamma_n>0$) for $n<N^*$ ($n>N^*$). Hence, at $n=N^*$ the hole effective mass
{\it diverges} ($\gamma_{N^*} =0$) and, by tuning the chemical potential,
$\mu$, via a back gate, to the hole region of the spectrum ($\mu<0$) one
should be able to observe an anomaly in the Shubnikov-de Haas (SdH)
oscillations. This is how zitterbewegung could manifest itself
in magnetotransport.

\subsection{Bilayer graphene: tight-binding approach}
\label{sec:bilayer}

\begin{figure}
\begin{center}
\includegraphics*[width=8cm]{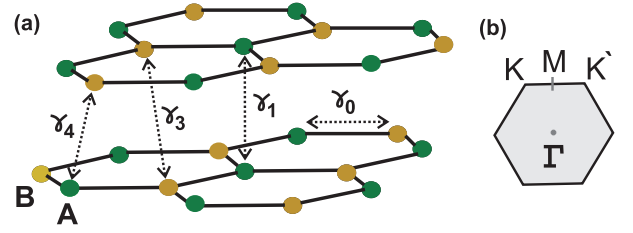}
\end{center}
\caption{(Color online)(a) Lattice structure of the bilayer with the various
hopping parameters according to the SWM model. The
  A-sublattices are indicated by the darker spheres. (b) Brillouin zone.
Adapted from \onlinecite{Malard07}.}
\label{fig_lattice}
\end{figure}

The tight-binding model developed for graphite can be easily extended to
stacks with a finite number of graphene layers. The simplest generalization
is a bilayer \cite{MF06}. A bilayer is interesting because the IQHE
shows anomalies, although different from those observed in a
single layer \cite{Ketal06}, and also a gap can open between the conduction and valence
band \cite{MF06}. The bilayer structure, with the AB stacking of 3D graphite,
is shown in Fig.\ref{fig_lattice}.

The tight-binding Hamiltonian for this problem can be written as:
\begin{eqnarray}
  \label{eq:HTB}
  \Hca_{\text{t.b.}} &=&
  -\gamma_0 \sum_{\substack{<i,j> \\ m, \sigma}}
  (a_{m,i,\sigma}^{\dag} b_{m,j,\sigma}^{\,} + \text{h.c.} )
  \nonumber \\
  &-& \gamma_1 \sum_{j,\sigma} (a_{1,j,\sigma}^{\dag}
  a_{2,j,\sigma}^{\,} + \text{h.c.}),
 \nonumber \\
&-& \gamma_3 \sum_{j,\sigma} (a_{1,j,\sigma}^{\dag}
  b_{2,j,\sigma}^{\,} + a_{2,j,\sigma}^{\dag}
  b_{1,j,\sigma}^{\,}+ \text{h.c.})
\nonumber \\
&-& \gamma_4 \sum_{j,\sigma} (b^{\dag}_{1,j,\sigma} b_{2,j,\sigma} + \text{h.c.}),
\end{eqnarray}
where $a_{m,i,\sigma}$ ($b_{m,i\sigma}$) annihilates an electron with spin $\sigma$, on sublattice A (B), in plane $m=1,2$, at site ${\bf R}_i$.
Here we use the graphite nomenclature for the hopping parameters:
$\gamma_0=t$ is the in-plane hopping energy and
$\gamma_1$ ($\gamma_1=\tp \approx 0.4$ eV in graphite \cite{BCP88,gic})
is the hopping energy between atom $\text{A}_1$ and atom $\text{A}_2$
(see Fig.~\ref{fig_lattice}), and $\gamma_3$ ($\gamma_3 \approx 0.3$ eV
in graphite \cite{BCP88,gic}) is the hopping energy between atom $\text{A}_1$ ($\text{A}_2$)
and atom $\text{B}_2$ ($\text{B}_1$), and
$\gamma_4$ ($\gamma_4 \approx -0.04$ eV in graphite \cite{BCP88,gic})
that connects $\text{B}_1$ and $\text{B}_2$.

In the continuum limit, by expanding the momentum close to the K point in the
BZ, the Hamiltonian reads,
\begin{eqnarray}
{\cal H} = \sum_{{\bf k}} \Psi_{{\bf k}}^{\dag} \cdot {\cal H}_K \cdot \Psi_{{\bf k}}
\end{eqnarray}
where  (ignoring $\gamma_4$ for the time being):
\begin{equation}
{\cal H}_K \!\!\equiv\!\! \left(\!\begin{array}{cccc} - V & v_F k
    &0 &3 \gamma_3 a k^* \\ v_F k^* &- V &\gamma_1 &0
    \\ 0 &\gamma_1 & V &v_F k \\ 3 \gamma_3 a k &0
    &v_F k^* &V \end{array} \!\right) \, ,
\label{bilayer}
\end{equation}
where $k = k_x + i k_y$ is a complex number, and we have added $V$
which is here half the shift
in electro-chemical potential between the two layers (this term will
appear if a potential bias is applied between the layers), and
\begin{eqnarray}
\Psi_{{\bf k}}^{\dag} = \left(a^{\dag}_{1}({\bf k}),a^{\dag}_{2}({\bf k}),b^{\dag}_{1}({\bf k}),b^{\dag}_{2}({\bf k})\right)
\end{eqnarray}
is a four component spinor.

\begin{figure}
\begin{center}
\includegraphics*[width=7cm]{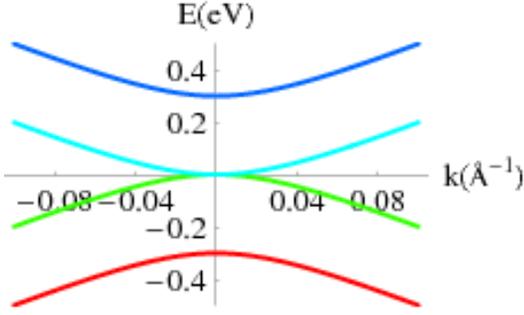}
\end{center}
\caption{(Color online) Band structure for bilayer graphene for $V=0$ and
$\gamma_3=0$. }
\label{bandsv0}
\end{figure}

If $V = 0$ and $\gamma_3 , v_F k \ll \gamma_1$, one can eliminate
the high energy states perturbatively and
write an effective Hamiltonian:
\begin{equation}
{\cal H}_K \equiv \left( \begin{array}{cc} 0 &\frac{v_F^2 k^2}{\gamma_1}
+ 3 \gamma_3 a k^* \\ \frac{v_F^2 (k^*)^2}{\gamma_1} + 3 \gamma_3 a k & 0
\end{array} \right) \, .
\label{hamil_bilayer_red}
\end{equation}
The hopping $\gamma_4$ leads to a $k$ dependent coupling between the
sublattices or a small
renormalization of $\gamma_1$. The same role is played by the inequivalence between
sublattices within a layer.

For $\gamma_3 = 0$, (\ref{hamil_bilayer_red}) gives two parabolic bands,
$\epsilon_{k,\pm} \approx \pm v_F^2 k^2 / t_{\perp}$ which touch at $\epsilon = 0$ (as shown in Fig.\ref{bandsv0}).
The spectrum is electron-hole symmetric. There are two
additional bands which start at $\pm t_{\perp}$. Within this approximation, the
bilayer is metallic, with a constant density of states. The term $\gamma_3$ changes qualitatively
the spectrum at low energies since it introduces a trigonal distortion, or warping, of the bands
(notice that this trigonal distortion, unlike the one introduced by large momentum in (\ref{Efull}),
occurs at low energies).
The electron-hole symmetry is preserved but,
instead of two bands touching at $k = 0$, we obtain three sets of
Dirac-like linear  bands. One Dirac point is at $\epsilon = 0$ and $k = 0$,
while the three
other Dirac points, also at $\epsilon = 0$, lie at three equivalent points
with a finite momentum. The stability of points where bands touch can be
understood using topological arguments \cite{MGV07}. The winding number
of a closed curve in the plane around a given point is an integer representing
the total number of times that the curve travels counterclockwise around the point so that
the wavefunction remains unaltered. The winding number of the
point where the two parabolic bands come together for $\gamma_3 = 0$ has winding number
$+ 2$. The trigonal warping term, $\gamma_3$, splits it into a Dirac point at $k
= 0$ and winding number $-1$, and three Dirac points at $k \ne 0$ and winding
numbers $+1$. An in-plane magnetic field, or a small rotation of one layer with
respect to the other splits the $\gamma_3 = 0$ degeneracy into two Dirac
points with winding number $+1$.

\begin{figure}
\begin{center}
\includegraphics*[width=7cm]{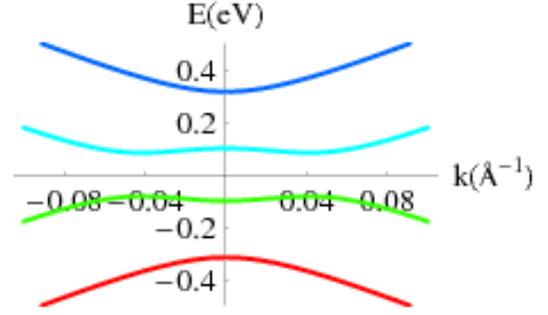}
\end{center}
\caption{(Color online) Band structure for bilayer graphene for $V \neq 0$ and
  $\gamma_3=0$.}
\label{bandsv}
\end{figure}

The term $V$ in (\ref{bilayer}) breaks the equivalence of the
two layers, or, alternatively, inversion symmetry. In this case,
the dispersion relation becomes:
\begin{eqnarray}
\epsilon^2_{\pm,{\bf k}} &=& V^2 + v_F^2 k^2 + \tp^2 / 2
\nonumber
\\
&\pm& \sqrt{4
    V^2  v_F^2 k^2 + t^2 v_F^2 k^2 + \tp^4 / 4} \, ,
\end{eqnarray}
given rise to the dispersion shown in Fig.~\ref{bandsv}, and to the opening
of a gap close, but not directly at, the K point.
For small momenta,
and $V \ll t$, the energy of the conduction band can be expanded:
\begin{eqnarray}
\epsilon_k \approx V - ( 2 V v_F^2 k^2 ) / \tp + ( v_F^4 k^4 ) / ( 2 \tp^2 V ) \, .
\end{eqnarray}
The dispersion for the valence band can be obtained by replacing $\epsilon_k$ by $-
\epsilon_k$. The bilayer has a gap at $k^2 \approx ( 2 V^2 ) /
v_F^2$. Notice, therefore, that the gap in the biased bilayer system depends
on the applied bias and hence can be measured experimentally \cite{MF06,MC06,Cetal06}.
The ability to open a gap makes bilayer graphene most interesting for technological
applications.

\subsection{Epitaxial graphene}
\label{epitaxial}

It has been known for a long time that monolayers of graphene could be grown
epitaxially on metal surfaces by using catalytic decomposition of hydrocarbons
or carbon oxide \cite{SPB74,EB79,CT89,ON97,SY06}. When such surfaces are
heated, oxygen or hydrogen desorbs, and the carbon atoms form a graphene
monolayer.
The resulting graphene structures could reach sizes up to a micrometer, with
few defects and were characterized by different surface-science techniques
and local scanning probes \cite{HCHEF82}. For example, graphene grown on ruthenium has zigzag edges and also
ripples associated with a $( 10 \times 10 )$ reconstruction \cite{Vetal07}.

Graphene can also be formed on the surface of SiC. Upon heating, the silicon
from the top layers desorbs, and a few layers of graphene are left on the
surface \cite{VCV75,FTD98,Cetal02,Betal04,Retal05,Hetal07b,Hetal07}. The
number of layers can be controlled by limiting time or temperature of
the heating treatment. The quality and the number of layers in the samples depends on the SiC
face used for their growth \cite{Hetal07} (the carbon terminated surface produces few layers
but with a low mobility, whereas the silicon terminated surface produces several layers but
with higher mobility). Epitaxially grown multilayers exhibit SdH oscillations with a Berry phase shift of $\pi$ \cite{Betal06}, which is the same as the phase shift for Dirac fermions observed in a single layer as well as for some subbands present in multilayer graphene (see further) and graphite \cite{LK04}.
The carbon layer directly on top of
the substrate is expected to be strongly bonded to it, and it shows no $\pi$
bands \cite{Vetal07b}. The next layer shows a $( 6 \sqrt{3} \times 6 \sqrt{3} )$
reconstruction due to the substrate, and has graphene properties.
An alternate
route to produce few layers graphene is based on synthesis from nanodiamonds \cite{Affoune01}.

Angle resolved photo-emission experiments (ARPES) show that
epitaxial graphene grown on SiC has linearly dispersing quasiparticles
(Dirac fermions) \cite{lanzara06,Oetal07,Betal07},
in agreement with the theoretical expectation. Nevertheless, these
experiments show that the electronic properties can
change locally in space indicating a certain degree of inhomogeneity due to the growth
method \cite{ZGFFHLGCNL07}. Similar inhomogeneities due to disorder in the c-axis
orientation of graphene planes is observed in graphite \cite{ZGL06}.
Moreover, graphene grown this way is heavily doped due to the
charge transfer from the substrate to the graphene layer (with the chemical
potential well above the Dirac point) and therefore all
samples have strong metallic character with large electronic mobilities
\cite{Betal06,Hetal07}. There is also evidence for strong interaction between
a substrate and the graphene layer leading to the appearance of gaps at the Dirac
point \cite{ZGFFHLGCNL07}. Indeed, gaps can be generated by the breaking of
the sublattice symmetry  and, as in the case of other carbon based systems
such as polyacethylene \cite{SSH79_PRL,SSH79_PRB}, it can lead to soliton-like excitations
\cite{JR76,HCM07}.
Multilayer graphene grown on SiC have also been studied
with ARPES \cite{Oetal06,Oetal07,BOMESHR07} and the results seem to agree
quite well with band structure calculations \cite{MO0704}. Spectroscopy
measurements also show the transitions associated with Landau levels \cite{SMPBH06},
and weak localization effects at low magnetic fields, also expected for
Dirac fermions \cite{WLSBH07}.  Local probes
reveal a rich structure of terraces \cite{Metal07} and interference patterns due to defects
at or below the graphene layers \cite{RCGLFS07}.

\begin{figure}[]
\begin{center}
\includegraphics[width=5cm]{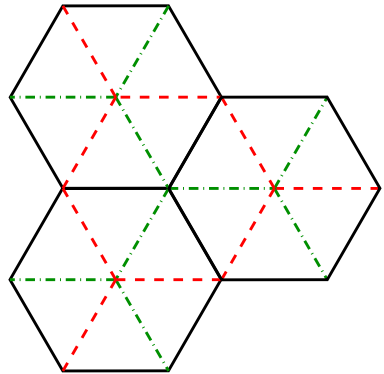}\\
\caption[fig]{\label{stacking} (Color online) Sketch of the three inequivalent orientations
  of graphene layers with respect to each other.}
\end{center}
\end{figure}

\subsection{Graphene stacks}
\label{stacks}

In stacks with more than one graphene layer, two consecutive layers are normally
oriented in such a way that the atoms in one of the two sublattices, $A_n$, of the
honeycomb structure of one layer are directly above one half of the atoms in
the neighboring layer, sublattice $A_{n \pm 1}$. The second set of atoms in
one layer sits on top of
the (empty) center of an hexagon in the other layer. The shortest distance
between carbon atoms in different layers is $d_{A_n A_{n \pm 1} } = c =
  3.4$\AA. The next distance is $d_{A_n B_{n \pm 1}} = \sqrt{c^2 +
    a^2}$. This is the most common arrangement of nearest neighbor layers observed
  in nature, although a stacking order in which all atoms in one layer occupy
  positions directly above the atoms in the neighboring layers (hexagonal
  stacking) has been considered theoretically \cite{CMGV91} and appears in
  graphite intercalated compounds \cite{gic}.

The relative position of two neighboring layers allows for two different
orientations of the third layer. If we label the positions of the two first
atoms as 1 and 2, the third layer can be of type 1, leading to the sequence
121, or it can fill a third position different from 1 and 2 (see
Fig.~\ref{stacking}), labeled 3. There are no more inequivalent positions
where a new layer can be placed, so that thicker stacks can be described in
terms of these three orientations.
In the most common version of bulk graphite the stacking order is $1212
\cdots$ (Bernal stacking). Regions with the stacking $123123 \cdots$ (rhombohedral stacking)
have also
been observed in different types of graphite \cite{B50,G67}. Finally, samples
with no discernible stacking order (turbostratic graphite) are also commonly
reported.

Beyond two layers, the stack ordering can be arbitrarily complex. Simple analytical
expressions for the electronic bands can be obtained for perfect Bernal (
$1212 \cdots$ ) and rhombohedral  ( $123123 \cdots$ )
stacking \cite{GNP06}. Even if we consider one interlayer hopping, $\tp =
\gamma_1$, the two stacking orders show rather different band structures near
$\epsilon = 0$. A Bernal stack with $N$ layers, $N$ even,  has $N/2$ electron like
and $N/2$ hole like parabolic subbands touching at $\epsilon = 0$. When $N$
is odd, an additional subband with linear (Dirac) dispersion
emerges. Rhombohedral systems have only two subbands that touch at
$\epsilon = 0$. These subbands disperse as $k^N$, and become surface states
localized at the top and bottom layer when $N \rightarrow \infty$. In this
limit, the remaining $2N-2$ subbands of a rhombohedral stack become Dirac
like, with the same Fermi velocity as a single graphene layer. The subband
structure of a tri-layer with the Bernal stacking includes two touching
parabolic bands, and one with Dirac dispersion, combining the features of
bilayer and monolayer graphene.

\begin{figure}[]
\begin{center}
\includegraphics*[width=4cm]{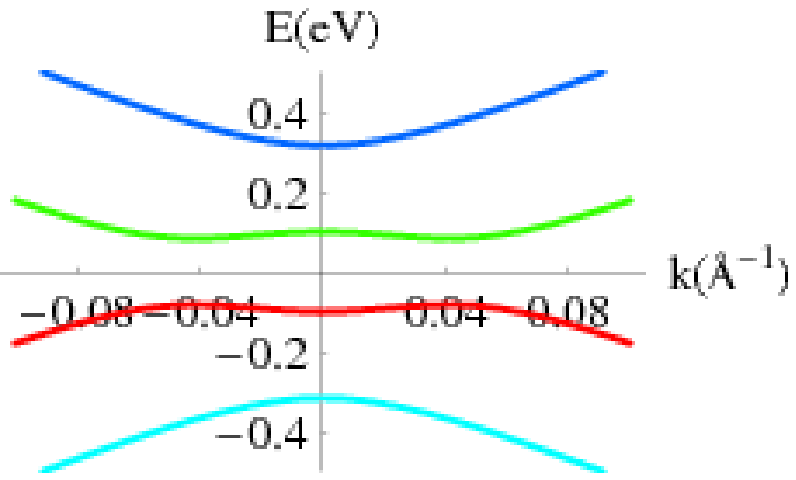}
\includegraphics*[width=4cm]{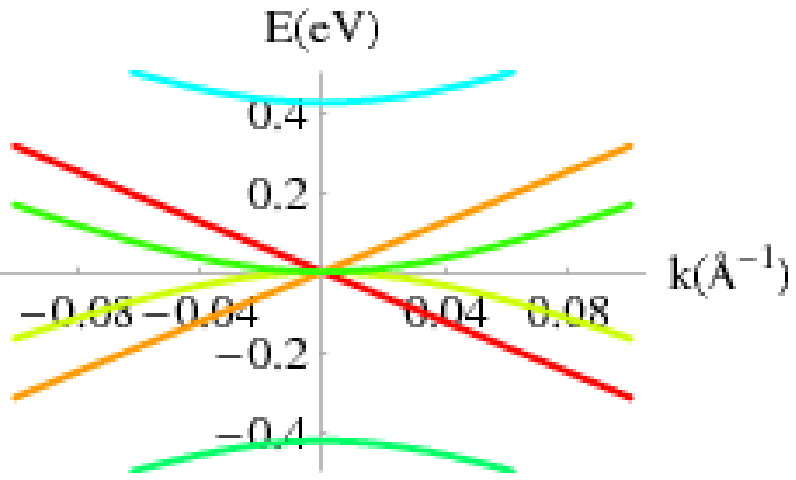}
\includegraphics*[width=4cm]{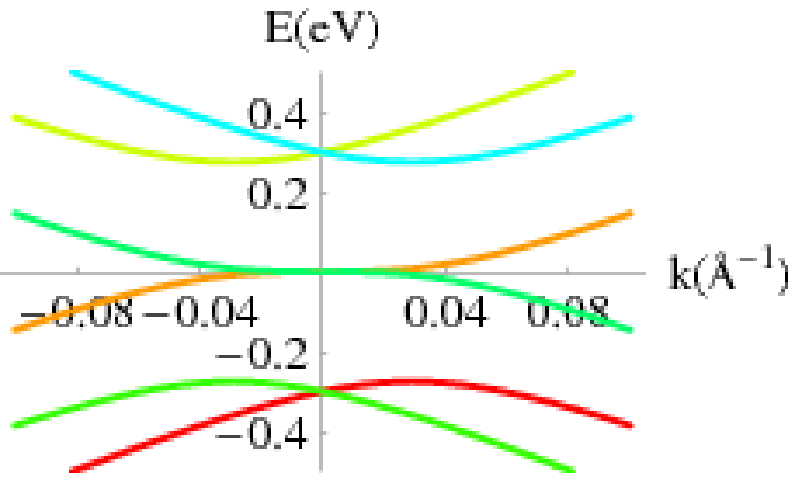}
\includegraphics*[width=4cm]{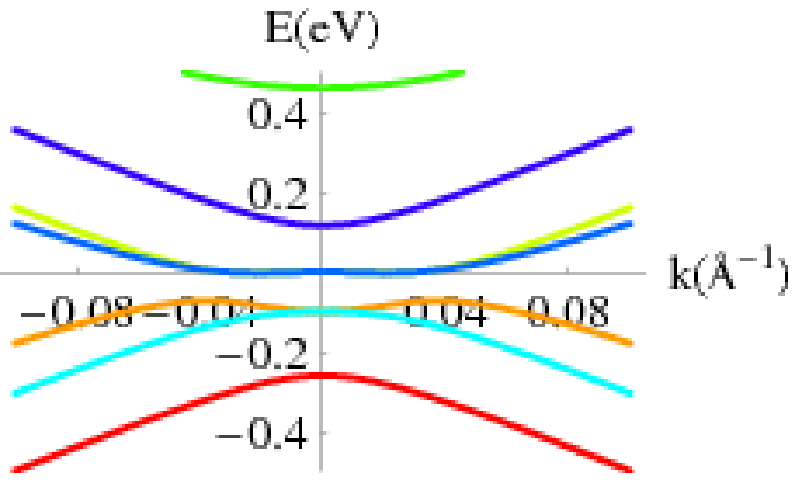}
\caption[fig]{\label{trilayer} (Color online) Electronic bands of graphene
  multilayers: top left: biased bilayer; top right: trilayer with Bernal
  stacking; bottom left: trilayer with orthorhombic stacking; bottom right:
  stack with four layers where the top and bottom layers are shifted in
  energy with respect to the two middle layers by $+0.1$ eV.}
\end{center}
\end{figure}

The low energy bands have different weights on the two sublattices of each
graphene layer. The states at a site directly coupled to the neighboring
planes are pushed to energies $\epsilon \approx \pm \tp$. The bands near
$\epsilon = 0$ are localized mostly at the sites without neighbors in the
next layers. For the Bernal stacking, this feature implies that the density
of states at $\epsilon = 0$ at sites without nearest neighbors in the
contiguous layers is finite, while it vanishes linearly at the other
sites. In stacks with rhombohedral stacking, all sites have one neighbor in
another plane, and the density of states vanishes at $\epsilon =
0$ \cite{GNP06}. This result is consistent with the well known fact that only
one of the two sublattices at a graphite surface can be resolved by scanning
tunneling microscopy (STM) \cite{Tetal87}.

As in the case of a bilayer, an inhomogeneous charge distribution can change
the electrostatic potential in the different layers. For more than two
layers, this breaking of the equivalence between layers can take place even
in the absence of an applied electric field. It is interesting to note that a
gap can open in a stack with Bernal ordering and four layers, if the
electronic charge at the two surface layers is different from that at the two
inner ones. Systems with a higher number of layers do not show a gap, even in
the presence of charge inhomogeneity. Four representative examples are shown
in Fig.~\ref{trilayer}. The band structure analyzed here will be modified by
the inclusion of the trigonal warping term, $\gamma_3$. Experimental studies
of graphene stacks have showed that, with increasing number of layers, the
system becomes increasingly metallic (concentration of charge carriers at
zero energy gradually increases), and there appear several types of electron-and-hole-like
carries \cite{Netal04,Metal05c}. An inhomogeneous charge distribution between
layers becomes very important in this case, leading to 2D electron and hole
systems that occupy only a few graphene layers near the surface and can completely
dominate transport properties of graphene stacks \cite{Metal05c}.

The degeneracies of the bands at
$\epsilon = 0$ can be studied using topological
arguments \cite{MGV07}. Multilayers with an
even number of layers and Bernal stacking have inversion symmetry, leading
to degeneracies with winding number +2, as in the case of a bilayer. The
trigonal lattice symmetry implies that these points can lead, at most, to
four Dirac points. In stacks with an odd number of layers, these degeneracies
can be completely removed. The winding number of the degeneracies found in
stacks with $N$ layers and orthorhombic ordering is $\pm N$. The inclusion of
trigonal warping terms will lead to the existence of many weaker degeneracies
near $\epsilon = 0$.

Furthermore, it is well known that in graphite the planes can be rotated
relative each other giving rise to Moir\'e patterns that are observed in STM
of graphite surfaces \cite{Rong93}.
The graphene layers can be rotated relative to each other due to the
weak coupling between planes that allows for the presence of many different
orientational states that are quasidegenerate in energy. For certain
angles the graphene layers become commensurate with each other leading to
a lowering of the electronic energy. Such phenomenon is quite similar to
the commensurate-incommensurate transitions observed in certain charge
density wave systems or adsorption of gases on graphite \cite{Bak82}.
This kind of dependence of the electronic structure on the relative rotation
angle between graphene layers leads to what is called super-lubricity in graphite
\cite{superlub}, namely, the vanishing of the friction between layers as
a function of the angle of rotation. In the case of bilayer graphene, a
rotation by a small commensurate angle leads to the effective decoupling
between layers and the recovery of the linear Dirac spectrum of the single
layer albeit with a modification on the value of the Fermi velocity \cite{LSPCN07}.

\subsubsection{Electronic structure of bulk graphite}
\label{graphite}

The tight-binding description of graphene described earlier can be extended
to systems with an infinite number of layers. The coupling between layers leads to
hopping terms between $\pi$ orbitals in different layers, leading to the so
called Slonczewski-Weiss-McClure model \cite{SW58}. This model describes the band
structure of bulk graphite with the Bernal stacking order in terms of seven
parameters, $\gamma_0 , \gamma_1 , \gamma_2 , \gamma_3 , \gamma_4 , \gamma_5$
and $\Delta$. The parameter $\gamma_0$ describes the hopping within each
layer, and it has been considered previously. The coupling between orbitals
in atoms that are nearest neighbors in successive layers is $\gamma_1$, which
we called $\tp$ earlier. The
parameters $\gamma_3$ and $\gamma_4$ describe the hopping between orbitals at
next nearest neighbors in successive layers and were discussed in the case
of the bilayer. The coupling between orbitals at
next nearest neighbor layers are $\gamma_2$ and $\gamma_5$. Finally, $\Delta$
is an on site energy which reflects the inequivalence between the two
sublattices in each graphene layer once the presence of neighboring layers is
taken into account. The values of these parameters, and their dependence with
pressure, or, equivalently, the interatomic distances, have been extensively
studied \cite{M57,N58,SMS64,M64,DM64,DSM77,BCP88}. A representative set of values
is shown in Table[\ref{hoppings}]. It is unknown, however, how these
parameters may vary in graphene stacks with a small number of layers.

The unit cell of graphite with Bernal stacking includes two layers, and two
atoms within each layer. The tight-binding Hamiltonian described previously can be
represented as a $4 \times 4$ matrix. In the continuum limit, the two
inequivalent corners of the BZ can be treated separately, and the
in plane terms can be described by the Dirac equation. The next terms in
importance for the low energy electronic spectrum are the nearest neighbor
couplings $\gamma_1$ and $\gamma_3$. The influence of the
parameter $\gamma_4$ on the low energy bands is much
smaller, as discussed below. Finally, the fine
details of  the spectrum of bulk graphite are determined by $\Delta$, which
breaks the electron-hole symmetry of the bands preserved by $\gamma_0 ,
\gamma_1$ and $\gamma_3$. It is
usually assumed to be much smaller than the other terms.

\begin{table}
\begin{tabular}{||c|c||}
\hline \hline
$\gamma_0$ &3.16 eV \\ \hline
$\gamma_1$ &0.39 eV \\ \hline
$\gamma_2$ & -0.020 eV \\ \hline
$\gamma_3$ & 0.315 eV \\ \hline
$\gamma_4$ & -0.044 eV \\ \hline
$\gamma_5$ & 0.038 eV \\ \hline
$\Delta$ & -0.008 eV \\
\hline \hline
\end{tabular}
\caption{Band structure parameters of graphite \protect{\cite{gic}}.}
\label{hoppings}
\end{table}

We label the two atoms from the unit cell in one layer as 1 and 2, and 3 and 4
correspond to the second layer. Atoms 2 and 3 are directly on top of each
other. Then, the matrix elements of the Hamiltonian can be written as:
\begin{eqnarray}
H_{11}^K &= &2 \gamma_2 \cos ( 2 \pi
    k_z / c ) \nonumber \\
H_{12}^K &= &v_F ( k_x + i k_y )
 \nonumber \\
H_{13}^K &= &\frac{3 \gamma_4 a}{2} \left( 1 + e^{i k_z c} \right)
    \left( k_x + i k_y \right)
 \nonumber \\
H_{14}^K &= &\frac{3 \gamma_3 a}{2} \left( 1 + e^{i k_z c} \right) \left(
    k_x - i k_y \right)
\nonumber \\
H_{22}^K &= &\Delta  + 2 \gamma_5 \cos ( 2 \pi
    k_z / c ) \nonumber \\
H_{23}^K &= &\gamma_1 \left( 1 + e^{i k_z c} \right) \nonumber \\
H_{24}^K &= &\frac{3 \gamma_4 a}{2} \left( 1 + e^{i k_z c} \right)
    \left( k_x + i k_y \right) \nonumber \\
H_{33}^K &= &\Delta  + 2 \gamma_5 \cos ( 2 \pi
    k_z / c ) \nonumber \\
H_{34}^K &= &v_F ( k_x + i k_y )
\nonumber \\
H_{44}^K &= &2 \gamma_2 \cos ( 2 \pi
    k_z / c )
\label{hamil_SW}
\end{eqnarray}
where $c$ is the lattice constant in the out of plane direction,
equal to twice the interlayer spacing. The matrix elements of
$H^{K'}$ can be obtained by replacing $k_x$ by $-k_x$ (other
conventions for the unit cell and the orientation of the lattice
lead to different phases).
Recent ARPES experiments
\cite{ZGL06,Oetal06,Zetal06,Betal07} performed
in epitaxially grown graphene stacks \cite{Betal04} confirm the main
features of this model, formulated mainly on the basis of Fermi
surface measurements \cite{M57,SMS64}. The electronic spectrum of the
model can also be calculated in a magnetic field \cite{G64,N76}, and
the results are also consistent with STM
on graphite surfaces \cite{Metal05,Ketal05,Netal06,LA07}, epitaxially
grown graphene stacks \cite{Metal07}, and with optical measurements
in the infrared range \cite{Letal06}.

%------------------------------------------------------------------------------
\subsection{Surface states in graphene}
\label{surface}
%------------------------------------------------------------------------------

\begin{figure}
\begin{center}
\includegraphics*[width=8cm]{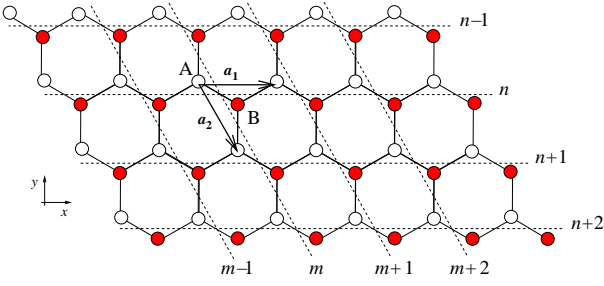}
\end{center}
\caption{\label{cap:ribbon}(Color online) Ribbon geometry with zigzag edges.}
\end{figure}

So far, we have discussed the basic bulk properties of graphene.
Nevertheless, graphene has very interesting surface (edge) states that
do not occur in other systems.
A semi-infinite graphene sheet with a zigzag edge has a band of
zero energy states localized at the surface \cite{FWNK96,WFA+99,Nakada96}. In section \ref{sec:nanoribbon}
we will discuss the existence of edge states using the Dirac equation.
Here will discuss the same problem using the tight-binding Hamiltonian.
To see why these edge states exist we consider the ribbon geometry
with zigzag
edges shown in Fig.~\ref{cap:ribbon}. The ribbon width is such that
it has $N$ unit cells in the transverse cross section ($y$ direction).
We will assume that the ribbon has infinite length in the longitudinal
direction ($x$ direction).

Let us rewrite (\ref{H1}), with $t'=0$, in terms of the integer
indices $m$ and $n$, introduced in Fig.
\ref{cap:ribbon}, and labeling the unit cells:
\begin{eqnarray}
H & = & -t\sum_{m,n,\sigma}[a_{\sigma}^{\dagger}(m,n)b_{\sigma}(m,n)+a_{\sigma}^{\dagger}(m,n)b_{\sigma}(m-1,n)
\nonumber
\\
 &  & +a_{\sigma}^{\dagger}(m,n)b_{\sigma}(m,n-1)+\textrm{h.c.}].\label{eq:Hribbon}\end{eqnarray}
Given that the ribbon is infinite in the $\mathbf{a}_{1}$ direction
one can introduce a  Fourier decomposition of the operators leading to
\begin{eqnarray}
H & = & -t\int \frac{\textrm{d}k}{2 \pi}
\sum_{n,\sigma}[a_{\sigma}^{\dagger}(k,n)b_{\sigma}(k,n)+e^{ika}a_{\sigma}^{\dagger}(k,n)b_{\sigma}(k,n)
\nonumber
\\
 &  & +a_{\sigma}^{\dagger}(k,n)b_{\sigma}(k,n-1)+\textrm{h.c.}] \, ,
\label{eq:Hribbonk}
\end{eqnarray}
where $c_{\sigma}^{\dagger}(k,n)\left|0\right\rangle =\left|c,\sigma,k,n\right\rangle $, and  $c=a,b$. The one-particle Hamiltonian can be written as:
\begin{eqnarray}
H^{1p} & = & -t\int\textrm{d}k\,\sum_{n,\sigma}[(1+e^{ika})\left|a,k,n,\sigma\right\rangle \left\langle b,k,n,\sigma\right|
\nonumber
\\
 &  & +\left|a,k,n,\sigma\right\rangle \left\langle b,k,n-1,\sigma\right|+\textrm{h.c.}].
\label{eq:Hribbonk1p}
\end{eqnarray}
The solution of the Schr\"odinger equation,
$H^{1p}\left|\mu,k,\sigma\right\rangle =E_{\mu,k}\left|\mu,k,\sigma\right\rangle $,
can be generally expressed as:
\begin{equation}
\left|\mu,k,\sigma\right\rangle =\sum_{n}[\alpha(k,n)\left|a,k,n,\sigma\right\rangle +\beta(k,n)\left|b,k,n,\sigma\right\rangle ],
\label{eq:solutionH1p}
\end{equation}
where the coefficients $\alpha$ and $\beta$ satisfy the following
equations:
\begin{eqnarray}
E_{\mu,k}\alpha(k,n) & = & -t[(1+e^{ika})\beta(k,n)+\beta(k,n-1)],
\label{eq:harper1noB}
\\
E_{\mu,k}\beta(k,n) & = & -t[(1+e^{-ika})\alpha(k,n)+\alpha(k,n+1)].
\label{eq:harper2noB}
\end{eqnarray}
As the ribbon has a finite width we have to be careful with the boundary
conditions. Since the ribbon only exists between $n=0$ and $n=N-1$
at the boundary Eqs.~(\ref{eq:harper1noB}) and~(\ref{eq:harper2noB})
read:
\begin{eqnarray}
E_{\mu,k}\alpha(k,0) & = & -t(1+e^{ika})\beta(k,0)\,,
\label{eq:harper1noBbc}
\\
E_{\mu,k}\beta(k,N-1) & = & -t(1+e^{-ika})\alpha(k,N-1).
\label{eq:harper2noBbc}
\end{eqnarray}
The surface (edge) states are solutions of Eqs.~(\ref{eq:harper1noB}-\ref{eq:harper2noBbc})
with $E_{\mu,k}=0$:
\begin{eqnarray}
0 & = & (1+e^{ika})\beta(k,n)+\beta(k,n-1)\,,
\label{eq:harper1noBes}
\\
0 & = & (1+e^{-ika})\alpha(k,n)+\alpha(k,n+1)\,,
\label{eq:harper2noBes}
\\
0 & = & \beta(k,0)\,,
\label{eq:harper1noBbces}
\\
0 & = & \alpha(k,N-1)\,.
\label{eq:harper2noBbces}
\end{eqnarray}
Equations~(\ref{eq:harper1noBes}) and~(\ref{eq:harper2noBbces}) are easily solved giving:
\begin{eqnarray}
\alpha(k,n) & = & [-2\cos(ka/2)]^{n}e^{i\frac{ka}{2}n}\alpha(k,0),
\label{eq:esA}
\\
\beta(k,n) & \!=\! & [\!-\!2\cos(ka/2)]^{N\!-\!1\!-\!n}e^{-i\frac{ka}{2}(N\!-\!1\!-\!n)}\beta(k,N\!-\!1).
\label{eq:esB}
\end{eqnarray}

Let us consider, for simplicity, a semi-infinite system with a
single edge. We must require the convergence condition $\left|-2\cos(ka/2)\right|<1$,
in (\ref{eq:esB}) because otherwise the wavefunction would diverge in the semi-infinite
graphene sheet. Therefore, the semi-infinite system has edge states
for $ka$ in the region $2\pi/3<ka<4\pi/3$, which corresponds to
$1/3$ of the possible momenta. Note that the amplitudes of the edge
states are given by,
\begin{eqnarray}
|\alpha(k,n)| & = & \sqrt{\frac{2}{\lambda(k)}}e^{-n/\lambda(k)},
\label{eq:modesA}
\\
|\beta(k,n)| & = & \sqrt{\frac{2}{\lambda(k)}}e^{-(N-1-n)/\lambda(k)},
\label{eq:modesB}
\end{eqnarray}
where the penetration length is given by:
\begin{eqnarray}
\lambda(k)=-1/\ln|2\cos(ka/2)|.
\end{eqnarray}
It is easily seen that the penetration length diverges when $ka$
approaches the limits of the region $]2\pi/3,4\pi/3[$.

Although the boundary conditions defined by Eqs.~(\ref{eq:harper1noBbces})
and~(\ref{eq:harper2noBbces}) are satisfied for solutions~(\ref{eq:esA})
and~(\ref{eq:esB}) in the semi-infinite system, they are not in
the ribbon geometry. In fact, Eqs.~(\ref{eq:modesA}) and~(\ref{eq:modesB})
are eigenstates only in the semi-infinite system. In the graphene
ribbon the two edge states, which come from both sides of the edge,
will overlap with each other. The bonding and anti-bonding states
formed by the two edge states will then be the ribbon eigenstates \cite{WFA+99} (note that at zero energy there are no other states
with which the edge states could hybridize). As bonding and anti-bonding
states result in a gap in energy the zero energy flat bands of edge states
will become slightly dispersive, depending on the ribbon width $N$.
The overlap between the two edge states is larger as $ka$ approaches
$2\pi/3$ and $4\pi/3$. This means that deviations from zero energy
flatness will be stronger near these points.

Edge states in graphene nanoribbons, just as the case of carbon nanotubes,
are predicted to be Luttinger liquids, that is,
interacting one-dimensional electron systems \cite{NGP06}. Hence, clean
nanoribbons must have 1D square root singularities in their density of states
\cite{Nakada96} that can be probed by Raman spectroscopy. Disorder
may smooth out these singularities, however.
In the
presence of a magnetic field, when the bulk states are gapped,
the edge states are responsible for the transport of spin and charge \cite{Abanin06,ALL07,ANZLGL07,AL0704}.

\subsection{Surface states in graphene stacks}
\label{surface_stack}

\begin{figure}[]
\begin{center}
\includegraphics[width=8cm]{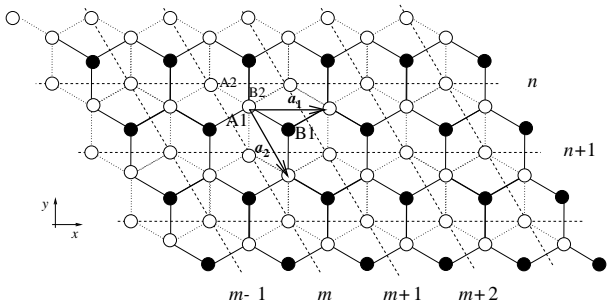}\\
\caption[fig]{\label{bilayer_zigzag} (Color online) Sketch of a zigzag termination of a
  graphene bilayer. As discussed in \protect{\cite{Cetal07}}, there is a band
  of surface states completely localized in the bottom layer,
and another surface band which alternates between the two.}
\end{center}
\end{figure}

Single layer graphene can be considered a zero gap semiconductor, which leads
to the extensively studied possibility of gap states, at $\epsilon = 0$, as
discussed in the previous section. The most studied such states are those
localized near a graphene zigzag edge \cite{FWNK96,WS00}. It can be shown
analytically \cite{Cetal07} that a bilayer zigzag edge, like that shown in
Fig.~\ref{bilayer_zigzag}, analyzed within the nearest neighbor tight-binding
approximation described before, has two bands of localized states, one
completely localized in the top layer and indistinguishable from similar
states in single layer graphene, and another band which alternates between
the two layers. These states, as they lie at $\epsilon = 0$, have finite
amplitudes on one half of the sites only.

These bands, as in single layer graphene, occupy one third of the
BZ of a stripe bounded by zigzag edges. They become
dispersive in a biased bilayer. As graphite can be described in
terms of effective bilayer systems, one for each value of the
perpendicular momentum, $k_z$, bulk graphite with a zigzag
termination should show one surface band per layer.

%------------------------------------------------------------------------------
\subsection{The spectrum of graphene nanoribbons}
\label{sec:nanoribbon}
%------------------------------------------------------------------------------

\begin{figure}
\begin{center}
\includegraphics*[width=7cm]{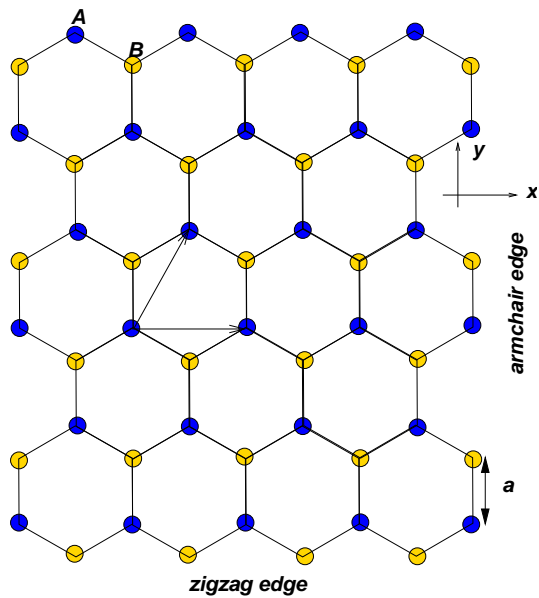}
\end{center}
\caption{\label{Fig:nanorA}(Color online) A piece of a honeycomb
lattice displaying both zigzag and armchair edges.}
\end{figure}

The spectrum of graphene nanoribbons depend very much on the
nature of their edges -- zigzag or armchair \cite{Brey106,Brey206,Nakada96}. In Fig.
\ref{Fig:nanorA}  we show a honeycomb lattice having zigzag edges
along the $x$ direction and armchair edges along the $y$ direction.
If we choose the ribbon to be infinite in the $x$ direction
we produce a graphene nanoribbon with zigzag edges; conversely
choosing the ribbon to be macroscopically large along the $y$ but finite
in the $x$ direction
we produce a graphene nanoribbon with armchair edges.

\begin{figure}
\begin{center}
\includegraphics*[width=9cm]{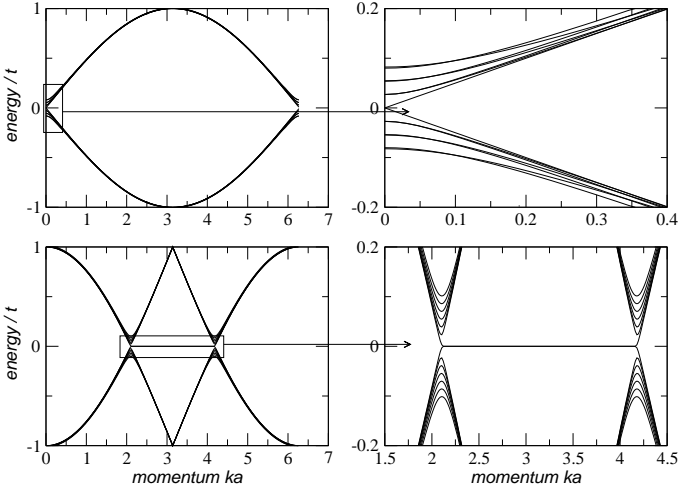}
\end{center}
\caption{\label{Fig:spectrum} (Color online)
Left: Energy spectrum, as calculated from the tight-binding equations,
for a nanoribbon with  armchair(top)
and zigzag(bottom) edges. The width of the nanoribbon
is $N=200$ unit cells. Only fourteen eigenstates are depicted.
Right: Zoom of the low energy states shown on the right. }
\end{figure}

In Fig.~\ref{Fig:spectrum} we show the fourteen energy levels,
calculated in the tight-binding approximation, closest
to zero energy for a nanoribbon with zigzag and armchair edges and
of width $N=200$ unit cells. We can see that they are both metallic,
and that the zigzag ribbon presents a band of zero energy modes that is
absent in the armchair case. This band at zero energy is the surface
states living  near the edge of the graphene ribbon.
More detailed {\it ab initio} calculations of the spectra of graphene nanoribbons
show that
interaction effects can lead to electronic gaps \cite{SCL06a} and magnetic states
close to the graphene edges, independent of their nature \cite{SCL06b,YPSCL07,YCL07}.

From the experimental point of view, however, graphene nanoribbons currently have a
high degree of roughness at the edges.
Such edge disorder can change significantly the properties of edge states \cite{AW07,GAW07},
leading to Anderson localization, and anomalies in the quantum Hall effect
\cite{NGP06,MB07} as well as Coulomb blockade effects \cite{SGN07b}. Such effects
have already been observed in lithographically engineered graphene
nanoribbons \cite{HOZK07,OJEALK07}. Furthermore, the problem of edge passivation by
hydrogen or other elements is not clearly understood experimentally
at this time. Passivation can be modeled in the tight-binding approach by modifications
of the hopping energies \cite{N07} or via additional phases in the boundary conditions
\cite{KM97}. Theoretical modeling of edge passivation indicate that those have
a strong effect on the electronic properties at the edge of graphene nanoribbons \cite{BHS06,HBPS07}.

In what follows we derive the spectrum for both zigzag
and armchair edges directly from the Dirac equation. This was
originally done both with and without a magnetic field \cite{Brey106,Brey206,Nakada96}.

%------------------------------------------------------------------------------
\subsubsection{Zigzag nanoribbons}
\label{zigzag}
%------------------------------------------------------------------------------

In the geometry of Fig.~\ref{Fig:nanorA} the
unit cell vectors are $\bm a_1=a_0(1,0)$
and $\bm a_2=a_0\left(1/2,\sqrt 3 /2\right)$, which generate the
unit vectors of the BZ given by
$\bm b_1=4\pi/(a_0\sqrt 3)\left(\sqrt 3/2,-1/2\right)$
and  $\bm b_2=4\pi/(a_0\sqrt 3)(0,1)$.
From these two vectors we find two inequivalent Dirac points
given by $\bm K = \left(4\pi/3a_0,0\right)=(K,0)$
and  $\bm K' = \left(- 4\pi/3a_0,0\right)=(-K,0)$,
with $a_0=\sqrt 3 a$. The Dirac Hamiltonian
around the Dirac point $\bm K$ reads in momentum space:
\begin{equation}
H_{\bm K} =v_F
\left(
\begin{array}{cc}
0 & p_x -ip_y \\
p_x+ip_y & 0
\end{array}
\right),
\label{eq:DiracK}
\end{equation}
and around the $\bm K'$ as:
\begin{equation}
H_{\bm K'} =v_F
\left(
\begin{array}{cc}
0 & p_x +ip_y \\
p_x-ip_y & 0
\end{array}
\right).
\label{eq:DiracKp}
\end{equation}
The wavefunction, in real space, for the sublattice $A$ is given by:
\begin{equation}
\Psi_A(\bm r)=e^{i\bm K\cdot \bm r}\psi_A(\bm r) +
e^{i\bm K'\cdot \bm r}\psi'_A(\bm r)\,,
\end{equation}
and for sublattice $B$ is given by
\begin{equation}
\Psi_B(\bm r)=e^{i\bm K\cdot \bm r}\psi_B(\bm r) +
e^{i\bm K'\cdot \bm r}\psi'_B(\bm r)\,,
\end{equation}
where $\psi_A$ and $\psi_B$ are the components of the
spinor wavefunction of Hamiltonian (\ref{eq:DiracK})
and  $\psi'_A$ and $\psi'_B$ have  identical meaning
but relatively to (\ref{eq:DiracKp}).
Let us assume that the edges of the nanoribbons are parallel
to the $x-$axis. In this case, the translational symmetry guarantees
that the spinor wavefunction can be written as:
\begin{equation}
\psi(\bm r)=e^{ik_x x}
\left(
\begin{array}{c}
\phi_A(y)\\
\phi_B(y)
\end{array}
\right),
\end{equation}
and a similar equation for the spinor of Hamiltonian
(\ref{eq:DiracKp}).
For zigzag
edges the boundary conditions at the edge of the
ribbon (located at $y=0$ and $y=L$, where $L$ is the ribbon width) are:
\begin{equation}
 \Psi_A(y=L)=0,\hspace{1cm}\Psi_B(y=0)=0\,,
\end{equation}
leading to:
\begin{eqnarray}
0&=&e^{i Kx}e^{ik_x x}\phi_A(L) +
e^{-iKx}e^{ik_x x}\phi'_A(L)\,,
\label{eq:zizzagBCa}
\\
0&=&e^{i Kx}e^{ik_x x}\phi_B(0) +
e^{-i Kx}e^{ik_x x}\phi'_B(0)\,.
\label{eq:zizzagBCb}
\end{eqnarray}
The boundary conditions (\ref{eq:zizzagBCa}) and
(\ref{eq:zizzagBCb}) are satisfied for any
$x$ by the choice:
\begin{equation}
\phi_A(L)=\phi'_A(L)=\phi_B(0)=\phi'_B(0)=0\,.
\end{equation}
We need now to find out the form of the envelope
functions. The eigenfunction around the point ${\bf K}$
has the form:
\begin{equation}
\left(
\begin{array}{cc}
0 & k_x -\partial_y \\
k_x +\partial_y& 0
\end{array}
\right)
\left(
\begin{array}{c}
\phi_A(y)\\
\phi_B(y)
\end{array}
\right)
=\tilde\epsilon
\left(
\begin{array}{c}
\phi_A(y)\\
\phi_B(y)
\end{array}
\right),
\end{equation}
with $\tilde \epsilon = \epsilon/v_F$ and $\epsilon$ the energy
eigenvalue. The eigenproblem can be written as  two linear
differential equations of the form:
\begin{equation}
\left\{
\begin{array}{c}
(k_x-\partial_y)\phi_B=\tilde\epsilon\phi_A\,,\\
(k_x+\partial_y)\phi_A=\tilde\epsilon\phi_B.
\end{array}
\right.
\label{1D}
\end{equation}
Applying the operator $(k_x+\partial_y)$ to the
first of Eqs. (\ref{1D}) leads to:
\begin{equation}
(-\partial^2_y+k_x^2)\phi_B=\tilde\epsilon^2\phi_B,
\label{2D}
\end{equation}
with $\phi_A$ given by:
\begin{equation}
\phi_A=\frac 1 {\tilde\epsilon}
(k_x-\partial_y)\phi_B\,.
\label{phiA1D}
\end{equation}
The solution of (\ref{2D}) has the form:
\begin{equation}
\phi_B=Ae^{zy}+Be^{-zy},
\end{equation}
leading to an eigenenergy  $\tilde\epsilon^2=k^2_x-z^2$.
The boundary conditions for  a zigzag edge require that
$\phi_A(y=L)=0$ and $\phi_B(y=0)=0$, leading to:
\begin{equation}
\left\{
\begin{array}{c}
  \phi_B(y=0)=0 \Leftrightarrow A+B=0\,,\\
\phi_A(y=L)=0\Leftrightarrow (k_x-z)Ae^{zL}+(k_x+z)Be^{-zL}=0
\end{array}
\right.,
\end{equation}
which leads to an eigenvalue equation of the form:
\begin{equation}
e^{-2zL}=\frac {k_x-z}{k_x+z}\,.
\label{EGV1}
\end{equation}
Equation (\ref{EGV1}) has real solutions for $z$, whenever
$k_x$ is positive; these solutions
correspond to surface waves (edge states)
existing near the edge of the graphene ribbon. In section
\ref{surface}
we discussed these states from the point of view of the tight-binding model. In addition
to real solutions for $z$, (\ref{EGV1}) also supports complex
ones, of the form $z=ik_n$, leading to:
\begin{equation}
k_x = \frac{k_n}{\tan(k_nL)}.
\label{EGV1b}
\end{equation}
The solutions of (\ref{EGV1b}) correspond to confined modes in the
graphene ribbon.

If we apply the same procedure to the Dirac equation around
the Dirac point $\bm K'$ we obtain a different eigenvalue equation
given by:
\begin{equation}
e^{-2zL}=\frac {k_x+z}{k_x-z}\,.
\label{EGV2}
\end{equation}
This equation supports real solutions for $z$ if $k_x$ is negative.
Therefore we have edge states for negative values $k_x$, with
momentum around $\bm K'$. As in the case of $\bm K$, the system
also supports confined modes, given by:
\begin{equation}
k_x = -\frac{k_n}{\tan(k_nL)}.
\label{EGV2b}
\end{equation}
One should note that the eigenvalue equations for $\bm K'$ are obtained
from those for $\bm K$ by inversion, $k_x\rightarrow -k_x$.

We finally notice that the edge states for zigzag nanoribbons are
dispersionless (localized in real space) when $t'=0$. When electron-hole
symmetry is broken ($t'\neq 0$) these states become dispersive with a
Fermi velocity $v_e \approx t' a$ \cite{NGP06}.

%------------------------------------------------------------------------------
\subsubsection{Armchair nanoribbons}
\label{armchair}
%------------------------------------------------------------------------------

Let us now consider an armchair nanoribbon with armchair edges
along the $y$ direction. The boundary conditions at the edges
of the ribbon (located at $x=0$ and $x=L$, where $L$ is the
width of the ribbon):
\begin{equation}
\Psi_A(x=0)=\Psi_B(x=0)=\Psi_A(x=L)=\Psi_B(x=L)=0\,.
\end{equation}
 Translational symmetry guarantees
that the spinor wavefunction of Hamiltonian
(\ref{eq:DiracK}) can be written as:
\begin{equation}
\psi(\bm r)=e^{ik_y y}
\left(
\begin{array}{c}
\phi_A(x)\\
\phi_B(x)
\end{array}
\right),
\end{equation}
and a similar equation for the spinor of the Hamiltonian
(\ref{eq:DiracKp}). The boundary conditions have the form:
\begin{eqnarray}
0&=&e^{ik_yy}\phi_A(0)+e^{ik_yy}\phi'_A(0)\,,
\\
0&=&e^{ik_yy}\phi_B(0)+e^{ik_yy}\phi'_B(0)\,,
\\
0&=&e^{iKL}e^{ik_yy}\phi_A(L)
+e^{-iKL}e^{ik_yy}\phi'_A(L)\,,
\\
0&=&e^{iKL}e^{ik_yy}\phi_B(L)
+e^{-iKL}e^{ik_yy}\phi'_B(L)\,,
\end{eqnarray}
and are satisfied for any $y$ if:
\begin{equation}
\phi_\mu(0)+\phi'_\mu(0)=0\,,
\label{eq:BCAC1}
\end{equation}
and
\begin{equation}
e^{iKL}\phi_\mu(L)+e^{-iKL}\phi'_\mu(L)=0\,,
\label{eq:BCAC2}
\end{equation}
with $\mu=A,B$. It is clear that these boundary conditions
mix states from the two Dirac points. Now we must  find the
form of the envelope functions obeying the boundary conditions
(\ref{eq:BCAC1}) and (\ref{eq:BCAC2}). As before, the functions
$\phi_B$ and $\phi'_B$ obey  the second order differential
equation (\ref{2D}) (with $y$ replaced by $x$) and the function $\phi_A$ and $\phi'_A$
are determined from (\ref{phiA1D}).
The solutions of (\ref{2D}) have the form:
\begin{eqnarray}
\phi_B&=&Ae^{ik_nx}+Be^{-ik_n x}\,,\\
\phi'_B&=&Ce^{ik_nx}+De^{-ik_n x}\,.
\end{eqnarray}
Applying the boundary conditions:
(\ref{eq:BCAC1}) and (\ref{eq:BCAC2}),
one obtains:
\begin{eqnarray}
0&=&A+B+C+D\,,\\
0&=&Ae^{i(k_n+K)L}+De^{-i(k_n+K)L}\nonumber\\
&+&Be^{-i(k_n-K)L}+Ce^{i(k_n-K)L}\,.
\end{eqnarray}
The boundary conditions are satisfied with the choice:
\begin{equation}
A=-D\,,\hspace{1cm} B=C=0\,,
\end{equation}
which leads to $\sin[(k_n+K)L]=0$. Therefore the allowed values of
$k_n$ are given by
\begin{equation}
k_n = \frac {n\pi}L - \frac {4\pi}{3a_0}\,,
\end{equation}
and the eigenenergies are given by:
\begin{equation}
\tilde\epsilon^2=k^2_y+k^2_n \, .
\end{equation}
No surface states exist in this case.

%------------------------------------------------------------------------------
\subsection{Dirac fermions in a magnetic field}
\label{sec:contmag}
%------------------------------------------------------------------------------

Let us now consider the problem of a uniform magnetic field $B$ applied
perpendicular to the graphene plane \footnote[2]{The problem of transverse magnetic and
electric fields can also be solved exactly. See: \cite{LSB06,PC07}.}. We use
the Landau gauge: ${\bf A} = B (-y,0)$. Notice that the
magnetic field introduces a new length scale in the problem:
\begin{eqnarray}
\ell_B = \sqrt{\frac{c}{e B}} \, ,
\label{lb}
\end{eqnarray}
which is the magnetic length. The
only other scale in the problem is the Fermi-Dirac velocity. Dimensional
analysis shows that the only quantity with dimensions of energy we
can make is $v_F/\ell_B$. In fact, this determines the cyclotron
frequency of the Dirac fermions:
\begin{eqnarray}
\omega_c = \sqrt{2} \, \frac{v_F}{\ell_B}
\label{wc}
\end{eqnarray}
(the $\sqrt{2}$ factor comes from the quantization of the problem,
see below). Eqs.~(\ref{wc}) and (\ref{lb}) show that the
cyclotron energy scales like $\sqrt{B}$, in clear contrast with
the non-relativistic problem where the cyclotron energy is linear in
$B$. This implies that the energy scale associated with
the Dirac fermions is rather different from the one find in the ordinary
2D electron gas. For instance, for fields of the order $B \approx 10$ T
the cyclotron energy in the 2D electron gas is of the order of 10 K.
In contrast, for the Dirac fermion problem, for the same fields,
the cyclotron energy is of the order of $1,000$ K, that is, two orders
of magnitude bigger. This has strong implications for the observation
of the quantum Hall effect at room temperature \cite{Netal07}. Furthermore, for $B = 10$ T the Zeeman
energy is relatively small, $g \mu_B B \approx 5$ K, and can be disregarded.

\begin{figure}[htb]
\centerline{\includegraphics[width=8cm]{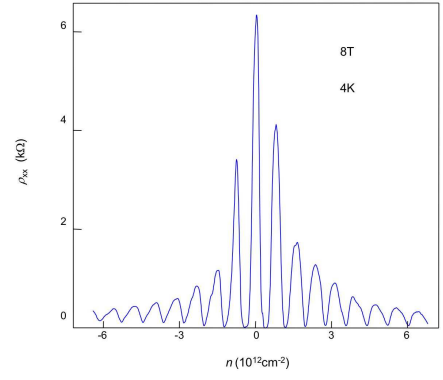}}
\caption{\label{sdhexp}
(Color online) SdH oscillations observed in longitudinal resistivity $\rho_{xx}$ of graphene
as a function of the charge carrier concentration $n$. Each peak corresponds to the
population of one Landau level. Note that the sequence is not interrupted when
passing through the Dirac point, between electrons and holes. The period of oscillations
$\Delta n = 4 B/\Phi_0$, where $B$ is the applied field and $\Phi_0$ is the flux
quantum \cite{Netal05}.}
\end{figure}

Let us now consider the Dirac equation in more detail. Using the minimal coupling
in (\ref{diraceq}) (i.e., replacing $- i \nabla $ by $-i \nabla +e {\bf A}/c$)
we find:
\begin{eqnarray}
v_F \left[\vec{\sigma} \cdot (- i \nabla + e {\bf A}/c) \right] \psi({\bf r}) &=& E \, \, \psi({\bf r}) \,  ,
\end{eqnarray}
in the Landau gauge the generic solution for the wavefunction has the form
$\psi(x,y) = e^{i k x} \phi(y)$,
and the Dirac equation reads:
\begin{eqnarray}
v_F\! \left[ \!\begin{array}{cc}
0 & \partial_y \!-\!k\!+\! B e y/c\\
- \partial_y \!-\! k \!+\! B e y/c& 0
\end{array}\!\right] \!\phi(y) \!\!&=&\!\! E \phi(y) \, ,
\label{diracshmag}
\end{eqnarray}
that can be rewritten as:
\begin{eqnarray}
\omega_c \left[\begin{array}{cc}
0 & {\cal O}\\
{\cal O}^{\dag}& 0 \end{array}\right] \phi(\xi) =  E \, \phi(\xi) \, ,
\end{eqnarray}
or equivalently:
\begin{eqnarray}
({\cal O} \sigma^+ + {\cal O}^{\dag} \sigma^-) \phi = (2 E/\omega_c) \phi \, ,
\label{diracmag}
\end{eqnarray}
where $\sigma^{\pm} = \sigma_x \pm i \sigma_y$, and we have defined
the dimensionless length scale:
\begin{eqnarray}
\xi &=& \frac{y}{\ell_B} - \ell_B k \, ,
\end{eqnarray}
and 1D harmonic oscillator operators:
\begin{eqnarray}
{\cal O} &=& \frac{1}{\sqrt{2}} \left(\partial_{\xi} + \xi\right) \, ,
\nonumber
\\
{\cal O}^{\dag} &=&  \frac{1}{\sqrt{2}} \left(- \partial_{\xi} + \xi\right) \, ,
\end{eqnarray}
that obey canonical commutation relations:
$[{\cal O},{\cal O}^{\dag}]=1$. The number operator is simply:
$N = {\cal O}^{\dag} {\cal O}$.

Firstly, we notice that (\ref{diracmag}) allows for a solution with
zero energy:
\begin{eqnarray}
({\cal O} \sigma^+ + {\cal O}^{\dag} \sigma^-) \phi_0 = 0 \, ,
\label{diracmagzero}
\end{eqnarray}
and since the Hilbert space generated by $\vec{\sigma}$ is of
dimension $2$, and the spectrum generated by ${\cal O}^{\dag}$ is
bounded from below, we just need to ensure that:
\begin{eqnarray}
{\cal O} \phi_0 &=& 0 \, ,
\nonumber
\\
\sigma^- \phi_0 &=& 0 \, ,
\end{eqnarray}
in order for (\ref{diracmagzero}) to be fulfilled.
The obvious zero mode solution is:
\begin{eqnarray}
\phi_0(\xi) =  \psi_0(\xi) \otimes |\Downarrow \rangle \, ,
\end{eqnarray}
where $|\Downarrow\rangle$ indicates the state localized on
sublattice $A$ and $|\Uparrow\rangle$ indicates the state
localized on sublattice $B$. Furthermore,
\begin{eqnarray}
{\cal O} \psi_0(\xi) = 0 \, ,
\end{eqnarray}
is the ground states of the 1D harmonic oscillator.
All the solutions can now be constructed from the zero mode:
\begin{eqnarray}
\phi_{N,\pm} (\xi) &=& \psi_{N-1}(\xi) \otimes |\Uparrow\rangle
\pm \psi_N(\xi) \otimes |\Downarrow \rangle
\nonumber
\\
&=& \left(\begin{array}{c}
\psi_{N-1}(\xi) \\
\pm \psi_N (\xi)
\end{array}\right) \, ,
\label{landaufunc}
\end{eqnarray}
and their energy is given by \cite{M56}:
\begin{eqnarray}
E_{\pm}(N) &=& \pm \omega_c \sqrt{N} \, ,
\label{landaulev}
\end{eqnarray}
where $N =0,1,2,...$ is a positive integer, $\psi_N(\xi)$ is the solution of the 1D Harmonic oscillator
(explicitly: $\psi_N(\xi) = 2^{-N/2} (N!)^{-1/2} \exp\{-\xi^2/2\} H_N(\xi)$ where $H_N(\xi)$ is
a Hermite polynomial). The Landau levels at the opposite Dirac point,
K', have exactly the same spectrum and hence each Landau level is doubly degenerate.
Of particular importance for the Dirac problem discussed here is the existence
of a zero energy state $N=0$ which is responsible, as we are going to show,
to the anomalies observed in the quantum Hall effect.
This particular Landau level structure has been observed by many different
experimental probes, from Shubnikov-de Haas oscillations in single layer graphene (see Fig. \ref{sdhexp})
\cite{Netal05,Zetal05b}, to infrared spectroscopy  \cite{Jetal07},
and to scanning tunneling spectroscopy \cite{LA07} (STS) on a graphite surface.

\subsection{The anomalous integer quantum Hall effect}
\label{iqhe}

In the presence of disorder Landau levels get broadened and mobility
edges appear \cite{laughlin}.
Notice that there will be a Landau level at zero energy that separates
states with hole character ($\mu <0$) from states with electron character
($\mu>0$).
The components of the resistivity and conductivity tensors are related by:
\begin{eqnarray}
\rho_{xx} &=& \frac{\sigma_{xx}}{\sigma_{xx}^2 + \sigma_{xy}^2} \, ,
\nonumber
\\
\rho_{xy} &=& \frac{\sigma_{xy}}{\sigma_{xx}^2 + \sigma_{xy}^2} \, ,
\label{rhosig}
\end{eqnarray}
where $\sigma_{xx}$ ($\rho_{xx}$)
is the longitudinal component and $\sigma_{xy}$ ($\rho_{xy}$)
is the Hall component of the conductivity (resistivity).
When the chemical potential is inside of a region of localized
states the longitudinal conductivity vanishes, $\sigma_{xx} =0$,
and hence: $\rho_{xx}=0$, $\rho_{xy}=1/\sigma_{xy}$. On the other
hand, when the chemical potential is a region of delocalized states,
when the chemical potential is crossing a Landau level,
we have $\sigma_{xx} \neq 0$ and $\sigma_{xy}$ varies continuously \cite{SSW06,SSHB07}.

\begin{figure}[htb]
\centerline{\includegraphics[width=9cm]{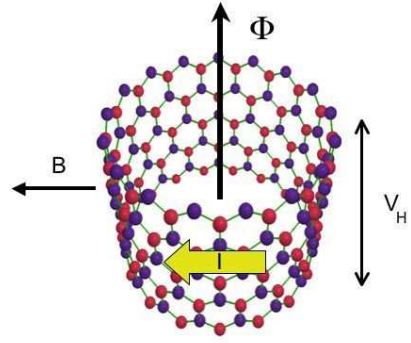}}
\caption{\label{laughlin}
(Color online) Geometry of Laughlin's thought experiment
with a graphene ribbon: a magnetic field $B$ is applied normal to the surface
of the ribbon, a current $I$ circles the loop, generating a Hall
voltage V$_{\rm H}$, and a magnetic flux $\Phi$.}
\end{figure}

The value
of $\sigma_{xy}$ in the region of localized states can be obtained
from Laughlin's gauge invariance argument \cite{laughlin}: one
imagines making a graphene ribbon such as the one in Fig.~\ref{laughlin}
with a magnetic field $B$ normal through its surface and a current $I$
circling its loop. Due to the Lorentz force the magnetic field produces a Hall voltage $V_H$
perpendicular to the field and current. The circulating current
generates a magnetic flux $\Phi$ that threads the loop. The current
is given by:
\begin{eqnarray}
I = c \frac{\delta E}{\delta \Phi} \, ,
\end{eqnarray}
where $E$ is the total energy of the system. The localized states
do not respond to changes in $\Phi$, only the delocalized ones.
When the flux is changed by a flux quantum $\delta \Phi=\Phi_0 = h c/e$
the extended states remain the same by gauge invariance. If the chemical
potential is in the region of localized states, all the extended
states below the chemical potential will be filled both before and
after the change of flux by $\Phi_0$. However, during the
change of flux an integer number of states enter the cylinder at one
edge and leave at the opposite edge.

\begin{figure}[htb]
\centerline{\includegraphics[width=10cm]{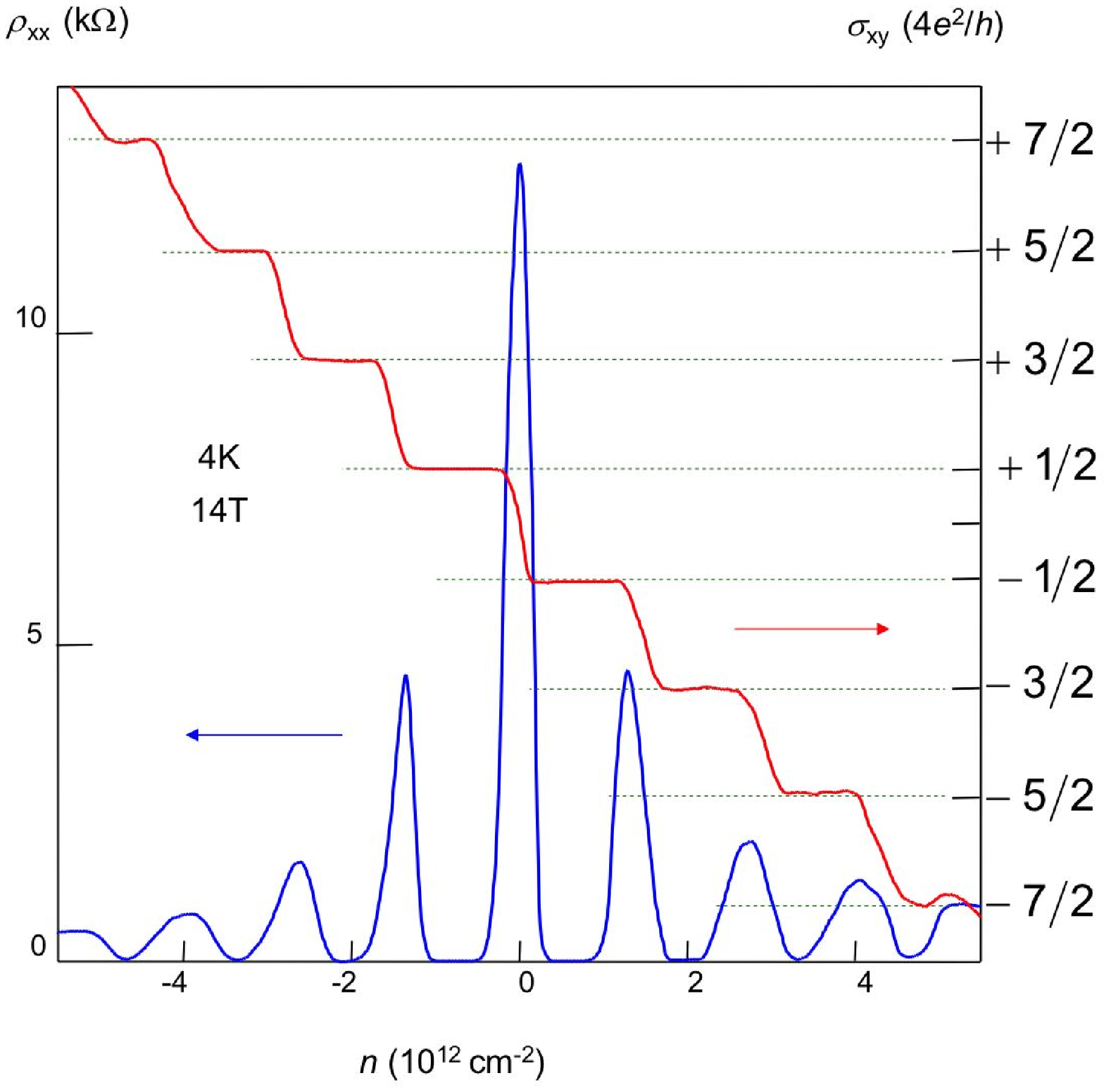}}
\caption{\label{iqheexp}
(Color online) Quantum Hall effect in graphene as a function of charge carrier
concentration. The peak at $n=0$ shows that in high magnetic fields there
appears a Landau level at zero energy where no states exist in zero field.
The field draws electronic states for this level from both conduction and
valence bands. The dashed line indicate plateaus in $\sigma_{xy}$ described
by Eq. (\ref{lagh}). Adapted from \cite{Netal05}.}
\end{figure}

The question is how many
occupied states are transferred between edges. Let us consider
a naive and as shown further incorrect
calculation in order to show the importance
of the zero mode in this problem. Each Landau
level contributes with one state times its degeneracy $g$.
In the case of graphene we have $g=4$ since there
are $2$ spin states and $2$ Dirac cones. Hence, we would expect
that when the flux changes by one flux quantum the change
in energy would be $\delta E_{\rm inc.} = \pm 4 N e V_H$, where
$N$ is an integer. The plus sign applies to electron states
(charge $+e$) and the minus sign to hole states (charge $-e$).
Hence, we would conclude that $I_{\rm inc.} = \pm 4 (e^2/h) V_H$
and hence $\sigma_{xy,{\rm inc.}} = I/V_H = \pm 4 N e^2/h$, which
is the naive expectation. The problem with this result is that
when the chemical potential is exactly at half-filling, that is,
at the Dirac point, it would predict a Hall plateau at $N=0$ with
$\sigma_{xy,{\rm inc.}} = 0$ which is not possible since there is
a $N=0$ Landau level, with extended states at this energy. The
solution for this paradox is rather simple: because of the presence
of the zero mode which is shared by the two Dirac points, there
are exactly $2 \times (2N+1)$ occupied states that are transferred
from one edge to another. Hence, the change in energy is
$\delta E = \pm 2 (2N+1) e V_H$ for a change of flux of
$\delta \Phi = hc/e$. Therefore, the Hall conductivity is \cite{Sch91,PGC06,PGN06,GS05,H06}:
\begin{eqnarray}
\sigma_{xy} = \frac{I}{V_H} = \frac{c}{V_H} \frac{\delta E}{\delta \Phi}
= \pm 2 (2 N+1) \frac{e^2}{h} \, ,
\label{lagh}
\end{eqnarray}
without any Hall plateau at $N=0$. This amazing result
has been observed experimentally \cite{Netal05,Zetal05b}
as shown in Fig.\ref{iqheexp}.

%------------------------------------------------------------------------------
\subsection{Tight-binding model in a magnetic field}
\label{tbfield}
%------------------------------------------------------------------------------

In the tight-binding approximation the hopping integrals are replaced
by a Peierls substitution:
\begin{equation}
e^{i e \int_{\mathbf{R}}^{\mathbf{R}'}
\mathbf{A}\cdot\textrm{d}\mathbf{r}}t_{\mathbf{R},\mathbf{R}'}
=
e^{i\frac{2\pi}{\Phi_0}\int_{\mathbf{R}}^{\mathbf{R}'}
\mathbf{A}\cdot\textrm{d}\mathbf{r}}t_{\mathbf{R},\mathbf{R}'}
\,,
\end{equation}
where $t_{\mathbf{R},\mathbf{R}'}$ represents the hopping integral
between the sites $\mathbf{R}$ and $\mathbf{R}'$,
with no field present.
The tight-binding
Hamiltonian for a single graphene layer, in a
constant magnetic field perpendicular to the plane,
is conveniently written as,
\begin{widetext}
\begin{equation}
H=-t\sum_{m,n,\sigma}[e^{i\pi \frac{\Phi}{\Phi_{0}}n\frac {1+z}2}
a_{\sigma}^{\dagger}(m,n)b_{\sigma}(m,n)+
e^{-i\pi\frac{\Phi}{\Phi_{0}}n}
a_{\sigma}^{\dagger}(m,n)b_{\sigma}(m-1,n-(1-z)/2)+
e^{i\pi \frac{\Phi}{\Phi_{0}}n\frac {z-1}2}
a_{\sigma}^{\dagger}(m,n)b_{\sigma}(m,n-z)+\textrm{h.c.}],
\label{eq:HzigzagB}
\end{equation}
\end{widetext}
holding  for a graphene stripe with a zigzag
($z=1$) and armchair ($z=-1$)  edges oriented
along the $x-$direction.
Fourier transforming along the $x$ direction gives,
\begin{widetext}
\begin{eqnarray*}
H  =  -t\sum_{k,n,\sigma}[e^{i\pi\frac{\Phi}{\Phi_{0}}n\frac {1+z}2}
a_{\sigma}^{\dagger}(k,n)b_{\sigma}(k,n)+
e^{-i\pi\frac{\Phi}{\Phi_{0}}n}e^{ika}
a_{\sigma}^{\dagger}(k,n)b_{\sigma}(k,n-(1-z)/2)+
e^{i\pi \frac{\Phi}{\Phi_{0}}n\frac {z-1}2}
a_{\sigma}^{\dagger}(k,n)b_{\sigma}(k,n-z)+\textrm{h.c.}].
\end{eqnarray*}
\end{widetext}

Let us now consider the case of zigzag edges. The eigenproblem
can be rewritten in terms of Harper's equations \cite{harper}, and for zigzag edges
we obtain \cite{rammal85}:
\begin{eqnarray}
E_{\mu,k}\alpha(k,n) & = & -t[
e^{ika/2}2\cos(\pi\frac{\Phi}{\Phi_{0}}n-\frac{ka}{2})\beta(k,n)
\nonumber\\
&+&
\beta(k,n-1)],\label{eq:harper1B}\\
E_{\mu,k}\beta(k,n) & = &
-t[e^{-ika/2}2\cos(\pi\frac{\Phi}{\Phi_{0}}n
-\frac{ka}{2})\alpha(k,n)
\nonumber\\
&+&\alpha(k,n+1)],
\label{eq:harper2B}
\end{eqnarray}
where the coefficients $\alpha(k,n)$ and $\beta(k,n)$
show up in Hamiltonian's eigenfunction, $\vert \psi (k)\rangle$,
written in terms of lattice-position-state states as:
\begin{equation}
\vert \psi(k) \rangle = \sum_{n,\sigma}\left(
\alpha(k,n) \vert a; k, n, \sigma \rangle
+
\beta(k,n) \vert b; k, n, \sigma  \rangle
\right)\,.
\end{equation}
Eqs.~(\ref{eq:harper1B}) and~(\ref{eq:harper2B}) hold in the bulk.
Considering that the zigzag ribbon has $N$ unit cells along
its width, from $n=0$ to $n=N-1$,
the boundary conditions  at the edges are obtained from
Eqs.~(\ref{eq:harper1B}) and~(\ref{eq:harper2B}), and  read
\begin{eqnarray}
E_{\mu,k}\alpha(k,0) =
-te^{ika/2}2\cos\left(
\frac{ka}{2}\right)\beta(k,0)\,,\label{eq:harper1Bbc}
\end{eqnarray}
\begin{eqnarray}
E_{\mu,k}\beta(k,N\!\!-\!\!1\!) \!\!=\!\!  -
2 t e^{-ika/2}\!\cos\!\left[\!\pi\frac{\Phi}{\Phi_{0}}(N\!\!-\!\!1)\!-\!\frac{ka}{2}\!\right]
\! \alpha(k,N\!\!-\!\!1).
\label{eq:harper2Bbc}
\end{eqnarray}
Similar equations hold for a graphene ribbon with armchair edges.

\begin{figure}
\begin{center}
\includegraphics*[width=8cm]{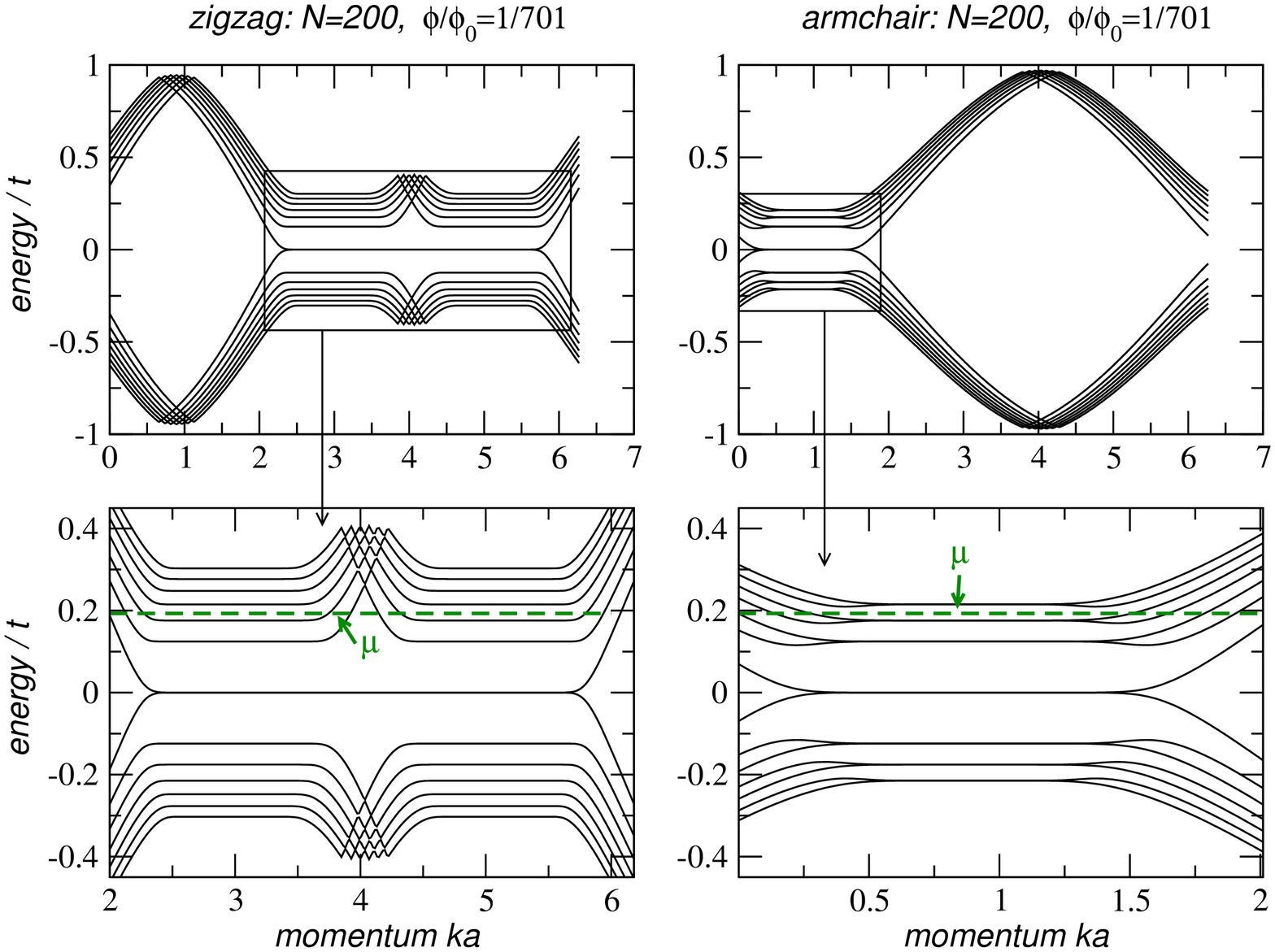}
\end{center}
\caption{\label{cap:LL_TB} (Color online)
Fourteen Energy levels of  tight-binding electrons in graphene
in the presence of a magnetic flux $\Phi=\Phi_0/701$,
for a finite stripe with $N=200$ unit cells. The bottom panels
are zoom in images of the top ones. The dashed line
represents the chemical potential $\mu$.}
\end{figure}

In Fig.~\ref{cap:LL_TB} we show  fourteen energy levels,
around zero energy, for both the zigzag and the armchair
cases. The formation of the Landau levels is signaled by the presence
of flat energy bands, following the bulk energy spectrum.
From Fig.~\ref{cap:LL_TB} it is straightforward to
obtain the value of the
Hall conductivity in the quantum Hall effect regime. Let us assume
that the chemical potential is in between two Landau levels
at positive energies, as
represented by the dashed line in Fig. \ref{cap:LL_TB}. The Landau
level structure shows  two zero energy modes, one of them is
electron-like (hole-like), since close to the edge of the sample its energy is
shifted upwards (downwards). The other Landau levels are doubly degenerate.
The determination of the values for the Hall conductivity is done by counting
how many energy levels (of electron-like nature) are below chemical
potential. This counting  produces the value ($2N+1$), with
$N=0,1,2,\ldots$ (for the case of Fig. \ref{cap:LL_TB} one has
$N=2$). From this counting the  Hall conductivity is given,
including an extra factor of two accounting for the spin degree of freedom, by
\begin{equation}
\sigma_{xy}= \pm 2\frac {e^2}h(2N+1)
=\pm 4\frac {e^2}h\left(N+\frac 1 2\right)\,.
\label{Eq:sigmahall}
\end{equation}
Eq.~(\ref{Eq:sigmahall}) leads to the anomalous integer
quantum Hall effect discussed previously, which is the hallmark of
single layer graphene.

\subsection{Landau levels in graphene stacks}
\label{llstack}

\begin{figure}[]
\begin{center}
\includegraphics*[width=4cm]{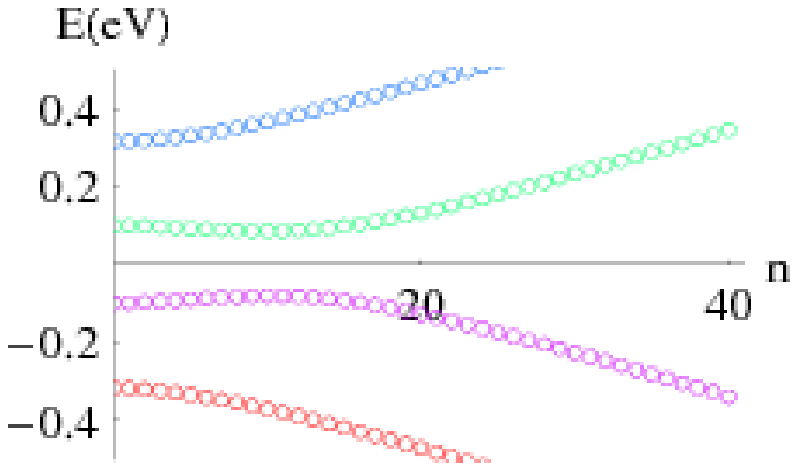}
\includegraphics*[width=4cm]{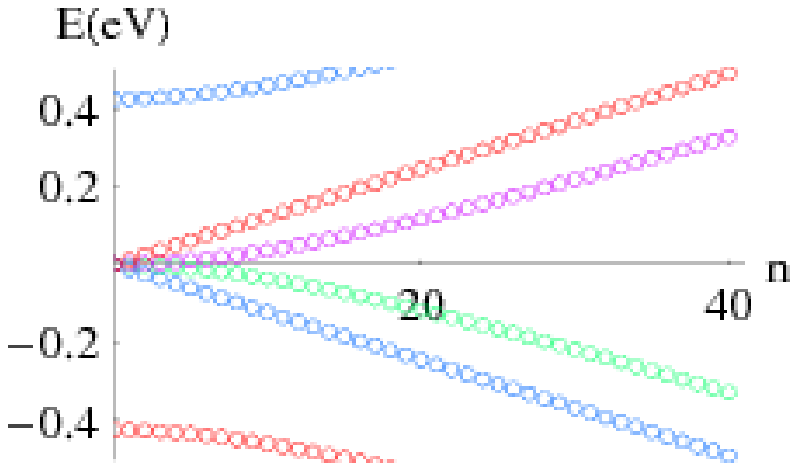}
\includegraphics*[width=4cm]{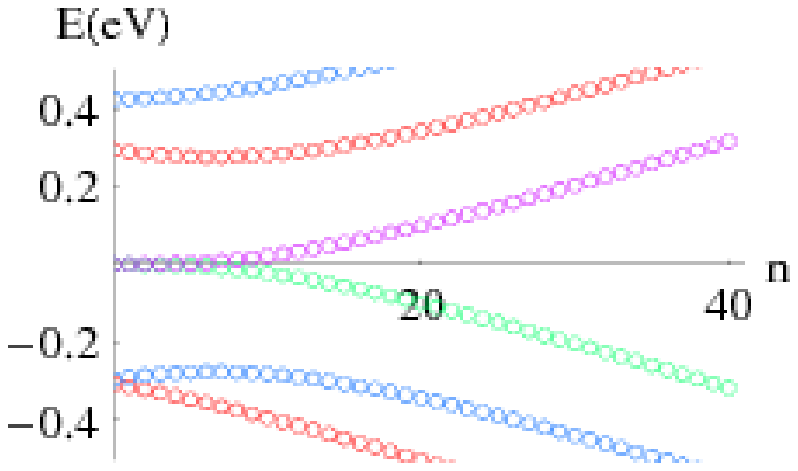}
\includegraphics*[width=4cm]{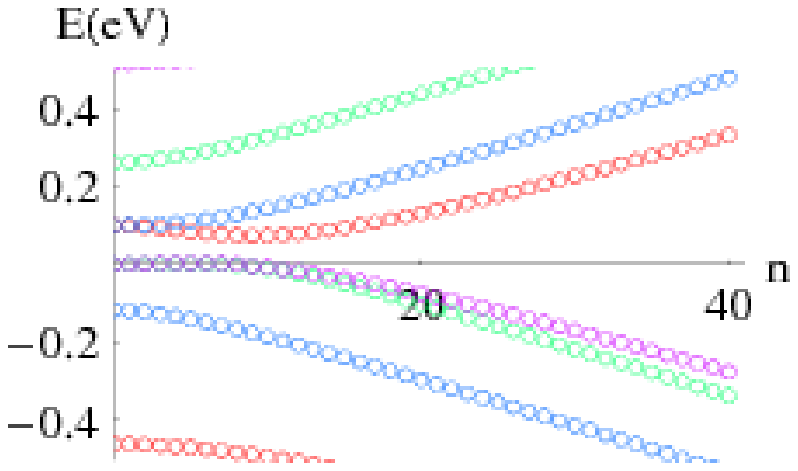}
\caption[fig]{\label{trilayer_mag} (Color online) Landau levels of the
graphene stacks shown in Fig.\protect{\ref{trilayer}}.
The applied magnetic field is 1 T.}
\end{center}
\end{figure}

The dependence of the Landau level energies  on the Landau index $N$ roughly
follows  the dispersion relation of the bands, as shown in
Fig.~\ref{trilayer_mag}. Note that, in a trilayer with Bernal stacking, two
sets of levels have a $\sqrt{N}$ dependence, while the energies of other two
depend linearly on $N$. In an infinite rhombohedral stack, the Landau levels
remain discrete and quasi-2D \cite{GNP06}. Note that, even in an
infinite stack with the Bernal structure, the Landau level closest to
$E = 0$ forms a band which does not overlap with the other Landau
levels, leading to the possibility of a 3D integer quantum
Hall effect \cite{BHRA07,Ketal03,KPSME06,LK04}.

\begin{figure}[]
\begin{center}
\includegraphics*[width=4cm]{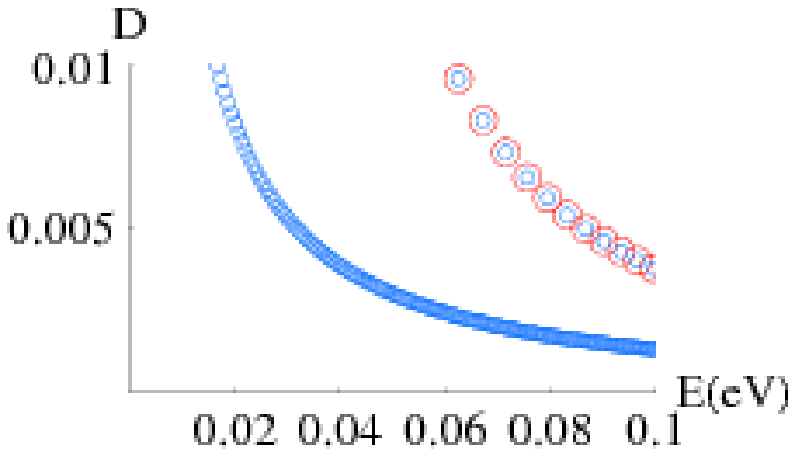}
\includegraphics*[width=4cm]{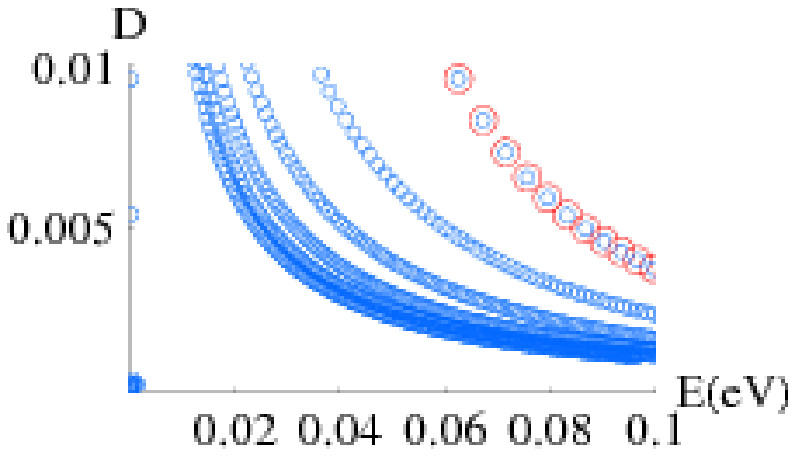}
\includegraphics*[width=4cm]{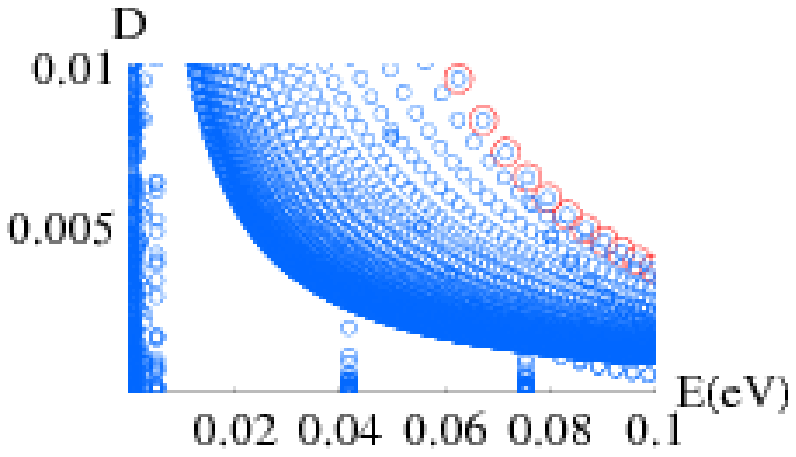}
\caption[fig]{\label{multilayer_LL} (Color online) Relative spectral
strength of the low energy optical transitions between Landau levels in
graphene stacks with Bernal ordering and an odd number of layers. The
applied magnetic field is 1 T. Top left: 3 layers. Top right: 11 layers.
Bottom: 51 layers. The large red circles are the transitions in a single layer.}
\end{center}
\end{figure}

The optical transitions between Landau levels can also be calculated. The
selection rules are the same as for a graphene single layer, and only
transitions between subbands with Landau level indices $M$ and $N$ such that
$| N | = | M \pm 1 |$ are allowed. The resulting transitions, with their
respective spectral strengths, are shown in Fig.~\ref{multilayer_LL}. The
transitions are grouped into subbands, which give rise to a continuum when
the number of layers tends to infinity. In Bernal stacks with an odd number
of layers, the transitions associated to Dirac subbands with linear
dispersion have the largest spectral strength, and they give a significant
contribution to the total absorption even if the number of layers is large,
$N_{\rm L} \lesssim 30-40$ \cite{SMPBH06}.

\subsection{Diamagnetism}
\label{sec_diamag}

Back in 1952 Mrozowski \cite{M1952} studied
diamagnetism of polycrystalline graphite and
other condensed-matter molecular-ring systems.
It was concluded that in such
ring systems diamagnetism has two contributions: (1) diamagnetism from the filled bands of electrons; (2)
Landau diamagnetism of free electrons and holes. For graphite
the second source of diamagnetism is by far the largest of the two.

McClure \cite{M56} computed diamagnetism of a
2D honeycomb lattice using the
theory  introduced by Wallace \cite{W47}, a calculation he
later generalized to three dimensional graphite \cite{M1960}.
For the honeycomb plane the magnetic susceptibility in units
of emu/g is
\begin{equation}
\chi=-\frac{0.0014}T\gamma_0^2\,\mathrm{sech}^2\left(\frac{\mu}{2k_B T}\right)\,,
\end{equation}
where $\mu$ is the Fermi energy, $T$ the temperature, and
$k_B$ the Boltzmann constant.
For graphite the magnetic susceptibility is anisotropic
and the difference between the susceptibility parallel to
the principal axis and that perpendicular to the principal
axis is -21.5$\times 10^{-6}$ emu/g. The susceptibility
perpendicular to the principal axis is about the
free-atom susceptibility of -0.5$\times 10^{-6}$ emu/g.
In the 2D model the susceptibility turns out to
be large due to the presence of fast moving electrons,  with a velocity of the
order of $v_F\simeq 10^6$ m/s, which in turn is a consequence
of the large value of the hopping parameter $\gamma_0$. In fact the susceptibility
turns out to be proportional to the square of $\gamma_0$. Later Sharma {\it et al.}
extended the calculation of McClure for graphite by including the
effect of trigonal warping  and showed that the low temperature diamagnetism
increases \cite{SJM1974}.

Safran and DiSalvo \cite{SD1979},
interested in the susceptibility of graphite intercalation compounds, recalculated, in a  tight-binding model,
the susceptibility perpendicular to a graphite plane using Fukuyama's theory
\cite{FK1971}, which includes interband transitions.
The results agree with those published by McClure \cite{M56}.
Later, Safran computed the susceptibility of a graphene
bilayer showing that this system is diamagnetic at small values of the
Fermi energy, but there appears a paramagnetic peak when the Fermi
energy is of the order of the interlayer coupling \cite{S1984}.

The magnetic susceptibility of other carbon based materials, as carbon
nanotubes and $C_{60}$ molecular solids was measured \cite{HOM1994}
showing a diamagnetic response at finite magnetic
fields different from that of graphite.
 The study of the magnetic response of nanographite ribbons with both
zig-zag and arm-chair edges was done by Wakabayashi {\it et al.}
using a numerical differentiation of the free
energy \cite{WFA+99}. From these two systems, the zig-zag edge state is of
particular interest since the system supports edge states even in the
presence of a magnetic field, leading to very high density of states
near the edge of the ribbon. The high temperature response of these
nanoribbons was found to be diamagnetic whereas the low temperature
susceptibility is paramagnetic.

The Dirac-like nature of the electronic quasiparticles in graphene
led \cite{GGC2007} to consider in general the
problem of the diamagnetism of nodal fermions  and Nakamura
to study the orbital magnetism of Dirac fermions in weak magnetic
fields\cite{N2007}. Koshino and Ando considered the diamagnetism of disordered
graphene in the self consistent Born approximation,
with a disorder potential of the form
$U(\bm r)=\bm 1 u_i\delta(\bm r-\bm R)$ \cite{KA0705}. At the neutrality point
and zero temperature the susceptibility of disordered graphene
is given by
\begin{equation}
\chi(0)=-\frac {g_vg_s}{3\pi^2} \, e^2\gamma_0^2 \, \frac{2W}{\Gamma_0}\,,
\end{equation}
where $g_s=g_v=2$ is the spin and valley degeneracies,
$W$ is a dimensionless parameter for the disorder strength,
defined as $W=n_iu^2_i/4\pi\gamma_0^2$, $n_i$ the impurity concentration, and $\Gamma_0$ is given
by $\Gamma_0=\epsilon_c\exp[-1/(2W)]$ with $\epsilon_c$  a parameter
defining a cut-off function used in the theory. At finite Fermi energy
$\epsilon_F$
and zero temperature the magnetic susceptibility is given by
\begin{equation}
\chi(\epsilon_F)=-\frac {g_vg_s}{3\pi} \, e^2\gamma_0^2 \,
\frac{W}{\vert\epsilon_F\vert}\,.
\end{equation}
In the clean limit the susceptibility is given by
\cite{M56,SD1979,KA0705}:
\begin{equation}
\chi(\epsilon_F)=-\frac {g_vg_s}{6\pi} \, e^2\gamma_0^2 \,
\delta(\epsilon_F)\,.
\label{chifree}
 \end{equation}

\subsection{Spin orbit coupling}

Spin-orbit coupling describes a process
in which an electron changes simultaneously its spin and angular momentum
or, in general, moves from one orbital wavefunction to another. The mixing of
the spin and the orbital motion is a relativistic effect, which can be
derived from Dirac's model of the electron. It is large in
heavy ions, where the average velocity of the electrons is higher. Carbon is
a light atom, and the spin orbit interaction is expected to be weak.

The symmetries of the spin orbit interaction, however, allow the
formation of a gap at the Dirac points in clean graphene. The spin orbit
interaction leads to a spin dependent shift of the orbitals, which is of
a different sign for the two sublattices, acting as an effective mass within
each Dirac point \cite{DD65,KM05,WC07}. The appearance of this gap leads to a non
trivial spin Hall conductance, in similar way to other models which study the
parity anomaly in relativistic field theory in (2+1)
dimensions \cite{H88}. When the inversion symmetry of the honeycomb
lattice is broken, either because the graphene layer is curved or because an
external electric field is applied (Rashba interaction) additional terms,
which modulate the nearest neighbor hopping, are induced \cite{A00}.
The intrinsic and extrinsic spin orbit interactions can be written as \cite{DD65,KM05}:
\begin{eqnarray}
{\cal H}_{SO;int} \!\!&\equiv &\!\! \Delta_{so}\int d^2 {\bf r} \hat{\Psi}^\dag ( {\bf r} )
\hat{s}_z \hat{\sigma}_z \hat{\tau}_z \hat{\Psi} ( {\bf r} ) \, ,
\nonumber \\
{\cal H}_{SO;ext} \!\!&\equiv &\!\! \lambda_R \int \!\!d^2 {\bf r} \hat{\Psi}^\dag (
 {\bf r} ) (
-\hat{s}_x \hat{\sigma}_y \!\!+\!\! \hat{s}_y \hat{\sigma}_x \hat{\tau}_z  )
 \hat{\Psi} ( {\bf r} ) \, ,
\end{eqnarray}
where $\hat{\sigma}$ and $\hat{\tau}$ are Pauli matrices which describe the
sublattice and valley degrees of freedom, and $\hat{s}$ are Pauli matrices
acting on actual spin space. $\Delta_{so}$ is the spin-orbit coupling and
$\lambda_R$ is the Rashba coupling.

The magnitude of the spin orbit coupling in graphene can be inferred from the known spin
orbit coupling in the carbon atom. This coupling allows for transitions
between the $p_z$ and $p_x$ and $p_y$ orbitals. An electric field induces
also transitions between the $p_z$ and $s$ orbitals. These intra-atomic
processes mix the $\pi$ and $\sigma$ bands in graphene. Using second order
perturbation theory, one obtains a coupling between the low energy states in
the $\pi$ band. Tight-binding \cite{HGB06,ZS07} and band structure
calculations \cite{MHSSKM06,YYQZF07} give estimates for the intrinsic and
extrinsic spin-orbit interactions in the range $0.01-0.2$ K, and hence
much smaller than the other energy scales in the problem (kinetic, interaction,
and disorder).

\section{Flexural phonons, elasticity, and crumpling}
\label{flex}

\begin{figure}[tbh]
(a)
\includegraphics[width=8cm, keepaspectratio]{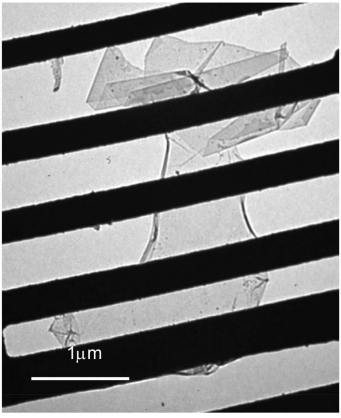}
(b)
\includegraphics[width=8cm, keepaspectratio]{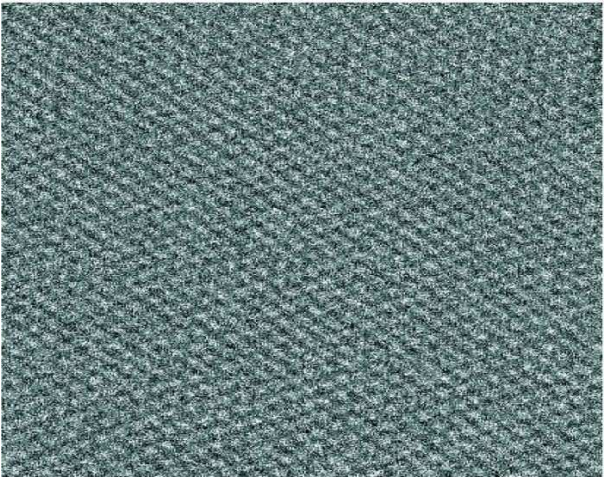}
\caption{(Color online) Suspended graphene sheet.
(a) Bright-field transmission-electron-microscope
image of a graphene membrane. Its central part (homogeneous
and featureless region) is monolayer
graphene. Adapted from \cite{meyer07}.
(b) Despite being only one atom thick, graphene remains a
perfect crystal as this atomic resolution image shows.
The image is obtained in a scanning transmission
electron microscope. The visible periodicity
is given by the lattice of benzene rings.
Adapted from \cite{Nairal07}.}
\label{suspend}
\end{figure}

Graphite, in the Bernal stacking configuration, is a layered crystalline solid
with 4 atoms per unit cell. Its basic structure is essentially a repetition of
the bilayer structure discussed earlier. The coupling between the layers, as
we discussed, is weak and, therefore, graphene has been always the starting point
for the discussion of phonons in graphite \cite{rubio}. Graphene has two atoms
per unit cell and if we consider graphene as strictly 2D it should have 2
acoustic modes (with dispersion $\omega_{{\rm ac}}(k) \propto k$ as $k \to 0$)
and 2 optical modes (with dispersion $\omega_{{\rm op}}(k) \propto {\rm constant}$,
as $k \to 0$) solely due to the in-plane translation and stretching of the graphene
lattice. Nevertheless, graphene exists in the 3D space and hence the atoms can
oscillate out-of-plane leading to 2 out-of-plane phonons (one acoustic and another optical)
called flexural modes. The acoustic flexural mode has dispersion
$\omega_{{\rm flex}}(k) \propto k^2$ as $k \to 0$ which represents the translation of
the whole graphene plane (essentially a one atom thick membrane)
in the perpendicular direction (free particle motion). The optical flexural
mode represents the out-of-phase out-of-plane oscillation of the neighboring atoms.
In first approximation,
if we neglect the coupling between graphene planes, graphite has essentially
the same phonon modes, albeit they are degenerate. The coupling between planes
has two main effects: (1) it lifts the degeneracy of the phonon modes, and (2) leads
to a strong suppression of the energy of the flexural modes. Theoretically, flexural
modes should become ordinary acoustic and optical modes in a fully covalent 3D solid,
but in practice, the flexural modes survive due to the fact the planes are coupled
by weak van der Waals-like forces. These modes have been measured experimentally in graphite
\cite{rubio}. Graphene can also be obtained as a suspended membrane, that is,
without a substrate, being supported only by a scaffold or bridging micron-scale gaps
\cite{bunch07,meyer07,MGKNORGZ07}. Figure \ref{suspend} shows a suspended graphene
sheet and an atomic resolution image of its crystal lattice.

Because the flexural modes disperse like $k^2$ they dominate the behavior of
structural fluctuations in graphene
at low energies (low temperatures) and long wavelengths. It is instructive to
understand how these modes appear from the point of view of elasticity theory \cite{nelson,chaikin}.
Consider, for instance, a graphene sheet in 3D and let us parameterize the
position of the sheet relative of a fixed coordinate frame
in terms of the in-plane vector ${\bf r}$ and the height variable $h({\bf r})$ so that
a position in the graphene is given by the vector ${\bf R}=({\bf r},h({\bf r}))$. The unit
vector normal to the surface is given by:
\begin{eqnarray}
{\bf N} = \frac{{\bf z}-\nabla h}{\sqrt{1+(\nabla h)^2}} \, ,
\label{normal}
\end{eqnarray}
where $\nabla = (\partial_x,\partial_y)$ is the 2D gradient operator, and ${\bf z}$ is the
unit vector in the third direction. In a flat graphene configuration all the normal vectors
are aligned and therefore $\nabla \cdot {\bf N} =0$. Deviations from the flat configuration
requires misalignment of the normal vectors and costs elastic energy. This elastic energy
can be written as:
\begin{eqnarray}
E_0 = \frac{\kappa}{2} \int d^2{\bf r} \left(\nabla \cdot {\bf N}\right)^2
\approx \frac{\kappa}{2} \int d^2{\bf r} \left(\nabla^2 h \right)^2
\label{e0}
\end{eqnarray}
where $\kappa$ is the bending stiffness of graphene, and the expression in terms
of $h({\bf r})$  is valid for
smooth distortions of the graphene sheet ($(\nabla h)^2 \ll 1$). The energy (\ref{e0}) is
valid in the absence of a surface tension or a substrate which break the rotational
and translational symmetry of the problem, respectively. In the presence of tension
there is an energy cost for solid rotations of the graphene sheet ($\nabla h \neq 0$)
and hence a new term has to be added to the elastic energy:
\begin{eqnarray}
E_{{\rm T}} = \frac{\gamma}{2} \int d^2 {\bf r} \left(\nabla h\right)^2 \, ,
\label{et}
\end{eqnarray}
where $\gamma$ is the interfacial stiffness. A substrate, described by a height
variable $s({\bf r})$, pins the graphene sheet through van der Waals and other
electrostatic potentials so that the graphene configuration tries to follow
the substrate $h({\bf r}) \sim s({\bf r})$. Deviations from this configuration
cost extra elastic energy that can be approximated by a harmonic potential \cite{swain99}:
\begin{eqnarray}
E_{{\rm S}} = \frac{v}{2} \int d^2 {\bf r} \left(s({\bf r})-h({\bf r})\right)^2 \, ,
\label{es}
\end{eqnarray}
where $v$ characterizes the strength of the interaction potential between substrate and graphene.

Firstly, let us consider the free floating graphene problem (\ref{e0}) that we can rewrite in
momentum space as:
\begin{eqnarray}
E_0 = \frac{\kappa}{2} \sum_{{\bf k}}  k^4 h_{-{\bf k}} h_{{\bf k}} \, .
\label{e0f}
\end{eqnarray}
We now canonically quantize the problem by introducing a momentum operator $P_{{\bf k}}$
that has the following commutator with $h_{{\bf k}}$:
\begin{eqnarray}
\left[h_{{\bf k}},P_{{\bf k'}}\right]= i \delta_{{\bf k},{\bf k'}} \, ,
\end{eqnarray}
and write the Hamiltonian as:
\begin{eqnarray}
H = \sum_{{\bf k}} \left\{ \frac{P_{-{\bf k}} P_{{\bf k}}}{2 \sigma} + \frac{\kappa k^4}{2}
h_{-{\bf k}} h_{{\bf k}}\right\} \, ,
\label{hph}
\end{eqnarray}
where $\sigma$ is graphene's 2D mass density. From the Heisenberg equations of
motion for the operators it is trivial to find that $h_{{\bf k}}$ oscillates harmonically
with a frequency given by:
\begin{eqnarray}
\omega_{{\rm flex}}({\bf k}) = \left(\frac{\kappa}{\sigma}\right)^{1/2} k^2 \, ,
\label{omegaflex}
\end{eqnarray}
which is the long wavelength dispersion of the flexural modes. In the presence of tension
it is easy to see that the dispersion is modified to:
\begin{eqnarray}
\omega({\bf k}) = k \, \sqrt{\frac{\kappa}{\sigma} k^2 + \frac{\gamma}{\sigma}}  \, ,
\label{omegat}
\end{eqnarray}
indicating that the dispersion of the flexural modes becomes linear in $k$,
as $k \to 0$, under tension.
That is what happens in graphite where the interaction between layers breaks the rotational
symmetry of the graphene layers.

Eq. (\ref{omegaflex}) also allows us to relate the bending
energy of graphene with the Young modulus, $Y$, of graphite.
The fundamental resonance frequency of a macroscopic graphite sample
of thickness $t$ is given by \cite{bunch07}:
\begin{eqnarray}
\nu({\bf k}) = \left(\frac{Y}{\rho}\right)^{1/2} \, t \, k^2 \, ,
\label{omgra}
\end{eqnarray}
where $\rho = \sigma/t$ is the 3D mass density. Assuming that (\ref{omgra}) works down
to the single plane level, that is, when $t$ is the distance between planes, we find:
\begin{eqnarray}
\kappa = Y t^3 \, ,
\label{kappay}
\end{eqnarray}
which provides a simple relationship between the bending stiffness and the Young modulus.
Given that $Y \approx 10^{12}$ N/m and $t \approx 3.4$ \AA \, we find, $\kappa \approx 1$ eV.
This result is in good agreement with {\it ab initio} calculations of the bending rigidity
\cite{lenosky92,tu02} and experiments in graphene resonators \cite{bunch07}.

The problem of structural order of a ``free floating'' graphene sheet can be fully understood from the
existence of the flexural modes. Consider, for instance, the number of flexural modes
per unit of area at a certain temperature $T$:
\begin{eqnarray}
N_{{\rm ph}} = \int \frac{d^2{\bf k}}{(2 \pi)^2} n_{{\bf k}} =
\frac{1}{2 \pi} \int_0^{\infty}dk \frac{k}{e^{\beta \sqrt{\kappa/\sigma} k^2}-1}
\label{nph}
\end{eqnarray}
where $n_{{\bf k}}$ is the Bose-Einstein occupation number ($\beta = 1/(k_B T)$).
For $T \neq 0$ the above integral is logarithmically divergent
in the infrared ($k \to 0$) indicating a divergent number of phonons in the
thermodynamic limit. For a system with finite size $L$ the smallest possible
wave vector is of the order of $k_{{\rm min}} \sim 2\pi/L$. Using $k_{{\rm min}}$ as a
lower cut-off in the integral (\ref{nph}) we find:
\begin{eqnarray}
N_{{\rm ph}} = \frac{\pi}{L^2_{{\rm T}}}
\ln\left(\frac{1}{1-e^{-L^2_{{\rm T}}/L^2}}\right) \, ,
\label{nphf}
\end{eqnarray}
where
\begin{eqnarray}
L_{{\rm T}} = \frac{2 \pi}{\sqrt{k_B T}} \left(\frac{\kappa}{\sigma}\right)^{1/4} \, ,
\end{eqnarray}
is the thermal wavelength of the flexural modes. Notice that  that when
$L \gg L_{{\rm T}}$ the number of flexural
phonons in (\ref{nphf}) diverges logarithmically with the size of the system:
\begin{eqnarray}
N_{{\rm ph}} \approx \frac{2 \pi}{L^2_{{\rm T}}}
\ln\left(\frac{L}{L_{{\rm T}}}\right) \, ,
\end{eqnarray}
indicating that the system cannot be structurally ordered at any finite temperature.
This is nothing but the crumpling instability of soft membranes \cite{nelson,chaikin}.
For $L \ll L_{{\rm T}}$ one finds that $N_{{\rm ph}}$ goes
to zero exponentially with the size of the system indicating that systems with finite
size can be flat at sufficiently low temperatures. Notice that for $\kappa \approx 1$ eV,
$\rho \approx 2200$ kg/m$^3$, $t=3.4$ \AA \,($\sigma \approx 7.5 \times 10^{-7}$ kg/m$^2$),
and $T \approx 300$ K, we find $L_{{\rm T}} \approx 1$ \AA \, indicating that free floating
graphene should always crumple at room temperature due to thermal fluctuations associated
with flexural phonons. Nevertheless, the previous discussion only involves the harmonic
(quadratic part) of the problem. Non-linear effects such as large
bending deformations \cite{PL85},
the coupling between flexural and in-plane modes (or phonon-phonon interactions \cite{BLMM07,RlD92})
and the presence of topological defects \cite{NP87} can lead to strong renormalizations
of the bending rigidity, driving the system toward a flat phase at low temperatures \cite{chaikin}.
This situation has been confirmed in numerical simulations of free graphene sheets \cite{FLK07,ANASNT07}.

 The situation is rather different if the system is under tension
or in the presence of a substrate or scaffold that can hold the graphene
sheet. In fact, static rippling of graphene flakes suspended on scaffolds
have been observed for single layer as well as bilayers \cite{meyer07,MGKNORGZ07}.
In this case
the dispersion, in accordance with (\ref{omegat}), is at least linear in $k$, and the integral
in (\ref{nph}) converges in the infrared ($k \to 0$) indicating that the number of flexural
phonons is finite and graphene does not crumple. We should notice that these thermal fluctuations
are dynamic and hence average to zero over time, therefore, the graphene sheet is expected to be flat under
these circumstances. Obviously, in the presence of a substrate or scaffold described by (\ref{es})
static deformations of the graphene sheet are allowed. Also, hydrocarbon
molecules that are often present on top of free hanging graphene membranes could quench
flexural fluctuations making them static.

Finally, one should notice that in the presence of a metallic gate the
electron-electron interactions lead to the coupling of the phonon modes
to the electronic excitations in the gate. This coupling could be partially responsible
to the damping of the phonon modes due to dissipative effects \cite{SGCN07}
as observed in graphene resonators \cite{bunch07}.

\section{Disorder in graphene}
\label{disorder}

Graphene is a remarkable material from the electronic point of view.
Because of the robustness and specificity of the sigma bonding, it
is very hard for alien atoms to replace the carbon atoms in the
honeycomb lattice. This is one of the reasons why the electron
mean free path in graphene can be so long, reaching up to one micrometer
in the existing samples. Nevertheless, graphene is not immune to disorder and
its electronic properties are controlled by extrinsic as well as
intrinsic effects that are unique to this system.
Among the intrinsic sources of disorder we can highlight: surface ripples
and topological defects. Extrinsic disorder can come about in many
different forms: adatoms, vacancies, charges on top of graphene or in
the substrate, and extended defects such as cracks and edges.

It is easy to see that from the point of view of single electron
physics (that is, terms that can be added to (\ref{H1})),
there are two main terms that disorder couples to. The
first one is a local change in the single site energy,
\begin{eqnarray}
H_{{\rm dd}} = \sum_{i} V_i \left(a^{\dag}_i a_i +
b^{\dag}_i b_i\right) \, ,
\label{diagdis}
\end{eqnarray}
where $V_i$ is the strength of the disorder potential on site ${\bf R}_i$,
which is diagonal in the sublattice indices and hence, from the point
of view of the Dirac Hamiltonian (\ref{H2}), can be written as:
\begin{eqnarray}
H_{{\rm dd}} = \int d^2 r \sum_{a=1,2} V_a({\bf r})
\hat \Psi^{\dag}_a({\bf r}) \hat \Psi_a({\bf r}) \, ,
\label{diagdisdir}
\end{eqnarray}
which acts as a chemical potential shift for the Dirac fermions, that is,
shifts locally the Dirac point.

Because of the vanishing of the density of states in single layer graphene,
and by consequence the lack of electrostatic screening, charge
potentials may be rather important in determining the spectroscopic and
transport properties \cite{A06,NMac07,AHGS07}.
Of particular importance is the Coulomb impurity
problem where,
\begin{eqnarray}
V_a(r) = \frac{e^2}{\epsilon_0} \frac{1}{r} \, ,
\end{eqnarray}
where $\epsilon_0$ is the dielectric constant of the medium.
The solution of the Dirac equation for the Coulomb potential in 2D can
be studied analytically  \cite{mele,SKL07,PNCN07,BSS07,Nov07}. Its solution has many
of the features of the 3D relativistic hydrogen atom problem \cite{Baym}.
Just as in the case of the 3D problem the nature of the eigenfunctions
depends strongly on graphene's dimensionless coupling constant:
\begin{eqnarray}
g = \frac{Z e^2}{\epsilon_0 v_F} \, .
\label{gcou}
\end{eqnarray}
Notice, therefore, that the coupling constant can be varied by
either changing the charge of the impurity, $Z$, or modifying
the dielectric environment and changing $\epsilon_0$.
For $g<g_c=1/2$ the solutions of this problem are given in terms of
Coulomb wavefunctions with logarithmic phase shifts. The local density
of states (LDOS) is affected close to the impurity due the electron-hole
asymmetry generated by the Coulomb potential. The local charge density
decays like $1/r^3$ plus fast oscillations of the order of the lattice
spacing (in the continuum limit this would give rise to a Dirac delta
function for the density \cite{KSBC06}).
Just like in 3D QED, the 2D problem becomes unstable for $g > g_c = 1/2$
leading to super-critical behavior and the so-called {\it fall of electron
to the center} \cite{LL81}. In this case the LDOS is strongly affected
by the presence of the Coulomb impurity with the appearance of bound states
outside the band and scattering resonances within the band \cite{PNCN07}
and the local electronic density decays monotonically like $1/r^2$ at large distances.

It has been argued \cite{Setal07} that without high vacuum environment
these Coulomb effects can be strongly suppressed by large effective dielectric
constants due to the presence of a nanometer thin layer of absorbed water \cite{SSFGCNS07}.
In fact, experiments in ultra-high vacuum conditions \cite{CJFWI07} display strong scattering
features in the transport that can be associated to charge impurities. Screening
effects that affect the strength and range of the Coulomb interaction,
are rather non-trivial in graphene \cite{Shk07,FNS07} and, therefore, important
for the interpretation of transport data \cite{BTBB07,NKR07,SPG07,LMCN07}.

Another type of disorder is the one that
changes the distance or angles between the $p_z$ orbitals. In this case,
the hopping energies between different sites are modified leading to
a new term to the original Hamiltonian (\ref{H1}):
\begin{eqnarray}
H_{{\rm od}} &=& \sum_{i,j}
\left\{
\delta t^{(ab)}_{ij} \left(a^{\dag}_i b_j +
{\rm h.c.} \right)
\right.
\nonumber
\\
&+& \left.
\delta t^{(aa)}_{ij} \left(a^{\dag}_i a_j +
b^{\dag}_i b_j \right)
\right\} \, ,
\label{odd}
\end{eqnarray}
or in Fourier space:
\begin{eqnarray}
H_{{\rm od}} &=& \sum_{{\bf k},{\bf k'}}
a^{\dag}_{{\bf k}} b_{{\bf k'}}
\sum_{i,\vec{\delta}_{ab}} \delta t^{(ab)}_{i} e^{i({\bf k}-{\bf k'}) \cdot
  {\bf R}_i - i \vec{\delta}_{aa} \cdot {\bf k'}}
+ {\rm h.c.}
\nonumber
\\
\!\!\!&+&\!\!\!
\left(\!a^{\dag}_{{\bf k}} a_{{\bf k'}} \!\!+\!\! b^{\dag}_{{\bf k}} b_{{\bf k'}}\!\right)
\sum_{i,\vec{\delta}_{aa}} \delta t^{(aa)}_{i} e^{i({\bf k}-{\bf k'}) \cdot
  {\bf R}_i - i \vec{\delta}_{ab} \cdot {\bf k'}}   ,
\label{offdiagdis}
\end{eqnarray}
where $\delta t^{(ab)}_{ij}$ ($\delta t^{(aa)}_{ij}$)
is the change of the hopping energy between orbitals on lattice sites
${\bf R}_i$ and ${\bf R}_j$ on the same (different)
sublattices (we have written ${\bf R}_j = {\bf R}_i + \vec{\delta}$ where
$\vec{\delta}_{ab}$ is the nearest neighbor vector, and $\vec{\delta}_{aa}$
is the next nearest neighbor vector).  Following the procedure of
Sec.~\ref{sec:continuous} we
project out the Fourier components of the operators close to the K and K'
points of the BZ using (\ref{projection}). If we assume that
$\delta t_{ij}$ is smooth over the lattice spacing scale, that is,
it does not have an Fourier component with momentum ${\bf K}-{\bf K'}$
(so the two Dirac cones are not coupled by disorder),
we can rewrite (\ref{offdiagdis}) in real space as:
\begin{eqnarray}
H_{{\rm od}} &=& \int d^2 r \left[{\cal A}({\bf r})
a^{\dag}_1({\bf r}) b_1({\bf r}) + {\rm h.c.}  \right.
\nonumber
\\
&+& \left. \phi({\bf r}) \left(a^{\dag}_1({\bf r}) a_1({\bf r}) + b^{\dag}_1({\bf r})
  b_1({\bf r}) \right) \right] \, ,
\label{offdiacon}
\end{eqnarray}
with a similar expression for the cone $2$ but with ${\cal A}$ replaced
by ${\cal A}^*$, where,
\begin{eqnarray}
{\cal A}({\bf r}) &=& \sum_{\vec{\delta}_{ab}} \delta t^{(ab)}({\bf r})
e^{-i \vec{\delta}_{ab} \cdot {\bf K}} \, ,
\label{defa}
\\
\phi({\bf r}) &=& \sum_{\vec{\delta}_{aa}} \delta t^{(aa)}({\bf r})
e^{-i \vec{\delta}_{aa} \cdot {\bf K}} \, .
\label{defphi}
\end{eqnarray}
Notice that whereas $\phi({\bf r}) = \phi^*({\bf r})$, because of the inversion
symmetry of the two triangular sublattices that make up the honeycomb
lattice, ${\cal A}$ is complex because of lack of inversion symmetry for
nearest neighbor hopping. Hence,
\begin{eqnarray}
{\cal A}({\bf r}) = {\cal A}_x({\bf r}) + i {\cal A}_y({\bf r}) \, .
\end{eqnarray}
In terms of the Dirac Hamiltonian (\ref{H2}) we can rewrite (\ref{offdiacon})
as:
\begin{eqnarray}
H_{{\rm od}} &=& \int d^2 r
\left[
\hat \Psi^{\dag}_1({\bf r}) \bm \sigma \cdot  \vec{{\cal A}}({\bf r}) \hat
\Psi_1 ({\bf r})
\right.
\nonumber
\\
&+& \left.
\phi({\bf r}) \hat \Psi^{\dag}_1({\bf r}) \hat \Psi_1({\bf r})
\right] \, ,
\label{hodir}
\end{eqnarray}
where $\vec{{\cal A}} = ({\cal A}_x,{\cal A}_y)$.
This result shows that changes in the hopping amplitude lead to the
appearance of vector, $\vec{{\cal A}}$, and scalar, $\Phi$,
potentials in the Dirac Hamiltonian. The presence of a vector potential in the
problem indicates that an effective magnetic field $\vec{B} = c/(e v_F) \nabla \times
\vec{{\cal A}}$ should also be present, naively implying a broken time reversal
symmetry, although the original problem was time reversal invariant.
This broken time reversal symmetry is not real since (\ref{hodir})
is the Hamiltonian around only one of the Dirac cones. The second Dirac
cone is related to the first by time reversal symmetry indicating that the
effective magnetic field is reversed in the second cone. Therefore, there is no
global broken symmetry but a compensation between the two cones.

\subsection{Ripples}
\label{ripples}

Graphene is a one atom thick system, the extreme case of a soft
membrane. Hence, just like soft membranes,
it is subject to distortions of its structure either due to
thermal fluctuations (as we discussed in Sec.~\ref{flex})
or interaction with a
substrate, scaffold, and absorbands \cite{swain99}. In the first case the fluctuations are time
dependent (although with time scales much longer than the electronic
ones), while in the second case the distortions act as quenched disorder.
In both cases, the disorder comes about because of the modification of
the distance and relative angle between the carbon atoms due to the
bending of the graphene sheet. This type of off-diagonal disorder does
not exist in ordinary 3D solids, or even in
quasi-1D or quasi-2D systems, where atomic chains and
atomic planes, respectively, are embedded in a 3D
crystalline structure. In fact, graphene is also very different from
other soft membranes because it is (semi) metallic, while previously studied
membranes were insulators.

\begin{figure}[tbh]
\centerline{\includegraphics[width=5cm, keepaspectratio]{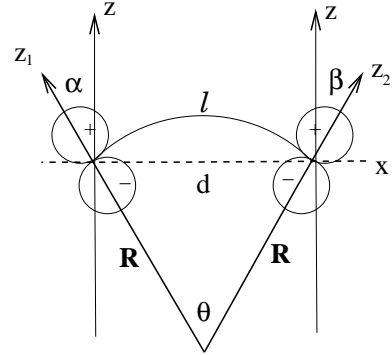}}
\caption{Bending of the surface of graphene by a radius $R$ and its
effect on the $p_z$ orbitals.}
\label{bending}
\end{figure}

The problem of the bending of graphitic systems and its effect on
the hybridization of the $\pi$ orbitals has been studied a great
deal in the context of classical minimal surfaces \cite{lenosky92}
and applied to fullerenes and carbon nanotubes
\cite{tersoff92,zhou97,zhou00,tu02,KM97}.
In order to understand the effect of bending on
graphene, consider the situation shown in Fig.\ref{bending}.
The bending of the graphene sheet has three main effects:
the decrease of the distance between carbon atoms, a rotation
of the $p_Z$ orbitals (compression or dilation of the lattice are
energetically costly due to the large spring constant of graphene
$\approx 57$ eV/\AA$^2$ \cite{zhou00}), and a re-hybridization
between $\pi$ and $\sigma$ orbitals \cite{NK07}.
Bending by a radius $R$ decreases
the distance between the orbitals from $\ell$ to $d = 2 R \sin[\ell/(2 R)]
\approx \ell - \ell^3/(24 R^2)$ for $R \gg \ell$. The decrease in the distance between
the orbitals increases the overlap between the two lobes of the $p_Z$
orbital \cite{harrison}:
$V_{ppa} \approx V_{ppa}^0 [1+\ell^2/(12 R^2)]$, where $a=\pi,\sigma$,
and $V_{ppa}^0$ is the overlap for a flat graphene sheet.
The rotation of the $p_Z$ orbitals can be understood within
the Slater-Koster formalism, namely, the rotation can be decomposed
into a $p_z-p_z$ ($\pi$ bond) plus a $p_x-p_x$ ($\sigma$ bond) hybridization
with energies $V_{pp\pi}$ and $V_{pp\sigma}$, respectively \cite{harrison}:
$V(\theta) = V_{pp\pi} \cos^2(\theta)-V_{pp\sigma} \sin^2(\theta)
\approx V_{pp\pi} - (V_{pp\pi}+V_{pp\sigma}) (\ell/(2 R))^2$, leading
to a decrease in the overlap. Furthermore, the rotation leads to re-hybridization
between $\pi$ and $\sigma$ orbitals leading to a further shift in energy
of the order of \cite{NK07}: $\delta \epsilon_{\pi} \approx (V^2_{sp\sigma}+V^2_{pp\sigma})/(\epsilon_{\pi}-\epsilon_a)$.

In the presence of a substrate, as we discussed in Sec.\ref{flex}, elasticity
theory predicts that graphene can be expected to follow the substrate in a smooth way.
Indeed, by minimizing the elastic energy (\ref{e0}), (\ref{et}), and (\ref{es})
with respect to the height $h$ we get \cite{swain99}:
\begin{eqnarray}
\kappa \nabla^4 h({\bf r}) - \gamma \nabla^2 h({\bf r}) + v h({\bf r}) = v s({\bf r}) \, ,
\end{eqnarray}
that can be solved by Fourier transform:
\begin{eqnarray}
h({\bf k}) = \frac{s({\bf k})}{1 + (\ell_t k)^2 + (\ell_c k)^4} \, ,
\label{hsk}
\end{eqnarray}
where
\begin{eqnarray}
\ell_t &=& \left(\frac{\gamma}{v}\right)^{1/2} \, ,
\nonumber
\\
\ell_c &=& \left(\frac{\kappa}{v}\right)^{1/4} \, .
\end{eqnarray}
Eq.~(\ref{hsk}) gives the height configuration in terms of the substrate
profile, and $\ell_t$ and $\ell_c$ provide the length scales for elastic
distortion of graphene on a substrate. Hence, disorder in the substrate
translates into disorder in the graphene sheet (albeit restricted by
elastic constraints). This picture has been confirmed by STM measurements
on graphene \cite{SRRMKBHHF07,Ishal07} in which strong correlations were found
between the roughness of the substrate and the graphene topography.
{\it Ab initio} band structure calculations also give support to this
scenario \cite{DW0703}.

The connection between the ripples and the electronic problem comes from the
relation between the height field $h({\bf r})$ and the local curvature of
the graphene sheet, $R$:
\begin{eqnarray}
\frac{2}{R({\bf r})} \approx \nabla^2 h({\bf r}) \, ,
\end{eqnarray}
and, hence we see that due to bending
the electrons are subject to a potential which depends on the structure
of a graphene sheet \cite{NK07}:
\begin{eqnarray}
V({\bf r}) \approx V^0 - \alpha \, (\nabla^2 h({\bf r}))^2 \, ,
\label{vripple}
\end{eqnarray}
where $\alpha$ ($\alpha \approx 10$ eV \AA$^2$) is the constant that depends on microscopic details.
The conclusion from (\ref{vripple}) is that the Dirac fermions
are scattered by ripples of the graphene sheet through a potential
which is proportional to the square of the local curvature. The coupling
between geometry and electron propagation is unique to graphene, and results
in additional scattering and resistivity due to ripples \cite{KG07}.

\subsection{Topological lattice defects}
\label{topo}

Structural defects of the honeycomb lattice like pentagons, heptagons and
their combinations such as Stone-Wales defect (a combination of two pentagon-heptagon
pairs) are also possible in graphene and can lead to scattering \cite{CV07,CV07b}.
These defects induce long range deformations, which modify the
electron trajectories.

Let us consider first a
disclination. This defect is equivalent to the deletion or inclusion
of a wedge in the lattice. The simplest one in the honeycomb lattice
is the absence of a $60^{\circ}$ wedge. The resulting edges can be glued
in such a way that all sites remain three-fold coordinated. The
honeycomb lattice is recovered everywhere, except at the apex of the
wedge, where a fivefold ring, a pentagon, is formed. One can imagine
a situation where the nearest neighbor hoppings are unchanged.
Nevertheless, the existence of a pentagon implies that the two
sublattices of the honeycomb structure can no longer be defined. A
trajectory around the pentagon after a closed circuit has to change
the sublattice index.

\begin{figure}[]
\begin{center}
\includegraphics[width=5cm]{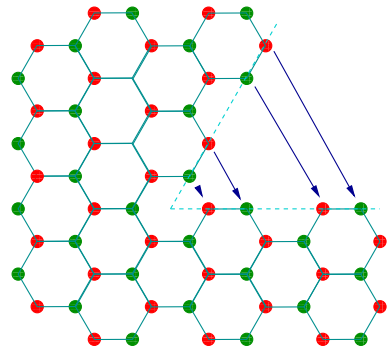}
\caption[fig]{\label{disclination} (Color online) Sketch of the boundary conditions
  associated to a disclination (pentagon) in the honeycomb lattice.}
\end{center}
\end{figure}

The boundary conditions imposed at the edges of a disclination are
sketched in Fig.~\ref{disclination}, showing the identification of
sites from different sublattices. In addition, the wavefunctions at
the $K$ and $K'$ points are exchanged when moving around the
pentagon.

Far away from the defect, a slow rotation of the components of the
spinorial wavefunction can be described by a gauge field which acts
on the valley and sublattice indices \cite{GGV92,GGV93b}. This gauge
field is technically non-abelian, although a transformation can be
defined which makes the resulting Dirac equation equivalent to one
with an effective abelian gauge field \cite{GGV93b}. The final
continuum equation gives a reasonable description of the electronic
spectrum of fullerenes of different sizes \cite{GGV92,GGV93b}, and
other structures which contain
pentagons \cite{LC00,OKP03,KO04,LC04,KO06}. It is easy to see that an
heptagon leads to the opposite effective field.

An in-plane dislocation, that is, the inclusion of a semi-infinite
row of sites, can be considered as induced by a pentagon and a
heptagon together. The non-abelian field described above is
canceled away from the core. A closed path implies a shift by one
(or more) lattice spacings. The wavefunctions at the $K$ and $K'$
points acquire phases, $e^{\pm 2 \pi i / 3}$, under a translation by
one lattice unit. Hence, the description of a dislocation in the
continuum limit requires an (abelian) vortex of charge $\pm 2 \pi /
3$ at its core. Dislocations separated over distances of the order of $d$
lead to an effective flux through an area of perimeter $l$ of
the order of \cite{MG06}:
\begin{equation}
\Phi \sim \frac{d}{\kf l^2}
\end{equation}
where $\kf$ is the Fermi vector of the electrons.

 In general, a
local rotation of the axes of the honeycomb lattice induces changes
in the hopping which lead to mixing of the $K$ and $K'$
wavefunctions, leading to a gauge field like the one induced by a
pentagon \cite{GGV01}. Graphene samples with disclinations and
dislocations are feasible in particular substrates \cite{CNBM08}, and gauge fields related to
the local curvature are then expected to play a crucial role in
such structures. The resulting electronic
structure can be analyzed using the theory of quantum mechanics in
curved space \cite{BD82,CV07,CV07b,JCV07}.

\subsection{Impurity states}
\label{impurity}

Point defects, similar to impurities and vacancies, can nucleate electronic
states in their vicinity. Hence, a concentration of $n_i$ impurities
per carbon atom leads to a change in the electronic density of the order of
$n_i$. The corresponding shift in the Fermi energy is $\epsilon_{\rm F}
\simeq v_{\rm F} \sqrt{n_{i}}$. In addition, impurities lead to a
finite elastic mean free path, $l_{\rm elas} \simeq a n_{i}^{-1/2}$, and
to an elastic scattering time $\tau_{\rm elas} \simeq ( v_{\rm F} n_i
)^{-1}$. Hence, the regions with few impurities can be considered
low-density metals in the dirty limit, as $\tau_{\rm elas}^{-1} \simeq \epsilon_{\rm F}$.

The Dirac equation allows for localized solutions that satisfy many possible
boundary conditions. It is known that small circular defects result in
localized and semi-localized states \cite{DHM98}, that is, states whose
wavefunction decays as $1/r$ as a function of the distance from the center of the
defect. A discrete version of these states can be realized in a nearest neighbor
tight-binding model with unitary scatterers such as vacancies \cite{Petal06}.
In the continuum, the
Dirac equation (\ref{diraceq}) for the wavefunction, $\psi({\bf r}) = (\phi_1({\bf r}),
\phi_2({\bf r}))$, can be written as:
\begin{eqnarray}
\partial_w \phi_1(w,w^*) &=& 0 \, ,
\nonumber
\\
\partial_{w^*} \phi_2(w,w^*) &=& 0 \, ,
\end{eqnarray}
where $w = x + i y$ is a complex number. These equations indicate that
at zero energy the components of the wavefunction can only be either
holomorphic or anti-holomorphic with respect to the variable $w$ (that is,
$\phi_1(w,w^*) = \phi_1(w^*)$ and $\phi_2(w,w^*) = \phi_2(w)$). Since the
boundary conditions require that the wavefunction vanishes at infinity
the only possible solutions have the form:
$\Psi_K ( {\bf \vec{r}} ) \propto ( 1 / ( x + i y )^n ,0 )$ or
$\Psi_{K'} ( {\bf \vec{r}} ) \propto ( 0, 1 / ( x - i y )^n )$. The
wavefunctions in the discrete lattice must be real, and at large distances
the actual solution
found near a vacancy tends to a superposition of two
solutions formed from wavefunctions from the two valleys with equal weight,
in a way similar to the mixing at armchair edges \cite{Brey206}.

The construction of the semi-localized state around a vacancy in the honeycomb
lattice can be extended to other discrete models which leads to the Dirac
equation in the continuum limit. A particular case is the nearest neighbor
square lattice with half flux per plaquette, or the nearest neighbor square
lattice with two flavors per site. The latter has been extensively studied in
relation to the effects of impurities on the electronic structure of d-wave
superconductors \cite{BVZ06}, and numerical results are in agreement with the
existence of this solution. As the state is localized on one sublattice
only, the solution can be generalized for the case of two vacancies.

\subsection{Localized states near edges, cracks, and voids}
\label{crack}

\begin{figure}[]
\begin{center}
\includegraphics[width=5cm]{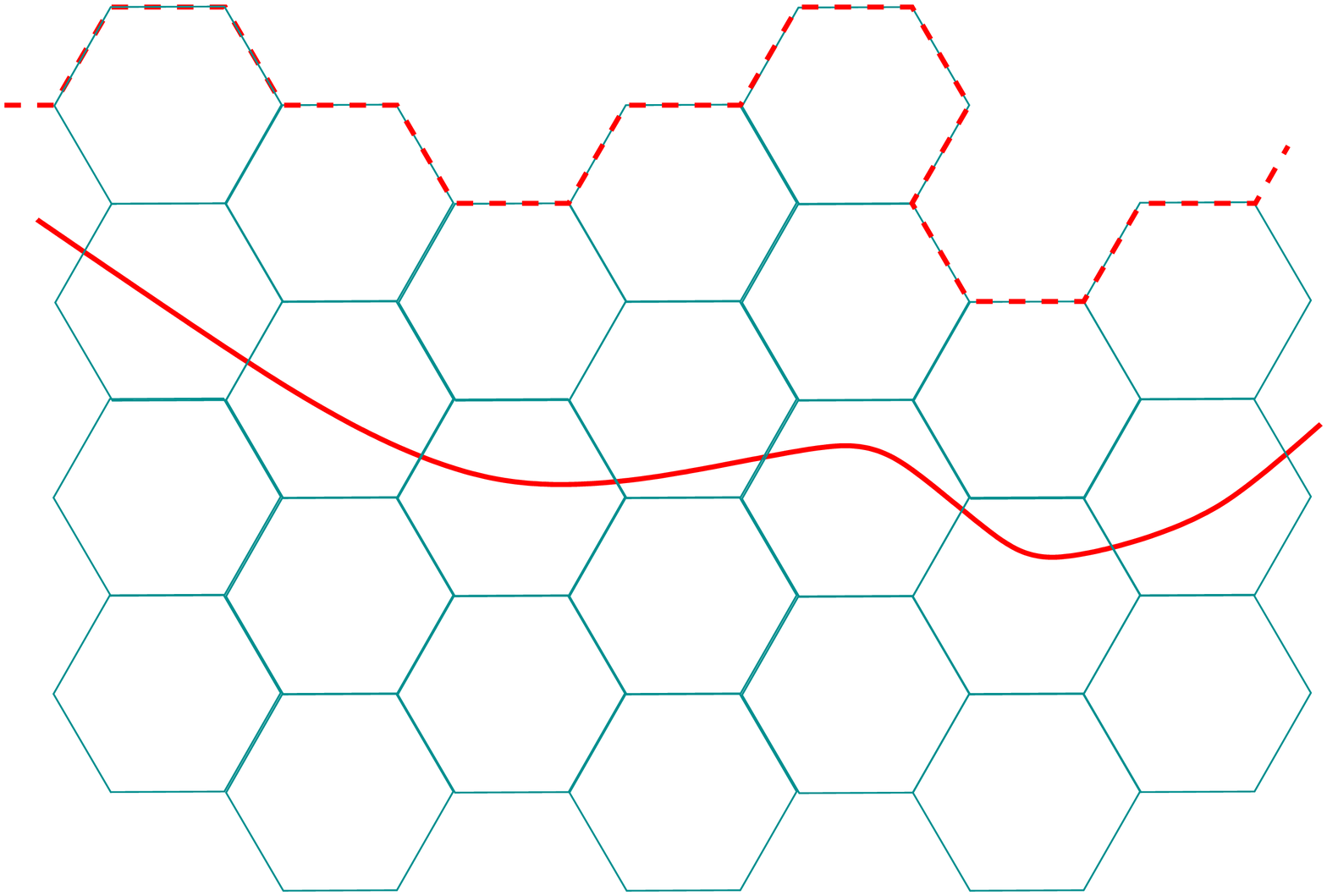}
\caption[fig]{\label{graphene_boundary} (Color online) Sketch of a rough
  graphene surface. The full line gives the boundary beyond which the lattice can be considered
  undistorted. The number of mid-gap states changes depending on a difference in
the number of removed sites for two sublattices.}
\end{center}
\end{figure}

Localized states can be defined at edges where the number of atoms in the two
sublattices is not compensated. The number of them depend on details of the
edge. The graphene edges can be strongly deformed, due to the bonding of other
atoms to carbon atoms at the edges. These atoms should not induce states in
the graphene $\pi$ band. In general, a boundary inside the graphene material
will exist, as sketched in Fig.~\ref{graphene_boundary}, beyond which the
$sp^2$ hybridization is well defined. If this is
the case, the number of mid-gap states near the edge is roughly
proportional to the difference in sites between the two sublattices near this
boundary.

Along a zigzag edge there is one localized state per three lattice units.
This implies that a precursor structure for
localized states at the Dirac energy can be found in ribbons or constrictions
of small lengths \cite{MJFP06}, which modifies the electronic structure and transport
properties.

Localized solutions can also be found near other defects that contain broken
bonds or vacancies. These states do not allow an analytical solution,
although, as discussed above, the continuum Dirac equation is compatible with
many boundary conditions, and it should describe well localized states that
vary slowly over distances comparable to the lattice spacing. The existence
of these states can be investigated by analyzing the scaling of the spectrum
near a defect as a function of the size of the system, $L$ \cite{VLSG05}. A
number of small voids and elongated cracks show states whose energy scales as
$L^{-2}$, while the energy of extended states scales as $L^{-1}$. A state
with energy scaling $L^{-2}$ is compatible with continuum states for which
the modulus of the wavefunction decays as $r^{-2}$ as a function of the
distance from the defect.

\subsection{Self-doping}
\label{self}

The band structure calculations discussed in previous sections
show that the electronic structure of a single
graphene plane is not strictly symmetrical in energy \cite{Retal02}.
The absence of electron-hole symmetry shifts the energy
of the states localized near impurities above or below the Fermi level,
leading to a transfer of charge from/to the clean regions. Hence,
the combination of localized defects and the lack of perfect electron-hole
symmetry around the Dirac points leads to the possibility of self-doping, in
addition to the usual scattering processes.

Extended lattice defects, like edges, grain boundaries, or micro-cracks, are
likely to induce a number of electronic states proportional
to their length, $L/a$, where $a$ is of the order of the lattice
constant. Hence, a distribution of extended defects of length $L$ at a
distance equal to $L$ itself gives rise to a concentration of $L/a$
carriers per carbon in regions of size of the order of $(L/a)^2$. The resulting system
can be considered a metal with a low density of carriers, $n_{\rm carrier} \propto a/L$ per
unit cell, and an elastic mean free path $l_{\rm elas} \simeq L$. Then, we
obtain:
\begin{eqnarray}
\epsilon_{\rm F} &\simeq &\frac{v_{\rm F}}{\sqrt{a L}} \nonumber \\
\frac{1}{\tau_{{\rm elas}}} &\simeq &\frac{v_{\rm F}}{L}
\end{eqnarray}
and, therefore, $(   \tau_{{\rm elas}} )^{-1} \ll \epsilon_{\rm F}$ when $a/L \ll
1$. Hence, the existence of extended defects leads to the possibility of
self-doping but maintaining most of the sample in the clean limit.
In this regime,
coherent oscillations of transport properties are expected,
although the observed electronic properties may correspond to a shifted
Fermi energy with respect to the nominally neutral defect--free system.

One can describe the effects that break electron-hole symmetry near the Dirac
points in terms of a finite next-nearest neighbor hopping between $\pi$
orbitals, $t'$, in (\ref{defphi}).
Consider, for instance, electronic structure of a
ribbon of width $L$ terminated by zigzag edges, which, as discussed, lead to
surface states for $t'=0$. The translational symmetry along the axis of the
ribbon allows us to define bands in terms of a wavevector parallel to this
axis. On the other hand, the localized surface bands, extending from $k_\parallel = ( 2 \pi )/3$
to $k_\parallel = - ( 2 \pi )/3$ acquire a dispersion of
order $t'$.
Hence, if the Fermi energy remains unchanged at the position of
the Dirac points ($\epsilon_{\rm Dirac} = - 3 t'$),
this band will be filled, and the ribbon will no longer be
charge neutral. In order to restore charge neutrality, the Fermi level needs
to be shifted by an amount
of the order of $t'$. As a consequence, some of the extended states near the Dirac
points are filled, leading to the phenomenon of self-doping. The local
charge is a function of distance to the edges, setting the Fermi energy so that
the ribbon is globally neutral. Note
that the charge transferred to the surface states is mostly localized near the
edges of the system.

\begin{figure}[htb]
\includegraphics*[width=7cm,angle=-90]{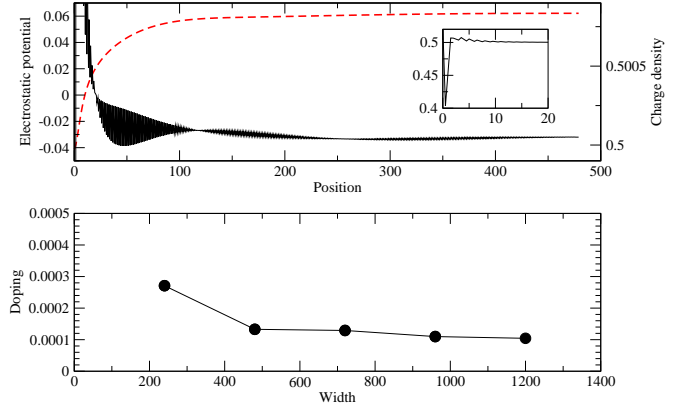}
\caption{\label{dopingfig} (Color online) Top: Self-consistent analysis of the displaced
charge density (in units of number of electrons per carbon) is shown
as a continuous line, and the corresponding electrostatic potential (in units of $t$) is
shown as a dashed line, for a graphene ribbon with periodic boundary
conditions along the zig-zag edge (with a length of $L=960 a$)
and with a circumference of size $W=80 \sqrt{3} a$.
The  inset shows the charge density near the edge.
Due to the presence of the
edge, there is a displaced charge in the bulk (bottom panel) that is
shown as a function of width $W$. Notice that the displaced
charge vanishes in the bulk limit ($W \to \infty$), in agreement with
(\ref{dispc}). Adapted from \onlinecite{PGC06}.}
\end{figure}

The charge transfer is suppressed by
electrostatic effects, as large deviations from charge neutrality have an
associated energy cost \cite{PGC06}. In order to study these charging effects we add to
the free-electron Hamiltonian (\ref{H1}) the Coulomb energy of interaction
between electrons:
\begin{eqnarray}
H_I = \sum_{i,j} U_{i,j} n_i n_j \, ,
\label{interact}
\end{eqnarray}
where $n_i = \sum_{\sigma} (a^{\dag}_{i,\sigma} a_{i,\sigma} +
b^{\dag}_{i,\sigma} b_{i,\sigma})$ is the number operator at site
${\bf R}_i$, and
\begin{eqnarray}
U_{i,j} = \frac{e^2}{\epsilon_0 |{\bf R}_i-{\bf R}_j|} \, ,
\end{eqnarray}
is the Coulomb interaction between electrons.
We expect, on physics grounds, that an
electrostatic potential builds up at the edges, shifting the position of the
surface states, and reducing the charge transferred to/from them. The potential
at the edge induced by a constant doping $\delta$ per carbon atom
is roughly, $\sim (\delta e^2/a) (W/a)$ ($\delta e^2/a$ is the Coulomb energy per carbon),
and  $W$ the width of the ribbon ($W/a$ is the number of atoms involved).
The charge transfer is stopped when the potential shifts the localized states to the Fermi energy,
that is, when $t' \approx (e^2/a) (W/a) \delta$. The resulting self-doping is
therefore
\begin{eqnarray}
\delta \sim \frac{t' a^2}{e^2 W} \, ,
\label{dispc}
\end{eqnarray}
that vanishes when $W \to \infty$.

We treat Hamiltonian (\ref{interact}) within the Hartree approximation
(that is, we replace $H_I$ by $H_{{\rm M.F.}} = \sum_i V_i n_i$
where $V_i = \sum_j U_{i,j} \langle n_j \rangle$, and solve the
problem self-consistently for $\langle n_i \rangle$).
Numerical results for graphene ribbons of
length $L = 80 \sqrt{3} a$ and different widths are shown in
Fig.~\ref{dopingfig} ($t'/t= 0.2$ and $e^2/a
= 0.5 t$). The largest width studied is $\sim 0.1 \mu$m, and the total number of
carbon atoms in the ribbon is $\approx 10^5$.  Notice that as $W$ increases, the
self-doping decreases indicating that, for a perfect graphene
plane ($W \rightarrow \infty$), the self-doping effect disappears.
For realistic parameters, we find that the amount of self-doping is $10^{-4} - 10^{-5}$ electrons
per unit cell for sizes $0.1 - 1 \mu$m.

\subsection{Vector potential and gauge field disorder}
\label{vector}

As discussed in Sec.~\ref{disorder},
lattice distortions modify the Dirac equation that describes the
low energy band structure of graphene. We consider here deformations
that change slowly on the lattice scale, so that they
do not mix the two inequivalent valleys.
As shown earlier, perturbations that hybridize the two sublattices lead to terms
that change the Dirac Hamiltonian from $v_F \bm \sigma \cdot {\bf k}$
into $\vf \bm \sigma \cdot {\bf k} + \bm \sigma \cdot {\bf A}$.
Hence, the vector ${\bf A}$ can be
thought of as if induced by an effective gauge field, ${\bf A}$.
In order
to preserve time reversal symmetry, this gauge field must have
opposite signs at the two Dirac cones, ${\bf A}_K = - {\bf A}_{K'}$.

\begin{figure}[]
\begin{center}
\includegraphics[width=6cm]{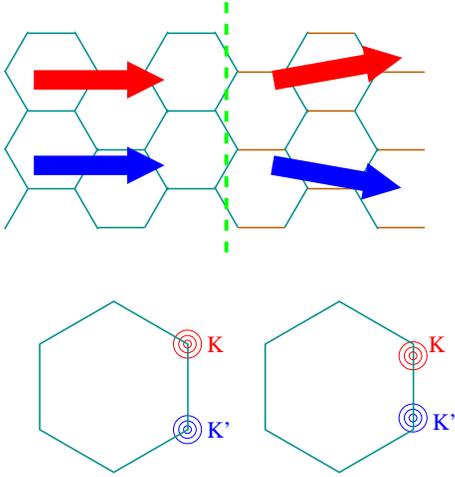}\\
\caption[fig]{(Color online) Gauge field induced by a simple
elastic strain. Top: The hopping along the horizontal bonds is assumed to be
changed on the right hand side of the graphene lattice, defining a straight
boundary between the unperturbed and perturbed regions (green dashed
line). Bottom: The modified hopping acts like a constant gauge field, which
displaces the Dirac cones in opposite directions at the $K$ and $K'$ points
of the Brillouin zone. The conservation of energy and momentum
parallel to the boundary leads to a deflection of electrons by the
boundary. }
\label{chiral_strain}
\end{center}
\end{figure}

A simple example is a distortion that changes the hopping between
all bonds along a given axis of the lattice. Let us assume that the
sites at the ends of those bonds define the unit cell, as sketched
in Fig.~\ref{chiral_strain}. If the distortion is constant, its
only effect is to displace the Dirac points away from the BZ corners.
The two inequivalent points are displaced in opposite
directions. This uniform distortion is the equivalent of a constant
gauge field, which does not change the electronic spectrum. The
situation changes if one considers a boundary that separates two
domains where the magnitude of the distortion is different. The
shift of the Dirac points leads to a deflection of the electronic
trajectories that cross the boundary, as also sketched in
Fig.~\ref{chiral_strain}. The modulation of the gauge field leads
to an effective magnetic field, which is of opposite sign for the
two valleys.

We have shown in Section \ref{topo} how topological lattice defects,
such as disclinations and dislocations, can be described by an
effective gauge field. Those defects can only exist in graphene sheets
that are intrinsically curved, and the gauge field only depends on
topology of the lattice. Changes in the nearest neighbor hopping also lead
to effective gauge fields. We consider next two physical processes that
induce effective gauge fields: ({\it i}) changes in the hopping induced
by hybridization between $\pi$ and $\sigma$ bands, which arise in curved
sheets, and ({\it ii}) changes in the hopping due to modulation in the bond
length, which is associated with elastic strain. The strength of these fields
depends on parameters that describe the value of the $\pi$-$\sigma$ hybridization,
and the dependence of hopping on the bond length.

A comparison of the relative strengths of the gauge fields induced by intrinsic
curvature, $\pi-\sigma$ hybridization (extrinsic curvature), and elastic strains,
arising from a ripple of typical height and size is given in Table \ref{table_gauge}.

\begin{table}
\begin{tabular}{||c||c|c||}
\hline \hline & &B \\
&$l_B$ &h=1 nm, l=10nm, a=0.1nm  \\ \hline \hline Intrinsic
curvature &$l \left( \frac{l}{h} \right)$ &0.06T  \\ \hline
Extrinsic curvature &$l \sqrt{\frac{t}{E}
\frac{l^3}{a h^2}}$ &0.006T
\\ \hline
Elastic strains &$l \sqrt{\frac{1}{\beta} \frac{a l}{h^2}}$
 &6T \\
\hline \hline
\end{tabular}
\caption{Estimates of the effective magnetic length, and effective
magnetic fields generated by the deformations considered in this
section. The intrinsic curvature entry also refers to the
contribution from topological defects.} \label{table_gauge}
\end{table}

\subsubsection{Gauge field induced by curvature}
\label{curvature}

As we discussed in Sec.~\ref{ripples}, when the $\pi$ orbitals are not parallel,
 the hybridization between them depends on their relative orientation. The angle
$\theta_i$ determines the relative orientation of neighboring
orbitals at some position ${\bf r}_i$ in the graphene sheet.
The value of $\theta_i$ depends on the local
curvature of the layer. The relative angle of rotation of two $p_z$ orbitals
at position ${\bf r}_i$ and ${\bf r}_j$ can be written as:
 $\cos(\theta_i-\theta_j) = {\bf N}_i \cdot {\bf N}_j$, where ${\bf N}_i$
is the unit vector perpendicular to the surface, defined in (\ref{normal}).
If ${\bf r}_j = {\bf r}_i + {\bf u}_{ij}$ we can write:
\begin{eqnarray}
{\bf N}_i \!\cdot\! {\bf N}_j \!\approx\! 1 \!+\! {\bf N}_i \!\cdot\! [({\bf u}_{ij} \!\cdot\! \nabla) {\bf N}_i] \!+\!
\frac{1}{2} {\bf N}_i \!\cdot\! [({\bf u}_{ij} \!\cdot\! \nabla)^2 {\bf N}_i] \, ,
\label{ninj}
\end{eqnarray}
where we assume smoothly varying ${\bf N}({\bf  r})$.
We use (\ref{normal}) in terms of the height field $h({\bf r})$
(${\bf N}({\bf r}) \approx {\bf z}-\nabla h({\bf r}) - (\nabla h)^2 {\bf
  z}/2$) to rewrite (\ref{ninj}) as:
\begin{eqnarray}
{\bf N}_i \cdot {\bf N}_j \approx 1 - \frac{1}{2} [({\bf u}_{ij} \cdot
\nabla) \nabla h({\bf r}_i)]^2 \, .
\label{ninjh}
\end{eqnarray}
Hence, bending of the graphene sheet leads to a modification of the
hopping amplitude between different sites in the form:
\begin{eqnarray}
\delta t_{ij} \approx - \frac{t_{ij}^0}{2} [({\bf u}_{ij} \cdot
\nabla) \nabla h({\bf r}_i)]^2 \, ,
\end{eqnarray}
where $t_{ij}^0$ is the bare hopping energy.
A similar effect leads to
changes the electronic states in a carbon nanotubes \cite{KM97}.
Using the results of Sec.~\ref{disorder}, namely (\ref{defa}),
we can now see  that a vector potential is generated for nearest neighbor
hopping (${\bf u} = \vec{\delta}_{ab}$) \cite{NK07}:
\begin{eqnarray}
{\cal A}^{(h)}_x &= &- \frac{3 E_{ab} a^2}{8} \left[ (
\partial_x^2 h )^2 -
( \partial_y^2 h )^2 \right] \nonumber \\
{\cal A}^{(h)}_y &= &\frac{3 E_{ab} a^2}{4} \left( \partial_x^2 h
+ \partial_y^2 h \right) \partial_x h \partial_y h
\label{field_curvature}
\end{eqnarray}
where the coupling constant $E_{ab}$ depends on microscopic details
\cite{NK07}. The
flux of effective magnetic field through a ripple of lateral
dimension $l$ and height $h$ is given approximately by:
\begin{equation}
\Phi \approx \frac{E_{ab} a^2 h^2}{\vf l^3}
\end{equation}
where the radius of curvature is $R^{-1} \approx h l^{-2}$. For a
ripple with $l \approx 20$nm, $h \approx 1$nm, taking $E_{ab} / \vf
\approx 10$ \AA$^{-1}$ , we find $\Phi \approx 10^{-3} \Phi_0$.

\subsubsection{Elastic strain}
\label{strain}

The elastic free energy for graphene can be written in terms of
the in-plane displacement ${\bf u}({\bf r})=(u_x,u_y)$ as:
\begin{eqnarray}
F[{\bf u}] \!=\! \frac{1}{2} \! \int \! d^2r \left[({\cal B}-{\cal G})
(\sum_{i=1,2} u_{ii})^2 + 2 {\cal G} \sum_{i,j=1,2} u^2_{ij}\right] \, ,
\end{eqnarray}
where ${\cal B}$ is the bulk modulus, ${\cal G}$ is the shear modulus, and
\begin{eqnarray}
u_{ij} = \frac{1}{2} \left(\frac{\partial u_i}{\partial x_j}+
\frac{\partial u_j}{\partial x_j}\right) \, ,
\end{eqnarray}
is the strain tensor ($x_1=x$ and $x_2=y$).

There are many types of static deformation of the honeycomb lattice
which can affect the propagation of Dirac fermions.
The simplest one is due to changes in the area of the unit
cell either due to dilation or contraction. Changes in the
unit cell area lead to local changes in the density of electrons
and, therefore, local changes in the chemical potential in the
system. In this case, their effect is similar to the one found
in (\ref{defphi}), and we must have:
\begin{eqnarray}
\phi_{{\rm dp}}({\bf r}) = g (u_{xx}+u_{yy}) \, ,
\label{phidp}
\end{eqnarray}
and their effect is diagonal in the sublattice index.

The nearest neighbor hopping depends on the length of the carbon
bond. Hence, elastic strains that modify the relative orientation
of the atoms also lead to an effective gauge field, which acts on
each $K$ point separately, as first discussed in relation to carbon
nanotubes \cite{SA02b,M07}. Consider two carbon atoms located in
two different sublattices in the same unit cell at ${\bf R}_i$.
The change in the local bond length can be written as:
\begin{eqnarray}
\delta u_i = \frac{\vec{\delta}_{ab}}{a} \cdot [{\bf u}_{{\rm A}}({\bf R}_i)-{\bf u}_{{\rm B}}({\bf R}_i+\vec{\delta}_{ab})] \, .
\label{dui}
\end{eqnarray}
The local displacements of the atoms in the unit cell can be related
to ${\bf u}({\bf r})$ by \cite{ando_opt}:
\begin{eqnarray}
(\vec{\delta}_{ab} \cdot \nabla) {\bf u} = \kappa^{-1} ({\bf u}_{{\rm A}} - {\bf u}_{{\rm B}}) \, ,
\label{uac}
\end{eqnarray}
where $\kappa$ is a dimensionless quantity that depends on microscopic details.
Changes in the bond length lead to changes in the hopping amplitude:
\begin{eqnarray}
t_{ij} \approx t_{ij}^0 + \frac{\partial t_{ij}}{\partial a} \delta u_i \, ,
\label{tijdel}
\end{eqnarray}
and we can write:
\begin{eqnarray}
\delta t^{(ab)}({\bf r}) \approx \beta \frac{\delta u({\bf r})}{a} \, ,
\label{dtab}
\end{eqnarray}
where
\begin{eqnarray}
\beta = \frac{\partial t^{(ab)}}{\partial \ln(a)} \, .
\label{beta}
\end{eqnarray}
Substituting (\ref{dui}) into (\ref{dtab}) and the final result into
(\ref{defa}), one finds \cite{ando_opt}:
\begin{eqnarray}
{\cal A}^{(s)}_x &= &\frac{3}{4} \beta \, \kappa \, ( u_{xx} - u_{yy} ) \, ,
\nonumber
\\
{\cal A}^{(s)}_y &= &\frac{3}{2} \beta \, \kappa \, u_{xy} \, .
\label{field_elastic}
\label{vecac}
\end{eqnarray}
We assume that the strains induced by a ripple of
dimension $l$ and height $h$ scale as  $u_{ij} \sim ( h
/ l )^2$. Then, using $\beta / \vf \approx a^{-1} \sim
1$\AA$^{-1}$, we find that the total flux through a ripple is:
\begin{equation}
\Phi \approx \frac{h^2}{a l} \, .
\end{equation}
For ripples such that $h \sim 1$nm and $l \sim 20$nm, this estimate
gives $\Phi \sim 10^{-1} \Phi_0$ in reasonable agreement
with the estimates in ref.~\cite{Metal06}.

The strain tensor must satisfy some additional constraints, as it is derived from a
displacement vector field. These constraints are called Saint Venant
compatibility conditions \cite{LL59}:
\begin{equation}
W_{ijkl}=\frac{\partial u_{ij}}{\partial x_k \partial x_l} +
\frac{\partial u_{kl}}{\partial x_i \partial x_j} - \frac{\partial
u_{il}}{\partial x_j \partial x_k} - \frac{\partial u_{jk}}{\partial
x_i \partial x_l} = 0 \, .
\label{saint_venant}
\end{equation}
An elastic deformation changes the distances in the crystal lattice
and can be considered as a change in the metric:
\begin{equation}
g_{ij} = \delta_{ij} + u_{ij} \label{metric}
\end{equation}
The compatibility equations (\ref{saint_venant}) are equivalent
to the condition that the curvature tensor derived from
(\ref{metric}) is zero. Hence, a purely elastic deformation
cannot induce intrinsic curvature in the sheet, which only arises
from topological defects.
 The effective fields associated with elastic strains can be
 large \cite{Metal06}, leading to significant changes in the
 electronic wavefunctions. An analysis of the resulting state, and
 the possible instabilities that may occur can be found
 in \cite{GKV07}.

\subsubsection{Random gauge fields}
\label{random}

The preceding discussion suggests that the effective fields
associated with lattice defects can modify significantly the
electronic properties. This is the case when the fields do not
change appreciably on scales comparable to the (effective) magnetic
length. The general problem of random gauge fields for Dirac
fermions has been extensively analyzed before the current interest
in graphene, as the topic is also relevant for the IQHE \cite{LFSG94} and d-wave
superconductivity \cite{NTW94}. The one electron nature of this two
dimensional problem makes it possible, at the Dirac energy, to map it
onto models of interacting electrons in one dimension, where many
exact results can be obtained \cite{CCFGM97}. The low energy density
of states, $\rho ( \omega )$, acquires an anomalous exponent, $\rho (
\omega ) \propto | \omega |^{1 - \Delta}$, where $\Delta > 0$. The
density of states is enhanced near the Dirac energy, reflecting the
tendency of disorder to close gaps. For sufficiently large values of
the random gauge field, a phase transition is also
possible \cite{CMW96,HD02}.

 Perturbation theory shows that random gauge fields are a marginal
 perturbation at the Dirac point, leading to logarithmic divergences.
 These divergences tend to have the opposite sign with respect to those induced
 by the Coulomb interaction (see Sec.~\ref{elec_elec}). As a result,
a renormalization group (RG)
 analysis of interacting electrons in a random gauge field suggests
 the possibility of non-trivial phases \cite{FA05,FA06,NKR07,SGV05,Khv07,A06c,AE06,DAnna06}, where
 interactions and disorder cancel each other.

\subsection{Coupling to magnetic impurities}

Magnetic impurities in graphene can be introduced chemically by
deposition and intercalation \cite{ULCN07,CM07}, or self-generated
by the introduction of defects \cite{KH06,KH07}.
The energy dependence of the density of states in graphene leads to
changes in the formation of a Kondo resonance between a magnetic
impurity and the graphene electrons. The vanishing of the density of
states at the Dirac energy implies that a Kondo singlet in the
ground state is not formed unless the exchange coupling exceeds a
critical value, of the order of the electron bandwidth, a problem
already studied in connection with magnetic impurities in d-wave
superconductors \cite{CF96,CF97,anatoli02,PSV01,FFV06}. For weak exchange couplings, the
magnetic impurity remains unscreened. An external gate changes the
chemical potential, allowing for a tuning of the Kondo
resonance \cite{SB07}. The situation changes significantly if the
scalar potential induced by the magnetic impurity is taken into
account. This potential that can be comparable to the bandwidth
allows the formation of mid-gap states and changes
the phase-shift associated to spin scattering \cite{HG07}. These
phase-shifts have a weak logarithmic dependence on the chemical
potential, and a Kondo resonance can exist, even close to the Dirac
energy.

The RKKY interaction between magnetic impurities is also modified in
graphene. At finite fillings, the absence of intra-valley
backscattering leads to a reduction of the Friedel oscillations,
which decay as $\sin ( 2 \kf r ) / | r |^3$ \cite{CF06,WSSG06,A06}.
This effect leads to an RKKY interaction, at finite fillings, which
oscillate and decay as $| r |^{-3}$. When intervalley scattering is
included, the interaction reverts to the usual dependence on
distance in two dimensions, $| r |^{-2}$ \cite{CF06}. At half-filling
extended defects lead to an RKKY interaction with an
$| r |^{-3}$ dependence \cite{VLSG05,DLB06}. This behavior is changed
when the impurity potential is localized on atomic
scales \cite{S07,BFS07}, or for highly symmetrical
couplings \cite{S07}.

\subsection{Weak and strong localization}
\label{localization}

In sufficiently clean systems, where the Fermi wavelength is much
shorter than the mean free path, $\kf l \gg 1$, electronic transport
can be described in classical terms, assuming that electrons follow
well defined trajectories. At low temperatures, when electrons
remain coherent over long distances, quantum effects lead to
interference corrections to the classical expressions for the
conductivity, the weak localization correction \cite{B84,CS86}. These
corrections are usually due to the positive interference between two
paths along closed loops, traversed  in opposite directions. As a
result, the probability that the electron goes back to the origin is
enhanced, so that quantum corrections decrease the conductivity.
These interferences are suppressed for paths longer than the
dephasing length, $l_\phi$, determined by interactions between the
electron and environment. Interference effects can also be
suppressed by magnetic fields that break down
time reversal symmetry and adds a random relative phase to the
process discussed above. Hence, in most metals, the conductivity
increases when a small magnetic field is applied (negative
magnetoresistance).

Graphene is special in this respect, due to the chirality of its electrons.
The motion along a closed path induces a change in the relative
weight of the two components of the wavefunction, leading to a new
phase, which contributes to the interference processes. If the
electron traverses a path without being scattered from one valley to
the other, this (Berry) phase changes the sign of the amplitude of one path
with respect to the time-reversed path. As a consequence, the two
paths interfere destructively, leading to a suppression of
backscattering \cite{SA02}. Similar processes take place in materials
with strong spin orbit coupling, as the spin direction changes along
the path of the electron \cite{B84,CS86}. Hence, if scattering
between valleys in graphene can be neglected, one expects a positive
magnetoresistance, i. e., weak anti-localization. In general, intra-
and intervalley elastic scattering can be described in terms of two
different scattering times, $\tau_{intra}$ and $\tau_{inter}$, so
that if $\tau_{intra} \ll \tau_{inter}$ one expects weak
anti-localization processes, while if $\tau_{inter} \ll \tau_{intra}$
ordinary weak localization will take place. Experimentally, localization
effects are always strongly suppressed close to the Dirac point but can
be partially or, in rare cases, completely recovered at high carrier
concentrations, depending on a particular single-layer sample \cite{Metal06,THGS07}.
Multilayer samples exhibit an additional positive magnetoresistance in higher
magnetic fields, which can be attribued to classical changes in the current
distribution due to a vertical gradient of concentration \cite{Metal06}
and anti-localization effects \cite{WLSBH07}.

The propagation of an electron in the absence of intervalley
scattering can be affected by the effective gauge fields induced by
lattice defects and curvature. These fields can suppress the interference
corrections to the conductivity \cite{Metal06,MG06}.
In addition, the description in terms of free Dirac electrons is only
valid near the neutrality point. The Fermi energy acquires a
trigonal distortion away from the Dirac point, and backward
scattering within each valley is no longer completely
suppressed \cite{Metal06b}, leading to a further suppression of
anti-localization effects at high dopings. Finally, the gradient of
external potentials induce a small asymmetry between the two
sublattices \cite{MG06}. This effect will also contribute to reduce
anti-localization, without giving rise to localization effects.

The above analysis has to be modified for a graphene bilayer. Although the description
of the electronic states requires a two component spinor, the total
phase around a closed loop is $2 \pi$, and backscattering is not
suppressed \cite{KFMA07}. This result is consistent with experimental
observations, which show the existence of weak localization
effects in a bilayer \cite{GTMHS07}.

When the Fermi energy is at the Dirac point, a
replica analysis shows that the conductivity approaches a universal
value of the order of $e^2 / h$ \cite{F86,F86b}. This result is valid when
intervalley scattering is neglected \cite{OGM06,OGM07,RMOF07}. Localization
is induced when these terms are included \cite{A06c,AE06}, as also
confirmed by numerical calculations \cite{LVGC07}. Interaction
effects tend to suppress the effects of disorder. The same result, namely
a conductance of the order of $e^2/h$, is obtained for disordered
graphene bilayers where a self-consistent calculation
leads to universal conductivity at the neutrality point \cite{NNGP06,NCNGP07,Kats07}.
In a biased graphene bilayer, the presence of impurities leads to the
appearance of impurity tails in the density of states due to the
creation of mid-gap states which are sensitive to the applied electric
field that opens the gap between the conduction and valence bands \cite{NCV06}.

One should point out that most of the calculations of transport properties
assume self-averaging, that is, that one can exchange a problem with lack
of translational invariance by an effective medium system with damping.
Obviously this procedure only works when the disorder is weak and the system
is in the metallic phase. Close to the localized phase this procedure breaks
down, the system divides itself into regions of different chemical
potential  and one has to think about transport in real
space, usually described in terms of percolation \cite{CFAA07,Shk07}.
Single electron transistor (SET) measurements of graphene show that this
seems to be the situation in graphene at half-filling \cite{MAULSKY07}.

Finally, we should point out that graphene stacks suffer from another source
of disorder, namely, c-axis disorder that is either due to impurities between
layers or rotation of graphene planes relative to each other. In either case
the in-plane and out-of-plane transport is directly affected.
This kind of disorder
has been observed experimentally by different techniques \cite{HVMSHBFMC07,BZYBOMHRC07}.
In the case
of the bilayer, the rotation of planes changes substantially the spectrum
restoring the Dirac fermion description \cite{LSPCN07}.
The transport properties in the out of plane direction are determined by the
interlayer current operator, $\hat{\bf j}_{n,n+1} = i t \sum ( c_{A,n,s}^\dag
c_{A,n+1,s} - c_{A,n+1,s}^\dag c_{A,n,s})$, where $n$ is a layer index, and
$A$ is a generic index that defines the sites coupled by the interlayer
hopping $t$. If we only consider hopping between nearest neighbor sites in
consecutive layers, these sites belong to one of the two sublattices in each
layer.

In a multilayer with Bernal stacking, these connected sites are the ones
where the density of states vanishes at zero energy, as discussed
above. Hence, even in a clean system, the number of conducting channels in
the direction perpendicular to the layers vanishes at zero
energy \cite{NNGP06,NCNGP07}. This situation is reminiscent of the in plane transport
properties of a single layer graphene. Similar to the latter case, a self-consistent
Born approximation for a
small concentration of impurities leads to a finite conductivity, which becomes
independent of the number of impurities.

\subsection{Transport near the Dirac point}
\label{transdirac}

In clean graphene, the number of channels available for electron
transport decreases as the chemical potential approaches the Dirac
energy. As a result, the conductance through a clean graphene ribbon
is, at most, $4 e^2 / h$, where the factor of 4 stands for the spin
and valley degeneracy. In addition, only one out of every three
possible clean graphene ribbons have a conduction channel at the
Dirac energy. The other two thirds are semiconducting, with a gap of
the order of $\vf / W$, where $W$ is the width. This result is a
consequence of the additional periodicity introduced by the
wavefunctions at the $K$ and $K'$ points of the Brillouin Zone,
irrespective of the boundary conditions.

A wide graphene ribbon allows for many channels, which can be
approximately classified by the momentum perpendicular to the axis
of the ribbon, $k_y$. At the Dirac energy, transport through these
channels is inhibited by the existence of a gap, $\Delta_{k_y} = \vf
k_y$. Transport through these channels is suppressed by a factor of
the order of $e^{- k_y L}$, where $L$ is the length of the ribbon. The
number of transverse channels increases as $W/a$, where $W$ is the
width of the ribbon and $a$ is a length of the order of the lattice
spacing. The allowed values of $k_y$ are $\propto n_y / W$,
where $n_y$ is an integer. Hence, for a ribbon such that $W \gg L$,
there are many channels which satisfy $k_y L \ll 1$. Transport
through these channels is not strongly inhibited, and their
contribution dominates when the Fermi energy lies near the Dirac
point. The conductance arising from these channels is given
approximately by \cite{K06,TTTRB06}:
\begin{equation}
G \sim \frac{e^2}{h} \frac{W}{2 \pi} \int d k_y e^{- k_y L} \sim
\frac{e^2}{h} \frac{W}{L} \, .
\label{evanescent}
\end{equation}
The transmission at normal incidence, $k_y = 0$, is
one, in agreement with the absence of
backscattering in graphene, for any barrier that does
not induce intervalley scattering \cite{KNG06}. The transmission of a
given channel scales as $T ( k_y ) = 1/\cosh^2 ( k_y L / 2)$.

 Eq.(\ref{evanescent}) shows that the
contribution from all transverse channels lead to a conductance
which scales, similar to a function of the length and width of the system, as
the conductivity of a diffusive metal. Moreover, the value of the
effective conductivity is of the order of $e^2 / h$. It can also be shown
that the shot noise depends on current in the same way as in a
diffusive metal. A detailed analysis of possible boundary conditions
at the contacts and their influence on evanescent waves can be
found in \cite{S06,RS06}.
The calculations leading to eq.(\ref{evanescent}) can be extended to
a graphene bilayer. The conductance is, again, a summation of terms
arising from evanescent waves between the two contacts, and it has
the dependence on sample dimensions of a 2D
conductivity of the order of $e^2 / h$ \cite{SB07b}, although there is a prefactor twice
bigger than the one in single layer graphene.

The calculation of the conductance of clean graphene in terms of
transmission coefficients, using the Landauer method leads to an
effective conductivity which is equal to the value obtained for bulk
graphene using diagrammatic methods, the Kubo formula \cite{PGN06},
in the limit of zero impurity concentration and zero doping.
Moreover, this correspondence remains valid for the case of a bilayer
without and with trigonal warping effects \cite{KA06,CCD07}.

Disorder at the Dirac energy changes the conductance of graphene
ribbons in two opposite directions \cite{LVGC07}: i) a sufficiently
strong disorder, with short range (intervalley) contributions, lead
to a localized regime, where the conductance depends exponentially
on the ribbon length, and ii) at the Dirac energy, disorder allows
mid-gap states that can enhance the conductance mediated by
evanescent waves discussed above. A fluctuating electrostatic
potential also reduces the effective gap for the transverse
channels, enhancing further the conductance. The resonant tunneling
regime mediated by mid-gap state was suggested by analytical
calculations \cite{T07}. The enhancement of the conductance by
potential fluctuations can also be studied semi-analytically. In the
absence of intervalley scattering, it leads to an effective
conductivity which grows with ribbon length \cite{SPG07}. In
fact, analytical and numerical studies \cite{BTBB07,NKR07,SPG07,LMCN07} show that the
conductivity obeys a universal scaling with the lattice size $L$:
\begin{eqnarray}
\sigma(L) = \frac{2 e^2}{h} \left(A \ln(L/\xi) + B \right) \, ,
\end{eqnarray}
where $\xi$ is a length scale associated with range of interactions
and $A$ and $B$ are numbers of the order of unit ($A\approx 0.17$ and $B \approx 0.23$
for a graphene lattice in the shape of a square of size $L$\cite{LMCN07}).
Notice, therefore, that the conductivity is always
of the order of $e^2/h$ and has a weak dependence on size.

%------------------------------------------------------------------------------
\subsection{Boltzmann Equation description of DC transport in doped graphene}
\begin{figure}[ht]
\begin{center}
\includegraphics*[width=10cm]{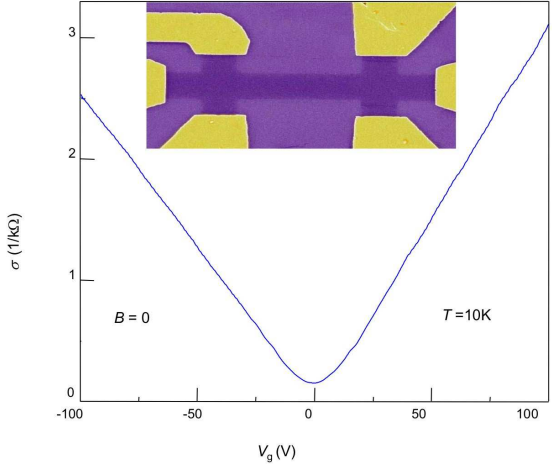}
\end{center}
\caption{\label{sigxx} (Color online) An example of changes in conductivity $\sigma$ of graphene
with varying gate voltage, $V_g$, and, therefore, carrier concentration $n$. Here $\sigma$ is
proportional to $n$ as discussed in the text. Note that samples with higher mobility ($>1$ m$^2$/Vs)
normally show a sublinear dependence, presumably indicating the presence of different types of
scatterers. Inset: scanning-electron micrograph of one of experimental devices (in false colors matching
those seen in visible optics. The scale of the micrograph is given by the width of the Hall bar, which is
one micrometer. Adapted from \cite{Netal05}.}
\end{figure}

%------------------------------------------------------------------------------
It was shown experimentally that the DC conductivity
of graphene depends linearly on the gate potential \cite{Netal04,Netal05,Netal05b},
except very close to the neutrality point (see Fig.\ref{sigxx}). Since the gate potential
depends linearly on the electronic density, $n$, one has a
conductivity $\sigma\propto n$. As shown  by Shon and  Ando \cite{Ando7}
if the scatterers are short range the DC conductivity
should be independent of the electronic density, at odds with the experimental
result. It has been shown \cite{A06,NMac06,NMac07} that by considering
a scattering mechanism based on screened charged impurities
it is possible to obtain from a Boltzmann equation approach
a conductivity varying linearly with the density, in agreement
with the experimental result \cite{A06,PLS07,Novapl07,TS07,KG07}.

The Boltzmann equation has the form \cite{Ziman}
\begin{equation}
-\bm v_{\bm k}\cdot\bm\nabla_{\bm r} f(\epsilon_{\bm k})
-e (\bm E+\bm v_{\bm k}\times H)\cdot\nabla_{\bm k}f(\epsilon_{\bm k})
=-\left.\frac {\partial f_{\bm k}}{\partial t}
\right\vert_{scatt.}\,.
\label{boltzmann}
\end{equation}
The solution of the Boltzmann equation in its general form is difficult and one needs therefore to rely  upon some approximation. The first
step in the usual approximation scheme is to write the distribution
as $f(\epsilon_{\bm k}) = f^0(\epsilon_{\bm k})+g(\epsilon_{\bm k})$
where $f^0(\epsilon_{\bm k})$ is the steady state distribution function
and $g(\epsilon_{\bm k})$ is assumed to be small. Inserting this ansatz
in (\ref{boltzmann}) and keeping
only terms that are linear in the external fields one obtains the
linearized Boltzmann equation \cite{Ziman} which reads
\begin{eqnarray}
&-&\frac {\partial f^0(\epsilon_{\bm k})}{\partial \epsilon_{\bm k}}
\bm v_{\bm k}\cdot
\left[
\left(
-\frac{\epsilon_{\bm k}-\zeta}{T}
\right)\bm\nabla_{\bm r}T
+e\left(
\bm E-\frac 1 e\bm\nabla_{\bm r}\zeta
\right)
\right] =\nonumber\\
&-&\left.\frac {\partial f_{\bm k}}{\partial t}
\right\vert_{scatt.} +
{\bm v_{\bm k}\cdot \bm \nabla_{\bm r}}g_{\bm k}+ e
(\bm v_{\bm k}\times \bm H )\cdot \bm \nabla_{\bm k} g_{\bm k}\,.
\end{eqnarray}
The second approximation has to do with the form of the
scattering term. The simplest approach  is to
introduce a relaxation time approximation:
\begin{equation}
-\left.\frac {\partial f_{\bm k}}{\partial t}
\right\vert_{scatt.}\rightarrow\frac {g_{\bm k}}{\tau_{\bm k}}\,,
\end{equation}
where $\tau_{\bm k}$ is the relaxation time, assumed to be
momentum dependent. This momentum dependence
is determined
phenomenologically in such way that the dependence of the
conductivity upon the electronic density agrees with
experimental data.
The Boltzmann equation is certainly not
valid at the Dirac point, but since many experiments are performed
at finite carrier density, controlled by an external
gate voltage,  we expect the Boltzmann equation to give
reliable results if an appropriate form for  $\tau_{\bm k}$
is used \cite{AHGS07}.

Let us compute the Boltzmann relaxation time, $\tau_{\bm k}$,
for two different
scattering potentials:(i) a Dirac delta function potential; (ii) a
unscreened Coulomb potential. The  relaxation time
$\tau_{\bm k}$ is defined as:
\begin{equation}
\frac 1{\tau_{\bm k}}= n_i \int d\,\theta
\int \frac{k'd\,k'}{(2\pi)^2}
S(\bm k,\bm k')(1-\cos\theta)\,,
\end{equation}
where $n_i$ is impurity concentration per unit of area,
and the transition rate $S(\bm k,\bm k')$ is given,
in the Born approximation, by:
\begin{equation}
S(\bm k,\bm k')=2\pi \vert H_{\bm k',\bm k}\vert^2
\frac {1}{v_F}
\delta( k'- k)\,,
\end{equation}
where the $v_F k$ is the dispersion of Dirac fermions in graphene
and $H_{\bm k',\bm k}$ is defined as
\begin{equation}
H_{\bm k',\bm k} = \int d\bm r\psi_{\bm k'}^\ast(\bm r)U_S(\bm r)
\psi_{\bm k}(\bm r)\,,
\end{equation}
with $U_S(\rm r)$  the scattering potential and $\psi_{\bm k}(\bm r)$
is the electronic spinor wavefunction of a clean graphene sheet. If the potential
is short range,\cite{Ando7} of the
form $U_S=v_0\delta(\bm r)$, the Boltzmann relaxation
 time turns out to be
\begin{equation}
\tau_{\bm k} =\frac {4 v_F}{n_iv_0^2}\frac 1 k\,.
\end{equation}
On the other hand, if the potential is the Coulomb
potential, given by $U_S(\bm r)=eQ/(4\pi\epsilon_0\epsilon r)$
for charged impurities of charge $Q$, the relaxation time is given by
\begin{equation}
\tau_{\bm k} =\frac {v_F}{u_0^2}k\,.
\label{taucoulomb}
\end{equation}
where $u_0^2=n_iQ^2e^2/(16\epsilon_0^2\epsilon^2)$.
As we argue below, the phenomenology of Dirac fermions
implies that the scattering in graphene must be of the form
(\ref{taucoulomb}).

Within the relaxation time approximation the solution of the
linearized Boltzmann equation when an electric field is applied to the
sample is
\begin{equation}
g_{\bm k} = -\frac {\partial f^0(\epsilon_{\bm k})}
{\partial \epsilon_{\bm k}}
e\tau_{\bm k}\bm v_{\bm k}\cdot\bm E\,,
\end{equation}
and the electric current reads (spin and valley indexes included)
\begin{equation}
\bm J=\frac {4}{A}\sum_{\bm k}e\bm v_{\bm k}g_{\bm k}\,.
\end{equation}
Since at low temperatures the following relation
$-\partial f^0(\epsilon_{\bm k})/\partial \epsilon_{\bm k}\rightarrow
\delta (\mu-v_F k)$ holds, one can easily see that assuming (\ref{taucoulomb})
where $k$ is measured relatively to the Dirac point,
the electronic conductivity turns out to be
\begin{equation}
\sigma_{xx} = 2 \frac {e^2}{h}\frac {\mu^2}{u_0^2}=2 \frac {e^2}{h}\frac
{\pi v_F^2}{u_0^2}n,
\label{sigmaxx}
\end{equation}
where $u_0$ is the strength of the scattering potential (with dimensions
of energy). The electronic  conductivity depends linearly on the
electron density, in agreement with the experimental data.
We stress that the Coulomb potential
is one possible mechanism of producing a scattering rate of the
form (\ref{taucoulomb}) but we do not exclude that other mechanisms may exist
(see, for instance, \cite{KG07}).

%------------------------------------------------------------------------------
\subsection{Magnetotransport and universal conductivity}
%------------------------------------------------------------------------------

The description of the magnetotransport properties of electrons
in a disordered honeycomb lattice is complex because of
the interference effects associated with the Hofstadter
problem \cite{Gumbs97}.
We shall simplify our problem by describing electrons in the
honeycomb lattice as Dirac fermions in the continuum approximation,
introduced in  Sec. \ref{sec:continuous}. Furthermore, we will only
focus on the problem of short range scattering in the unitary limit
since in this regime many analytical results are obtained \cite{PGC06,MGKO07,Petal06,SL06,SL07,KH06,PLSCN07}.
The problem of magnetotransport in the presence of Coulomb impurities,
as discussed in the previous section is still an open research problem.
A similar approach was considered by Abrikosov
in the quantum magnetoresistance study of non-stoichiometric
chalcogenides \cite{Abrikosov98}.
In the case of graphene, the effective Hamiltonian describing Dirac fermions
in a magnetic field (including disorder) can be written as:
$H=H_0+H_i$ where $H_0$ is given by (\ref{H1})
and $H_i$ is the impurity potential reading \cite{PGC06}:
\begin{equation}
H_i=V {\sum_{j=1}^{N_i}\delta(\bm r-\bm r_j) \bm I}\,
\label{llH1}
\end{equation}
The formulation of the problem in second quantization requires
the solution of $H_0$, which was done in Section \ref{sec:contmag}.
The field operators, close to the K point,
are defined as (the spin index is omitted for simplicity):
\begin{eqnarray}
\Psi(\bm r)&=&\sum_{k}\frac {e^{ikx}}{\sqrt L}
\left(
\begin{array}{c}
 0\\
\phi_0(y)
\end{array}
\right)c_{k,-1} \nonumber\\
&+&
\sum_{n,k,\alpha}
\frac {e^{ikx}}{\sqrt {2L}}
\left(
\begin{array}{c}
 \phi_{n}(y-kl_B^2)\\
\phi_{n+1}(y-kl_B^2)
\end{array}
\right)c_{k,n,\alpha}\, ,
\end{eqnarray}
where $c_{k,n,\alpha}$ destroys an electron in band $\alpha = \pm 1$,
with energy level $n$ and guiding center $kl_B^2$;
$c_{k,-1}$ destroys an electron in the zero Landau level;
the cyclotron frequency is given by (\ref{wc}).
The sum over $n=0,1,2,\ldots,$ is cut off at $n_0$ given by $E(1,n_0)=W$,
where $W$ is of the order of the electronic bandwidth.
In this representation $H_0$ becomes diagonal, leading to
Green's functions of the form (in Matsubara representation):
\begin{equation}
G_0(k,n,\alpha;i\omega)
=\frac 1{i\omega - E(\alpha,n)} \,,
\end{equation}
where
\begin{eqnarray}
E(\alpha,n) = \alpha \omega_c \sqrt{n}
\end{eqnarray}
are the Landau levels for this problem ($\alpha = \pm 1$ labels the two bands).
Notice that $G_0(k,n,\alpha;i\omega)$ is effectively $k$-independent, and
$E(\alpha,-1)=0$ is the zero energy Landau level.
When expressed in the Landau basis, the scattering Hamiltonian
(\ref{llH1}) connects Landau
levels of negative and positive energy.

%%%%%%%%%%%%%%%%%%%%%%%%%%%%%%%%%%%%%%%%%%
% Figure 10                               %
%%%%%%%%%%%%%%%%%%%%%%%%%%%%%%%%%%%%%%%%%%
\begin{figure}[ht]
\begin{center}
\includegraphics*[width=8cm]{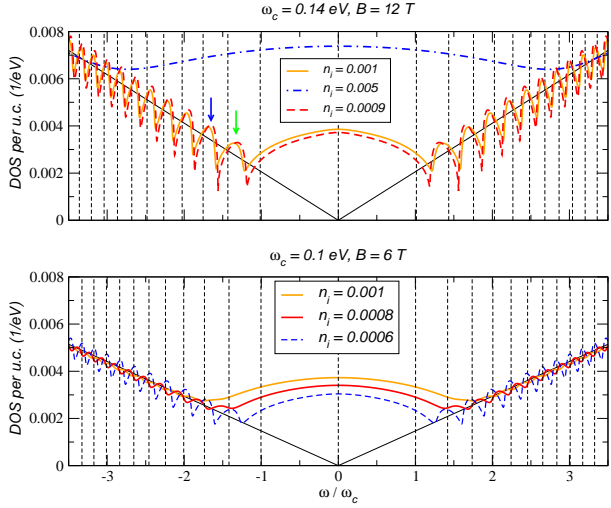}
\end{center}
\caption{\label{fig_dos_landau} (Color online)
Top: Electronic density of states (DOS), $\rho(\omega)$, as a function of $\omega/\omega_c$
($\omega_c = 0.14$ eV) in a magnetic field $B=12$ T
for different impurity concentrations $n_i$. Bottom: $\rho(\omega)$, as a function of $\omega/\omega_c$
($\omega_c = 0.1$ eV is the cyclotron frequency) in a magnetic field $B=6$ T.
The solid line shows
the DOS in the absence of disorder. The position of the Landau levels in the
absence of disorder are shown as vertical lines. The two arrows in the
top panel show the position of the renormalized Landau levels
(see Fig.\ref{fig_self_energy}) given by
the solution of Eq. (\ref{rpole}). Adapted from \onlinecite{PGC06}.}
\end{figure}
% END OF FIGURE 10 %%%%%%%%%%%%%%%%%%%%%%%%

%%%%%%%%%%%%%%%%%%%%%%%%%%%%%%%%%%%%%%%%%%%%%%%%%%%%%%%%%%%%%%%%%%%%%

\subsubsection{The full self-consistent Born approximation (FSBA)}
%%%%%%%%%%%%%%%%%%%%%%%

In order to describe the effect of impurity scattering
on the magnetoresistance of graphene, the Green's function for
Landau levels in the presence of disorder needs to be computed. In the
context of the 2D electron gas, an equivalent study was performed
by Ohta and Ando, \cite{Ohta1,Ohta2,Ando1,Ando2,Ando4,Ando5,Ando6}
using the averaging procedure over impurity positions of Duke \cite{Duke68}.
Below the averaging procedure over impurity positions
is performed in the
standard way, namely, having determined the Green's function for a given
impurity configuration $(\bm r_1,\ldots\bm r_{N_i})$, the position averaged Green's function
is determined from:
\begin{eqnarray}
\langle G(p,n,\alpha;i\omega;\bm r_1,\ldots\bm r_{N_i})\rangle
\equiv G(p,n,\alpha;i\omega) \nonumber\\
= L^{-2N_i}\left[\prod_{j=1}^{N_i}
\int d\bm r_j\right] G(p,n,\alpha;i\omega;\bm r_1,\ldots\bm r_{N_i})\,.
\end{eqnarray}
In
the presence of
Landau levels the average over impurity positions
involves the wavefunctions of the one-dimensional
harmonic oscillator.
After a lengthy algebra, the Green's function in the presence of
vacancies, in the FSBA, can be written as:
\begin{eqnarray}
\label{gn}
G(p,n,\alpha;\omega+0^+)&=&[\omega-E(n,\alpha)-\Sigma_1(\omega)]^{-1}\,,
\\
\label{g0}
G(p,-1;\omega+0^+)&=&[\omega-\Sigma_2(\omega)]^{-1}\,,
\end{eqnarray}
where
\begin{eqnarray}
\label{S1}
\Sigma_1(\omega)&=&- n_i[Z(\omega)]^{-1}\,,\\
\label{S2}
\Sigma_2(\omega)&=&- n_i[g_cG(p,-1;\omega+0^+)/2 + Z(\omega)]^{-1}\,,\\
\label{Z}
Z(\omega)&=&g_cG(p,-1;\omega+0^+)/2\nonumber\\
&+&g_c\sum_{n,\alpha}G(p,n,\alpha;\omega+0^+)/2\,,
\end{eqnarray}
and $g_c=A_c/(2\pi l_B^2)$ is the degeneracy of a Landau level per
unit cell. One should notice that the Green's functions do not depend
upon $p$ explicitly.
The self-consistent solution of Eqs. (\ref{gn}),
(\ref{g0}), (\ref{S1}), (\ref{S2}) and (\ref{Z}) gives
the density of states, the electron self-energy, and
the change of Landau level energy position due to disorder.

The effect of disorder  in the density of states of Dirac fermions
in a magnetic field is shown in Fig.~\ref{fig_dos_landau}.
For reference we note that $E(1,1)=0.14$ eV, for $B=14$ T,
and $E(1,1)=0.1$ eV, for $B=6$ T.
From Fig.~\ref{fig_dos_landau} we see that, for a given $n_i$,
the effect of broadening due to impurities is less effective
as the magnetic field increases. It is also clear that the
position of Landau levels is renormalized relatively to
the non-disordered case. The renormalization of the Landau
level position can be determined from poles of (\ref{gn}) and
(\ref{g0}):
\begin{equation}
\omega-E(\alpha,n)-{\rm Re}\Sigma(\omega)=0\,.
\label{rpole}
\end{equation}
Of course, due to the importance of scattering at low energies,
the solution to Eq. (\ref{rpole}) does not represent exact
eigenstates of system since the imaginary part of the self-energy
is non-vanishing. However, these energies do determine
the form of the density of states, as we discuss below.

\begin{figure}[ht]
\begin{center}
\includegraphics*[width=8cm]{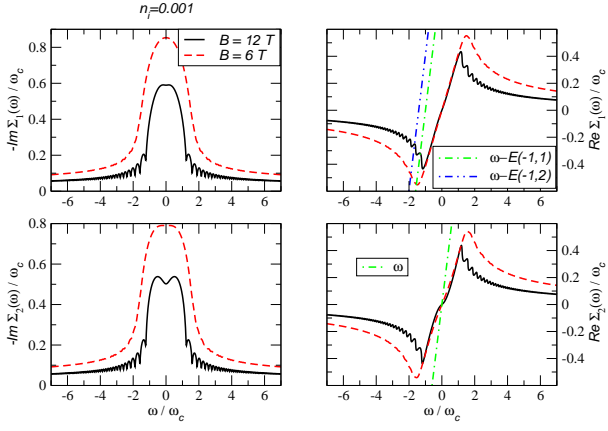}
\end{center}
\caption{\label{fig_self_energy} (Color online) Imaginary (right) and real (left) parts of $\Sigma_1(\omega)$ (top) and $\Sigma_2(\omega)$(bottom),
in units of $\omega_c$, as a function of $\omega/\omega_c$.
The right panels also show the intercept of $\omega-E(\alpha,n)$ with ${\rm Re}\Sigma(\omega)$
as required by Eq. (\ref{rpole}).  Adapted from \onlinecite{PGC06}.}
\end{figure}

In Fig.~\ref{fig_self_energy}, the graphical solution to
Eq. (\ref{rpole}) is given for two different energies
($E(-1,n)$, with $n=1,2$), its is clear that the renormalization
is important for the first Landau level. This result is due to the
increase in scattering at low energies, which is present
already in the case of zero magnetic field. The values of $\omega$
satisfying Eq. (\ref{rpole}) show up in the density of states
as the energy values where the oscillations due to the Landau level
quantization have a maximum. In Fig.~\ref{fig_dos_landau},
the position of the renormalized Landau levels
is shown in the upper panel (marked by two arrows), corresponding to the
bare energies $E(-1,n)$, with $n=1,2$. The importance
of this renormalization decreases with the reduction of
the number of impurities. This is clear in  Fig. \ref{fig_dos_landau}
where a visible shift toward low energies is evident
when $n_i$ has a small 10$\%$ change, from $n_i=10^{-3}$ to $n_i=9 \times 10^{-4}$.

%%%%%%%%%%%%%%%%%%%%%%%%%%%%%%%%%%%%%%%%%%

%%%%%%%%%%%%%%%%%%%%%%%%%%%%%%%%%%%%%%%%%%
% Figure 11                              %
%%%%%%%%%%%%%%%%%%%%%%%%%%%%%%%%%%%%%%%%%%
% END OF FIGURE 11%%%%%%%%%%%%%%%%%%%%%%%%
%%%%%%%%%%%%%%%%%%%%%%%%%%%%%%%%%%%%%%%%%%

The study of the magnetoresistance properties of the system requires the
calculation of the conductivity tensor.
We compute the current-current correlation function
and from it the conductivity tensor is derived.
The details of the calculations are presented in \cite{PGC06}.
If we however neglect the real part of the self-energy,  assume for
 ${\rm Im}\Sigma_{i}(\omega)=\Gamma$  ($i=1,2$)  a constant value,
and consider that  $E(1,1)\gg \Gamma$,
these results reduce to those of \cite{Getal02}.

It is instructive to consider
first the case $\omega,T\rightarrow 0$, which leads to
($\sigma_{xx}(0,0)=\sigma_0$):
\begin{eqnarray}
\sigma_0&=&\frac {e^2}h \frac 4{\pi}
\left[\frac {{\rm Im} \Sigma_1(0)/{\rm Im}\Sigma_2(0)-1}{1+({\rm Im}\Sigma_1(0)/\omega_c)^2
}\right.\nonumber\\
&+&\left.\frac {n_0+1}{n_0+1 + ({\rm Im}\Sigma_1(0)/\omega_c)^2}
\right]\,,
\label{s0B}
\end{eqnarray}
where we include a factor $2$ due to the valley degeneracy. In the absence
of a magnetic field ($\omega_c \to 0$) the above expression reduces to:
\begin{eqnarray}
\sigma_0&=&\frac {e^2}h \frac 4{\pi}
\left[1-\frac{[{\rm Im}\Sigma_1(0)]^2}{(v_F \Lambda)^2+[{\rm Im}\Sigma_1(0)]^2}
\right]\,,
\label{s0B0}
\end{eqnarray}
where we have introduced the energy cut-off, $v_F \Lambda$.
Either when ${\rm Im} \Sigma_1(0)\simeq{\rm Im}\Sigma_2(0)$
and $\omega_c\gg {\rm Im}\Sigma_1(0)$ (or $n_0\gg {\rm
  Im}\Sigma_1(0)/\omega_c$, $\omega_c=E(0,1)=\sqrt 2 v_F /l^2_B$), or
when $\Lambda v_F \gg {\rm Im}\Sigma_1(0)$, in the absence of an applied
field, Eqs. (\ref{s0B}) and (\ref{s0B0}) reduce to:
\begin{eqnarray}
\sigma_0 = \frac{4}{\pi} \frac{e^2}{h} \, ,
\label{unisig}
\end{eqnarray}
which is the so-called universal conductivity of graphene
\cite{F86,F86b,PLee93,LFSG94,NTW94,Ziegler98,Yang02,PGC06,TTTRB06,K06}.
This result was obtained previously by Ando and collaborators
using the second order self-consistent Born approximation \cite{Ando7,Ando8}.

%%%%%%%%%%%%%%%%%%%%%%%%%%%%%%%%%%%%%%%%%%
% Figure 12                              %
%%%%%%%%%%%%%%%%%%%%%%%%%%%%%%%%%%%%%%%%%%
\begin{figure}[ht]
\begin{center}
\includegraphics*[width=8cm]{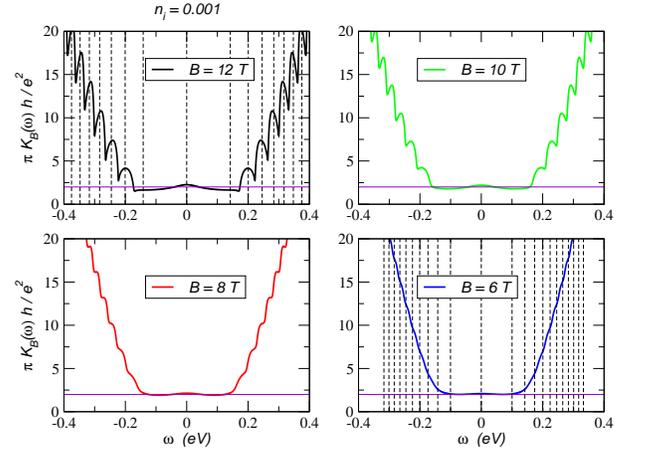}
\end{center}
\caption{\label{fig_kernel_landau} (Color online) Conductivity kernel, $K(\omega)$ (in units of $e^2/(\pi h)$), as a function of energy $\omega$ for
different magnetic fields and for $n_i=10^{-3}$. The horizontal lines mark the universal limit of the conductivity per cone, $\sigma_0=2 e^2/(\pi h)$. The vertical lines show the position of the Landau
levels in the absence of disorder.  Adapted from \onlinecite{PGC06}.}
\end{figure}
% END OF FIGURE 12 %%%%%%%%%%%%%%%%%%%%%%%
%%%%%%%%%%%%%%%%%%%%%%%%%%%%%%%%%%%%%%%%%%

Because the DC magnetotransport properties of graphene
are normally measured with the possibility of tuning its
electronic density by a gate potential \cite{Netal04},
it is  important to compute the conductivity kernel,
since this has direct experimental relevance. In the
the case  $\omega\rightarrow 0$ we write the conductivity
$\sigma_{xx}(0,T)$ as:
\begin{equation}
\sigma_{xx}(0,T)= \frac {e^2}{\pi h}\int_{-\infty}^{\infty}
d\epsilon \frac {\partial f(\epsilon)}{\partial \epsilon}
K_B(\epsilon)\,,
\label{sxxb}
\end{equation}
where the conductivity kernel $K_B(\epsilon)$ is given
in the  Appendix of Ref. \cite{PGC06}. The magnetic field dependence of
kernel $K_B(\epsilon)$
is shown in Fig. \ref{fig_kernel_landau}. Observe
that the effect of disorder is the creation of a region
where $K_B(\epsilon)$ remains constant before
it starts to increase in energy with superimposed oscillations
coming from the Landau levels.
The same effect, but with the absence of the oscillations,
was identified at the
level of the self-consistent density of states plotted
in Fig. \ref{fig_dos_landau}. Together with $\sigma_{xx}(0,T)$,
the Hall conductivity $\sigma_{xy}(0,T)$ allows the calculation
of the resistivity tensor (\ref{rhosig}).

\begin{figure}[ht]
\begin{center}
\includegraphics*[width=8cm]{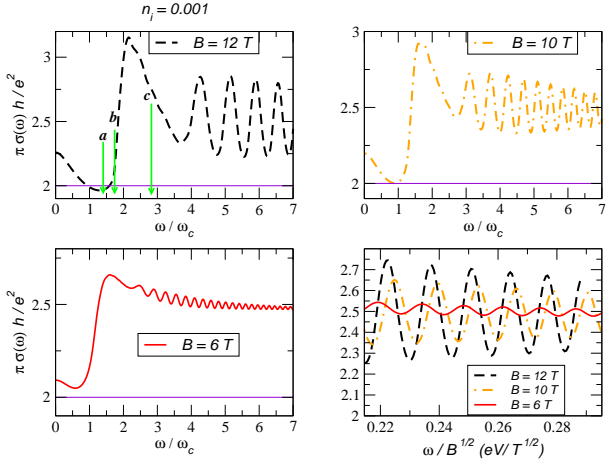}
\end{center}
\caption{\label{fig_sigomega_landau} (Color online) Frequency dependent
conductivity per cone, $\sigma(\omega)$ (in units of $e^2/(\pi h)$) at $T=10$ K
and $n_i = 10^{-3}$, as a function of the energy $\omega$
(in units of $\omega_c$) for different values of the magnetic field $B$.
The vertical arrows in the upper left panel, labeled {\bf a}, {\bf b},
and {\bf c}, show the positions of the transitions between different
Landau levels: $E(1,1)-E(-1,0)$, $E(2,1)-E(-1,0)$, and $E(1,1)-E(1,-1)$,
respectively. The horizontal continuous lines show the value of the
universal conductivity. The lower right panel shows the conductivity
for different values of magnetic field as a function of $\omega/\sqrt{B}$.
 Adapted from \onlinecite{PGC06}.}
\end{figure}

Let us now focus on the optical conductivity, $\sigma_{xx}(\omega)$ \cite{PGC06,GSC07}.
This quantity can be probed by reflectivity experiments in the subterahertz
to mid-infrared frequency range \cite{B05}. This quantity is
 depicted in Fig. \ref{fig_sigomega_landau} for
different magnetic fields.
It is clear that the first peak is controlled
by the $E(1,1)-E(1,-1)$, and we have checked that it does not
obey any particular scaling form as a function of
$\omega/B$. On the other hand, as the effect of scattering
becomes less important the high energy conductivity
oscillations
start obeying the scaling $\omega/\sqrt{B}$, as we show
in the lower right panel of Fig.~\ref{fig_sigomega_landau}.

%%%%%%%%%%%%%%%%%%%%%%%%%%%%%%%%%%%%%%%%%%
% Figure 13                           %
%%%%%%%%%%%%%%%%%%%%%%%%%%%%%%%%%%%%%%%%%%
% END OF FIGURE 13 %%%%%%%%%%%%%%%%%%%%%%%
%%%%%%%%%%%%%%%%%%%%%%%%%%%%%%%%%%%%%%%%%%
%%%%%%%%%%%%%%%%%%%%%%%%%%%%%%%%%%%%%%%%%%

\section{Many-body effects}
\label{manybody}

\subsection{Electron-phonon interactions}
\label{elec_pho}

In Sec.~\ref{curvature} and Sec.~\ref{strain} we discussed how static deformations
of the graphene sheet due to bending and strain couple to the Dirac fermions
via vector potentials. Just as bending has to do with the flexural modes of the
graphene sheet (as discussed in Sec.~\ref{flex}), strain fields are related
to optical and acoustic modes \cite{rubio}. Given the local displacements of the atoms
in each sublattice, ${\bf u}_A$ and ${\bf u}_B$, the electron-phonon coupling
has essentially the form discussed previously for static fields.

The coupling to acoustic modes is the most straightforward one, since it already
appears in the elastic theory. If ${\bf u}_{{\rm ac}}$ is the acoustic phonon
displacement, then the relation between this displacement and the atom displacement
is given by equation (\ref{uac}), and its coupling to electrons is given by
the vector potential (\ref{vecac}) in the Dirac equation (\ref{hodir}).

For optical modes the situation is slightly different since the optical mode
displacement is \cite{ando_opt,ando_opt_bi}:
\begin{eqnarray}
{\bf u}_{{\rm op}} = \frac{1}{\sqrt{2}} ({\bf u}_A - {\bf u}_B) \, ,
\label{uop}
\end{eqnarray}
that is, the bond length deformation vector. To calculate the coupling to the
electrons we can proceed as previously and compute the change in the nearest
neighbor hopping energy due to the lattice distortion through (\ref{tijdel}),
(\ref{dtab}), (\ref{dui}), and (\ref{uop}). Once again the electron-phonon interaction
becomes a problem of the coupling of the electrons with a vector potential
as in (\ref{hodir}) where the components of the vector potential are:
\begin{eqnarray}
{\cal A}^{({\rm op})}_x &=& - \sqrt{\frac{3}{2}} \frac{\beta}{a^2} u^{{\rm op}}_y  \, ,
\nonumber
\\
{\cal A}^{({\rm op})}_y &=&  - \sqrt{\frac{3}{2}} \frac{\beta}{a^2} u^{{\rm op}}_x \, ,
\label{aopt}
\end{eqnarray}
where $\beta = \partial t/\partial \ln(a)$ was defined in (\ref{beta}). Notice
that we can write:
$\vec{{\cal A}}^{{\rm op}} = - \sqrt{3/2} (\beta/a^2) \, \vec{\sigma} \times {\bf u}_{{\rm op}}$.
A similar expression is valid close to the K' point with $\vec{{\cal A}}$ replaced by
$-\vec{{\cal A}}$.

Optical phonons are particularly important in graphene research because of
Raman spectroscopy. The latter has played a particularly important role in the
study of carbon nanotubes \cite{nanotubes} because of the 1D character of these
systems, namely, the presence of van Hove singularities in the 1D spectrum
lead to colossal enhancements of the Raman signal that can be easily
detected, even for a single isolated carbon nanotube. In graphene the
situation is rather different since its 2D character leads to a much smoother
density of states (except for the van Hove singularity at high energies of
the order of the hopping energy $t \approx 2.8$ eV). Nevertheless, graphene is an
open surface and hence is readily accessible by Raman spectroscopy. In fact,
it has played a very important role
because it allows the identification of the number of planes
\cite{FMSCLMPJNG06,GCJTE06,YZKP07,PLCNGFM07,GMESJHW07},
and the study of the optical phonon modes in graphene, particularly the ones
in the center of the BZ with momentum $q \approx 0$. Similar studies
have been performed in graphite ribbons \cite{Cancado04}.

\begin{figure}[htb]
\begin{center}
\includegraphics[width=8cm]{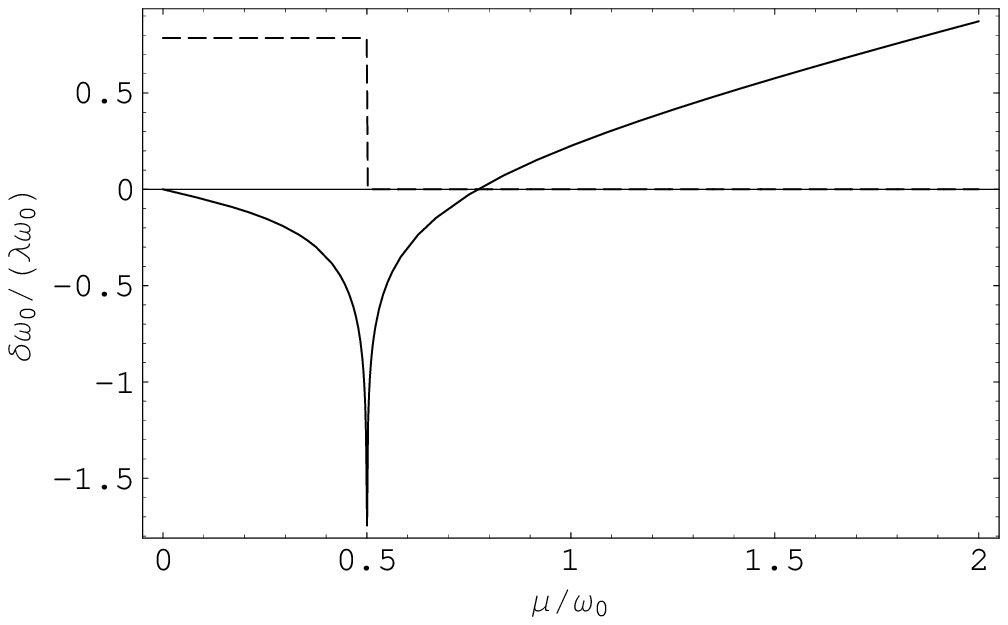}
\includegraphics[width=3cm]{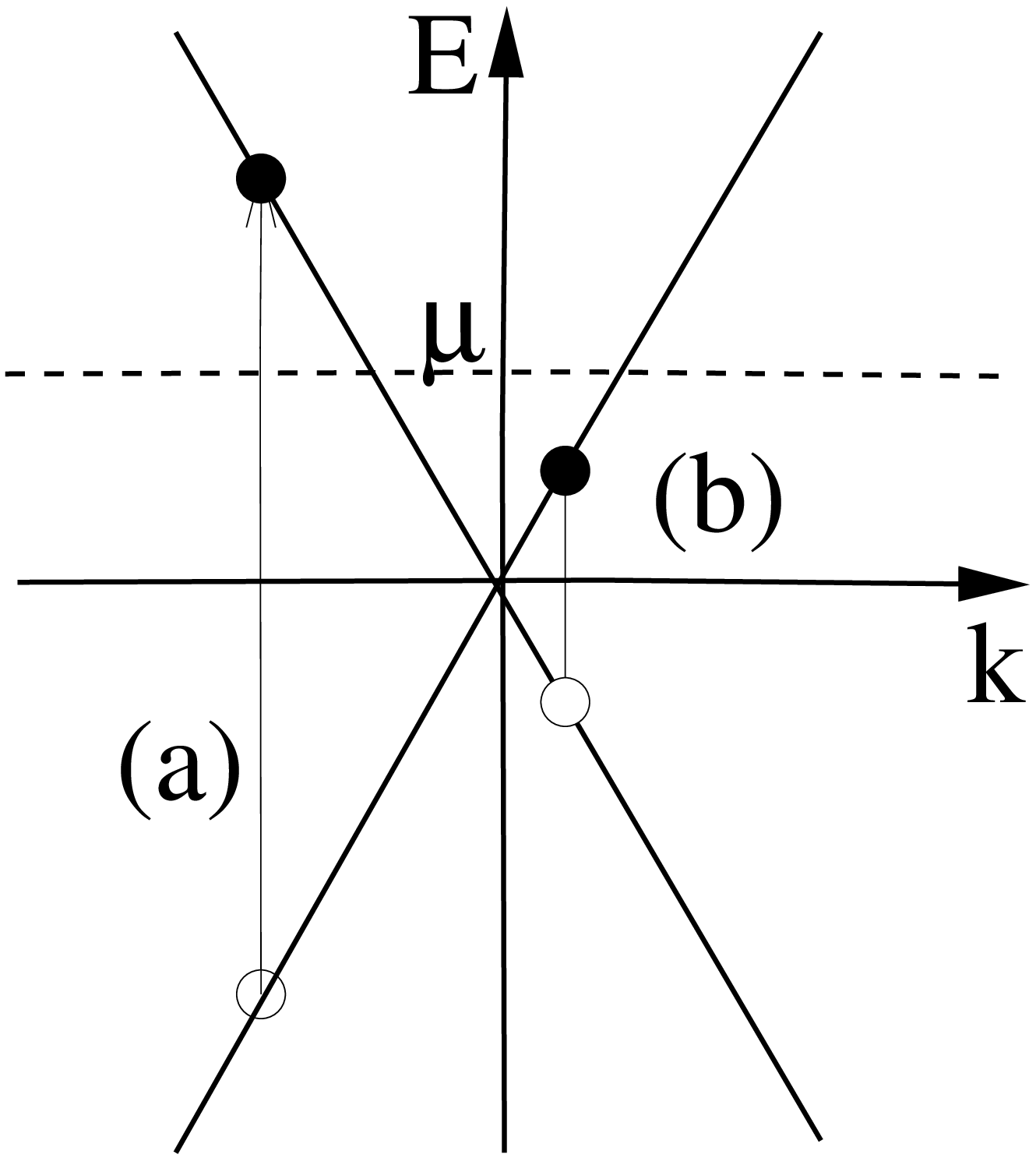}
\caption[fig]{\label{wshift} (Color online) Top: The continuous line is the
relative phonon frequency shift
as a function of $\mu/\omega_0$, and the dashed line is the damping of
the phonon due to electron-hole pair creation;
Bottom: (a) Electron-hole process that
leads to phonon softening ($\omega_0>2 \mu$), and (b) electron-hole process
that leads to phonon hardening ($\omega_0<2 \mu$).}
\end{center}
\end{figure}

Let us consider the effect of the Dirac fermions on the optical modes. If one
treats the vector potential, electron-phonon coupling, (\ref{hodir}) and
(\ref{aopt}) up to second order perturbation theory, its main effect is
the polarization of the electron system by creating electron-hole pairs.
In the QED language, the creation electron-hole  pairs is called pair
(electron/anti-electron) production \cite{CN07}. Pair production is
equivalent to a renormalization of the phonon propagator by a self-energy
that is proportional to the polarization function of the Dirac fermions.

The renormalized phonon frequency, $\Omega_0({\bf q})$, is given
by \cite{ando_opt,ando_opt_bi,CMG07,LM06,SWKS07}:
\begin{eqnarray}
\Omega_0({\bf q}) \approx \omega_0 -\frac{2 \beta^2}{a^2 \omega_0}
\chi({\bf q},\omega_0) \, ,
\label{fren}
\end{eqnarray}
where $\omega_0$ is the bare phonon frequency, and the electron-phonon
polarization function is given by:
\begin{eqnarray}
\chi({\bf q},\omega) \!=\!\!\!
\sum_{s,s'=\pm 1} \!\! \int \!\frac{d^2 {\bf k}}{(2 \pi)^2} \!
\frac{f[E_{s}({\bf k}+{\bf q})]\!-\!f[E_{s'}({\bf k})]}{\omega_0\!-\!E_{s}({\bf
    k}+{\bf q})\!+\!E_{s'}({\bf k})\!+\!i \eta} \, ,
\label{chipho}
\end{eqnarray}
where $E_s({\bf q})$ is the Dirac fermion dispersion ($s=+1$ for the upper
band, and $s=-1$ for the lower band), and
$f[E]$ is the Fermi-Dirac distribution function. For Raman spectroscopy, the response
of interest is at $q=0$ where clearly only the interband processes
such that $s s' = -1$ (that is, processes between the lower and upper cones)
contribute. The electron-phonon polarization function can be easily
calculated using the linearized Dirac fermion dispersion (\ref{eq:conical})
and the low energy density of states (\ref{rho0}):
\begin{eqnarray}
\chi(0,\omega_0) \!&=&\! \frac{6 \sqrt{3}}{\pi v_F^2}
\int_0^{v_F \Lambda}\!\!\!\!dE E \left(f[-E]\!-\!f[E]\right)
\left(\frac{1}{\omega_0\!+\!2E\!+\!i \eta}
\right.
\nonumber
\\
&-& \left. \frac{1}{\omega\!-\!2E\!+\!i \eta}\right) \, ,
\label{cho0pho}
\end{eqnarray}
where we have introduced the cut-off momentum $\Lambda$ ($\approx 1/a$)
so that the integral converges in the ultraviolet. At zero temperature,
$T=0$, we have $f[E] = \theta(\mu-E)$ and we assume electron doping, $\mu
>0$, so that $f[-E]=1$ (for the case of hole doping, $\mu<0$, is obtained by
electron-hole symmetry). The integration in (\ref{cho0pho}) gives:
\begin{eqnarray}
\chi(0,\omega_0) &=& \frac{6 \sqrt{3}}{\pi v_F^2}
\left[
v_F \Lambda - \mu + \frac{\omega_0}{4}
\left(\ln\left|\frac{\omega_0/2+\mu}{\omega_0/2-\mu}\right|
\right.\right.
\nonumber
\\
&+& \left. \left. i \pi \theta\left(\omega_0/2-\mu\right)\right)\right] \, ,
\end{eqnarray}
where the cut-off dependent term is a contribution coming from the
occupied states in the lower $\pi$ band and hence is independent
of the value of the chemical potential. This contribution can be
fully incorporated into the bare value of $\omega_0$ in (\ref{fren}).
Hence the relative shift in the phonon frequency can be written as:
\begin{eqnarray}
\frac{\delta \omega_0}{\omega_0} \approx - \frac{\lambda}{4}
\left(-\frac{\mu}{\omega_0}+\ln\left|\frac{\omega_0/2+\mu}{\omega_0/2-\mu}
\right| +i \pi \theta\left(\omega_0/2-\mu\right)\right) \, ,
\label{w0shift}
\end{eqnarray}
where
\begin{eqnarray}
\lambda = \frac{36 \sqrt{3}}{\pi} \frac{\beta^2}{8 M a^2 \omega_0} \, ,
\label{epc}
\end{eqnarray}
is the dimensionless electron-phonon coupling. Notice that (\ref{w0shift})
has a real and imaginary part. The real part represents the actual shift in
frequency, while the imaginary part gives the damping of the phonon mode due to
pair production (see Fig.~\ref{wshift}). There is a clear change in behavior
depending whether $\mu$ is larger or smaller than $\omega_0/2$. For $\mu <
\omega_0/2$ there is a decrease in the phonon frequency implying that the
lattice is softening, while for $\mu > \omega_0/2$ the lattice hardens. The
interpretation for this effect is also given in Fig.~\ref{wshift}. On the one
hand, if the
frequency of the phonon is larger than twice the chemical potential, real
electron-hole pairs are produced, leading to stronger screening of the
electron-ion interaction and hence, to a softer phonon mode. At the same
time the phonons become damped and decay. On the other hand,
if the frequency of the phonon is smaller than the twice the chemical
potential, the production of electron-hole pairs is halted by the Pauli
principle and only virtual excitations can be generated leading to
polarization and lattice hardening. In this case, there is no damping and
the phonon is long lived. This amazing result has been observed
experimentally by Raman spectroscopy \cite{YZKP07,PLCNGFM07}. Electron-phonon
coupling has also been investigated theoretically in the case of a finite
magnetic field \cite{A07,GFKF07}. In this case, resonant coupling occurs
due to the large degeneracy of the Landau levels and different Raman
transitions are expected as compared with the zero-field case. The coupling
of electrons to flexural modes on a free standing graphene sheet was discussed
in ref.~\cite{mVO07}.

\subsection{Electron-electron interactions}
\label{elec_elec}

Of all disciplines of condensed matter physics, the study of
electron-electron interactions is probably one of the most
complex since it involves the understanding of the behavior
of a macroscopic number of variables. Hence, the problem of
interacting systems is a field in constant motion and we
shall not try to give here a comprehensive survey of the problem for graphene.
Instead, we will focus on a small number of topics that are
of current discussion in the literature.

Since graphene is a truly 2D system, it is informative
to compare it with the more standard 2DEG that has been studied
extensively in the last 25 years since the development of heterostructures
and the discovery of the quantum Hall effect (for a review, see \cite{iqhe_si}). At
the simplest level, metallic systems have two main kind of excitations:
electron-hole pairs and collective modes such as plasmons.

\begin{figure}[]
\begin{center}
\includegraphics[width=8cm]{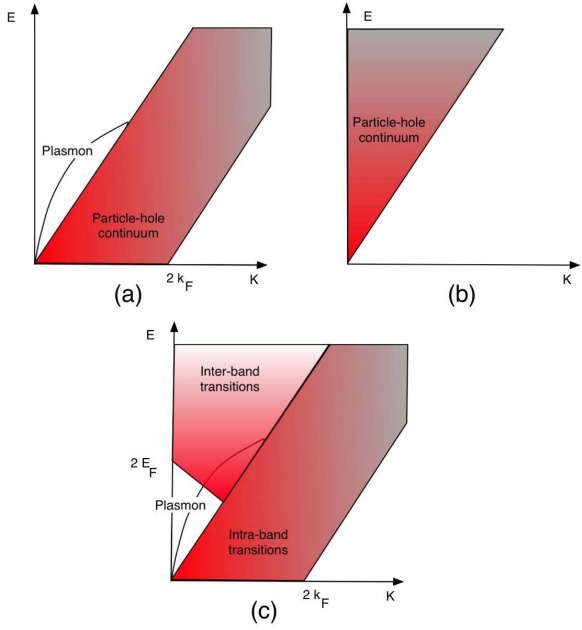}
\caption[fig]{\label{phc} (Color online) Electron-hole continuum and collective modes of:
(a) a 2DEG; (b) undoped graphene; (c) doped graphene.}
\end{center}
\end{figure}

Electron-hole pairs are incoherent excitations of the Fermi sea and a
direct result of Pauli's exclusion principle: an electron inside the
Fermi sea with momentum ${\bf k}$ is excited outside the Fermi sea to
a new state with momentum ${\bf k}+{\bf q}$, leaving a hole behind.
The energy associated with such an excitation is simply:
$\omega = \epsilon_{{\bf k}+{\bf q}} - \epsilon_{\bf k}$ and for
states close to the Fermi surface (${\bf k} \approx {\bf k}_F$) their
energy scales linearly with the excitation momentum, $\omega_q \approx v_F
q$. In a system with non-relativistic dispersion such as normal metals
and semiconductors, the electron-hole continuum is made out of intra-band
transitions only and exists even at zero energy since it is always
possible to produce electron-hole pairs with arbitrarily low energy
close to the Fermi surface, as shown in Fig.~\ref{phc}(a). Besides that,
the 2DEG can also sustain collective excitations such as plasmons that
have dispersion: $\omega_{{\rm plasmon}}(q) \propto \sqrt{q}$, and exist
outside the electron-hole continuum at sufficiently long wavelengths
\cite{S86}.

In systems with relativistic-like dispersion, such as graphene, these
excitations change considerably, especially when the Fermi energy
is at the Dirac point. In this case the Fermi surface shrinks to
a point and hence intra-band excitations disappear and only interband
transitions between the lower and upper cones can exist (see
Fig.\ref{phc}(b)). Therefore, neutral graphene has no electron-hole
excitations at low energy, instead each electron-hole pair costs energy
and hence the electron-hole occupies the upper part of the energy versus
momentum diagram. In this case, plasmons are suppressed and no coherent
collective excitations can exist. If the chemical potential is moved
away from the Dirac point then intra-band excitations are restored
and the electron-hole continuum of graphene shares features of the
2DEG and undoped graphene. The full electron-hole continuum of doped
graphene is shown in Fig. \ref{phc}(c), and in this case plasmon modes
are allowed. As the chemical potential is raised away from the Dirac
point, graphene resembles more and more the 2DEG.

These features in the elementary excitations of graphene reflect its
screening properties as well. In fact, the polarization and dielectric
functions of undoped graphene are rather different from the ones of the
2DEG (Lindhard function). In the random phase approximation (RPA),
the polarization function can be calculated analytically \cite{S86,GGV93,GGV94}:
\begin{eqnarray}
\Pi(q,\omega) = \frac{q^2}{4 \sqrt{v_F^2 q^2-\omega^2}} \, ,
\end{eqnarray}
and hence, for $\omega > v_F q$ the polarization function is
imaginary indicating the damping of electron-hole pairs. Notice
that the static polarization function ($\omega=0$) vanishes
linearly with $q$, indicating the lack of screening in the system.
This polarization function has been also calculated in the
presence of a finite chemical potential \cite{A06,S86,S86b,WSSG06,HdS06}.

Undoped, clean graphene is a semimetal, with a
vanishing density of states at the Fermi level. As a result the {\it linear}
Fermi Thomas screening length diverges, and the long range Coulomb
interaction is not screened. At finite electron density $n$,
the Thomas-Fermi screening length reads \label{A06,NMac06}:
\begin{eqnarray}
\lambda_{TF} \approx \frac{1}{4 \alpha} \frac{1}{k_F} = \frac{1}{4 \alpha} \frac{1}{\sqrt{\pi n}} \, ,
\label{ltf}
\end{eqnarray}
where
\begin{eqnarray}
\alpha = \frac{e^2}{\epsilon_0  v_F} \,,
\label{alfacou}
\end{eqnarray}
is the dimensionless coupling constant in the problem
(the analogue of (\ref{gcou}) in the Coulomb impurity problem).
Going beyond the linear Thomas-Fermi regime, it has been shown that the Coulomb law is modified
\cite{FNS07,Kscreening,ZF07}.

The Dirac Hamiltonian in the presence of interactions can be
written as:
\begin{eqnarray}
{\cal H} &\equiv& - i \vf \int d^2 {\bf r} \, \hat{\Psi}^{\dag} ({\bf r} )
\bm \sigma \cdot \nabla \hat{\Psi} ( {\bf r} )
\nonumber
\\
&&+ \frac{e^2}{2 \epsilon_0} \int d^2
{\bf r} d^2 {\bf r}'  \frac{1}{| {\bf r}
- {\bf r}' |} \hat{\rho}({\bf r})\hat{\rho}({\bf r'}) \, ,
\end{eqnarray}
where
\begin{eqnarray}
\hat{\rho}({\bf r}) = \hat{\Psi}^{\dag} ( {\bf r} ) \hat{\Psi}
( {\bf r} ) \, ,
\end{eqnarray}
is the electronic density. Observe that Coulomb interaction, unlike in QED,
is assumed to be instantaneous since $v_F/c \approx 1/300$ and hence
retardation effects are very small. Moreover, the photons propagate
in 3D space whereas the electrons are confined to the 2D graphene sheet.
Hence, the Coulomb interaction breaks the Lorentz invariance of the problem
and makes the many-body situation rather different from the one in QED \cite{BC76}.
Furthermore, the problem depends
on two parameters, $\vf$ and $e^2/\epsilon_0$. Under
a dimensional scaling, ${\bf r} \rightarrow \lambda {\bf r} , t
\rightarrow \lambda t , \Psi \rightarrow \lambda^{-1} \Psi$, both
parameters remain invariant. In RG language, the
Coulomb interaction is a marginal variable, whose strength relative
to the kinetic energy does not change upon a change in scale. If the
units are chosen in such a way that $\vf$ is dimensionless, the
value of $e^2/\epsilon_0$ will also be rendered dimensionless. This is the case
in theories considered renormalizable in quantum field theory.

The Fermi velocity in graphene is comparable to that in
half-filled metals. In solids with lattice constant $a$, the
total kinetic energy per site, $1/ ( m a^2 )$, where $m$ is
the bare mass of the electron, is of the same order of magnitude as
the electrostatic energy, $e^2 / (\epsilon_0 a)$. The Fermi velocity for fillings
of the order of unity is $\vf \sim 1/ ( m a )$. Hence, $e^2 / (\epsilon_0 \vf) \sim
1$. This estimate is also valid in graphene. Hence, unlike in
QED, where $\alpha_{{\rm QED}} = 1 / 137$, the coupling
constant in graphene is $\alpha \sim 1$.

Despite the fact that the coupling constant is of the order of unity, a
perturbative RG analysis can be applied. RG
techniques allow us to identify stable fixed points of the model,
which may be attractive over a broader range than the one where a
perturbative treatment can be rigorously justified. Alternatively,
an RG scheme can be reformulated as the process of piecewise
integration of high energy excitations \cite{S94}. This procedure
leads to changes in the effective low energy couplings. The scheme is
valid if the energy of the renormalized modes is much larger than
the scales of interest.

\begin{figure}[]
\begin{center}
\includegraphics[width=5cm]{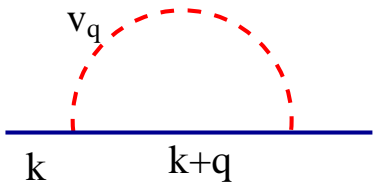}
\caption[fig]{\label{diagram_HF} Hartree-Fock self-energy diagram
which leads to a logarithmic renormalization of the Fermi velocity.}
\end{center}
\end{figure}

The Hartree-Fock correction due to Coulomb interactions between electrons (given by
the diagram in Fig.~\ref{diagram_HF}) gives a logarithmic correction to the
electron self-energy \cite{GGV94}:
\begin{equation}
\Sigma_{HF} ( {\bf k} ) = \frac{\alpha}{4} k \ln \left(
\frac{\Lambda}{k} \right)
\label{gonza}
\end{equation}
where $\Lambda$ is a momentum cutoff which sets the range of
validity of the Dirac equation. This result remains true even
to higher order in perturbation theory \cite{Mish07} and
is also obtained in large $N$ expansions \cite{RWP89,RWP91,DTS07} ($N$ is the number of flavors of
Dirac fermions), with the only modification
being the prefactor in (\ref{gonza}).
This result implies that the Fermi
velocity is renormalized towards higher values. As a consequence,
the density of states near the Dirac energy is reduced, in agreement
with the general trend of repulsive interactions to induce or
increase gaps.

This result can be understood from the RG point of view by
studying the effect of reducing the cut-off from $\Lambda$
to $\Lambda-d\Lambda$ and its effect on the effective
coupling. It can be shown that $\alpha$ obeys the
equation \cite{GGV94}:
\begin{equation}
\Lambda \frac{\partial \alpha}{\partial \Lambda} = -
\frac{\alpha}{4} \, .
\end{equation}
Therefore, the Coulomb interaction becomes marginally
irrelevant. These features are confirmed by a full
relativistic calculation, although the Fermi velocity cannot,
obviously, surpass the velocity of light \cite{GGV94}. This
result indicates that strongly correlated electronic phases, such as
ferromagnetism \cite{PGN05b} and Wigner
crystals \cite{DJBB06} are suppressed in clean graphene.

A calculation of higher order self-energy terms leads to a
wavefunction renormalization, and to a finite quasiparticle
lifetime, which grows linearly with quasiparticle
energy \cite{GGV94,GGV96}. The wavefunction renormalization implies
that the quasiparticle weight tends to zero as its energy is
reduced. A strong coupling expansion is also possible, assuming that
the number of electronic flavors justifies an RPA expansion, keeping
only electron-hole bubble diagrams \cite{GGV99}. This analysis
confirms that the Coulomb interaction is renormalized towards lower
values.

The enhancement in the Fermi velocities leads to a widening of the
electronic spectrum. This is consistent with measurements of the
gaps in narrow single wall nanotubes, which show deviations from the
scaling with $R^{-1}$, where $R$ is the radius, expected from the
Dirac equation \cite{KM04}. The linear dependence of the inverse
quasiparticle lifetime with energy is consistent with photo-emission
experiments in graphite, for energies larger with respect to the
interlayer interactions \cite{Xetal96,Zetal06,ZGL06,Betal07,SSSTS07}.
Note that in graphite, band structure effects modify the lifetimes at
low energies \cite{SCABEL01}. The vanishing of the quasiparticle peak
at low energies can lead to an energy dependent renormalization of
the interlayer hopping \cite{VLG02,VLG03}. Other thermodynamic
properties of undoped and doped graphene can also be
calculated \cite{BPPAM07,V07}.

Non-perturbative calculations of the effects of the long range
interactions in undoped graphene show that a transition to a gapped
phase is also possible, when the number of electronic flavors is
large \cite{K01,LK04,KS06}. The broken symmetry phase is similar to
the excitonic transition found in materials where it becomes
favorable to create electron-hole pairs that then form bound
excitons (excitonic transition).

Undoped graphene cannot have well defined plasmons, as their
energies fall within the electron-hole continuum, and therefore have a
significant Landau damping. At finite temperatures, however,
thermally excited quasiparticles screen the Coulomb interaction, and
an acoustic collective charge excitation can exist \cite{V06}.

Doped graphene shows a finite density of states at the Fermi level,
and the long range Coulomb interaction is screened.
Accordingly, there are collective plasma interactions near ${\bf q} \rightarrow
0$, which disperse as $\omega_p \sim \sqrt{| {\bf q} |}$, since
the system is 2D \cite{S86,S86b,CT89}. The fact that the
electronic states are described by the massless Dirac equation
implies that $\omega_P \propto n^{1/4}$, where $n$ is the carrier
density. The static dielectric constant has a continuous derivative
at $2 \kf$, unlike in the case of the 2D electron gas \cite{A06,WSSG06,SHT07}.
This fact is associated with the
suppressed backward scattering in graphene. The simplicity of the
band structure of graphene allows analytical calculation of
the energy and momentum dependence of the dielectric
function \cite{WSSG06,SHT07}.
The screening of the long-range Coulomb interaction implies that the
low energy quasiparticles show a quadratic dependence on energy with
respect to the Fermi energy \cite{HHS07}.

One way to probe the strength of the electron-electron interactions
is via the electronic compressibility. Measurements of the compressibility
using a single electron transistor (SET) show very little sign of interactions
in the system, being well fitted by the non-interacting result that,
contrary to the two-dimensional electron gas (2DEG) \cite{EPW94,giovani},
is positively divergent \cite{MAULSKY07,PABPM07}. Bilayer graphene, on the
other hand, shares characteristics of the single layer and the 2DEG
with a non-monotonic dependence of the compressibility on the carrier density
\cite{KNCN07}. In fact, bilayer graphene very close to
half-filling has been predicted to be unstable towards Wigner crystallization
\cite{DWBZB07}, just like the 2DEG. Furthermore, according to Hartree-Fock
calculations, clean bilayer graphene is unstable towards ferromagnetism
\cite{NCNPG05}.

\subsubsection{Screening in graphene stacks}
\label{screen_stack}

The electron-electron interaction leads to the screening of external
potentials. In a doped stack, the charge tends to accumulate
near the surfaces, and its distribution is determined by the
dielectric function of the stack in the out-of-plane direction. The
same polarizability describes the screening of an external field
perpendicular to the layers, like the one induced by a gate in
electrically doped systems \cite{Netal04}. The self-consistent
distribution of charge in a biased graphene bilayer has been studied
in ref. \cite{MC06}. From the observed charge distribution and
self-consistent calculations, an estimate of the band structure
parameters and their relation with the induced gap can be
obtained \cite{Cetal06}.

In the absence of interlayer hopping, the polarizability of a set of
stacks of 2D electron gases can be written as a sum of the
screening by the individual layers. Using the accepted values for
the effective masses and carrier densities of graphene, this scheme
gives a first approximation to screening in graphite \cite{VF71}. The
screening length in the out of plane direction is of about 2 graphene
layers \cite{Metal05c}. This model is easily generalizable to a stack of semimetals
described by the 2D Dirac equation \cite{GGV01}. At half
filling, the screening length in all directions diverges, and the
screening effects are weak.

Interlayer hopping modifies this picture significantly. The hopping
leads to coherence \cite{G06}. The out of plane electronic dispersion
is similar to that of a one dimensional conductor. The out of plane
polarizability of a multilayer contains intra- and interband
contributions. The subbands in a system with the Bernal stacking
have a parabolic dispersion, when only the nearest neighbor hopping
terms are included. This band structure leads to an interband
susceptibility described by a sum of terms like those in
(\ref{susc_bilayer}), which diverges at half-filling. In an
infinite system, this divergence is more pronounced at $k_\perp =
\pi / c$, that is, for a wave vector equal to twice the distance
between layers. This effect greatly enhances Friedel like
oscillations in the charge distribution in the out of plane
direction, which can lead to the changes in the sign of the charge
in neighboring layers \cite{G06}. Away from half-filling a graphene
bilayer behaves, from the point of view of screening, in a way very
similar to the 2DEG \cite{WC07b}.

%------------------------------------------------------------------------------
\subsection{Short range interactions}
\label{short}
%------------------------------------------------------------------------------

In this section we discuss the effect of short range
Coulomb interactions on the physics of graphene.
The simplest carbon system with a hexagonal shape is the
benzene molecule. The value of the Hubbard interaction among $\pi$-electrons
was, for this system,
computed long ago by Parr {\it et al.} \cite{Parr50}, yielding
a value of $U=16.93$ eV. For comparison purposes, in polyacetylene the value for the Hubbard
interaction is $U\simeq$10 eV and the hopping energy is
$t\approx$ 2.5 eV \cite{Baeriswyl86}. These two examples just show that
the value of the on-site Coulomb interaction is fairly large for
$\pi-$electrons. As a first guess for graphene, one can take
$U$ to be of the same order as for polyacethylene, with the hopping
integral $t\simeq$ 2.8 eV. Of course in pure graphene the
electron-electron interaction is not screened, since
the density of states is zero at the Dirac point, and one should work
out the effect of Coulomb interactions by considering the bare Coulomb
potential. On the other hand, as we have seen before,
defects induce a finite density of states at the Dirac point, which
could lead to an effective screening of the long-range Coulomb
interaction. Let us assume that the bare Coulomb interaction
is screened in graphene and that Coulomb interactions are represented
by the Hubbard interaction. This means that we must add to the
Hamiltonian (\ref{H1}) a term of the form:
\begin{eqnarray}
H_U&=& U \, \sum_{\bm R_i}\left[a^\dag_{\uparrow}(\bm R_i)a_{\uparrow}(\bm R_i)
a^\dag_{\downarrow}(\bm R_i)a_{\downarrow}(\bm R_i)
\right.
\nonumber\\
&+&\left.b^\dag_{\uparrow}(\bm R_i)b_{\uparrow}(\bm R_i)
b^\dag_{\downarrow}(\bm R_i)b_{\downarrow}(\bm R_i)\right]
\end{eqnarray}
The simplest question one can ask is whether this system shows a tendency
toward some kind of magnetic order driven by the interaction
$U$. Within the simplest Hartree-Fock approximation \cite{PAD04},
the instability line toward ferromagnetism is given by:
\begin{equation}
U_F(\mu)=\frac 2 {\rho(\mu)}\,,
\end{equation}
which is nothing but the Stoner criterion. Similar results are obtained
in more sophisticated calculations \cite{IH06}. Clearly, at half-filling
the value for the density of states is $\rho(0)=0$ and the critical
value for $U_F$ is arbitrarily large. Therefore we do not expect a
ferromagnetic ground state at the neutrality point
of one electron per carbon atom.
For other electronic densities, $\rho(\mu)$
becomes finite producing a finite value for $U_F$. We note that
the inclusion of $t'$ does not change these findings, since the
density of states remains zero at the neutrality point.

The line toward an antiferromagnetic ground state is given by \cite{PAD04}
\begin{equation}
U_{AF}(\mu)=\frac 2 {\frac 1 N \sum_{\bm k,\mu>0} \frac {1}
{\vert E_+(\bm k) \vert}}\,,
\end{equation}
where $E_+(\bm k)$ is given in (\ref{E1}). This result gives
a finite $U_{AF}$ at the neutrality point \cite{sorella92,martelo97}:
\begin{equation}
U_{AF}(0)=2.23t\,.
\end{equation}
Quantum Monte Carlo calculations \cite{sorella92,paiva05}, raise however
its value to:
\begin{equation}
U_{AF}(0)\simeq 5t\,.
\end{equation}
Taking for graphene the same value for $U$  as in polyacetylene
and $t=2.8$ eV, one obtains $U/t\simeq 3.6$, which put the system
far from the transition toward an antiferromagnet
ground state.
Yet another possibility is that the system may be in  a sort of
a quantum spin liquid \cite{sungsik05} (as originally proposed by
Pauling \cite{P72} in 1956) since mean field calculations
give a critical value for $U$ to be of the order
of $U/t\simeq 1.7$. Whether this type of ground
state really exists and whether quantum fluctuations pushes this
value of $U$ toward larger values is not known.

\subsubsection{Bilayer graphene: exchange}
\label{exchange}

The exchange interaction can be large in an unbiased graphene
bilayer with a small concentration of carriers. It was shown
previously that the exchange contribution to the electronic energy
of a single graphene layer does not lead to a ferromagnetic
instability \cite{PGN05b}. The reason for this is a significant
contribution from the interband  exchange, which is a term usually
neglected in doped semiconductors. This contribution depends on the
overlap of the conduction and valence wavefunctions, and it is
modified in a bilayer. The interband exchange energy is reduced in a
bilayer \cite{NNPG06}, and a positive contribution that depends
logarithmically on the bandwidth in graphene is absent in its bilayer.
As a result, the exchange energy becomes negative, and scales as
$n^{3/2}$, where $n$ is the carrier density, similar to the 2DEG. The
quadratic dispersion at low energies implies that the kinetic energy
scales as $n^2$, again as in the 2DEG. This expansion leads to:
\begin{equation}
E = E_{kin} + E_{exc} \approx \frac{\pi v_F^2 n^2}{8 t_{\perp}} - \frac{e^2
  n^{3/2}}{27 \sqrt{\pi} \epsilon_0}
\label{exchange_bilayer}
\end{equation}
Writing $n_{\uparrow} = ( n + s ) / 2 , n_{\downarrow} = ( n - s ) / 2$,
where $s$ is the magnetization, (\ref{exchange_bilayer}) predicts a second
order transition to a ferromagnetic state for $n = ( 4 e^4 t^2 ) / ( 81 \pi^3
v_F^4 \epsilon_0)$. Higher order corrections to (\ref{exchange_bilayer}) lead to a
first order transition at slightly higher densities \cite{NNPG06}. For a ratio
$\gamma_1 / \gamma_0 \approx 0.1$, this analysis implies that a graphene
bilayer should be ferromagnetic for carrier densities such that $| n |
\lesssim 4 \times 10^{10}$cm$^{-2}$.

A bilayer is also the unit cell of Bernal graphite, and the exchange
instability can also be studied in an infinite system. Taking into account
nearest neighbor interlayer hopping only, bulk graphite should also show an
exchange instability at low doping. In fact, there is some experimental
evidence for a ferromagnetic instability in strongly disordered graphite \cite{Eetal02,Eetal03b,KE07}.

The analysis described above can be extended to the biased bilayer, where a
gap separates the conduction and valence bands \cite{SPGN07}. The analysis of
this case is somewhat different, as the Fermi surface at low doping is a
ring, and the exchange interaction can change its bounds. The presence of a
gap reduces further the mixing of the valence and conduction band, leading to
an enhancement of the exchange instability. At all doping levels, where the Fermi
surface is ring shaped, the biased bilayer is unstable towards ferromagnetism.

\subsubsection{Bilayer graphene: short range interactions}
\label{bi_short}

\begin{figure}[]
\begin{center}
\includegraphics[width=5cm,angle=0]{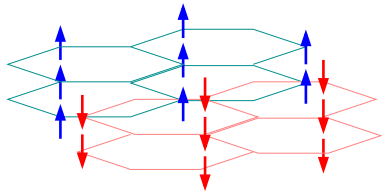}\\
\caption[fig]{\label{bilayer_AF} (Color online) Sketch of the expected magnetization of a
  graphene bilayer at half-filling.}
\end{center}
\end{figure}

The band structure of a graphene bilayer, at half filling, leads to
logarithmic divergences in different response functions at ${\bf q} =
0$. The two parabolic bands that are tangent at $\vk = 0$ lead to a
susceptibility which is proportional to:
\begin{equation}
\chi ( \vq , \omega ) \propto \int_{| {\vq} | < \Lambda} d^2 {\bf k} \frac{1}{
  \omega - ( v_F^2 / t ) | {\vk} |^2} \propto \log \left( \frac{\Lambda}{\sqrt{(
  \omega t ) / v_F^2}} \right)
  \label{susc_bilayer}
\end{equation}
where $\Lambda \sim \sqrt{t^2 / v_F^2}$ is a high momentum cutoff.
These logarithmic divergences are similar to the ones which show up
when the Fermi surface of a 2D  metal is near a saddle
point in the dispersion relation \cite{GGV96}.  A full
treatment of these divergences requires a RG
approach \cite{S94}. Within  a simpler mean field treatment, however,
it is easy to notice that the divergence of the bilayer
susceptibility gives rise to an instability towards an
antiferromagnetic phase, where the carbon atoms which are not
connected to the neighboring layers acquire a finite magnetization,
while the magnetization of the atoms with neighbors in the
contiguous layers remain zero. A scheme of the expected ordered
state is shown in Fig.~\ref{bilayer_AF}.

\subsection{Interactions in high magnetic fields}
\label{high_field}

The formation of Landau levels enhances the effect of
interactions due to the quenching of the kinetic energy.
This effect is most pronounced at low
fillings, when only the lowest levels are occupied. New phases may
appear at low temperatures. We consider here phases different from
the fractional quantum Hall effect, which has not been observed in
graphene so far. The existence of new phases can be inferred from the
splitting of the valley or spin degeneracy of the Landau levels,
which can be observed in spectroscopy
measurements \cite{SMPBH06,Jetal07}, or in the appearance of new
quantum Hall plateaus \cite{Zetal06b,JZSK07,GZKPGM07,GJC07,ANZLGL07}.

Interactions can lead to new phases when their effect overcomes that
of disorder. An analysis of the competition between disorder and
interactions is found in ref.~\cite{NMac07}. The energy splitting of the
different broken symmetry phases, in a clean system, is determined
by lattice effects, so that it is reduced by factors of order $a /
l_B$, where $a$ is a length of the order of the lattice spacing, and
$l_B$ is the magnetic length \cite{AF06,GMD06,AF07,WSSH07}. The combination of
disorder and a magnetic field may also lift the degeneracy between the
two valleys, favoring valley polarized phases \cite{ALL07}.

In addition to phases with enhanced ferromagnetism or with broken valley
symmetry, interactions at high magnetic fields can lead to excitonic
instabilities \cite{GMSS06} and Wigner crystal phases \cite{ZJ0703}.
When only the $n=0$ state is occupied,
the Landau levels have all their weight in a given sublattice. Then,
the breaking of valley degeneracy can be associated with a charge
density wave, which opens a gap \cite{FL07}. It is interesting to
note that in these phases new collective excitations are possible \cite{DMS07}.

Interactions modify the edge states in the quantum Hall regime. A
novel phase can appear when the $n=0$ is the last filled level. The
Zeeman splitting shifts the electron and hole like chiral states,
which disperse in opposite directions near the boundary of the
sample. The resulting level crossing between an electron like level
with spin anti-parallel to the field, and a hole like level with spin
parallel to the field, may lead to Luttinger liquid features in the
edge states \cite{FB06,ANZLGL07}.

\section{Conclusions}
\label{conclusions}

Graphene is a unique system in many ways. It is truly 2D, has unusual
electronic excitations described in terms of Dirac fermions that move
in a curved space, it is an interesting
 mix of a semiconductor (zero density of states) and a metal
(gaplessness), and has properties of soft matter. The electrons
in graphene seem to be almost insensitive to disorder and electron-electron
interactions and have very
long mean free paths. Hence, graphene's
properties are rather different from what is found in usual metals
and semiconductors. Graphene has also a very robust but flexible structure
with unusual phonon modes that do not exist in ordinary 3D solids. In some
sense, graphene brings together issues in quantum gravity and particle
physics, and also from soft and hard condensed matter. Interestingly enough,
these properties can be easily  modified with the application of electric
and magnetic fields, addition of layers, by control of its geometry, and
chemical doping. Moreover, graphene can be directly and relatively easily
probed by various scanning probe techniques from mesoscopic
down to atomic scales, because it is not buried inside a 3D
structure. This makes graphene one of the most versatile
systems in condensed matter research.

Besides the unusual basic properties, graphene has the potential for a
large number of applications \cite{geim_review}, from chemical sensors
\cite{Setal07,CLRA07} to transistors \cite{OHLMV07,NCNGP06}.
Graphene can be chemically and/or structurally modified in order to change
its functionality and henceforth its potential applications. Moreover,
graphene can be easily obtained from graphite, a material that is
abundant on earth's surface. This particular characteristic
makes graphene one of the most readily available
materials for basic research since it frees economically challenged
research institutions in developing countries
from the dependence of expensive sample growing techniques.

Many of graphene's properties
are currently subject of intense research and debate. The understanding of
the nature of the disorder and how it affects the transport properties
(a problem of fundamental importance for applications), the effect
of phonons on electronic transport, the nature of electron-electron
interactions and how they modify its physical properties
are research areas that are still in their infancy. In this review, we have
touched only the surface of a very deep sea that still has to be explored.

Whereas hundreds of papers have been written on monolayer graphene in the
last few years, only a small fraction actually deals with multilayers. The
majority of the theoretical and experimental efforts have been concentrated on
the single layer, perhaps because of its simplicity and
the natural attraction that a one atom
thick material, which can be produced by simple methods in almost any
laboratory in the world, creates for human imagination. Nevertheless, few
layer graphene is equally interesting and unusual with a technological
potential perhaps bigger than the single layer. Indeed, the theoretical
understanding and experimental exploration of multilayers is far behind the
single layer. This is a fertile and open field of research for the future.

Finally, we have focused entirely on pure carbon graphene where the band
structure is dominated by the Dirac description. Nevertheless, chemical
modification of graphene can lead to entirely new physics. Depending on the
nature of chemical dopants and how they are introduced into the graphene
lattice (adsorption, substitution, or intercalation) the results can be
many. Small concentrations of adsorbed alkali metal can be used to change the
chemical potential while adsorbed transition elements can lead to strong
hybridization effects that affect the electronic structure. In fact, the
introducion of d- and f-electron atoms in the graphene lattice may
produce a significant enhancement of the electron-electron interactions.
Hence, it is easy to envision a plethora of many-body effects that can be
induced by doping and have to be studied in the context of Dirac electrons:
Kondo effect, ferromagnetism, antiferromagnetism, charge
and spin density waves. The study of chemically
induced many-body effects in graphene would add a new chapter to the
short but fascinating history of this material. Only future will tell but
the potential for more amazement is lurking on the horizon.

\section{Acknowledgments}

We have benefited immensely from discussions with
many colleagues and friends in the last few years but we would like
to thank especially, Boris Altshuler, Eva Andrei, Alexander Balatsky, Carlo Beenakker, Sankar Das Sarma,
Walt de Heer, Millie Dresselhaus, Vladimir Falko, Andrea Ferrari, Herb Fertig, Eduardo Fradkin, Ernie Hill,
Mihail Katsnelson, Eun-Ah Kim, Philip Kim, Valery Kotov, Alessandra Lanzara, Leonid Levitov,
Allan MacDonald, Serguey Morozov, Johan Nilsson, Vitor Pereira, Philip Phillips,
Ramamurti Shankar, Jo\~ao Lopes dos Santos, Shan-Wen Tsai, Bruno Uchoa,
and Maria Vozmediano.

N.M.R.P. acknowledges financial support from POCI 2010 via project
PTDC/FIS/64404/2006. F.G. was
supported by MEC (Spain) grant No.~FIS2005-05478-C02-01 and EU
contract 12881 (NEST).
A.~H.~C.~N was supported through NSF grant DMR-0343790.
K.S.N. and A. K. G. were supported by EPSRC (UK) and the Royal Society.

%------------------------------------------------------------------------------
%------------------------------------------------------------------------------
\bibliographystyle{apsrmp} %apsrev
\bibliography{graphene}

\newcommand{\npb}{Nucl. Phys.}\newcommand{\adv}{Adv.
  Phys.}\newcommand{\epl}{Europhys. Lett.}
\begin{thebibliography}{450}
\expandafter\ifx\csname natexlab\endcsname\relax\def\natexlab#1{#1}\fi
\expandafter\ifx\csname bibnamefont\endcsname\relax
  \def\bibnamefont#1{#1}\fi
\expandafter\ifx\csname bibfnamefont\endcsname\relax
  \def\bibfnamefont#1{#1}\fi
\expandafter\ifx\csname citenamefont\endcsname\relax
  \def\citenamefont#1{#1}\fi
\expandafter\ifx\csname url\endcsname\relax
  \def\url#1{\texttt{#1}}\fi
\expandafter\ifx\csname urlprefix\endcsname\relax\def\urlprefix{URL }\fi
\providecommand{\bibinfo}[2]{#2}
\providecommand{\eprint}[2][]{\url{#2}}

\bibitem[{\citenamefont{Abanin} \emph{et~al.}(2006)\citenamefont{Abanin, Lee,
  and Levitov}}]{Abanin06}
\bibinfo{author}{\bibnamefont{Abanin}, \bibfnamefont{D.~A.}},
  \bibinfo{author}{\bibfnamefont{P.~A.} \bibnamefont{Lee}}, and
  \bibinfo{author}{\bibfnamefont{L.~S.} \bibnamefont{Levitov}},
  \bibinfo{year}{2006}, \bibinfo{journal}{Phys. Rev. Lett.}
  \textbf{\bibinfo{volume}{96}}, \bibinfo{pages}{176803}.

\bibitem[{\citenamefont{Abanin}
  \emph{et~al.}(2007{\natexlab{a}})\citenamefont{Abanin, Lee, and
  Levitov}}]{ALL07}
\bibinfo{author}{\bibnamefont{Abanin}, \bibfnamefont{D.~A.}},
  \bibinfo{author}{\bibfnamefont{P.~A.} \bibnamefont{Lee}}, and
  \bibinfo{author}{\bibfnamefont{L.~S.} \bibnamefont{Levitov}},
  \bibinfo{year}{2007}{\natexlab{a}}, \bibinfo{journal}{Solid State Comm.}
  \textbf{\bibinfo{volume}{143}}, \bibinfo{pages}{77}.

\bibitem[{\citenamefont{Abanin and Levitov}(2007)}]{AL0704}
\bibinfo{author}{\bibnamefont{Abanin}, \bibfnamefont{D.~A.}}, and
  \bibinfo{author}{\bibfnamefont{L.~S.} \bibnamefont{Levitov}},
  \bibinfo{year}{2007}, \bibinfo{journal}{Science}
  \textbf{\bibinfo{volume}{317}}, \bibinfo{pages}{641}.

\bibitem[{\citenamefont{Abanin}
  \emph{et~al.}(2007{\natexlab{b}})\citenamefont{Abanin, Novoselov, Zeitler,
  Lee, Geim, and Levitov}}]{ANZLGL07}
\bibinfo{author}{\bibnamefont{Abanin}, \bibfnamefont{D.~A.}},
  \bibinfo{author}{\bibfnamefont{K.~S.} \bibnamefont{Novoselov}},
  \bibinfo{author}{\bibfnamefont{U.}~\bibnamefont{Zeitler}},
  \bibinfo{author}{\bibfnamefont{P.~A.} \bibnamefont{Lee}},
  \bibinfo{author}{\bibfnamefont{A.~K.} \bibnamefont{Geim}}, and
  \bibinfo{author}{\bibfnamefont{L.~S.} \bibnamefont{Levitov}},
  \bibinfo{year}{2007}{\natexlab{b}}, \bibinfo{journal}{Phys. Rev. Lett.}
  \textbf{\bibinfo{volume}{98}}, \bibinfo{pages}{196806}.

\bibitem[{\citenamefont{Abergel} \emph{et~al.}(2007)\citenamefont{Abergel,
  Russell, and Fal'ko}}]{ARF07}
\bibinfo{author}{\bibnamefont{Abergel}, \bibfnamefont{D.~S.~L.}},
  \bibinfo{author}{\bibfnamefont{A.}~\bibnamefont{Russell}}, and
  \bibinfo{author}{\bibfnamefont{V.~I.} \bibnamefont{Fal'ko}},
  \bibinfo{year}{2007}, \bibinfo{journal}{Appl. Phys. Lett.}
  \textbf{\bibinfo{volume}{91}}, \bibinfo{pages}{063125}.

\bibitem[{\citenamefont{Abrikosov}(1998)}]{Abrikosov98}
\bibinfo{author}{\bibnamefont{Abrikosov}, \bibfnamefont{A.~A.}},
  \bibinfo{year}{1998}, \bibinfo{journal}{Phys. Rev. B}
  \textbf{\bibinfo{volume}{58}}, \bibinfo{pages}{2788}.

\bibitem[{\citenamefont{Adam} \emph{et~al.}(2007)\citenamefont{Adam, Hwang,
  Galitski, and {das Sarma}}}]{AHGS07}
\bibinfo{author}{\bibnamefont{Adam}, \bibfnamefont{S.}},
  \bibinfo{author}{\bibfnamefont{E.~H.} \bibnamefont{Hwang}},
  \bibinfo{author}{\bibfnamefont{V.~M.} \bibnamefont{Galitski}}, and
  \bibinfo{author}{\bibfnamefont{S.}~\bibnamefont{{das Sarma}}},
  \bibinfo{year}{2007}, \bibinfo{journal}{Proc. Natl. Acad. Sci. USA}
  \textbf{\bibinfo{volume}{104}}, \bibinfo{pages}{18392}.

\bibitem[{\citenamefont{Adebpour} \emph{et~al.}(2007)\citenamefont{Adebpour,
  Neek-Amal, a~nd F.~Shahbazi, Nafari, and {Reza Rahimi Tabar}}}]{ANASNT07}
\bibinfo{author}{\bibnamefont{Adebpour}, \bibfnamefont{N.}},
  \bibinfo{author}{\bibfnamefont{M.}~\bibnamefont{Neek-Amal}},
  \bibinfo{author}{\bibfnamefont{R.~A.} \bibnamefont{a~nd F.~Shahbazi}},
  \bibinfo{author}{\bibfnamefont{N.}~\bibnamefont{Nafari}}, and
  \bibinfo{author}{\bibfnamefont{M.}~\bibnamefont{{Reza Rahimi Tabar}}},
  \bibinfo{year}{2007}, \bibinfo{journal}{Phys. Rev. B}
  \textbf{\bibinfo{volume}{76}}, \bibinfo{pages}{195407}.

\bibitem[{\citenamefont{Affoune} \emph{et~al.}(2001)\citenamefont{Affoune,
  Prasad, Saito, Enoki, Kaburagi, and Hishiyama}}]{Affoune01}
\bibinfo{author}{\bibnamefont{Affoune}, \bibfnamefont{A.~M.}},
  \bibinfo{author}{\bibfnamefont{B.~L.~V.} \bibnamefont{Prasad}},
  \bibinfo{author}{\bibfnamefont{H.}~\bibnamefont{Saito}},
  \bibinfo{author}{\bibfnamefont{T.}~\bibnamefont{Enoki}},
  \bibinfo{author}{\bibfnamefont{Y.}~\bibnamefont{Kaburagi}}, and
  \bibinfo{author}{\bibfnamefont{Y.}~\bibnamefont{Hishiyama}},
  \bibinfo{year}{2001}, \bibinfo{journal}{Chem. Phys. Lett.}
  \textbf{\bibinfo{volume}{348}}, \bibinfo{pages}{17}.

\bibitem[{\citenamefont{Akhmerov and Beenakker}(2007)}]{AkBe07}
\bibinfo{author}{\bibnamefont{Akhmerov}, \bibfnamefont{A.~R.}}, and
  \bibinfo{author}{\bibfnamefont{C.~W.~J.} \bibnamefont{Beenakker}},
  \bibinfo{year}{2007}, \eprint{arXiv:0710.2723}.

\bibitem[{\citenamefont{Aleiner and Efetov}(2006)}]{AE06}
\bibinfo{author}{\bibnamefont{Aleiner}, \bibfnamefont{I.~L.}}, and
  \bibinfo{author}{\bibfnamefont{K.~B.} \bibnamefont{Efetov}},
  \bibinfo{year}{2006}, \bibinfo{journal}{Phys. Rev. Lett.}
  \textbf{\bibinfo{volume}{97}}, \bibinfo{pages}{236801}.

\bibitem[{\citenamefont{Alicea and Fisher}(2006)}]{AF06}
\bibinfo{author}{\bibnamefont{Alicea}, \bibfnamefont{J.}}, and
  \bibinfo{author}{\bibfnamefont{M.~P.~A.} \bibnamefont{Fisher}},
  \bibinfo{year}{2006}, \bibinfo{journal}{Phys. Rev. B}
  \textbf{\bibinfo{volume}{74}}, \bibinfo{pages}{075422}.

\bibitem[{\citenamefont{Alicea and Fisher}(2007)}]{AF07}
\bibinfo{author}{\bibnamefont{Alicea}, \bibfnamefont{J.}}, and
  \bibinfo{author}{\bibfnamefont{M.~P.~A.} \bibnamefont{Fisher}},
  \bibinfo{year}{2007}, \bibinfo{journal}{Solid State Comm.}
  \textbf{\bibinfo{volume}{143}}, \bibinfo{pages}{504}.

\bibitem[{\citenamefont{Altland}(2006)}]{A06c}
\bibinfo{author}{\bibnamefont{Altland}, \bibfnamefont{A.}},
  \bibinfo{year}{2006}, \bibinfo{journal}{Phys. Rev. Lett.}
  \textbf{\bibinfo{volume}{97}}, \bibinfo{pages}{236802}.

\bibitem[{\citenamefont{Ando}(1974{\natexlab{a}})}]{Ando2}
\bibinfo{author}{\bibnamefont{Ando}, \bibfnamefont{T.}},
  \bibinfo{year}{1974}{\natexlab{a}}, \bibinfo{journal}{J. Phys. Soc. Jpn.}
  \textbf{\bibinfo{volume}{36}}, \bibinfo{pages}{1521}.

\bibitem[{\citenamefont{Ando}(1974{\natexlab{b}})}]{Ando4}
\bibinfo{author}{\bibnamefont{Ando}, \bibfnamefont{T.}},
  \bibinfo{year}{1974}{\natexlab{b}}, \bibinfo{journal}{J. Phys. Soc. Jpn.}
  \textbf{\bibinfo{volume}{37}}, \bibinfo{pages}{622}.

\bibitem[{\citenamefont{Ando}(1974{\natexlab{c}})}]{Ando5}
\bibinfo{author}{\bibnamefont{Ando}, \bibfnamefont{T.}},
  \bibinfo{year}{1974}{\natexlab{c}}, \bibinfo{journal}{J. Phys. Soc. Jpn.}
  \textbf{\bibinfo{volume}{37}}, \bibinfo{pages}{1233}.

\bibitem[{\citenamefont{Ando}(1975)}]{Ando6}
\bibinfo{author}{\bibnamefont{Ando}, \bibfnamefont{T.}}, \bibinfo{year}{1975},
  \bibinfo{journal}{J. Phys. Soc. Jpn.} \textbf{\bibinfo{volume}{38}},
  \bibinfo{pages}{989}.

\bibitem[{\citenamefont{Ando}(2000)}]{A00}
\bibinfo{author}{\bibnamefont{Ando}, \bibfnamefont{T.}}, \bibinfo{year}{2000},
  \bibinfo{journal}{J. Phys. Soc. Jpn.} \textbf{\bibinfo{volume}{69}},
  \bibinfo{pages}{1757}.

\bibitem[{\citenamefont{Ando}(2006{\natexlab{a}})}]{ando_opt}
\bibinfo{author}{\bibnamefont{Ando}, \bibfnamefont{T.}},
  \bibinfo{year}{2006}{\natexlab{a}}, \bibinfo{journal}{J. Phys. Soc. Jpn.}
  \textbf{\bibinfo{volume}{75}}, \bibinfo{pages}{124701}.

\bibitem[{\citenamefont{Ando}(2006{\natexlab{b}})}]{A06}
\bibinfo{author}{\bibnamefont{Ando}, \bibfnamefont{T.}},
  \bibinfo{year}{2006}{\natexlab{b}}, \bibinfo{journal}{J. Phys. Soc. Jpn.}
  \textbf{\bibinfo{volume}{75}}, \bibinfo{pages}{074716}.

\bibitem[{\citenamefont{Ando}(2007{\natexlab{a}})}]{A07}
\bibinfo{author}{\bibnamefont{Ando}, \bibfnamefont{T.}},
  \bibinfo{year}{2007}{\natexlab{a}}, \bibinfo{journal}{J. Phys. Soc. Jpn.}
  \textbf{\bibinfo{volume}{76}}, \bibinfo{pages}{024712}.

\bibitem[{\citenamefont{Ando}(2007{\natexlab{b}})}]{ando_opt_bi}
\bibinfo{author}{\bibnamefont{Ando}, \bibfnamefont{T.}},
  \bibinfo{year}{2007}{\natexlab{b}}, \bibinfo{journal}{J. Phys. Soc. Jpn.}
  \textbf{\bibinfo{volume}{76}}, \bibinfo{pages}{104711}.

\bibitem[{\citenamefont{Ando} \emph{et~al.}(1998)\citenamefont{Ando, Nakanishi,
  and Saito}}]{ANS98}
\bibinfo{author}{\bibnamefont{Ando}, \bibfnamefont{T.}},
  \bibinfo{author}{\bibfnamefont{T.}~\bibnamefont{Nakanishi}}, and
  \bibinfo{author}{\bibfnamefont{R.}~\bibnamefont{Saito}},
  \bibinfo{year}{1998}, \bibinfo{journal}{J. Phys. Soc. Jpn.}
  \textbf{\bibinfo{volume}{67}}, \bibinfo{pages}{2857}.

\bibitem[{\citenamefont{Ando and Uemura}(1974)}]{Ando1}
\bibinfo{author}{\bibnamefont{Ando}, \bibfnamefont{T.}}, and
  \bibinfo{author}{\bibfnamefont{Y.}~\bibnamefont{Uemura}},
  \bibinfo{year}{1974}, \bibinfo{journal}{J. Phys. Soc. Jpn.}
  \textbf{\bibinfo{volume}{36}}, \bibinfo{pages}{959}.

\bibitem[{\citenamefont{Ando} \emph{et~al.}(2002)\citenamefont{Ando, Zheng, and
  Suzuura}}]{Ando8}
\bibinfo{author}{\bibnamefont{Ando}, \bibfnamefont{T.}},
  \bibinfo{author}{\bibfnamefont{Y.}~\bibnamefont{Zheng}}, and
  \bibinfo{author}{\bibfnamefont{H.}~\bibnamefont{Suzuura}},
  \bibinfo{year}{2002}, \bibinfo{journal}{J. Phys. Soc. Jpn.}
  \textbf{\bibinfo{volume}{71}}, \bibinfo{pages}{1318}.

\bibitem[{\citenamefont{Andreoni}(2000)}]{fullerenes}
\bibinfo{author}{\bibnamefont{Andreoni}, \bibfnamefont{W.}},
  \bibinfo{year}{2000}, \emph{\bibinfo{title}{The Physics of Fullerene-Based
  and Fullerene-related materials}} (\bibinfo{publisher}{Springer}).

\bibitem[{\citenamefont{Areshkin and White}(2007)}]{AW07}
\bibinfo{author}{\bibnamefont{Areshkin}, \bibfnamefont{D.~A.}}, and
  \bibinfo{author}{\bibfnamefont{C.~T.} \bibnamefont{White}},
  \bibinfo{year}{2007}, \bibinfo{journal}{Nano Lett.}
  \textbf{\bibinfo{volume}{7}}, \bibinfo{pages}{204}.

\bibitem[{\citenamefont{Ashcroft and Mermin}(1976)}]{Ashcroft}
\bibinfo{author}{\bibnamefont{Ashcroft}, \bibfnamefont{N.~W.}}, and
  \bibinfo{author}{\bibfnamefont{N.~D.} \bibnamefont{Mermin}},
  \bibinfo{year}{1976}, \emph{\bibinfo{title}{Solid State Physics}}
  (\bibinfo{publisher}{Saunders College}, \bibinfo{address}{Philadelphia, PA}).

\bibitem[{\citenamefont{Bacon}(1950)}]{B50}
\bibinfo{author}{\bibnamefont{Bacon}, \bibfnamefont{G.~E.}},
  \bibinfo{year}{1950}, \bibinfo{journal}{Acta Crystalographica}
  \textbf{\bibinfo{volume}{4}}, \bibinfo{pages}{320}.

\bibitem[{\citenamefont{Baeriswyl} \emph{et~al.}(1986)\citenamefont{Baeriswyl,
  Campbell, and Mazumdar}}]{Baeriswyl86}
\bibinfo{author}{\bibnamefont{Baeriswyl}, \bibfnamefont{D.}},
  \bibinfo{author}{\bibfnamefont{D.~K.} \bibnamefont{Campbell}}, and
  \bibinfo{author}{\bibfnamefont{S.}~\bibnamefont{Mazumdar}},
  \bibinfo{year}{1986}, \bibinfo{journal}{Phys. Rev. Lett.}
  \textbf{\bibinfo{volume}{56}}, \bibinfo{pages}{1509}.

\bibitem[{\citenamefont{Bak}(1982)}]{Bak82}
\bibinfo{author}{\bibnamefont{Bak}, \bibfnamefont{P.}}, \bibinfo{year}{1982},
  \bibinfo{journal}{Rep. Prog. Phys.} \textbf{\bibinfo{volume}{45}},
  \bibinfo{pages}{587}.

\bibitem[{\citenamefont{Balatsky} \emph{et~al.}(2006)\citenamefont{Balatsky,
  Vekhter, and Zhu}}]{BVZ06}
\bibinfo{author}{\bibnamefont{Balatsky}, \bibfnamefont{A.~V.}},
  \bibinfo{author}{\bibfnamefont{I.}~\bibnamefont{Vekhter}}, and
  \bibinfo{author}{\bibfnamefont{J.-X.} \bibnamefont{Zhu}},
  \bibinfo{year}{2006}, \bibinfo{journal}{Rev. Mod. Phys.}
  \textbf{\bibinfo{volume}{78}}, \bibinfo{pages}{373}.

\bibitem[{\citenamefont{Bar} \emph{et~al.}(2007)\citenamefont{Bar, Zhang,
  Yayon, Bostwick, Ohta, McChesney, Horn, Rotenberg, and
  Crommie}}]{BZYBOMHRC07}
\bibinfo{author}{\bibnamefont{Bar}, \bibfnamefont{V.~W.}},
  \bibinfo{author}{\bibfnamefont{Y.}~\bibnamefont{Zhang}},
  \bibinfo{author}{\bibfnamefont{Y.}~\bibnamefont{Yayon}},
  \bibinfo{author}{\bibfnamefont{A.}~\bibnamefont{Bostwick}},
  \bibinfo{author}{\bibfnamefont{T.}~\bibnamefont{Ohta}},
  \bibinfo{author}{\bibfnamefont{J.~L.} \bibnamefont{McChesney}},
  \bibinfo{author}{\bibfnamefont{K.}~\bibnamefont{Horn}},
  \bibinfo{author}{\bibfnamefont{E.}~\bibnamefont{Rotenberg}}, and
  \bibinfo{author}{\bibfnamefont{M.~F.} \bibnamefont{Crommie}},
  \bibinfo{year}{2007}, \bibinfo{journal}{Appl. Phys. Lett.}
  \textbf{\bibinfo{volume}{91}}, \bibinfo{pages}{122102}.

\bibitem[{\citenamefont{Bardarson} \emph{et~al.}(2007)\citenamefont{Bardarson,
  Tworzydlo, Brouwer, and Beenakker}}]{BTBB07}
\bibinfo{author}{\bibnamefont{Bardarson}, \bibfnamefont{J.~H.}},
  \bibinfo{author}{\bibfnamefont{J.}~\bibnamefont{Tworzydlo}},
  \bibinfo{author}{\bibfnamefont{P.~W.} \bibnamefont{Brouwer}}, and
  \bibinfo{author}{\bibfnamefont{C.~W.~J.} \bibnamefont{Beenakker}},
  \bibinfo{year}{2007}, \bibinfo{journal}{Phys. Rev. Lett.}
  \textbf{\bibinfo{volume}{99}}, \bibinfo{pages}{106801}.

\bibitem[{\citenamefont{Barlas} \emph{et~al.}(2007)\citenamefont{Barlas,
  Pereg-Barnea, Polini, Asgari, and MacDonald}}]{BPPAM07}
\bibinfo{author}{\bibnamefont{Barlas}, \bibfnamefont{Y.}},
  \bibinfo{author}{\bibfnamefont{T.}~\bibnamefont{Pereg-Barnea}},
  \bibinfo{author}{\bibfnamefont{M.}~\bibnamefont{Polini}},
  \bibinfo{author}{\bibfnamefont{R.}~\bibnamefont{Asgari}}, and
  \bibinfo{author}{\bibfnamefont{A.~H.} \bibnamefont{MacDonald}},
  \bibinfo{year}{2007}, \bibinfo{journal}{Phys. Rev. Lett.}
  \textbf{\bibinfo{volume}{98}}, \bibinfo{pages}{236601}.

\bibitem[{\citenamefont{Barone} \emph{et~al.}(2006)\citenamefont{Barone, Hod,
  and Scuseria}}]{BHS06}
\bibinfo{author}{\bibnamefont{Barone}, \bibfnamefont{V.}},
  \bibinfo{author}{\bibfnamefont{O.}~\bibnamefont{Hod}}, and
  \bibinfo{author}{\bibfnamefont{G.~E.} \bibnamefont{Scuseria}},
  \bibinfo{year}{2006}, \bibinfo{journal}{Nano Letters}
  \textbf{\bibinfo{volume}{6}}, \bibinfo{pages}{2748}.

\bibitem[{\citenamefont{Baym}(1969)}]{Baym}
\bibinfo{author}{\bibnamefont{Baym}, \bibfnamefont{G.}}, \bibinfo{year}{1969},
  \emph{\bibinfo{title}{Lectures on quantum mechanics}}
  (\bibinfo{publisher}{Benjamin, New York}).

\bibitem[{\citenamefont{Baym and Chin}(1976)}]{BC76}
\bibinfo{author}{\bibnamefont{Baym}, \bibfnamefont{G.}}, and
  \bibinfo{author}{\bibfnamefont{S.~A.} \bibnamefont{Chin}},
  \bibinfo{year}{1976}, \bibinfo{journal}{Nuclear Physics}
  \textbf{\bibinfo{volume}{A262}}, \bibinfo{pages}{527}.

\bibitem[{\citenamefont{Bena and Kivelson}(2005)}]{BK05}
\bibinfo{author}{\bibnamefont{Bena}, \bibfnamefont{C.}}, and
  \bibinfo{author}{\bibfnamefont{S.~A.} \bibnamefont{Kivelson}},
  \bibinfo{year}{2005}, \bibinfo{journal}{Phys. Rev. B}
  \textbf{\bibinfo{volume}{72}}, \bibinfo{pages}{125432}.

\bibitem[{\citenamefont{Berger} \emph{et~al.}(2006)\citenamefont{Berger, Song,
  Li, Wu, Brown, Naud, Mayou, Li, Hass, Marchenkov, Conrad, First}
  \emph{et~al.}}]{Betal06}
\bibinfo{author}{\bibnamefont{Berger}, \bibfnamefont{C.}},
  \bibinfo{author}{\bibfnamefont{Z.}~\bibnamefont{Song}},
  \bibinfo{author}{\bibfnamefont{X.}~\bibnamefont{Li}},
  \bibinfo{author}{\bibfnamefont{X.}~\bibnamefont{Wu}},
  \bibinfo{author}{\bibfnamefont{N.}~\bibnamefont{Brown}},
  \bibinfo{author}{\bibfnamefont{C.}~\bibnamefont{Naud}},
  \bibinfo{author}{\bibfnamefont{D.}~\bibnamefont{Mayou}},
  \bibinfo{author}{\bibfnamefont{T.}~\bibnamefont{Li}},
  \bibinfo{author}{\bibfnamefont{J.}~\bibnamefont{Hass}},
  \bibinfo{author}{\bibfnamefont{A.~N.} \bibnamefont{Marchenkov}},
  \bibinfo{author}{\bibfnamefont{E.~H.} \bibnamefont{Conrad}},
  \bibinfo{author}{\bibfnamefont{P.~N.} \bibnamefont{First}}, \emph{et~al.},
  \bibinfo{year}{2006}, \bibinfo{journal}{Science}
  \textbf{\bibinfo{volume}{312}}, \bibinfo{pages}{1191}.

\bibitem[{\citenamefont{Berger} \emph{et~al.}(2004)\citenamefont{Berger, Song,
  Li, Li, Ogbazghi, Feng, Dai, Marchenkov, Conrad, First, and
  de~Heer}}]{Betal04}
\bibinfo{author}{\bibnamefont{Berger}, \bibfnamefont{C.}},
  \bibinfo{author}{\bibfnamefont{Z.~M.} \bibnamefont{Song}},
  \bibinfo{author}{\bibfnamefont{T.~B.} \bibnamefont{Li}},
  \bibinfo{author}{\bibfnamefont{X.~B.} \bibnamefont{Li}},
  \bibinfo{author}{\bibfnamefont{A.~Y.} \bibnamefont{Ogbazghi}},
  \bibinfo{author}{\bibfnamefont{R.}~\bibnamefont{Feng}},
  \bibinfo{author}{\bibfnamefont{Z.~T.} \bibnamefont{Dai}},
  \bibinfo{author}{\bibfnamefont{A.~N.} \bibnamefont{Marchenkov}},
  \bibinfo{author}{\bibfnamefont{E.~H.} \bibnamefont{Conrad}},
  \bibinfo{author}{\bibfnamefont{P.~N.} \bibnamefont{First}}, and
  \bibinfo{author}{\bibfnamefont{W.~A.} \bibnamefont{de~Heer}},
  \bibinfo{year}{2004}, \bibinfo{journal}{J. Phys. Chem. B}
  \textbf{\bibinfo{volume}{108}}, \bibinfo{pages}{19912}.

\bibitem[{\citenamefont{Bergman}(1984)}]{B84}
\bibinfo{author}{\bibnamefont{Bergman}, \bibfnamefont{G.}},
  \bibinfo{year}{1984}, \bibinfo{journal}{Phys. Rep.}
  \textbf{\bibinfo{volume}{107}}, \bibinfo{pages}{1}.

\bibitem[{\citenamefont{Bernevig} \emph{et~al.}(2007)\citenamefont{Bernevig,
  Hughes, Raghu, and Arovas}}]{BHRA07}
\bibinfo{author}{\bibnamefont{Bernevig}, \bibfnamefont{B.~A.}},
  \bibinfo{author}{\bibfnamefont{T.~L.} \bibnamefont{Hughes}},
  \bibinfo{author}{\bibfnamefont{S.}~\bibnamefont{Raghu}}, and
  \bibinfo{author}{\bibfnamefont{D.~P.} \bibnamefont{Arovas}},
  \bibinfo{year}{2007}, \eprint{cond-mat/0701436}.

\bibitem[{\citenamefont{Berry and Modragon}(1987)}]{BM87}
\bibinfo{author}{\bibnamefont{Berry}, \bibfnamefont{M.~V.}}, and
  \bibinfo{author}{\bibfnamefont{R.~J.} \bibnamefont{Modragon}},
  \bibinfo{year}{1987}, \bibinfo{journal}{Proc. R. Soc. Lond. A}
  \textbf{\bibinfo{volume}{412}}, \bibinfo{pages}{53}.

\bibitem[{\citenamefont{Bhattacharjee and Sengupta}(2006)}]{BS06}
\bibinfo{author}{\bibnamefont{Bhattacharjee}, \bibfnamefont{S.}}, and
  \bibinfo{author}{\bibfnamefont{K.}~\bibnamefont{Sengupta}},
  \bibinfo{year}{2006}, \bibinfo{journal}{Phys. Rev. Lett.}
  \textbf{\bibinfo{volume}{97}}, \bibinfo{pages}{217001}.

\bibitem[{\citenamefont{Birrell and Davies}(1982)}]{BD82}
\bibinfo{author}{\bibnamefont{Birrell}, \bibfnamefont{N.~D.}}, and
  \bibinfo{author}{\bibfnamefont{P.~C.~W.} \bibnamefont{Davies}},
  \bibinfo{year}{1982}, \emph{\bibinfo{title}{Quantum Fields in Curved Space}}
  (\bibinfo{publisher}{Cambridge Univ. Press, Cambridge}).

\bibitem[{\citenamefont{Biswas} \emph{et~al.}(2007)\citenamefont{Biswas,
  Sachdev, and Son}}]{BSS07}
\bibinfo{author}{\bibnamefont{Biswas}, \bibfnamefont{R.~B.}},
  \bibinfo{author}{\bibfnamefont{S.}~\bibnamefont{Sachdev}}, and
  \bibinfo{author}{\bibfnamefont{D.~T.} \bibnamefont{Son}},
  \bibinfo{year}{2007}, \bibinfo{journal}{Phys. Rev. B}
  \textbf{\bibinfo{volume}{76}}, \bibinfo{pages}{205122}.

\bibitem[{\citenamefont{Blake} \emph{et~al.}(2007)\citenamefont{Blake,
  Novoselov, {Castro Neto}, Jiang, Yang, Booth, Geim, and Hill}}]{blake07}
\bibinfo{author}{\bibnamefont{Blake}, \bibfnamefont{P.}},
  \bibinfo{author}{\bibfnamefont{K.~S.} \bibnamefont{Novoselov}},
  \bibinfo{author}{\bibfnamefont{A.~H.} \bibnamefont{{Castro Neto}}},
  \bibinfo{author}{\bibfnamefont{D.}~\bibnamefont{Jiang}},
  \bibinfo{author}{\bibfnamefont{R.}~\bibnamefont{Yang}},
  \bibinfo{author}{\bibfnamefont{T.~J.} \bibnamefont{Booth}},
  \bibinfo{author}{\bibfnamefont{A.~K.} \bibnamefont{Geim}}, and
  \bibinfo{author}{\bibfnamefont{E.~W.} \bibnamefont{Hill}},
  \bibinfo{year}{2007}, \bibinfo{journal}{Appl. Phys. Lett.}
  \textbf{\bibinfo{volume}{91}}, \bibinfo{pages}{063124}.

\bibitem[{\citenamefont{Bliokh}(2005)}]{B05}
\bibinfo{author}{\bibnamefont{Bliokh}, \bibfnamefont{K.~Y.}},
  \bibinfo{year}{2005}, \bibinfo{journal}{Phys. Lett. A}
  \textbf{\bibinfo{volume}{344}}, \bibinfo{pages}{127}.

\bibitem[{\citenamefont{Bommel} \emph{et~al.}(1975)\citenamefont{Bommel,
  Crombeen, and Tooren}}]{VCV75}
\bibinfo{author}{\bibnamefont{Bommel}, \bibfnamefont{A.~J.~V.}},
  \bibinfo{author}{\bibfnamefont{J.~E.} \bibnamefont{Crombeen}}, and
  \bibinfo{author}{\bibfnamefont{A.~V.} \bibnamefont{Tooren}},
  \bibinfo{year}{1975}, \bibinfo{journal}{Surf. Sci.}
  \textbf{\bibinfo{volume}{48}}, \bibinfo{pages}{463}.

\bibitem[{\citenamefont{Bonini} \emph{et~al.}(2007)\citenamefont{Bonini,
  Lazzeri, Marzari, and Mauri}}]{BLMM07}
\bibinfo{author}{\bibnamefont{Bonini}, \bibfnamefont{N.}},
  \bibinfo{author}{\bibfnamefont{M.}~\bibnamefont{Lazzeri}},
  \bibinfo{author}{\bibfnamefont{N.}~\bibnamefont{Marzari}}, and
  \bibinfo{author}{\bibfnamefont{F.}~\bibnamefont{Mauri}},
  \bibinfo{year}{2007}, \bibinfo{journal}{Phys. Rev. Lett.}
  \textbf{\bibinfo{volume}{99}}, \bibinfo{pages}{176802}.

\bibitem[{\citenamefont{Bostwick}
  \emph{et~al.}(2007{\natexlab{a}})\citenamefont{Bostwick, Ohta, McChesney,
  Emtsev, Seyller, Horn, and Rotenberg}}]{BOMESHR07}
\bibinfo{author}{\bibnamefont{Bostwick}, \bibfnamefont{A.}},
  \bibinfo{author}{\bibfnamefont{T.}~\bibnamefont{Ohta}},
  \bibinfo{author}{\bibfnamefont{J.~L.} \bibnamefont{McChesney}},
  \bibinfo{author}{\bibfnamefont{K.~V.} \bibnamefont{Emtsev}},
  \bibinfo{author}{\bibfnamefont{T.}~\bibnamefont{Seyller}},
  \bibinfo{author}{\bibfnamefont{K.}~\bibnamefont{Horn}}, and
  \bibinfo{author}{\bibfnamefont{E.}~\bibnamefont{Rotenberg}},
  \bibinfo{year}{2007}{\natexlab{a}}, \bibinfo{journal}{New J. Phys.}
  \textbf{\bibinfo{volume}{9}}, \bibinfo{pages}{385}.

\bibitem[{\citenamefont{Bostwick}
  \emph{et~al.}(2007{\natexlab{b}})\citenamefont{Bostwick, Ohta, Seyller, Horn,
  and Rotenberg}}]{Betal07}
\bibinfo{author}{\bibnamefont{Bostwick}, \bibfnamefont{A.}},
  \bibinfo{author}{\bibfnamefont{T.}~\bibnamefont{Ohta}},
  \bibinfo{author}{\bibfnamefont{T.}~\bibnamefont{Seyller}},
  \bibinfo{author}{\bibfnamefont{K.}~\bibnamefont{Horn}}, and
  \bibinfo{author}{\bibfnamefont{E.}~\bibnamefont{Rotenberg}},
  \bibinfo{year}{2007}{\natexlab{b}}, \bibinfo{journal}{Nature Physics}
  \textbf{\bibinfo{volume}{3}}, \bibinfo{pages}{36}.

\bibitem[{\citenamefont{Boyle and Nozi\`eres}(1958)}]{BN58}
\bibinfo{author}{\bibnamefont{Boyle}, \bibfnamefont{W.~S.}}, and
  \bibinfo{author}{\bibfnamefont{P.}~\bibnamefont{Nozi\`eres}},
  \bibinfo{year}{1958}, \bibinfo{journal}{Phys. Rev.}
  \textbf{\bibinfo{volume}{111}}, \bibinfo{pages}{782}.

\bibitem[{\citenamefont{Brandt} \emph{et~al.}(1988)\citenamefont{Brandt,
  Chudinov, and Ponomarev}}]{BCP88}
\bibinfo{author}{\bibnamefont{Brandt}, \bibfnamefont{N.~B.}},
  \bibinfo{author}{\bibfnamefont{S.~M.} \bibnamefont{Chudinov}}, and
  \bibinfo{author}{\bibfnamefont{Y.~G.} \bibnamefont{Ponomarev}},
  \bibinfo{year}{1988}, in \emph{\bibinfo{booktitle}{Modern Problems in
  Condensed Matter Sciences}}, edited by \bibinfo{editor}{\bibfnamefont{V.~M.}
  \bibnamefont{Agranovich}} and \bibinfo{editor}{\bibfnamefont{A.~A.}
  \bibnamefont{Maradudin}} (\bibinfo{publisher}{North Holland (Amsterdam)}),
  volume \bibinfo{volume}{20.1}.

\bibitem[{\citenamefont{Brey and Fertig}(2006{\natexlab{a}})}]{Brey106}
\bibinfo{author}{\bibnamefont{Brey}, \bibfnamefont{L.}}, and
  \bibinfo{author}{\bibfnamefont{H.}~\bibnamefont{Fertig}},
  \bibinfo{year}{2006}{\natexlab{a}}, \bibinfo{journal}{Phys. Rev. B}
  \textbf{\bibinfo{volume}{73}}, \bibinfo{pages}{195408}.

\bibitem[{\citenamefont{Brey and Fertig}(2006{\natexlab{b}})}]{Brey206}
\bibinfo{author}{\bibnamefont{Brey}, \bibfnamefont{L.}}, and
  \bibinfo{author}{\bibfnamefont{H.}~\bibnamefont{Fertig}},
  \bibinfo{year}{2006}{\natexlab{b}}, \bibinfo{journal}{Phys. Rev. B}
  \textbf{\bibinfo{volume}{73}}, \bibinfo{pages}{235411}.

\bibitem[{\citenamefont{Brey} \emph{et~al.}(2007)\citenamefont{Brey, Fertig,
  and Sarma}}]{BFS07}
\bibinfo{author}{\bibnamefont{Brey}, \bibfnamefont{L.}},
  \bibinfo{author}{\bibfnamefont{H.~A.} \bibnamefont{Fertig}}, and
  \bibinfo{author}{\bibfnamefont{S.~D.} \bibnamefont{Sarma}},
  \bibinfo{year}{2007}, \bibinfo{journal}{Phys. Rev. Lett.}
  \textbf{\bibinfo{volume}{99}}, \bibinfo{pages}{116802}.

\bibitem[{\citenamefont{Bunch} \emph{et~al.}(2007)\citenamefont{Bunch, van~der
  Zande, Verbridge, Frank, Tanenbaum, Parpia, Craighead, and McEuen}}]{bunch07}
\bibinfo{author}{\bibnamefont{Bunch}, \bibfnamefont{J.~S.}},
  \bibinfo{author}{\bibfnamefont{A.~M.} \bibnamefont{van~der Zande}},
  \bibinfo{author}{\bibfnamefont{S.~S.} \bibnamefont{Verbridge}},
  \bibinfo{author}{\bibfnamefont{I.~W.} \bibnamefont{Frank}},
  \bibinfo{author}{\bibfnamefont{D.~M.} \bibnamefont{Tanenbaum}},
  \bibinfo{author}{\bibfnamefont{J.~M.} \bibnamefont{Parpia}},
  \bibinfo{author}{\bibfnamefont{H.~G.} \bibnamefont{Craighead}}, and
  \bibinfo{author}{\bibfnamefont{P.~L.} \bibnamefont{McEuen}},
  \bibinfo{year}{2007}, \bibinfo{journal}{Science}
  \textbf{\bibinfo{volume}{315}}, \bibinfo{pages}{490}.

\bibitem[{\citenamefont{Calandra and Mauri}(2007)}]{CM07}
\bibinfo{author}{\bibnamefont{Calandra}, \bibfnamefont{M.}}, and
  \bibinfo{author}{\bibfnamefont{F.}~\bibnamefont{Mauri}},
  \bibinfo{year}{2007}, \bibinfo{journal}{Phys. Rev. B}
  \textbf{\bibinfo{volume}{76}}, \bibinfo{pages}{199901}.

\bibitem[{\citenamefont{Calizo} \emph{et~al.}(2007)\citenamefont{Calizo, Bao,
  Miao, Lau, and Balandin}}]{CBMLB07}
\bibinfo{author}{\bibnamefont{Calizo}, \bibfnamefont{I.}},
  \bibinfo{author}{\bibfnamefont{W.}~\bibnamefont{Bao}},
  \bibinfo{author}{\bibfnamefont{F.}~\bibnamefont{Miao}},
  \bibinfo{author}{\bibfnamefont{C.~N.} \bibnamefont{Lau}}, and
  \bibinfo{author}{\bibfnamefont{A.~A.} \bibnamefont{Balandin}},
  \bibinfo{year}{2007}, \bibinfo{journal}{Appl. Phys. Lett.}
  \textbf{\bibinfo{volume}{91}}, \bibinfo{pages}{201904}.

\bibitem[{\citenamefont{Calogeracos and Dombey}(1999)}]{calogeracos99}
\bibinfo{author}{\bibnamefont{Calogeracos}, \bibfnamefont{A.}}, and
  \bibinfo{author}{\bibfnamefont{N.}~\bibnamefont{Dombey}},
  \bibinfo{year}{1999}, \bibinfo{journal}{Contemp. Phys.}
  \textbf{\bibinfo{volume}{40}}, \bibinfo{pages}{313}.

\bibitem[{\citenamefont{Campagnoli and Tosatti}(1989)}]{CT89}
\bibinfo{author}{\bibnamefont{Campagnoli}, \bibfnamefont{G.}}, and
  \bibinfo{author}{\bibfnamefont{E.}~\bibnamefont{Tosatti}},
  \bibinfo{year}{1989}, in \emph{\bibinfo{booktitle}{Progress on Electron
  Properties of Metals}}, edited by \bibinfo{editor}{\bibnamefont{{R. Girlanda
  {\it et al}}}} (\bibinfo{publisher}{Kluwer Academic Publishing}), p.
  \bibinfo{pages}{337}.

\bibitem[{\citenamefont{Cancado} \emph{et~al.}(2004)\citenamefont{Cancado,
  Pimenta, Neves, Medeiros-Ribeiro, Enoki, Kobayashi, Takai, Fukui,
  Dresselhaus, Saito, and Jorio}}]{Cancado04}
\bibinfo{author}{\bibnamefont{Cancado}, \bibfnamefont{L.~G.}},
  \bibinfo{author}{\bibfnamefont{M.~A.} \bibnamefont{Pimenta}},
  \bibinfo{author}{\bibfnamefont{R.~B.~R.} \bibnamefont{Neves}},
  \bibinfo{author}{\bibfnamefont{G.}~\bibnamefont{Medeiros-Ribeiro}},
  \bibinfo{author}{\bibfnamefont{T.}~\bibnamefont{Enoki}},
  \bibinfo{author}{\bibfnamefont{Y.}~\bibnamefont{Kobayashi}},
  \bibinfo{author}{\bibfnamefont{K.}~\bibnamefont{Takai}},
  \bibinfo{author}{\bibfnamefont{K.-I.} \bibnamefont{Fukui}},
  \bibinfo{author}{\bibfnamefont{M.~S.} \bibnamefont{Dresselhaus}},
  \bibinfo{author}{\bibfnamefont{R.}~\bibnamefont{Saito}}, and
  \bibinfo{author}{\bibfnamefont{A.}~\bibnamefont{Jorio}},
  \bibinfo{year}{2004}, \bibinfo{journal}{Phys. Rev. Lett.}
  \textbf{\bibinfo{volume}{93}}, \bibinfo{pages}{047403}.

\bibitem[{\citenamefont{Casiraghi} \emph{et~al.}(2007)\citenamefont{Casiraghi,
  Hartschuh, Lidorikis, Qian, Harutyunyan, Gokus, Novoselov, and
  Ferrari}}]{CHLQHGNF07}
\bibinfo{author}{\bibnamefont{Casiraghi}, \bibfnamefont{C.}},
  \bibinfo{author}{\bibfnamefont{A.}~\bibnamefont{Hartschuh}},
  \bibinfo{author}{\bibfnamefont{E.}~\bibnamefont{Lidorikis}},
  \bibinfo{author}{\bibfnamefont{H.}~\bibnamefont{Qian}},
  \bibinfo{author}{\bibfnamefont{H.}~\bibnamefont{Harutyunyan}},
  \bibinfo{author}{\bibfnamefont{T.}~\bibnamefont{Gokus}},
  \bibinfo{author}{\bibfnamefont{K.~S.} \bibnamefont{Novoselov}}, and
  \bibinfo{author}{\bibfnamefont{A.~C.} \bibnamefont{Ferrari}},
  \bibinfo{year}{2007}, \bibinfo{journal}{Nano Lett.}
  \textbf{\bibinfo{volume}{7}}, \bibinfo{pages}{2711}.

\bibitem[{\citenamefont{Cassanello and Fradkin}(1996)}]{CF96}
\bibinfo{author}{\bibnamefont{Cassanello}, \bibfnamefont{C.~R.}}, and
  \bibinfo{author}{\bibfnamefont{E.}~\bibnamefont{Fradkin}},
  \bibinfo{year}{1996}, \bibinfo{journal}{Phys. Rev. B}
  \textbf{\bibinfo{volume}{53}}, \bibinfo{pages}{15079}.

\bibitem[{\citenamefont{Cassanello and Fradkin}(1997)}]{CF97}
\bibinfo{author}{\bibnamefont{Cassanello}, \bibfnamefont{C.~R.}}, and
  \bibinfo{author}{\bibfnamefont{E.}~\bibnamefont{Fradkin}},
  \bibinfo{year}{1997}, \bibinfo{journal}{Phys. Rev. B}
  \textbf{\bibinfo{volume}{56}}, \bibinfo{pages}{11246}.

\bibitem[{\citenamefont{Castillo} \emph{et~al.}(1997)\citenamefont{Castillo,
  de~C.~Chamon, Fradkin, Goldbart, and Mudry}}]{CCFGM97}
\bibinfo{author}{\bibnamefont{Castillo}, \bibfnamefont{H.~E.}},
  \bibinfo{author}{\bibfnamefont{C.}~\bibnamefont{de~C.~Chamon}},
  \bibinfo{author}{\bibfnamefont{E.}~\bibnamefont{Fradkin}},
  \bibinfo{author}{\bibfnamefont{P.~M.} \bibnamefont{Goldbart}}, and
  \bibinfo{author}{\bibfnamefont{C.}~\bibnamefont{Mudry}},
  \bibinfo{year}{1997}, \bibinfo{journal}{Phys. Rev. B}
  \textbf{\bibinfo{volume}{56}}, \bibinfo{pages}{10668}.

\bibitem[{\citenamefont{Castro}
  \emph{et~al.}(2007{\natexlab{a}})\citenamefont{Castro, Novoselov, Morozov,
  Peres, {Lopes dos Santos}, Nilsson, Guinea, Geim, and {Castro
  Neto}}}]{Cetal06}
\bibinfo{author}{\bibnamefont{Castro}, \bibfnamefont{E.~V.}},
  \bibinfo{author}{\bibfnamefont{K.~S.} \bibnamefont{Novoselov}},
  \bibinfo{author}{\bibfnamefont{S.~V.} \bibnamefont{Morozov}},
  \bibinfo{author}{\bibfnamefont{N.~M.~R.} \bibnamefont{Peres}},
  \bibinfo{author}{\bibfnamefont{J.}~\bibnamefont{{Lopes dos Santos}}},
  \bibinfo{author}{\bibfnamefont{J.}~\bibnamefont{Nilsson}},
  \bibinfo{author}{\bibfnamefont{F.}~\bibnamefont{Guinea}},
  \bibinfo{author}{\bibfnamefont{A.~K.} \bibnamefont{Geim}}, and
  \bibinfo{author}{\bibfnamefont{A.~H.} \bibnamefont{{Castro Neto}}},
  \bibinfo{year}{2007}{\natexlab{a}}, \bibinfo{journal}{Phys. Rev. Lett.}
  \textbf{\bibinfo{volume}{99}}, \bibinfo{pages}{216802}.

\bibitem[{\citenamefont{Castro}
  \emph{et~al.}(2007{\natexlab{b}})\citenamefont{Castro, Peres, {Lopes dos
  Santos}, {Castro Neto}, and Guinea}}]{Cetal07}
\bibinfo{author}{\bibnamefont{Castro}, \bibfnamefont{E.~V.}},
  \bibinfo{author}{\bibfnamefont{N.~M.~R.} \bibnamefont{Peres}},
  \bibinfo{author}{\bibfnamefont{J.~M.~B.} \bibnamefont{{Lopes dos Santos}}},
  \bibinfo{author}{\bibfnamefont{A.~H.} \bibnamefont{{Castro Neto}}}, and
  \bibinfo{author}{\bibfnamefont{F.}~\bibnamefont{Guinea}},
  \bibinfo{year}{2007}{\natexlab{b}}, \bibinfo{journal}{Phys. Rev. Lett.}
  \textbf{\bibinfo{volume}{100}}, \bibinfo{pages}{026802}.

\bibitem[{\citenamefont{{Castro Neto}}(2007)}]{CN07}
\bibinfo{author}{\bibnamefont{{Castro Neto}}, \bibfnamefont{A.~H.}},
  \bibinfo{year}{2007}, \bibinfo{journal}{Nature Materials}
  \textbf{\bibinfo{volume}{6}}, \bibinfo{pages}{176}.

\bibitem[{\citenamefont{{Castro Neto} and Guinea}(2007)}]{CMG07}
\bibinfo{author}{\bibnamefont{{Castro Neto}}, \bibfnamefont{A.~H.}}, and
  \bibinfo{author}{\bibfnamefont{F.}~\bibnamefont{Guinea}},
  \bibinfo{year}{2007}, \bibinfo{journal}{Phys. Rev. B}
  \textbf{\bibinfo{volume}{75}}, \bibinfo{pages}{045404}.

\bibitem[{\citenamefont{{Castro Neto}}
  \emph{et~al.}(2006{\natexlab{a}})\citenamefont{{Castro Neto}, Guinea, and
  Peres}}]{pw}
\bibinfo{author}{\bibnamefont{{Castro Neto}}, \bibfnamefont{A.~H.}},
  \bibinfo{author}{\bibfnamefont{F.}~\bibnamefont{Guinea}}, and
  \bibinfo{author}{\bibfnamefont{N.~M.~R.} \bibnamefont{Peres}},
  \bibinfo{year}{2006}{\natexlab{a}}, \bibinfo{journal}{Physics World}
  \textbf{\bibinfo{volume}{19}}, \bibinfo{pages}{33}.

\bibitem[{\citenamefont{{Castro Neto}}
  \emph{et~al.}(2006{\natexlab{b}})\citenamefont{{Castro Neto}, Guinea, and
  Peres}}]{NGP06}
\bibinfo{author}{\bibnamefont{{Castro Neto}}, \bibfnamefont{A.~H.}},
  \bibinfo{author}{\bibfnamefont{F.}~\bibnamefont{Guinea}}, and
  \bibinfo{author}{\bibfnamefont{N.~M.~R.} \bibnamefont{Peres}},
  \bibinfo{year}{2006}{\natexlab{b}}, \bibinfo{journal}{Phys. Rev. B}
  \textbf{\bibinfo{volume}{73}}, \bibinfo{pages}{205408}.

\bibitem[{\citenamefont{Chaikin and Lubensky}(1995)}]{chaikin}
\bibinfo{author}{\bibnamefont{Chaikin}, \bibfnamefont{P.}}, and
  \bibinfo{author}{\bibfnamefont{T.~C.} \bibnamefont{Lubensky}},
  \bibinfo{year}{1995}, \emph{\bibinfo{title}{Introduction to Condensed Matter
  Physics}} (\bibinfo{publisher}{Cambridge University Press}).

\bibitem[{\citenamefont{Chakravarty and Schmid}(1986)}]{CS86}
\bibinfo{author}{\bibnamefont{Chakravarty}, \bibfnamefont{S.}}, and
  \bibinfo{author}{\bibfnamefont{A.}~\bibnamefont{Schmid}},
  \bibinfo{year}{1986}, \bibinfo{journal}{Phys. Rep.}
  \textbf{\bibinfo{volume}{140}}, \bibinfo{pages}{193}.

\bibitem[{\citenamefont{Chamon} \emph{et~al.}(1996)\citenamefont{Chamon, Mudry,
  and Wen}}]{CMW96}
\bibinfo{author}{\bibnamefont{Chamon}, \bibfnamefont{C.~C.}},
  \bibinfo{author}{\bibfnamefont{C.}~\bibnamefont{Mudry}}, and
  \bibinfo{author}{\bibfnamefont{X.-G.} \bibnamefont{Wen}},
  \bibinfo{year}{1996}, \bibinfo{journal}{\prb} \textbf{\bibinfo{volume}{53}},
  \bibinfo{pages}{R7638}.

\bibitem[{\citenamefont{Charlier} \emph{et~al.}(2007)\citenamefont{Charlier,
  Blase, and Roche}}]{nanotube_review}
\bibinfo{author}{\bibnamefont{Charlier}, \bibfnamefont{J.-C.}},
  \bibinfo{author}{\bibfnamefont{X.}~\bibnamefont{Blase}}, and
  \bibinfo{author}{\bibfnamefont{S.}~\bibnamefont{Roche}},
  \bibinfo{year}{2007}, \bibinfo{journal}{Rev. Mod. Phys.}
  \textbf{\bibinfo{volume}{79}}, \bibinfo{pages}{677}.

\bibitem[{\citenamefont{Charlier} \emph{et~al.}(1991)\citenamefont{Charlier,
  Michenaud, Gonze, and Vigneron}}]{CMGV91}
\bibinfo{author}{\bibnamefont{Charlier}, \bibfnamefont{J.~C.}},
  \bibinfo{author}{\bibfnamefont{J.~P.} \bibnamefont{Michenaud}},
  \bibinfo{author}{\bibfnamefont{X.}~\bibnamefont{Gonze}}, and
  \bibinfo{author}{\bibfnamefont{J.~P.} \bibnamefont{Vigneron}},
  \bibinfo{year}{1991}, \bibinfo{journal}{Phys. Rev. B}
  \textbf{\bibinfo{volume}{44}}, \bibinfo{pages}{13237}.

\bibitem[{\citenamefont{Cheianov}
  \emph{et~al.}(2007{\natexlab{a}})\citenamefont{Cheianov, Fal'ko, and
  Altshuler}}]{CFA07}
\bibinfo{author}{\bibnamefont{Cheianov}, \bibfnamefont{V.~V.}},
  \bibinfo{author}{\bibfnamefont{V.}~\bibnamefont{Fal'ko}}, and
  \bibinfo{author}{\bibfnamefont{B.~L.} \bibnamefont{Altshuler}},
  \bibinfo{year}{2007}{\natexlab{a}}, \bibinfo{journal}{Science}
  \textbf{\bibinfo{volume}{315}}, \bibinfo{pages}{252}.

\bibitem[{\citenamefont{Cheianov and Fal'ko}(2006)}]{CF06}
\bibinfo{author}{\bibnamefont{Cheianov}, \bibfnamefont{V.~V.}}, and
  \bibinfo{author}{\bibfnamefont{V.~I.} \bibnamefont{Fal'ko}},
  \bibinfo{year}{2006}, \bibinfo{journal}{Phys. Rev. B}
  \textbf{\bibinfo{volume}{74}}, \bibinfo{pages}{041403}.

\bibitem[{\citenamefont{Cheianov}
  \emph{et~al.}(2007{\natexlab{b}})\citenamefont{Cheianov, Fal'ko, Altshuler,
  and Aleiner}}]{CFAA07}
\bibinfo{author}{\bibnamefont{Cheianov}, \bibfnamefont{V.~V.}},
  \bibinfo{author}{\bibfnamefont{V.~I.} \bibnamefont{Fal'ko}},
  \bibinfo{author}{\bibfnamefont{B.~L.} \bibnamefont{Altshuler}}, and
  \bibinfo{author}{\bibfnamefont{I.~L.} \bibnamefont{Aleiner}},
  \bibinfo{year}{2007}{\natexlab{b}}, \bibinfo{journal}{Phys. Rev. Lett.}
  \textbf{\bibinfo{volume}{99}}, \bibinfo{pages}{176801}.

\bibitem[{\citenamefont{Chen}
  \emph{et~al.}(2007{\natexlab{a}})\citenamefont{Chen, Apalkov, and
  Chakraborty}}]{CAC07}
\bibinfo{author}{\bibnamefont{Chen}, \bibfnamefont{H.-Y.}},
  \bibinfo{author}{\bibfnamefont{V.}~\bibnamefont{Apalkov}}, and
  \bibinfo{author}{\bibfnamefont{T.}~\bibnamefont{Chakraborty}},
  \bibinfo{year}{2007}{\natexlab{a}}, \bibinfo{journal}{Phys. Rev. Lett.}
  \textbf{\bibinfo{volume}{98}}, \bibinfo{pages}{186803}.

\bibitem[{\citenamefont{Chen}
  \emph{et~al.}(2007{\natexlab{b}})\citenamefont{Chen, Jang, Fuhrer, Williams,
  and Ishigami}}]{CJFWI07}
\bibinfo{author}{\bibnamefont{Chen}, \bibfnamefont{J.~H.}},
  \bibinfo{author}{\bibfnamefont{C.}~\bibnamefont{Jang}},
  \bibinfo{author}{\bibfnamefont{M.~S.} \bibnamefont{Fuhrer}},
  \bibinfo{author}{\bibfnamefont{E.~D.} \bibnamefont{Williams}}, and
  \bibinfo{author}{\bibfnamefont{M.}~\bibnamefont{Ishigami}},
  \bibinfo{year}{2007}{\natexlab{b}}, \eprint{arXiv:0708.2408}.

\bibitem[{\citenamefont{Chen}
  \emph{et~al.}(2007{\natexlab{c}})\citenamefont{Chen, Lin, Rooks, and
  Avouris}}]{CLRA07}
\bibinfo{author}{\bibnamefont{Chen}, \bibfnamefont{Z.}},
  \bibinfo{author}{\bibfnamefont{Y.-M.} \bibnamefont{Lin}},
  \bibinfo{author}{\bibfnamefont{M.~J.} \bibnamefont{Rooks}}, and
  \bibinfo{author}{\bibfnamefont{P.}~\bibnamefont{Avouris}},
  \bibinfo{year}{2007}{\natexlab{c}}, \bibinfo{journal}{Physica E}
  \textbf{\bibinfo{volume}{40/2}}, \bibinfo{pages}{228}.

\bibitem[{\citenamefont{Cho} \emph{et~al.}(2007)\citenamefont{Cho, Chen, and
  Fuhrer}}]{CCF07}
\bibinfo{author}{\bibnamefont{Cho}, \bibfnamefont{S.}},
  \bibinfo{author}{\bibfnamefont{Y.-F.} \bibnamefont{Chen}}, and
  \bibinfo{author}{\bibfnamefont{M.~S.} \bibnamefont{Fuhrer}},
  \bibinfo{year}{2007}, \bibinfo{journal}{Appl. Phys. Lett.}
  \textbf{\bibinfo{volume}{91}}, \bibinfo{pages}{123105}.

\bibitem[{\citenamefont{Coey} \emph{et~al.}(2002)\citenamefont{Coey,
  Venkatesan, Fitzgerald, Douvalis, and Sanders}}]{Cetal02}
\bibinfo{author}{\bibnamefont{Coey}, \bibfnamefont{J.~M.~D.}},
  \bibinfo{author}{\bibfnamefont{M.}~\bibnamefont{Venkatesan}},
  \bibinfo{author}{\bibfnamefont{C.~B.} \bibnamefont{Fitzgerald}},
  \bibinfo{author}{\bibfnamefont{A.~P.} \bibnamefont{Douvalis}}, and
  \bibinfo{author}{\bibfnamefont{I.~S.} \bibnamefont{Sanders}},
  \bibinfo{year}{2002}, \bibinfo{journal}{Nature}
  \textbf{\bibinfo{volume}{420}}, \bibinfo{pages}{156}.

\bibitem[{\citenamefont{Cortijo and Vozmediano}(2007{\natexlab{a}})}]{CV07b}
\bibinfo{author}{\bibnamefont{Cortijo}, \bibfnamefont{A.}}, and
  \bibinfo{author}{\bibfnamefont{M.~A.~H.} \bibnamefont{Vozmediano}},
  \bibinfo{year}{2007}{\natexlab{a}}, \bibinfo{journal}{Nucl. Phys. B}
  \textbf{\bibinfo{volume}{763}}, \bibinfo{pages}{293}.

\bibitem[{\citenamefont{Cortijo and Vozmediano}(2007{\natexlab{b}})}]{CV07}
\bibinfo{author}{\bibnamefont{Cortijo}, \bibfnamefont{A.}}, and
  \bibinfo{author}{\bibfnamefont{M.~A.~H.} \bibnamefont{Vozmediano}},
  \bibinfo{year}{2007}{\natexlab{b}}, \bibinfo{journal}{Europhys. Lett.}
  \textbf{\bibinfo{volume}{77}}, \bibinfo{pages}{47002}.

\bibitem[{\citenamefont{Couraux} \emph{et~al.}(2008)\citenamefont{Couraux,
  {N'Diaye}, Busse, and Michely}}]{CNBM08}
\bibinfo{author}{\bibnamefont{Couraux}, \bibfnamefont{J.}},
  \bibinfo{author}{\bibfnamefont{A.~T.} \bibnamefont{{N'Diaye}}},
  \bibinfo{author}{\bibfnamefont{C.}~\bibnamefont{Busse}}, and
  \bibinfo{author}{\bibfnamefont{T.}~\bibnamefont{Michely}},
  \bibinfo{year}{2008}, \bibinfo{journal}{Nano Letters}
  \bibinfo{note}{DOI:10.1021/nl0728874}.

\bibitem[{\citenamefont{Cserti}
  \emph{et~al.}(2007{\natexlab{a}})\citenamefont{Cserti, Csord\'as, and
  D\'avid}}]{CCD07}
\bibinfo{author}{\bibnamefont{Cserti}, \bibfnamefont{J.}},
  \bibinfo{author}{\bibfnamefont{A.}~\bibnamefont{Csord\'as}}, and
  \bibinfo{author}{\bibfnamefont{G.}~\bibnamefont{D\'avid}},
  \bibinfo{year}{2007}{\natexlab{a}}, \bibinfo{journal}{Phys. Rev. Lett.}
  \textbf{\bibinfo{volume}{99}}, \bibinfo{pages}{066802}.

\bibitem[{\citenamefont{Cserti}
  \emph{et~al.}(2007{\natexlab{b}})\citenamefont{Cserti, Palyi, and
  Peterfalvi}}]{CPP07}
\bibinfo{author}{\bibnamefont{Cserti}, \bibfnamefont{J.}},
  \bibinfo{author}{\bibfnamefont{A.}~\bibnamefont{Palyi}}, and
  \bibinfo{author}{\bibfnamefont{C.}~\bibnamefont{Peterfalvi}},
  \bibinfo{year}{2007}{\natexlab{b}}, \bibinfo{journal}{Phys. Rev. Lett.}
  \textbf{\bibinfo{volume}{99}}, \bibinfo{pages}{246801}.

\bibitem[{\citenamefont{Dahal} \emph{et~al.}(2006)\citenamefont{Dahal,
  Joglekar, Bedell, and Balatsky}}]{DJBB06}
\bibinfo{author}{\bibnamefont{Dahal}, \bibfnamefont{H.}},
  \bibinfo{author}{\bibfnamefont{Y.~N.} \bibnamefont{Joglekar}},
  \bibinfo{author}{\bibfnamefont{K.}~\bibnamefont{Bedell}}, and
  \bibinfo{author}{\bibfnamefont{A.~V.} \bibnamefont{Balatsky}},
  \bibinfo{year}{2006}, \bibinfo{journal}{Phys. Rev. B}
  \textbf{\bibinfo{volume}{74}}, \bibinfo{pages}{233405}.

\bibitem[{\citenamefont{Dahal} \emph{et~al.}(2007)\citenamefont{Dahal, Wehling,
  Bedell, Zhu, and Balatsky}}]{DWBZB07}
\bibinfo{author}{\bibnamefont{Dahal}, \bibfnamefont{H.~P.}},
  \bibinfo{author}{\bibfnamefont{T.~O.} \bibnamefont{Wehling}},
  \bibinfo{author}{\bibfnamefont{K.~S.} \bibnamefont{Bedell}},
  \bibinfo{author}{\bibfnamefont{J.-X.} \bibnamefont{Zhu}}, and
  \bibinfo{author}{\bibfnamefont{A.~V.} \bibnamefont{Balatsky}},
  \bibinfo{year}{2007}, \eprint{cond-mat/0706.1689}.

\bibitem[{\citenamefont{Das} \emph{et~al.}(2007)\citenamefont{Das, Chakraborty,
  and Sood}}]{DCS07}
\bibinfo{author}{\bibnamefont{Das}, \bibfnamefont{A.}},
  \bibinfo{author}{\bibfnamefont{B.}~\bibnamefont{Chakraborty}}, and
  \bibinfo{author}{\bibfnamefont{A.~K.} \bibnamefont{Sood}},
  \bibinfo{year}{2007}, \eprint{arXiv:0710.4160}.

\bibitem[{\citenamefont{Deacon} \emph{et~al.}(2007)\citenamefont{Deacon,
  Chuang, Nicholas, Novoselov, and Geim}}]{DCNNG07}
\bibinfo{author}{\bibnamefont{Deacon}, \bibfnamefont{R.~S.}},
  \bibinfo{author}{\bibfnamefont{K.-C.} \bibnamefont{Chuang}},
  \bibinfo{author}{\bibfnamefont{R.~J.} \bibnamefont{Nicholas}},
  \bibinfo{author}{\bibfnamefont{K.~S.} \bibnamefont{Novoselov}}, and
  \bibinfo{author}{\bibfnamefont{A.~K.} \bibnamefont{Geim}},
  \bibinfo{year}{2007}, \bibinfo{journal}{Phys. Rev. B}
  \textbf{\bibinfo{volume}{76}}, \bibinfo{pages}{081406(R)}.

\bibitem[{\citenamefont{{Dell'Anna}}(2006)}]{DAnna06}
\bibinfo{author}{\bibnamefont{{Dell'Anna}}, \bibfnamefont{L.}},
  \bibinfo{year}{2006}, \bibinfo{journal}{Nucl.Phys.B}
  \textbf{\bibinfo{volume}{758}}, \bibinfo{pages}{255}.

\bibitem[{\citenamefont{Dharma-Wardana}(2007)}]{DW0703}
\bibinfo{author}{\bibnamefont{Dharma-Wardana}, \bibfnamefont{M.~W.~C.}},
  \bibinfo{year}{2007}, \bibinfo{journal}{J. Phys.:Condens. Matter}
  \textbf{\bibinfo{volume}{19}}, \bibinfo{pages}{386228}.

\bibitem[{\citenamefont{Dienwiebel}
  \emph{et~al.}(2004)\citenamefont{Dienwiebel, Verhoeven, Pradeep, Frenken,
  Heimberg, and Zandbergen}}]{superlub}
\bibinfo{author}{\bibnamefont{Dienwiebel}, \bibfnamefont{M.}},
  \bibinfo{author}{\bibfnamefont{G.~S.} \bibnamefont{Verhoeven}},
  \bibinfo{author}{\bibfnamefont{N.}~\bibnamefont{Pradeep}},
  \bibinfo{author}{\bibfnamefont{J.~W.~M.} \bibnamefont{Frenken}},
  \bibinfo{author}{\bibfnamefont{J.~A.} \bibnamefont{Heimberg}}, and
  \bibinfo{author}{\bibfnamefont{H.~W.} \bibnamefont{Zandbergen}},
  \bibinfo{year}{2004}, \bibinfo{journal}{Phys. Rev. Lett.}
  \textbf{\bibinfo{volume}{92}}, \bibinfo{pages}{126101}.

\bibitem[{\citenamefont{Dillon} \emph{et~al.}(1977)\citenamefont{Dillon, Spain,
  and McClure}}]{DSM77}
\bibinfo{author}{\bibnamefont{Dillon}, \bibfnamefont{R.~O.}},
  \bibinfo{author}{\bibfnamefont{I.~L.} \bibnamefont{Spain}}, and
  \bibinfo{author}{\bibfnamefont{J.~W.} \bibnamefont{McClure}},
  \bibinfo{year}{1977}, \bibinfo{journal}{J. Phys. Chem. Sol.}
  \textbf{\bibinfo{volume}{38}}, \bibinfo{pages}{635}.

\bibitem[{\citenamefont{DiVincenzo and Mele}(1984)}]{mele}
\bibinfo{author}{\bibnamefont{DiVincenzo}, \bibfnamefont{D.~P.}}, and
  \bibinfo{author}{\bibfnamefont{E.~J.} \bibnamefont{Mele}},
  \bibinfo{year}{1984}, \bibinfo{journal}{Phys.Rev.B}
  \textbf{\bibinfo{volume}{29}}, \bibinfo{pages}{1685}.

\bibitem[{\citenamefont{Dong} \emph{et~al.}(1998)\citenamefont{Dong, Hou, and
  Ma}}]{DHM98}
\bibinfo{author}{\bibnamefont{Dong}, \bibfnamefont{S.-H.}},
  \bibinfo{author}{\bibfnamefont{X.-W.} \bibnamefont{Hou}}, and
  \bibinfo{author}{\bibfnamefont{Z.-Q.} \bibnamefont{Ma}},
  \bibinfo{year}{1998}, \bibinfo{journal}{Phys. Rev. A}
  \textbf{\bibinfo{volume}{58}}, \bibinfo{pages}{2160}.

\bibitem[{\citenamefont{Doretto and {Morais Smith}}(2007)}]{DMS07}
\bibinfo{author}{\bibnamefont{Doretto}, \bibfnamefont{R.~L.}}, and
  \bibinfo{author}{\bibfnamefont{C.}~\bibnamefont{{Morais Smith}}},
  \bibinfo{year}{2007}, \eprint{cond-mat/0704.3671}.

\bibitem[{\citenamefont{Dresselhaus and Dresselhaus}(1965)}]{DD65}
\bibinfo{author}{\bibnamefont{Dresselhaus}, \bibfnamefont{G.}}, and
  \bibinfo{author}{\bibfnamefont{M.~S.} \bibnamefont{Dresselhaus}},
  \bibinfo{year}{1965}, \bibinfo{journal}{Phys. Rev.}
  \textbf{\bibinfo{volume}{140}}, \bibinfo{pages}{A401}.

\bibitem[{\citenamefont{Dresselhaus and Dresselhaus}(2002)}]{gic}
\bibinfo{author}{\bibnamefont{Dresselhaus}, \bibfnamefont{M.~S.}}, and
  \bibinfo{author}{\bibfnamefont{G.}~\bibnamefont{Dresselhaus}},
  \bibinfo{year}{2002}, \bibinfo{journal}{Advances in Physics}
  \textbf{\bibinfo{volume}{51}}, \bibinfo{pages}{1}.

\bibitem[{\citenamefont{Dresselhaus}
  \emph{et~al.}(1983)\citenamefont{Dresselhaus, Dresselhaus, Fischer, and
  Moran}}]{DDFM83}
\bibinfo{author}{\bibnamefont{Dresselhaus}, \bibfnamefont{M.~S.}},
  \bibinfo{author}{\bibfnamefont{G.}~\bibnamefont{Dresselhaus}},
  \bibinfo{author}{\bibfnamefont{J.~E.} \bibnamefont{Fischer}}, and
  \bibinfo{author}{\bibfnamefont{M.~J.} \bibnamefont{Moran}},
  \bibinfo{year}{1983}, \emph{\bibinfo{title}{Intercalated graphite}}
  (\bibinfo{publisher}{North-Holland, New York}).

\bibitem[{\citenamefont{Dresselhaus and Mavroides}(1964)}]{DM64}
\bibinfo{author}{\bibnamefont{Dresselhaus}, \bibfnamefont{M.~S.}}, and
  \bibinfo{author}{\bibfnamefont{J.~G.} \bibnamefont{Mavroides}},
  \bibinfo{year}{1964}, \bibinfo{journal}{IBM J. Res. Dev}
  \textbf{\bibinfo{volume}{8}}, \bibinfo{pages}{262}.

\bibitem[{\citenamefont{Dugaev} \emph{et~al.}(2006)\citenamefont{Dugaev,
  Litvinov, and Barnas}}]{DLB06}
\bibinfo{author}{\bibnamefont{Dugaev}, \bibfnamefont{V.~K.}},
  \bibinfo{author}{\bibfnamefont{V.~I.} \bibnamefont{Litvinov}}, and
  \bibinfo{author}{\bibfnamefont{J.}~\bibnamefont{Barnas}},
  \bibinfo{year}{2006}, \bibinfo{journal}{Phys. Rev. B}
  \textbf{\bibinfo{volume}{74}}, \bibinfo{pages}{224438}.

\bibitem[{\citenamefont{Duke}(1968)}]{Duke68}
\bibinfo{author}{\bibnamefont{Duke}, \bibfnamefont{C.~B.}},
  \bibinfo{year}{1968}, \bibinfo{journal}{Phys. Rev.}
  \textbf{\bibinfo{volume}{168}}, \bibinfo{pages}{816}.

\bibitem[{\citenamefont{Eisenstein}
  \emph{et~al.}(1994)\citenamefont{Eisenstein, Pfeiffer, and West}}]{EPW94}
\bibinfo{author}{\bibnamefont{Eisenstein}, \bibfnamefont{J.~P.}},
  \bibinfo{author}{\bibfnamefont{L.~N.} \bibnamefont{Pfeiffer}}, and
  \bibinfo{author}{\bibfnamefont{K.~W.} \bibnamefont{West}},
  \bibinfo{year}{1994}, \bibinfo{journal}{Phys. Rev. B}
  \textbf{\bibinfo{volume}{50}}, \bibinfo{pages}{1760}.

\bibitem[{\citenamefont{Eizenberg and Blakely}(1979)}]{EB79}
\bibinfo{author}{\bibnamefont{Eizenberg}, \bibfnamefont{M.}}, and
  \bibinfo{author}{\bibfnamefont{J.~M.} \bibnamefont{Blakely}},
  \bibinfo{year}{1979}, \bibinfo{journal}{Surf. Sci.}
  \textbf{\bibinfo{volume}{82}}, \bibinfo{pages}{228}.

\bibitem[{\citenamefont{Esquinazi} \emph{et~al.}(2002)\citenamefont{Esquinazi,
  Setzer, H\"ohne, Semmelhack, Kopelevich, Spemann, Butz, Kohlstrunk, and
  L\"osch}}]{Eetal02}
\bibinfo{author}{\bibnamefont{Esquinazi}, \bibfnamefont{P.}},
  \bibinfo{author}{\bibfnamefont{A.}~\bibnamefont{Setzer}},
  \bibinfo{author}{\bibfnamefont{R.}~\bibnamefont{H\"ohne}},
  \bibinfo{author}{\bibfnamefont{C.}~\bibnamefont{Semmelhack}},
  \bibinfo{author}{\bibfnamefont{Y.}~\bibnamefont{Kopelevich}},
  \bibinfo{author}{\bibfnamefont{D.}~\bibnamefont{Spemann}},
  \bibinfo{author}{\bibfnamefont{T.}~\bibnamefont{Butz}},
  \bibinfo{author}{\bibfnamefont{B.}~\bibnamefont{Kohlstrunk}}, and
  \bibinfo{author}{\bibfnamefont{M.}~\bibnamefont{L\"osch}},
  \bibinfo{year}{2002}, \bibinfo{journal}{\prb} \textbf{\bibinfo{volume}{66}},
  \bibinfo{pages}{024429}.

\bibitem[{\citenamefont{Esquinazi} \emph{et~al.}(2003)\citenamefont{Esquinazi,
  Spemann, H\"ohne, Setzer, Han, and Butz}}]{Eetal03b}
\bibinfo{author}{\bibnamefont{Esquinazi}, \bibfnamefont{P.}},
  \bibinfo{author}{\bibfnamefont{D.}~\bibnamefont{Spemann}},
  \bibinfo{author}{\bibfnamefont{R.}~\bibnamefont{H\"ohne}},
  \bibinfo{author}{\bibfnamefont{A.}~\bibnamefont{Setzer}},
  \bibinfo{author}{\bibfnamefont{K.-H.} \bibnamefont{Han}}, and
  \bibinfo{author}{\bibfnamefont{T.}~\bibnamefont{Butz}}, \bibinfo{year}{2003},
  \bibinfo{journal}{Phys. Rev. Lett.} \textbf{\bibinfo{volume}{91}},
  \bibinfo{pages}{227201}.

\bibitem[{\citenamefont{{Eun-Ah Kim} and {Castro Neto}}(2007)}]{NK07}
\bibinfo{author}{\bibnamefont{{Eun-Ah Kim}}}, and
  \bibinfo{author}{\bibfnamefont{A.~H.} \bibnamefont{{Castro Neto}}},
  \bibinfo{year}{2007}, \eprint{cond-mat/0702562}.

\bibitem[{\citenamefont{Fasolino} \emph{et~al.}(2007)\citenamefont{Fasolino,
  Los, and Katsnelson}}]{FLK07}
\bibinfo{author}{\bibnamefont{Fasolino}, \bibfnamefont{A.}},
  \bibinfo{author}{\bibfnamefont{J.~H.} \bibnamefont{Los}}, and
  \bibinfo{author}{\bibfnamefont{M.~I.} \bibnamefont{Katsnelson}},
  \bibinfo{year}{2007}, \bibinfo{journal}{Nat. Mat.}
  \textbf{\bibinfo{volume}{6}}, \bibinfo{pages}{858}.

\bibitem[{\citenamefont{Faugeras} \emph{et~al.}(2007)\citenamefont{Faugeras,
  Nerriere, Potemski, Mahmood, Dujardin, Berger, and de~Heer}}]{FNPMDBH07}
\bibinfo{author}{\bibnamefont{Faugeras}, \bibfnamefont{C.}},
  \bibinfo{author}{\bibfnamefont{A.}~\bibnamefont{Nerriere}},
  \bibinfo{author}{\bibfnamefont{M.}~\bibnamefont{Potemski}},
  \bibinfo{author}{\bibfnamefont{A.}~\bibnamefont{Mahmood}},
  \bibinfo{author}{\bibfnamefont{E.}~\bibnamefont{Dujardin}},
  \bibinfo{author}{\bibfnamefont{C.}~\bibnamefont{Berger}}, and
  \bibinfo{author}{\bibfnamefont{W.~A.} \bibnamefont{de~Heer}},
  \bibinfo{year}{2007}, \eprint{arXiv:0709.2538}.

\bibitem[{\citenamefont{Fauser} \emph{et~al.}(2007)\citenamefont{Fauser,
  Tolksdorf, and Zeidler}}]{qgravity}
\bibinfo{author}{\bibnamefont{Fauser}, \bibfnamefont{B.}},
  \bibinfo{author}{\bibfnamefont{J.}~\bibnamefont{Tolksdorf}}, and
  \bibinfo{author}{\bibfnamefont{E.}~\bibnamefont{Zeidler}},
  \bibinfo{year}{2007}, \emph{\bibinfo{title}{Quantum gravity}}
  (\bibinfo{publisher}{Birkh\"auser}).

\bibitem[{\citenamefont{Ferrari} \emph{et~al.}(2006)\citenamefont{Ferrari,
  Meyer, Scardaci, Casiraghi, Lazzeri, Mauri, Piscanec, Jiang, Novoselov, and
  S.~Roth}}]{FMSCLMPJNG06}
\bibinfo{author}{\bibnamefont{Ferrari}, \bibfnamefont{A.~C.}},
  \bibinfo{author}{\bibfnamefont{J.~C.} \bibnamefont{Meyer}},
  \bibinfo{author}{\bibfnamefont{V.}~\bibnamefont{Scardaci}},
  \bibinfo{author}{\bibfnamefont{C.}~\bibnamefont{Casiraghi}},
  \bibinfo{author}{\bibfnamefont{M.}~\bibnamefont{Lazzeri}},
  \bibinfo{author}{\bibfnamefont{F.}~\bibnamefont{Mauri}},
  \bibinfo{author}{\bibfnamefont{S.}~\bibnamefont{Piscanec}},
  \bibinfo{author}{\bibfnamefont{D.}~\bibnamefont{Jiang}},
  \bibinfo{author}{\bibfnamefont{K.~S.} \bibnamefont{Novoselov}}, and
  \bibinfo{author}{\bibfnamefont{A.~K.~G.} \bibnamefont{S.~Roth}},
  \bibinfo{year}{2006}, \bibinfo{journal}{Phys. Rev. Lett.}
  \textbf{\bibinfo{volume}{97}}, \bibinfo{pages}{187401}.

\bibitem[{\citenamefont{Fertig and Brey}(2006)}]{FB06}
\bibinfo{author}{\bibnamefont{Fertig}, \bibfnamefont{H.~A.}}, and
  \bibinfo{author}{\bibfnamefont{L.}~\bibnamefont{Brey}}, \bibinfo{year}{2006},
  \bibinfo{journal}{Phys. Rev. Lett.} \textbf{\bibinfo{volume}{97}},
  \bibinfo{pages}{116805}.

\bibitem[{\citenamefont{Fogler}
  \emph{et~al.}(2007{\natexlab{a}})\citenamefont{Fogler, Glazman, Novikov, and
  Shklovskii}}]{FGNS07}
\bibinfo{author}{\bibnamefont{Fogler}, \bibfnamefont{M.~M.}},
  \bibinfo{author}{\bibfnamefont{L.~I.} \bibnamefont{Glazman}},
  \bibinfo{author}{\bibfnamefont{D.~S.} \bibnamefont{Novikov}}, and
  \bibinfo{author}{\bibfnamefont{B.~I.} \bibnamefont{Shklovskii}},
  \bibinfo{year}{2007}{\natexlab{a}}, \eprint{arXiv:0710.2150}.

\bibitem[{\citenamefont{Fogler}
  \emph{et~al.}(2007{\natexlab{b}})\citenamefont{Fogler, Novikov, and
  Shklovskii}}]{FNS07}
\bibinfo{author}{\bibnamefont{Fogler}, \bibfnamefont{M.~M.}},
  \bibinfo{author}{\bibfnamefont{D.~S.} \bibnamefont{Novikov}}, and
  \bibinfo{author}{\bibfnamefont{B.~I.} \bibnamefont{Shklovskii}},
  \bibinfo{year}{2007}{\natexlab{b}}, \bibinfo{journal}{Phys. Rev. B}
  \textbf{\bibinfo{volume}{76}}, \bibinfo{pages}{233402}.

\bibitem[{\citenamefont{Forbeaux} \emph{et~al.}(1998)\citenamefont{Forbeaux,
  Themlin, and Debever}}]{FTD98}
\bibinfo{author}{\bibnamefont{Forbeaux}, \bibfnamefont{I.}},
  \bibinfo{author}{\bibfnamefont{J.-M.} \bibnamefont{Themlin}}, and
  \bibinfo{author}{\bibfnamefont{J.-M.} \bibnamefont{Debever}},
  \bibinfo{year}{1998}, \bibinfo{journal}{Phys. Rev. B}
  \textbf{\bibinfo{volume}{58}}, \bibinfo{pages}{16396}.

\bibitem[{\citenamefont{Foster and Ludwig}(2006{\natexlab{a}})}]{FA05}
\bibinfo{author}{\bibnamefont{Foster}, \bibfnamefont{M.~S.}}, and
  \bibinfo{author}{\bibfnamefont{A.~W.~W.} \bibnamefont{Ludwig}},
  \bibinfo{year}{2006}{\natexlab{a}}, \bibinfo{journal}{Phys. Rev. B}
  \textbf{\bibinfo{volume}{73}}, \bibinfo{pages}{155104}.

\bibitem[{\citenamefont{Foster and Ludwig}(2006{\natexlab{b}})}]{FA06}
\bibinfo{author}{\bibnamefont{Foster}, \bibfnamefont{M.~S.}}, and
  \bibinfo{author}{\bibfnamefont{A.~W.~W.} \bibnamefont{Ludwig}},
  \bibinfo{year}{2006}{\natexlab{b}}, \bibinfo{journal}{Phys. Rev. B}
  \textbf{\bibinfo{volume}{74}}, \bibinfo{pages}{241102}.

\bibitem[{\citenamefont{Fradkin}(1986{\natexlab{a}})}]{F86}
\bibinfo{author}{\bibnamefont{Fradkin}, \bibfnamefont{E.}},
  \bibinfo{year}{1986}{\natexlab{a}}, \bibinfo{journal}{\prb}
  \textbf{\bibinfo{volume}{33}}, \bibinfo{pages}{3257}.

\bibitem[{\citenamefont{Fradkin}(1986{\natexlab{b}})}]{F86b}
\bibinfo{author}{\bibnamefont{Fradkin}, \bibfnamefont{E.}},
  \bibinfo{year}{1986}{\natexlab{b}}, \bibinfo{journal}{\prb}
  \textbf{\bibinfo{volume}{33}}, \bibinfo{pages}{3263}.

\bibitem[{\citenamefont{Fritz} \emph{et~al.}(2006)\citenamefont{Fritz, Florens,
  and Vojta}}]{FFV06}
\bibinfo{author}{\bibnamefont{Fritz}, \bibfnamefont{L.}},
  \bibinfo{author}{\bibfnamefont{S.}~\bibnamefont{Florens}}, and
  \bibinfo{author}{\bibfnamefont{M.}~\bibnamefont{Vojta}},
  \bibinfo{year}{2006}, \bibinfo{journal}{Phys. Rev. B}
  \textbf{\bibinfo{volume}{74}}, \bibinfo{pages}{144410}.

\bibitem[{\citenamefont{Fuchs and Lederer}(2007)}]{FL07}
\bibinfo{author}{\bibnamefont{Fuchs}, \bibfnamefont{J.-N.}}, and
  \bibinfo{author}{\bibfnamefont{P.}~\bibnamefont{Lederer}},
  \bibinfo{year}{2007}, \bibinfo{journal}{Phys. Rev. Lett.}
  \textbf{\bibinfo{volume}{98}}, \bibinfo{pages}{016803}.

\bibitem[{\citenamefont{Fujita} \emph{et~al.}(1996)\citenamefont{Fujita,
  Wakabayashi, Nakada, and Kusakabe}}]{FWNK96}
\bibinfo{author}{\bibnamefont{Fujita}, \bibfnamefont{M.}},
  \bibinfo{author}{\bibfnamefont{K.}~\bibnamefont{Wakabayashi}},
  \bibinfo{author}{\bibfnamefont{K.}~\bibnamefont{Nakada}}, and
  \bibinfo{author}{\bibfnamefont{K.}~\bibnamefont{Kusakabe}},
  \bibinfo{year}{1996}, \bibinfo{journal}{J. Phys. Soc. Jpn.}
  \textbf{\bibinfo{volume}{65}}, \bibinfo{pages}{1920}.

\bibitem[{\citenamefont{Fukuyama}(1971)}]{FK1971}
\bibinfo{author}{\bibnamefont{Fukuyama}, \bibfnamefont{H.}},
  \bibinfo{year}{1971}, \bibinfo{journal}{Prog. Theor. Phys.}
  \textbf{\bibinfo{volume}{45}}, \bibinfo{pages}{704}.

\bibitem[{\citenamefont{Gasparoux}(1967)}]{G67}
\bibinfo{author}{\bibnamefont{Gasparoux}, \bibfnamefont{H.}},
  \bibinfo{year}{1967}, \bibinfo{journal}{Carbon} \textbf{\bibinfo{volume}{5}},
  \bibinfo{pages}{441}.

\bibitem[{\citenamefont{Geim and MacDonald}(2007)}]{GMac07}
\bibinfo{author}{\bibnamefont{Geim}, \bibfnamefont{A.~K.}}, and
  \bibinfo{author}{\bibfnamefont{A.~H.} \bibnamefont{MacDonald}},
  \bibinfo{year}{2007}, \bibinfo{journal}{Physics Today}
  \textbf{\bibinfo{volume}{60}}, \bibinfo{pages}{35}.

\bibitem[{\citenamefont{Geim and Novoselov}(2007)}]{geim_review}
\bibinfo{author}{\bibnamefont{Geim}, \bibfnamefont{A.~K.}}, and
  \bibinfo{author}{\bibfnamefont{K.~S.} \bibnamefont{Novoselov}},
  \bibinfo{year}{2007}, \bibinfo{journal}{Nature Materials}
  \textbf{\bibinfo{volume}{6}}, \bibinfo{pages}{183}.

\bibitem[{\citenamefont{de~Gennes}(1964)}]{G64}
\bibinfo{author}{\bibnamefont{de~Gennes}, \bibfnamefont{P.~G.}},
  \bibinfo{year}{1964}, \bibinfo{journal}{Rev. Mod. Phys.}
  \textbf{\bibinfo{volume}{36}}, \bibinfo{pages}{225}.

\bibitem[{\citenamefont{Ghosal} \emph{et~al.}(2007)\citenamefont{Ghosal,
  Goswami, and Chakravarty}}]{GGC2007}
\bibinfo{author}{\bibnamefont{Ghosal}, \bibfnamefont{A.}},
  \bibinfo{author}{\bibfnamefont{P.}~\bibnamefont{Goswami}}, and
  \bibinfo{author}{\bibfnamefont{S.}~\bibnamefont{Chakravarty}},
  \bibinfo{year}{2007}, \bibinfo{journal}{Phys. Rev.}
  \textbf{\bibinfo{volume}{75}}, \bibinfo{pages}{115123}.

\bibitem[{\citenamefont{Giesbers} \emph{et~al.}(2007)\citenamefont{Giesbers,
  Zeitler, Katsnelson, Ponomarenko, Mohiuddin, and Maan}}]{GZKPGM07}
\bibinfo{author}{\bibnamefont{Giesbers}, \bibfnamefont{A.~J.~M.}},
  \bibinfo{author}{\bibfnamefont{U.}~\bibnamefont{Zeitler}},
  \bibinfo{author}{\bibfnamefont{M.~I.} \bibnamefont{Katsnelson}},
  \bibinfo{author}{\bibfnamefont{L.~A.} \bibnamefont{Ponomarenko}},
  \bibinfo{author}{\bibfnamefont{T.~M.~G.} \bibnamefont{Mohiuddin}}, and
  \bibinfo{author}{\bibfnamefont{J.~C.} \bibnamefont{Maan}},
  \bibinfo{year}{2007}, \bibinfo{journal}{Phys. Rev. Lett.}
  \textbf{\bibinfo{volume}{99}}, \bibinfo{pages}{206803}.

\bibitem[{\citenamefont{Giovannetti}
  \emph{et~al.}(2007)\citenamefont{Giovannetti, Khomyakov, Brocks, Kelly, and
  {van der Brink}}}]{GKBKB07}
\bibinfo{author}{\bibnamefont{Giovannetti}, \bibfnamefont{G.}},
  \bibinfo{author}{\bibfnamefont{P.~A.} \bibnamefont{Khomyakov}},
  \bibinfo{author}{\bibfnamefont{G.}~\bibnamefont{Brocks}},
  \bibinfo{author}{\bibfnamefont{P.~J.} \bibnamefont{Kelly}}, and
  \bibinfo{author}{\bibfnamefont{J.}~\bibnamefont{{van der Brink}}},
  \bibinfo{year}{2007}, \bibinfo{journal}{Phys. Rev. B}
  \textbf{\bibinfo{volume}{76}}, \bibinfo{pages}{073103}.

\bibitem[{\citenamefont{Giuliani and Vignale}(2005)}]{giovani}
\bibinfo{author}{\bibnamefont{Giuliani}, \bibfnamefont{G.~F.}}, and
  \bibinfo{author}{\bibfnamefont{G.}~\bibnamefont{Vignale}},
  \bibinfo{year}{2005}, \emph{\bibinfo{title}{Quantum theory of the electron
  liquid}} (\bibinfo{publisher}{Cambridge Press}).

\bibitem[{\citenamefont{Goerbig} \emph{et~al.}(2007)\citenamefont{Goerbig,
  Fuchs, Kechedzhi, and Fal'ko}}]{GFKF07}
\bibinfo{author}{\bibnamefont{Goerbig}, \bibfnamefont{M.~O.}},
  \bibinfo{author}{\bibfnamefont{J.-N.} \bibnamefont{Fuchs}},
  \bibinfo{author}{\bibfnamefont{K.}~\bibnamefont{Kechedzhi}}, and
  \bibinfo{author}{\bibfnamefont{V.~I.} \bibnamefont{Fal'ko}},
  \bibinfo{year}{2007}, \bibinfo{journal}{Phys. Rev. Lett.}
  \textbf{\bibinfo{volume}{99}}, \bibinfo{pages}{087402}.

\bibitem[{\citenamefont{Goerbig} \emph{et~al.}(2006)\citenamefont{Goerbig,
  Moessner, and Dou\u{c}ot}}]{GMD06}
\bibinfo{author}{\bibnamefont{Goerbig}, \bibfnamefont{M.~O.}},
  \bibinfo{author}{\bibfnamefont{R.}~\bibnamefont{Moessner}}, and
  \bibinfo{author}{\bibfnamefont{B.}~\bibnamefont{Dou\u{c}ot}},
  \bibinfo{year}{2006}, \bibinfo{journal}{Phys. Rev. B}
  \textbf{\bibinfo{volume}{74}}, \bibinfo{pages}{161407}.

\bibitem[{\citenamefont{Gonz\'alez}
  \emph{et~al.}(1992)\citenamefont{Gonz\'alez, Guinea, and Vozmediano}}]{GGV92}
\bibinfo{author}{\bibnamefont{Gonz\'alez}, \bibfnamefont{J.}},
  \bibinfo{author}{\bibfnamefont{F.}~\bibnamefont{Guinea}}, and
  \bibinfo{author}{\bibfnamefont{M.~A.~H.} \bibnamefont{Vozmediano}},
  \bibinfo{year}{1992}, \bibinfo{journal}{\prl} \textbf{\bibinfo{volume}{69}},
  \bibinfo{pages}{172}.

\bibitem[{\citenamefont{Gonz\'alez}
  \emph{et~al.}(1993{\natexlab{a}})\citenamefont{Gonz\'alez, Guinea, and
  Vozmediano}}]{GGV93}
\bibinfo{author}{\bibnamefont{Gonz\'alez}, \bibfnamefont{J.}},
  \bibinfo{author}{\bibfnamefont{F.}~\bibnamefont{Guinea}}, and
  \bibinfo{author}{\bibfnamefont{M.~A.~H.} \bibnamefont{Vozmediano}},
  \bibinfo{year}{1993}{\natexlab{a}}, \bibinfo{journal}{Mod. Phys. Lett.}
  \textbf{\bibinfo{volume}{B7}}, \bibinfo{pages}{1593}.

\bibitem[{\citenamefont{Gonz\'alez}
  \emph{et~al.}(1993{\natexlab{b}})\citenamefont{Gonz\'alez, Guinea, and
  Vozmediano}}]{GGV93b}
\bibinfo{author}{\bibnamefont{Gonz\'alez}, \bibfnamefont{J.}},
  \bibinfo{author}{\bibfnamefont{F.}~\bibnamefont{Guinea}}, and
  \bibinfo{author}{\bibfnamefont{M.~A.~H.} \bibnamefont{Vozmediano}},
  \bibinfo{year}{1993}{\natexlab{b}}, \bibinfo{journal}{Nucl. Phys. B}
  \textbf{\bibinfo{volume}{406 [FS]}}, \bibinfo{pages}{771}.

\bibitem[{\citenamefont{Gonz\'alez}
  \emph{et~al.}(1994)\citenamefont{Gonz\'alez, Guinea, and Vozmediano}}]{GGV94}
\bibinfo{author}{\bibnamefont{Gonz\'alez}, \bibfnamefont{J.}},
  \bibinfo{author}{\bibfnamefont{F.}~\bibnamefont{Guinea}}, and
  \bibinfo{author}{\bibfnamefont{M.~A.~H.} \bibnamefont{Vozmediano}},
  \bibinfo{year}{1994}, \bibinfo{journal}{Nucl. Phys. B}
  \textbf{\bibinfo{volume}{424}}, \bibinfo{pages}{596}.

\bibitem[{\citenamefont{Gonz\'alez}
  \emph{et~al.}(1996)\citenamefont{Gonz\'alez, Guinea, and Vozmediano}}]{GGV96}
\bibinfo{author}{\bibnamefont{Gonz\'alez}, \bibfnamefont{J.}},
  \bibinfo{author}{\bibfnamefont{F.}~\bibnamefont{Guinea}}, and
  \bibinfo{author}{\bibfnamefont{M.~A.~H.} \bibnamefont{Vozmediano}},
  \bibinfo{year}{1996}, \bibinfo{journal}{Phys. Rev. Lett.}
  \textbf{\bibinfo{volume}{77}}, \bibinfo{pages}{3589}.

\bibitem[{\citenamefont{Gonz\'alez}
  \emph{et~al.}(1999)\citenamefont{Gonz\'alez, Guinea, and Vozmediano}}]{GGV99}
\bibinfo{author}{\bibnamefont{Gonz\'alez}, \bibfnamefont{J.}},
  \bibinfo{author}{\bibfnamefont{F.}~\bibnamefont{Guinea}}, and
  \bibinfo{author}{\bibfnamefont{M.~A.~H.} \bibnamefont{Vozmediano}},
  \bibinfo{year}{1999}, \bibinfo{journal}{\prb} \textbf{\bibinfo{volume}{59}},
  \bibinfo{pages}{R2474}.

\bibitem[{\citenamefont{Gonz\'alez}
  \emph{et~al.}(2001)\citenamefont{Gonz\'alez, Guinea, and Vozmediano}}]{GGV01}
\bibinfo{author}{\bibnamefont{Gonz\'alez}, \bibfnamefont{J.}},
  \bibinfo{author}{\bibfnamefont{F.}~\bibnamefont{Guinea}}, and
  \bibinfo{author}{\bibfnamefont{M.~A.~H.} \bibnamefont{Vozmediano}},
  \bibinfo{year}{2001}, \bibinfo{journal}{\prb} \textbf{\bibinfo{volume}{63}},
  \bibinfo{pages}{134421}.

\bibitem[{\citenamefont{Gorbachev} \emph{et~al.}(2007)\citenamefont{Gorbachev,
  Tikhonenko, Mayorov, Horsell, and Savchenko}}]{GTMHS07}
\bibinfo{author}{\bibnamefont{Gorbachev}, \bibfnamefont{R.~V.}},
  \bibinfo{author}{\bibfnamefont{F.~V.} \bibnamefont{Tikhonenko}},
  \bibinfo{author}{\bibfnamefont{A.~S.} \bibnamefont{Mayorov}},
  \bibinfo{author}{\bibfnamefont{D.~W.} \bibnamefont{Horsell}}, and
  \bibinfo{author}{\bibfnamefont{A.~K.} \bibnamefont{Savchenko}},
  \bibinfo{year}{2007}, \bibinfo{journal}{Phys. Rev. Lett.}
  \textbf{\bibinfo{volume}{98}}, \bibinfo{pages}{176805}.

\bibitem[{\citenamefont{Gorbar} \emph{et~al.}(2002)\citenamefont{Gorbar,
  Gusynin, Miransky, and Shovkovy}}]{Getal02}
\bibinfo{author}{\bibnamefont{Gorbar}, \bibfnamefont{E.~V.}},
  \bibinfo{author}{\bibfnamefont{V.~P.} \bibnamefont{Gusynin}},
  \bibinfo{author}{\bibfnamefont{V.~A.} \bibnamefont{Miransky}}, and
  \bibinfo{author}{\bibfnamefont{I.~A.} \bibnamefont{Shovkovy}},
  \bibinfo{year}{2002}, \bibinfo{journal}{\prb} \textbf{\bibinfo{volume}{66}},
  \bibinfo{pages}{045108}.

\bibitem[{\citenamefont{Goswami} \emph{et~al.}(2007)\citenamefont{Goswami, Jia,
  and Chakravarty}}]{GJC07}
\bibinfo{author}{\bibnamefont{Goswami}, \bibfnamefont{P.}},
  \bibinfo{author}{\bibfnamefont{X.}~\bibnamefont{Jia}}, and
  \bibinfo{author}{\bibfnamefont{S.}~\bibnamefont{Chakravarty}},
  \bibinfo{year}{2007}, \bibinfo{journal}{Phys. Rev. B}
  \textbf{\bibinfo{volume}{76}}, \bibinfo{pages}{205408}.

\bibitem[{\citenamefont{Graf} \emph{et~al.}(2007)\citenamefont{Graf, Molitor,
  Ensslin, Stampfer, Jungen, Hierold, and Wirtz}}]{GMESJHW07}
\bibinfo{author}{\bibnamefont{Graf}, \bibfnamefont{D.}},
  \bibinfo{author}{\bibfnamefont{F.}~\bibnamefont{Molitor}},
  \bibinfo{author}{\bibfnamefont{K.}~\bibnamefont{Ensslin}},
  \bibinfo{author}{\bibfnamefont{C.}~\bibnamefont{Stampfer}},
  \bibinfo{author}{\bibfnamefont{A.}~\bibnamefont{Jungen}},
  \bibinfo{author}{\bibfnamefont{C.}~\bibnamefont{Hierold}}, and
  \bibinfo{author}{\bibfnamefont{L.}~\bibnamefont{Wirtz}},
  \bibinfo{year}{2007}, \bibinfo{journal}{Nano Letters}
  \textbf{\bibinfo{volume}{7}}, \bibinfo{pages}{238}.

\bibitem[{\citenamefont{Guinea}(2007)}]{G06}
\bibinfo{author}{\bibnamefont{Guinea}, \bibfnamefont{F.}},
  \bibinfo{year}{2007}, \bibinfo{journal}{Phys. Rev. B}
  \textbf{\bibinfo{volume}{75}}, \bibinfo{pages}{235433}.

\bibitem[{\citenamefont{Guinea} \emph{et~al.}(2006)\citenamefont{Guinea,
  {Castro Neto}, and Peres}}]{GNP06}
\bibinfo{author}{\bibnamefont{Guinea}, \bibfnamefont{F.}},
  \bibinfo{author}{\bibfnamefont{A.~H.} \bibnamefont{{Castro Neto}}}, and
  \bibinfo{author}{\bibfnamefont{N.~M.~R.} \bibnamefont{Peres}},
  \bibinfo{year}{2006}, \bibinfo{journal}{Phys. Rev. B}
  \textbf{\bibinfo{volume}{73}}, \bibinfo{pages}{245426}.

\bibitem[{\citenamefont{Guinea} \emph{et~al.}(2007)\citenamefont{Guinea,
  Katsnelson, and Vozmediano}}]{GKV07}
\bibinfo{author}{\bibnamefont{Guinea}, \bibfnamefont{F.}},
  \bibinfo{author}{\bibfnamefont{M.~I.} \bibnamefont{Katsnelson}}, and
  \bibinfo{author}{\bibfnamefont{M.~A.~H.} \bibnamefont{Vozmediano}},
  \bibinfo{year}{2007}, \bibinfo{journal}{Phys. Rev. B}
  \textbf{\bibinfo{volume}{76}}, \bibinfo{pages}{235309}.

\bibitem[{\citenamefont{Gumbs and Fekete}(1997)}]{Gumbs97}
\bibinfo{author}{\bibnamefont{Gumbs}, \bibfnamefont{G.}}, and
  \bibinfo{author}{\bibfnamefont{P.}~\bibnamefont{Fekete}},
  \bibinfo{year}{1997}, \bibinfo{journal}{Phy. Rev. B}
  \textbf{\bibinfo{volume}{56}}, \bibinfo{pages}{3787}.

\bibitem[{\citenamefont{Gunlycke} \emph{et~al.}(2007)\citenamefont{Gunlycke,
  Areshkin, and White}}]{GAW07}
\bibinfo{author}{\bibnamefont{Gunlycke}, \bibfnamefont{D.}},
  \bibinfo{author}{\bibfnamefont{D.~A.} \bibnamefont{Areshkin}}, and
  \bibinfo{author}{\bibfnamefont{C.~T.} \bibnamefont{White}},
  \bibinfo{year}{2007}, \bibinfo{journal}{Appl. Phys. Lett.}
  \textbf{\bibinfo{volume}{90}}, \bibinfo{pages}{12104}.

\bibitem[{\citenamefont{Gupta} \emph{et~al.}(2006)\citenamefont{Gupta, Chen,
  Joshi, Tadigadapa, and Eklund}}]{GCJTE06}
\bibinfo{author}{\bibnamefont{Gupta}, \bibfnamefont{A.}},
  \bibinfo{author}{\bibfnamefont{G.}~\bibnamefont{Chen}},
  \bibinfo{author}{\bibfnamefont{P.}~\bibnamefont{Joshi}},
  \bibinfo{author}{\bibfnamefont{S.}~\bibnamefont{Tadigadapa}}, and
  \bibinfo{author}{\bibfnamefont{P.~C.} \bibnamefont{Eklund}},
  \bibinfo{year}{2006}, \bibinfo{journal}{Nano Letters}
  \textbf{\bibinfo{volume}{12}}, \bibinfo{pages}{2667}.

\bibitem[{\citenamefont{Gusynin} \emph{et~al.}(2006)\citenamefont{Gusynin,
  Miransky, Sharapov, and Shovkovy}}]{GMSS06}
\bibinfo{author}{\bibnamefont{Gusynin}, \bibfnamefont{V.~P.}},
  \bibinfo{author}{\bibfnamefont{V.~A.} \bibnamefont{Miransky}},
  \bibinfo{author}{\bibfnamefont{S.~G.} \bibnamefont{Sharapov}}, and
  \bibinfo{author}{\bibfnamefont{I.~A.} \bibnamefont{Shovkovy}},
  \bibinfo{year}{2006}, \bibinfo{journal}{Phys. Rev. B}
  \textbf{\bibinfo{volume}{74}}, \bibinfo{pages}{195429}.

\bibitem[{\citenamefont{Gusynin and Sharapov}(2005)}]{GS05}
\bibinfo{author}{\bibnamefont{Gusynin}, \bibfnamefont{V.~P.}}, and
  \bibinfo{author}{\bibfnamefont{S.~G.} \bibnamefont{Sharapov}},
  \bibinfo{year}{2005}, \bibinfo{journal}{Phys. Rev. Lett.}
  \textbf{\bibinfo{volume}{95}}, \bibinfo{pages}{146801}.

\bibitem[{\citenamefont{Gusynin} \emph{et~al.}(2007)\citenamefont{Gusynin,
  Sharapov, and Carbotte}}]{GSC07}
\bibinfo{author}{\bibnamefont{Gusynin}, \bibfnamefont{V.~P.}},
  \bibinfo{author}{\bibfnamefont{S.~G.} \bibnamefont{Sharapov}}, and
  \bibinfo{author}{\bibfnamefont{J.~P.} \bibnamefont{Carbotte}},
  \bibinfo{year}{2007}, \bibinfo{journal}{J. Phys.: Condens.Matter}
  \textbf{\bibinfo{volume}{19}}, \bibinfo{pages}{026222}.

\bibitem[{\citenamefont{Haldane}(1988)}]{H88}
\bibinfo{author}{\bibnamefont{Haldane}, \bibfnamefont{F.~D.~M.}},
  \bibinfo{year}{1988}, \bibinfo{journal}{Phys. Rev. Lett.}
  \textbf{\bibinfo{volume}{61}}, \bibinfo{pages}{2015}.

\bibitem[{\citenamefont{Han} \emph{et~al.}(2007)\citenamefont{Han, \"Ozyilmaz,
  Zhang, and Kim}}]{HOZK07}
\bibinfo{author}{\bibnamefont{Han}, \bibfnamefont{M.~Y.}},
  \bibinfo{author}{\bibfnamefont{B.}~\bibnamefont{\"Ozyilmaz}},
  \bibinfo{author}{\bibfnamefont{Y.}~\bibnamefont{Zhang}}, and
  \bibinfo{author}{\bibfnamefont{P.}~\bibnamefont{Kim}}, \bibinfo{year}{2007},
  \bibinfo{journal}{Phys. Rev. Lett.} \textbf{\bibinfo{volume}{98}},
  \bibinfo{pages}{206805}.

\bibitem[{\citenamefont{Harper}(1955)}]{harper}
\bibinfo{author}{\bibnamefont{Harper}, \bibfnamefont{P.~G.}},
  \bibinfo{year}{1955}, \bibinfo{journal}{Proc. Phys. Soc. London A}
  \textbf{\bibinfo{volume}{68}}, \bibinfo{pages}{874}.

\bibitem[{\citenamefont{Harrison}(1980)}]{harrison}
\bibinfo{author}{\bibnamefont{Harrison}, \bibfnamefont{W.~A.}},
  \bibinfo{year}{1980}, \emph{\bibinfo{title}{Solid State Theory}}
  (\bibinfo{publisher}{Dover}).

\bibitem[{\citenamefont{Hass}
  \emph{et~al.}(2007{\natexlab{a}})\citenamefont{Hass, Feng, Mill\'an-Otoya,
  Li, Sprinkle, First, de~Heer, and Conrad}}]{Hetal07b}
\bibinfo{author}{\bibnamefont{Hass}, \bibfnamefont{J.}},
  \bibinfo{author}{\bibfnamefont{R.}~\bibnamefont{Feng}},
  \bibinfo{author}{\bibfnamefont{J.~E.} \bibnamefont{Mill\'an-Otoya}},
  \bibinfo{author}{\bibfnamefont{X.}~\bibnamefont{Li}},
  \bibinfo{author}{\bibfnamefont{M.}~\bibnamefont{Sprinkle}},
  \bibinfo{author}{\bibfnamefont{P.~N.} \bibnamefont{First}},
  \bibinfo{author}{\bibfnamefont{W.~A.} \bibnamefont{de~Heer}}, and
  \bibinfo{author}{\bibfnamefont{E.~H.} \bibnamefont{Conrad}},
  \bibinfo{year}{2007}{\natexlab{a}}, \bibinfo{journal}{Phys. Rev. B}
  \textbf{\bibinfo{volume}{75}}, \bibinfo{pages}{214109}.

\bibitem[{\citenamefont{Hass}
  \emph{et~al.}(2007{\natexlab{b}})\citenamefont{Hass, Varchon, Millan-Otoya,
  Sprinkle, {de Heer}, Berger, First, Magaud, and Conrad}}]{HVMSHBFMC07}
\bibinfo{author}{\bibnamefont{Hass}, \bibfnamefont{J.}},
  \bibinfo{author}{\bibfnamefont{F.}~\bibnamefont{Varchon}},
  \bibinfo{author}{\bibfnamefont{J.~E.} \bibnamefont{Millan-Otoya}},
  \bibinfo{author}{\bibfnamefont{M.}~\bibnamefont{Sprinkle}},
  \bibinfo{author}{\bibfnamefont{W.~A.} \bibnamefont{{de Heer}}},
  \bibinfo{author}{\bibfnamefont{C.}~\bibnamefont{Berger}},
  \bibinfo{author}{\bibfnamefont{P.~N.} \bibnamefont{First}},
  \bibinfo{author}{\bibfnamefont{L.}~\bibnamefont{Magaud}}, and
  \bibinfo{author}{\bibfnamefont{E.~H.} \bibnamefont{Conrad}},
  \bibinfo{year}{2007}{\natexlab{b}}, \eprint{cond-mat/0706.2134}.

\bibitem[{\citenamefont{de~Heer} \emph{et~al.}(2007)\citenamefont{de~Heer,
  Berger, Wu, First, Conrad, Li, Li, Sprinkle, Hass, Sadowski, Potemski, and
  Martinez}}]{Hetal07}
\bibinfo{author}{\bibnamefont{de~Heer}, \bibfnamefont{W.~A.}},
  \bibinfo{author}{\bibfnamefont{C.}~\bibnamefont{Berger}},
  \bibinfo{author}{\bibfnamefont{X.}~\bibnamefont{Wu}},
  \bibinfo{author}{\bibfnamefont{P.~N.} \bibnamefont{First}},
  \bibinfo{author}{\bibfnamefont{E.~H.} \bibnamefont{Conrad}},
  \bibinfo{author}{\bibfnamefont{X.}~\bibnamefont{Li}},
  \bibinfo{author}{\bibfnamefont{T.}~\bibnamefont{Li}},
  \bibinfo{author}{\bibfnamefont{M.}~\bibnamefont{Sprinkle}},
  \bibinfo{author}{\bibfnamefont{J.}~\bibnamefont{Hass}},
  \bibinfo{author}{\bibfnamefont{M.~L.} \bibnamefont{Sadowski}},
  \bibinfo{author}{\bibfnamefont{M.}~\bibnamefont{Potemski}}, and
  \bibinfo{author}{\bibfnamefont{G.}~\bibnamefont{Martinez}},
  \bibinfo{year}{2007}, \bibinfo{journal}{Sol. State Comm.}
  \textbf{\bibinfo{volume}{143}}, \bibinfo{pages}{92}.

\bibitem[{\citenamefont{Heersche} \emph{et~al.}(2007)\citenamefont{Heersche,
  Jarillo-Herrero, Oostinga, Vandersypen, and Morpurgo}}]{Hetal06}
\bibinfo{author}{\bibnamefont{Heersche}, \bibfnamefont{H.~B.}},
  \bibinfo{author}{\bibfnamefont{P.}~\bibnamefont{Jarillo-Herrero}},
  \bibinfo{author}{\bibfnamefont{J.~B.} \bibnamefont{Oostinga}},
  \bibinfo{author}{\bibfnamefont{L.~M.~K.} \bibnamefont{Vandersypen}}, and
  \bibinfo{author}{\bibfnamefont{A.}~\bibnamefont{Morpurgo}},
  \bibinfo{year}{2007}, \bibinfo{journal}{Nature}
  \textbf{\bibinfo{volume}{446}}, \bibinfo{pages}{56}.

\bibitem[{\citenamefont{Hentschel and Guinea}(2007)}]{HG07}
\bibinfo{author}{\bibnamefont{Hentschel}, \bibfnamefont{M.}}, and
  \bibinfo{author}{\bibfnamefont{F.}~\bibnamefont{Guinea}},
  \bibinfo{year}{2007}, \eprint{cond-mat/0705.0522}.

\bibitem[{\citenamefont{Herbut}(2006)}]{IH06}
\bibinfo{author}{\bibnamefont{Herbut}, \bibfnamefont{I.~F.}},
  \bibinfo{year}{2006}, \bibinfo{journal}{Phys. Rev. Lett.}
  \textbf{\bibinfo{volume}{97}}, \bibinfo{pages}{146401}.

\bibitem[{\citenamefont{Herbut}(2007)}]{H06}
\bibinfo{author}{\bibnamefont{Herbut}, \bibfnamefont{I.~F.}},
  \bibinfo{year}{2007}, \bibinfo{journal}{Phys. Rev. B}
  \textbf{\bibinfo{volume}{75}}, \bibinfo{pages}{165411}.

\bibitem[{\citenamefont{Heremans} \emph{et~al.}(1994)\citenamefont{Heremans,
  Olk, and Morelli}}]{HOM1994}
\bibinfo{author}{\bibnamefont{Heremans}, \bibfnamefont{J.}},
  \bibinfo{author}{\bibfnamefont{C.~H.} \bibnamefont{Olk}}, and
  \bibinfo{author}{\bibfnamefont{D.~T.} \bibnamefont{Morelli}},
  \bibinfo{year}{1994}, \bibinfo{journal}{Phys. Rev. B}
  \textbf{\bibinfo{volume}{49}}, \bibinfo{pages}{15122}.

\bibitem[{\citenamefont{Hill} \emph{et~al.}(2007)\citenamefont{Hill, Geim,
  Novoselov, Schedin, and Blake}}]{HGNSB07}
\bibinfo{author}{\bibnamefont{Hill}, \bibfnamefont{E.~W.}},
  \bibinfo{author}{\bibfnamefont{A.~K.} \bibnamefont{Geim}},
  \bibinfo{author}{\bibfnamefont{K.}~\bibnamefont{Novoselov}},
  \bibinfo{author}{\bibfnamefont{F.}~\bibnamefont{Schedin}}, and
  \bibinfo{author}{\bibfnamefont{P.}~\bibnamefont{Blake}},
  \bibinfo{year}{2007}, \bibinfo{journal}{IEEE Trans. Magn.}
  \textbf{\bibinfo{volume}{42}}, \bibinfo{pages}{2694}.

\bibitem[{\citenamefont{Himpsel} \emph{et~al.}(1982)\citenamefont{Himpsel,
  Christmann, Heimann, Eastman, and Feibelman}}]{HCHEF82}
\bibinfo{author}{\bibnamefont{Himpsel}, \bibfnamefont{F.~J.}},
  \bibinfo{author}{\bibfnamefont{K.}~\bibnamefont{Christmann}},
  \bibinfo{author}{\bibfnamefont{P.}~\bibnamefont{Heimann}},
  \bibinfo{author}{\bibfnamefont{D.~E.} \bibnamefont{Eastman}}, and
  \bibinfo{author}{\bibfnamefont{P.~J.} \bibnamefont{Feibelman}},
  \bibinfo{year}{1982}, \bibinfo{journal}{Surf. Sci.}
  \textbf{\bibinfo{volume}{115}}, \bibinfo{pages}{L159}.

\bibitem[{\citenamefont{Hobson and Nierenberg}(1953)}]{HN53}
\bibinfo{author}{\bibnamefont{Hobson}, \bibfnamefont{J.~P.}}, and
  \bibinfo{author}{\bibfnamefont{W.~A.} \bibnamefont{Nierenberg}},
  \bibinfo{year}{1953}, \bibinfo{journal}{Phys. Rev.}
  \textbf{\bibinfo{volume}{89}}, \bibinfo{pages}{662}.

\bibitem[{\citenamefont{Hod} \emph{et~al.}(2007)\citenamefont{Hod, Barone,
  Peralta, and Scuseria}}]{HBPS07}
\bibinfo{author}{\bibnamefont{Hod}, \bibfnamefont{O.}},
  \bibinfo{author}{\bibfnamefont{V.}~\bibnamefont{Barone}},
  \bibinfo{author}{\bibfnamefont{J.~E.} \bibnamefont{Peralta}}, and
  \bibinfo{author}{\bibfnamefont{G.~E.} \bibnamefont{Scuseria}},
  \bibinfo{year}{2007}, \bibinfo{journal}{Nano Letters}
  \textbf{\bibinfo{volume}{7}}, \bibinfo{pages}{2295}.

\bibitem[{\citenamefont{Horovitz and Doussal}(2002)}]{HD02}
\bibinfo{author}{\bibnamefont{Horovitz}, \bibfnamefont{B.}}, and
  \bibinfo{author}{\bibfnamefont{P.~L.} \bibnamefont{Doussal}},
  \bibinfo{year}{2002}, \bibinfo{journal}{\prb} \textbf{\bibinfo{volume}{65}},
  \bibinfo{pages}{125323}.

\bibitem[{\citenamefont{Hou} \emph{et~al.}(2007)\citenamefont{Hou, Chamon, and
  Mudry}}]{HCM07}
\bibinfo{author}{\bibnamefont{Hou}, \bibfnamefont{C.-Y.}},
  \bibinfo{author}{\bibfnamefont{C.}~\bibnamefont{Chamon}}, and
  \bibinfo{author}{\bibfnamefont{C.}~\bibnamefont{Mudry}},
  \bibinfo{year}{2007}, \bibinfo{journal}{Phys. Rev. Lett.}
  \textbf{\bibinfo{volume}{98}}, \bibinfo{pages}{186809}.

\bibitem[{\citenamefont{Huard} \emph{et~al.}(2007)\citenamefont{Huard,
  Sulpizio, Stander, Todd, Yang, and Goldhaber-Gordon}}]{HSSTYG07}
\bibinfo{author}{\bibnamefont{Huard}, \bibfnamefont{B.}},
  \bibinfo{author}{\bibfnamefont{J.~A.} \bibnamefont{Sulpizio}},
  \bibinfo{author}{\bibfnamefont{N.}~\bibnamefont{Stander}},
  \bibinfo{author}{\bibfnamefont{K.}~\bibnamefont{Todd}},
  \bibinfo{author}{\bibfnamefont{B.}~\bibnamefont{Yang}}, and
  \bibinfo{author}{\bibfnamefont{D.}~\bibnamefont{Goldhaber-Gordon}},
  \bibinfo{year}{2007}, \bibinfo{journal}{Phys. Rev. Lett.}
  \textbf{\bibinfo{volume}{98}}, \bibinfo{pages}{236803}.

\bibitem[{\citenamefont{Huertas-Hernando}
  \emph{et~al.}(2006)\citenamefont{Huertas-Hernando, Guinea, and
  Brataas}}]{HGB06}
\bibinfo{author}{\bibnamefont{Huertas-Hernando}, \bibfnamefont{D.}},
  \bibinfo{author}{\bibfnamefont{F.}~\bibnamefont{Guinea}}, and
  \bibinfo{author}{\bibfnamefont{A.}~\bibnamefont{Brataas}},
  \bibinfo{year}{2006}, \bibinfo{journal}{Phys. Rev. B}
  \textbf{\bibinfo{volume}{74}}, \bibinfo{pages}{155426}.

\bibitem[{\citenamefont{Hwang and {Das Sarma}}(2007)}]{HdS06}
\bibinfo{author}{\bibnamefont{Hwang}, \bibfnamefont{E.~H.}}, and
  \bibinfo{author}{\bibfnamefont{S.}~\bibnamefont{{Das Sarma}}},
  \bibinfo{year}{2007}, \bibinfo{journal}{Phys. Rev. B}
  \textbf{\bibinfo{volume}{75}}, \bibinfo{pages}{205418}.

\bibitem[{\citenamefont{Hwang} \emph{et~al.}(2007)\citenamefont{Hwang, Hu, and
  Sarma}}]{HHS07}
\bibinfo{author}{\bibnamefont{Hwang}, \bibfnamefont{E.~H.}},
  \bibinfo{author}{\bibfnamefont{B.~Y.-K.} \bibnamefont{Hu}}, and
  \bibinfo{author}{\bibfnamefont{S.~D.} \bibnamefont{Sarma}},
  \bibinfo{year}{2007}, \bibinfo{journal}{Phys. Rev. B}
  \textbf{\bibinfo{volume}{76}}, \bibinfo{pages}{115434}.

\bibitem[{\citenamefont{Ishigami} \emph{et~al.}(2007)\citenamefont{Ishigami,
  Chen, Cullan, Fuhrer, and Williams}}]{Ishal07}
\bibinfo{author}{\bibnamefont{Ishigami}, \bibfnamefont{M.}},
  \bibinfo{author}{\bibfnamefont{J.~H.} \bibnamefont{Chen}},
  \bibinfo{author}{\bibfnamefont{D.~W.~G.} \bibnamefont{Cullan}},
  \bibinfo{author}{\bibfnamefont{M.~S.} \bibnamefont{Fuhrer}}, and
  \bibinfo{author}{\bibfnamefont{E.~D.} \bibnamefont{Williams}},
  \bibinfo{year}{2007}, \bibinfo{journal}{Nano Letters}
  \textbf{\bibinfo{volume}{7}}, \bibinfo{pages}{1643}.

\bibitem[{\citenamefont{Itzykson and Zuber}(2006)}]{zuber06}
\bibinfo{author}{\bibnamefont{Itzykson}, \bibfnamefont{C.}}, and
  \bibinfo{author}{\bibfnamefont{J.-B.} \bibnamefont{Zuber}},
  \bibinfo{year}{2006}, \emph{\bibinfo{title}{Quantum Field Theory}}
  (\bibinfo{publisher}{Dover}).

\bibitem[{\citenamefont{Jackiw and Rebbi}(1976)}]{JR76}
\bibinfo{author}{\bibnamefont{Jackiw}, \bibfnamefont{R.}}, and
  \bibinfo{author}{\bibfnamefont{C.}~\bibnamefont{Rebbi}},
  \bibinfo{year}{1976}, \bibinfo{journal}{Phys. Rev. D}
  \textbf{\bibinfo{volume}{13}}, \bibinfo{pages}{3398}.

\bibitem[{\citenamefont{Jiang}
  \emph{et~al.}(2007{\natexlab{a}})\citenamefont{Jiang, Henriksen, Tung, Wang,
  Schwartz, Han, Kim, and Stormer}}]{Jetal07}
\bibinfo{author}{\bibnamefont{Jiang}, \bibfnamefont{Z.}},
  \bibinfo{author}{\bibfnamefont{E.~A.} \bibnamefont{Henriksen}},
  \bibinfo{author}{\bibfnamefont{L.~C.} \bibnamefont{Tung}},
  \bibinfo{author}{\bibfnamefont{Y.-J.} \bibnamefont{Wang}},
  \bibinfo{author}{\bibfnamefont{M.~E.} \bibnamefont{Schwartz}},
  \bibinfo{author}{\bibfnamefont{M.~Y.} \bibnamefont{Han}},
  \bibinfo{author}{\bibfnamefont{P.}~\bibnamefont{Kim}}, and
  \bibinfo{author}{\bibfnamefont{H.~L.} \bibnamefont{Stormer}},
  \bibinfo{year}{2007}{\natexlab{a}}, \bibinfo{journal}{Phys. Rev. Lett.}
  \textbf{\bibinfo{volume}{98}}, \bibinfo{pages}{197403}.

\bibitem[{\citenamefont{Jiang}
  \emph{et~al.}(2007{\natexlab{b}})\citenamefont{Jiang, Zhang, Stormer, and
  Kim}}]{JZSK07}
\bibinfo{author}{\bibnamefont{Jiang}, \bibfnamefont{Z.}},
  \bibinfo{author}{\bibfnamefont{Y.}~\bibnamefont{Zhang}},
  \bibinfo{author}{\bibfnamefont{H.~L.} \bibnamefont{Stormer}}, and
  \bibinfo{author}{\bibfnamefont{P.}~\bibnamefont{Kim}},
  \bibinfo{year}{2007}{\natexlab{b}}, \bibinfo{journal}{Phys. Rev. Lett.}
  \textbf{\bibinfo{volume}{99}}, \bibinfo{pages}{106802}.

\bibitem[{\citenamefont{de~Juan} \emph{et~al.}(2007)\citenamefont{de~Juan,
  Cortijo, and Vozmediano}}]{JCV07}
\bibinfo{author}{\bibnamefont{de~Juan}, \bibfnamefont{F.}},
  \bibinfo{author}{\bibfnamefont{A.}~\bibnamefont{Cortijo}}, and
  \bibinfo{author}{\bibfnamefont{M.~A.~H.} \bibnamefont{Vozmediano}},
  \bibinfo{year}{2007}, \bibinfo{journal}{Phys. Rev. B}
  \textbf{\bibinfo{volume}{76}}, \bibinfo{pages}{165409}.

\bibitem[{\citenamefont{Kane and Mele}(1997)}]{KM97}
\bibinfo{author}{\bibnamefont{Kane}, \bibfnamefont{C.~L.}}, and
  \bibinfo{author}{\bibfnamefont{E.~J.} \bibnamefont{Mele}},
  \bibinfo{year}{1997}, \bibinfo{journal}{Phys. Rev. Lett.}
  \textbf{\bibinfo{volume}{78}}, \bibinfo{pages}{1932}.

\bibitem[{\citenamefont{Kane and Mele}(2004)}]{KM04}
\bibinfo{author}{\bibnamefont{Kane}, \bibfnamefont{C.~L.}}, and
  \bibinfo{author}{\bibfnamefont{E.~J.} \bibnamefont{Mele}},
  \bibinfo{year}{2004}, \bibinfo{journal}{\prl} \textbf{\bibinfo{volume}{93}},
  \bibinfo{pages}{197402}.

\bibitem[{\citenamefont{Kane and Mele}(2005)}]{KM05}
\bibinfo{author}{\bibnamefont{Kane}, \bibfnamefont{C.~L.}}, and
  \bibinfo{author}{\bibfnamefont{E.~J.} \bibnamefont{Mele}},
  \bibinfo{year}{2005}, \bibinfo{journal}{Phys. Rev. Lett.}
  \textbf{\bibinfo{volume}{95}}, \bibinfo{pages}{226801}.

\bibitem[{\citenamefont{Katsnelson}(2006{\natexlab{a}})}]{Kscreening}
\bibinfo{author}{\bibnamefont{Katsnelson}, \bibfnamefont{M.~I.}},
  \bibinfo{year}{2006}{\natexlab{a}}, \bibinfo{journal}{Phys. Rev. B}
  \textbf{\bibinfo{volume}{74}}, \bibinfo{pages}{201401}.

\bibitem[{\citenamefont{Katsnelson}(2006{\natexlab{b}})}]{K06}
\bibinfo{author}{\bibnamefont{Katsnelson}, \bibfnamefont{M.~I.}},
  \bibinfo{year}{2006}{\natexlab{b}}, \bibinfo{journal}{Eur. J. Phys. B}
  \textbf{\bibinfo{volume}{51}}, \bibinfo{pages}{157}.

\bibitem[{\citenamefont{Katsnelson}(2007{\natexlab{a}})}]{Kats0703}
\bibinfo{author}{\bibnamefont{Katsnelson}, \bibfnamefont{M.~I.}},
  \bibinfo{year}{2007}{\natexlab{a}}, \bibinfo{journal}{Eur. Phys. J B}
  \textbf{\bibinfo{volume}{57}}, \bibinfo{pages}{225}.

\bibitem[{\citenamefont{Katsnelson}(2007{\natexlab{b}})}]{KMT}
\bibinfo{author}{\bibnamefont{Katsnelson}, \bibfnamefont{M.~I.}},
  \bibinfo{year}{2007}{\natexlab{b}}, \bibinfo{journal}{Materials Today,}
  \textbf{\bibinfo{volume}{10}}, \bibinfo{pages}{20}.

\bibitem[{\citenamefont{Katsnelson}(2007{\natexlab{c}})}]{Kats07}
\bibinfo{author}{\bibnamefont{Katsnelson}, \bibfnamefont{M.~I.}},
  \bibinfo{year}{2007}{\natexlab{c}}, \bibinfo{journal}{Phys. Rev. B}
  \textbf{\bibinfo{volume}{76}}, \bibinfo{pages}{073411}.

\bibitem[{\citenamefont{Katsnelson and Geim}(2008)}]{KG07}
\bibinfo{author}{\bibnamefont{Katsnelson}, \bibfnamefont{M.~I.}}, and
  \bibinfo{author}{\bibfnamefont{A.~K.} \bibnamefont{Geim}},
  \bibinfo{year}{2008}, \bibinfo{journal}{Phil. Trans. R. Soc. A}
  \textbf{\bibinfo{volume}{366}}, \bibinfo{pages}{195}.

\bibitem[{\citenamefont{Katsnelson and Novoselov}(2007)}]{KN0703}
\bibinfo{author}{\bibnamefont{Katsnelson}, \bibfnamefont{M.~I.}}, and
  \bibinfo{author}{\bibfnamefont{K.~S.} \bibnamefont{Novoselov}},
  \bibinfo{year}{2007}, \bibinfo{journal}{Sol. Stat. Comm.}
  \textbf{\bibinfo{volume}{143}}, \bibinfo{pages}{3}.

\bibitem[{\citenamefont{Katsnelson}
  \emph{et~al.}(2006)\citenamefont{Katsnelson, Novoselov, and Geim}}]{KNG06}
\bibinfo{author}{\bibnamefont{Katsnelson}, \bibfnamefont{M.~I.}},
  \bibinfo{author}{\bibfnamefont{K.~S.} \bibnamefont{Novoselov}}, and
  \bibinfo{author}{\bibfnamefont{A.~K.} \bibnamefont{Geim}},
  \bibinfo{year}{2006}, \bibinfo{journal}{Nature Physics}
  \textbf{\bibinfo{volume}{2}}, \bibinfo{pages}{620}.

\bibitem[{\citenamefont{Kechedzhi} \emph{et~al.}(2007)\citenamefont{Kechedzhi,
  Fal'ko, McCann, and Altshuler}}]{KFMA07}
\bibinfo{author}{\bibnamefont{Kechedzhi}, \bibfnamefont{K.}},
  \bibinfo{author}{\bibfnamefont{V.~I.} \bibnamefont{Fal'ko}},
  \bibinfo{author}{\bibfnamefont{E.}~\bibnamefont{McCann}}, and
  \bibinfo{author}{\bibfnamefont{B.~L.} \bibnamefont{Altshuler}},
  \bibinfo{year}{2007}, \bibinfo{journal}{Phys. Rev. Lett.}
  \textbf{\bibinfo{volume}{98}}, \bibinfo{pages}{176806}.

\bibitem[{\citenamefont{Khveshchenko}(2001)}]{K01}
\bibinfo{author}{\bibnamefont{Khveshchenko}, \bibfnamefont{D.~V.}},
  \bibinfo{year}{2001}, \bibinfo{journal}{\prl} \textbf{\bibinfo{volume}{87}},
  \bibinfo{pages}{246802}.

\bibitem[{\citenamefont{Khveshchenko}(2007)}]{Khv07}
\bibinfo{author}{\bibnamefont{Khveshchenko}, \bibfnamefont{D.~V.}},
  \bibinfo{year}{2007}, \eprint{cond-mat/0705.4105}.

\bibitem[{\citenamefont{Khveshchenko and Shively}(2006)}]{KS06}
\bibinfo{author}{\bibnamefont{Khveshchenko}, \bibfnamefont{D.~V.}}, and
  \bibinfo{author}{\bibfnamefont{W.~F.} \bibnamefont{Shively}},
  \bibinfo{year}{2006}, \bibinfo{journal}{Phys. Rev. B}
  \textbf{\bibinfo{volume}{73}}, \bibinfo{pages}{115104}.

\bibitem[{\citenamefont{Kobayashi} \emph{et~al.}(2005)\citenamefont{Kobayashi,
  Fukui, Enoki, Kusakabe, and Kaburagi}}]{Ketal05}
\bibinfo{author}{\bibnamefont{Kobayashi}, \bibfnamefont{Y.}},
  \bibinfo{author}{\bibfnamefont{K.}~\bibnamefont{Fukui}},
  \bibinfo{author}{\bibfnamefont{T.}~\bibnamefont{Enoki}},
  \bibinfo{author}{\bibfnamefont{K.}~\bibnamefont{Kusakabe}}, and
  \bibinfo{author}{\bibfnamefont{Y.}~\bibnamefont{Kaburagi}},
  \bibinfo{year}{2005}, \bibinfo{journal}{Phys. Rev. B}
  \textbf{\bibinfo{volume}{71}}, \bibinfo{pages}{193406}.

\bibitem[{\citenamefont{Kolesnikov and Osipov}(2004)}]{KO04}
\bibinfo{author}{\bibnamefont{Kolesnikov}, \bibfnamefont{D.~V.}}, and
  \bibinfo{author}{\bibfnamefont{V.~A.} \bibnamefont{Osipov}},
  \bibinfo{year}{2004}, \bibinfo{journal}{JETP Letters}
  \textbf{\bibinfo{volume}{79}}, \bibinfo{pages}{532}.

\bibitem[{\citenamefont{Kolesnikov and Osipov}(2006)}]{KO06}
\bibinfo{author}{\bibnamefont{Kolesnikov}, \bibfnamefont{D.~V.}}, and
  \bibinfo{author}{\bibfnamefont{V.~A.} \bibnamefont{Osipov}},
  \bibinfo{year}{2006}, \bibinfo{journal}{Eur. J. Phys. B}
  \textbf{\bibinfo{volume}{49}}, \bibinfo{pages}{465}.

\bibitem[{\citenamefont{Kolezhuk} \emph{et~al.}(2006)\citenamefont{Kolezhuk,
  Sachdev, Biswas, and Chen}}]{KSBC06}
\bibinfo{author}{\bibnamefont{Kolezhuk}, \bibfnamefont{A.}},
  \bibinfo{author}{\bibfnamefont{S.}~\bibnamefont{Sachdev}},
  \bibinfo{author}{\bibfnamefont{R.~B.} \bibnamefont{Biswas}}, and
  \bibinfo{author}{\bibfnamefont{P.}~\bibnamefont{Chen}}, \bibinfo{year}{2006},
  \bibinfo{journal}{Phys. Rev. B} \textbf{\bibinfo{volume}{74}},
  \bibinfo{pages}{165114}.

\bibitem[{\citenamefont{Kopelevich and Esquinazi}(2006)}]{KE07}
\bibinfo{author}{\bibnamefont{Kopelevich}, \bibfnamefont{Y.}}, and
  \bibinfo{author}{\bibfnamefont{P.}~\bibnamefont{Esquinazi}},
  \bibinfo{year}{2006}, \eprint{cond-mat/0609497}.

\bibitem[{\citenamefont{Kopelevich}
  \emph{et~al.}(2006)\citenamefont{Kopelevich, {Medina Pantoja}, {da Silva},
  Mrowka, and Esquinazi}}]{KPSME06}
\bibinfo{author}{\bibnamefont{Kopelevich}, \bibfnamefont{Y.}},
  \bibinfo{author}{\bibfnamefont{J.~C.} \bibnamefont{{Medina Pantoja}}},
  \bibinfo{author}{\bibfnamefont{R.~R.} \bibnamefont{{da Silva}}},
  \bibinfo{author}{\bibfnamefont{F.}~\bibnamefont{Mrowka}}, and
  \bibinfo{author}{\bibfnamefont{P.}~\bibnamefont{Esquinazi}},
  \bibinfo{year}{2006}, \bibinfo{journal}{Physics Letters A}
  \textbf{\bibinfo{volume}{355}}, \bibinfo{pages}{233}.

\bibitem[{\citenamefont{Kopelevich}
  \emph{et~al.}(2003)\citenamefont{Kopelevich, Torres, da~Silva, Mrowka, Kempa,
  and Esquinazi}}]{Ketal03}
\bibinfo{author}{\bibnamefont{Kopelevich}, \bibfnamefont{Y.}},
  \bibinfo{author}{\bibfnamefont{J.~H.~S.} \bibnamefont{Torres}},
  \bibinfo{author}{\bibfnamefont{R.~R.} \bibnamefont{da~Silva}},
  \bibinfo{author}{\bibfnamefont{F.}~\bibnamefont{Mrowka}},
  \bibinfo{author}{\bibfnamefont{H.}~\bibnamefont{Kempa}}, and
  \bibinfo{author}{\bibfnamefont{P.}~\bibnamefont{Esquinazi}},
  \bibinfo{year}{2003}, \bibinfo{journal}{\prl} \textbf{\bibinfo{volume}{90}},
  \bibinfo{pages}{156402}.

\bibitem[{\citenamefont{Koshino and Ando}(2006)}]{KA06}
\bibinfo{author}{\bibnamefont{Koshino}, \bibfnamefont{M.}}, and
  \bibinfo{author}{\bibfnamefont{T.}~\bibnamefont{Ando}}, \bibinfo{year}{2006},
  \bibinfo{journal}{Phys. Rev. B} \textbf{\bibinfo{volume}{73}},
  \bibinfo{pages}{245403}.

\bibitem[{\citenamefont{Koshino and Ando}(2007)}]{KA0705}
\bibinfo{author}{\bibnamefont{Koshino}, \bibfnamefont{M.}}, and
  \bibinfo{author}{\bibfnamefont{T.}~\bibnamefont{Ando}}, \bibinfo{year}{2007},
  \bibinfo{journal}{Phys. Rev. B} \textbf{\bibinfo{volume}{75}},
  \bibinfo{pages}{235333}.

\bibitem[{\citenamefont{Kumazaki and Hirashima}(2006)}]{KH06}
\bibinfo{author}{\bibnamefont{Kumazaki}, \bibfnamefont{H.}}, and
  \bibinfo{author}{\bibfnamefont{D.~S.} \bibnamefont{Hirashima}},
  \bibinfo{year}{2006}, \bibinfo{journal}{J. Phys. Soc. Jpn.}
  \textbf{\bibinfo{volume}{75}}, \bibinfo{pages}{053707}.

\bibitem[{\citenamefont{Kumazaki and Hirashima}(2007)}]{KH07}
\bibinfo{author}{\bibnamefont{Kumazaki}, \bibfnamefont{H.}}, and
  \bibinfo{author}{\bibfnamefont{D.~S.} \bibnamefont{Hirashima}},
  \bibinfo{year}{2007}, \bibinfo{journal}{J. Phys. Soc. Jpn.}
  \textbf{\bibinfo{volume}{76}}, \bibinfo{pages}{064713}.

\bibitem[{\citenamefont{Kusminskiy}
  \emph{et~al.}(2007)\citenamefont{Kusminskiy, Nilsson, Campbell, and {Castro
  Neto}}}]{KNCN07}
\bibinfo{author}{\bibnamefont{Kusminskiy}, \bibfnamefont{S.~V.}},
  \bibinfo{author}{\bibfnamefont{J.}~\bibnamefont{Nilsson}},
  \bibinfo{author}{\bibfnamefont{D.~K.} \bibnamefont{Campbell}}, and
  \bibinfo{author}{\bibfnamefont{A.~H.} \bibnamefont{{Castro Neto}}},
  \bibinfo{year}{2007}, \eprint{cond-mat/0706.2359}.

\bibitem[{\citenamefont{Lammert and Crespi}(2004)}]{LC04}
\bibinfo{author}{\bibnamefont{Lammert}, \bibfnamefont{P.~E.}}, and
  \bibinfo{author}{\bibfnamefont{V.~H.} \bibnamefont{Crespi}},
  \bibinfo{year}{2004}, \bibinfo{journal}{Phys. Rev. B}
  \textbf{\bibinfo{volume}{69}}, \bibinfo{pages}{035406}.

\bibitem[{\citenamefont{Landau and Lifshitz}(1959)}]{LL59}
\bibinfo{author}{\bibnamefont{Landau}, \bibfnamefont{L.~D.}}, and
  \bibinfo{author}{\bibfnamefont{E.~M.} \bibnamefont{Lifshitz}},
  \bibinfo{year}{1959}, \emph{\bibinfo{title}{Theory of Elasticity}}
  (\bibinfo{publisher}{Pergamon Press (London)}).

\bibitem[{\citenamefont{Landau and Lifshitz}(1981)}]{LL81}
\bibinfo{author}{\bibnamefont{Landau}, \bibfnamefont{L.~D.}}, and
  \bibinfo{author}{\bibfnamefont{E.~M.} \bibnamefont{Lifshitz}},
  \bibinfo{year}{1981}, \emph{\bibinfo{title}{Quantum Mechanics:
  Non-Relativistic Theory}} (\bibinfo{publisher}{Pergamon Press (London)}).

\bibitem[{\citenamefont{Laughlin}(1981)}]{laughlin}
\bibinfo{author}{\bibnamefont{Laughlin}, \bibfnamefont{R.~B.}},
  \bibinfo{year}{1981}, \bibinfo{journal}{Phys. Rev. B}
  \textbf{\bibinfo{volume}{23}}, \bibinfo{pages}{5632}.

\bibitem[{\citenamefont{Lazzeri and Mauri}(2006)}]{LM06}
\bibinfo{author}{\bibnamefont{Lazzeri}, \bibfnamefont{M.}}, and
  \bibinfo{author}{\bibfnamefont{F.}~\bibnamefont{Mauri}},
  \bibinfo{year}{2006}, \bibinfo{journal}{Phys. Rev. Lett.}
  \textbf{\bibinfo{volume}{97}}, \bibinfo{pages}{266407}.

\bibitem[{\citenamefont{LeClair}(2000)}]{LC00}
\bibinfo{author}{\bibnamefont{LeClair}, \bibfnamefont{A.}},
  \bibinfo{year}{2000}, \bibinfo{journal}{Phys. Rev. Lett.}
  \textbf{\bibinfo{volume}{84}}, \bibinfo{pages}{1292}.

\bibitem[{\citenamefont{Lee}(1993)}]{PLee93}
\bibinfo{author}{\bibnamefont{Lee}, \bibfnamefont{P.~A.}},
  \bibinfo{year}{1993}, \bibinfo{journal}{Phys. Rev. Lett.}
  \textbf{\bibinfo{volume}{71}}, \bibinfo{pages}{1887}.

\bibitem[{\citenamefont{Lee and Ramakrishnan}(1985)}]{localization}
\bibinfo{author}{\bibnamefont{Lee}, \bibfnamefont{P.~A.}}, and
  \bibinfo{author}{\bibfnamefont{T.~V.} \bibnamefont{Ramakrishnan}},
  \bibinfo{year}{1985}, \bibinfo{journal}{Rev. Mod. Phys.}
  \textbf{\bibinfo{volume}{57}}, \bibinfo{pages}{287}.

\bibitem[{\citenamefont{Lee and Lee}(2005)}]{sungsik05}
\bibinfo{author}{\bibnamefont{Lee}, \bibfnamefont{S.-S.}}, and
  \bibinfo{author}{\bibfnamefont{P.~A.} \bibnamefont{Lee}},
  \bibinfo{year}{2005}, \bibinfo{journal}{Phys. Rev. Lett.}
  \textbf{\bibinfo{volume}{95}}, \bibinfo{pages}{036403}.

\bibitem[{\citenamefont{Leenaerts} \emph{et~al.}(2007)\citenamefont{Leenaerts,
  Partoens, and Peeters}}]{LPP07}
\bibinfo{author}{\bibnamefont{Leenaerts}, \bibfnamefont{O.}},
  \bibinfo{author}{\bibfnamefont{B.}~\bibnamefont{Partoens}}, and
  \bibinfo{author}{\bibfnamefont{F.~M.} \bibnamefont{Peeters}},
  \bibinfo{year}{2007}, \eprint{arXiv:0710.1757}.

\bibitem[{\citenamefont{Lemme} \emph{et~al.}(2007)\citenamefont{Lemme,
  Echtermeyer, Baus, and Kurz}}]{LEBK07}
\bibinfo{author}{\bibnamefont{Lemme}, \bibfnamefont{M.~C.}},
  \bibinfo{author}{\bibfnamefont{T.~J.} \bibnamefont{Echtermeyer}},
  \bibinfo{author}{\bibfnamefont{M.}~\bibnamefont{Baus}}, and
  \bibinfo{author}{\bibfnamefont{H.}~\bibnamefont{Kurz}}, \bibinfo{year}{2007},
  \bibinfo{journal}{IEEE Electron Device Letters}
  \textbf{\bibinfo{volume}{28}}, \bibinfo{pages}{282}.

\bibitem[{\citenamefont{Lenosky} \emph{et~al.}(1992)\citenamefont{Lenosky,
  Gonze, Teter, and Elser}}]{lenosky92}
\bibinfo{author}{\bibnamefont{Lenosky}, \bibfnamefont{T.}},
  \bibinfo{author}{\bibfnamefont{X.}~\bibnamefont{Gonze}},
  \bibinfo{author}{\bibfnamefont{M.}~\bibnamefont{Teter}}, and
  \bibinfo{author}{\bibfnamefont{V.}~\bibnamefont{Elser}},
  \bibinfo{year}{1992}, \bibinfo{journal}{Nature}
  \textbf{\bibinfo{volume}{355}}, \bibinfo{pages}{333}.

\bibitem[{\citenamefont{Lewenkopf} \emph{et~al.}(2007)\citenamefont{Lewenkopf,
  Mucciolo, and {Castro Neto}}}]{LMCN07}
\bibinfo{author}{\bibnamefont{Lewenkopf}, \bibfnamefont{C.~H.}},
  \bibinfo{author}{\bibfnamefont{E.~R.} \bibnamefont{Mucciolo}}, and
  \bibinfo{author}{\bibfnamefont{A.~H.} \bibnamefont{{Castro Neto}}},
  \bibinfo{year}{2007}, \eprint{arXiv:0711.3202}.

\bibitem[{\citenamefont{Li and Andrei}(2007)}]{LA07}
\bibinfo{author}{\bibnamefont{Li}, \bibfnamefont{G.}}, and
  \bibinfo{author}{\bibfnamefont{E.~Y.} \bibnamefont{Andrei}},
  \bibinfo{year}{2007}, \bibinfo{journal}{Nature Physics}
  \textbf{\bibinfo{volume}{3}}, \bibinfo{pages}{623}.

\bibitem[{\citenamefont{Li} \emph{et~al.}(2006)\citenamefont{Li, Tsai, Padilla,
  Dordevic, Burch, Wang, and Basov}}]{Letal06}
\bibinfo{author}{\bibnamefont{Li}, \bibfnamefont{Z.~Q.}},
  \bibinfo{author}{\bibfnamefont{S.-W.} \bibnamefont{Tsai}},
  \bibinfo{author}{\bibfnamefont{W.~J.} \bibnamefont{Padilla}},
  \bibinfo{author}{\bibfnamefont{S.~V.} \bibnamefont{Dordevic}},
  \bibinfo{author}{\bibfnamefont{K.~S.} \bibnamefont{Burch}},
  \bibinfo{author}{\bibfnamefont{Y.~J.} \bibnamefont{Wang}}, and
  \bibinfo{author}{\bibfnamefont{D.~N.} \bibnamefont{Basov}},
  \bibinfo{year}{2006}, \bibinfo{journal}{Phys. Rev. B}
  \textbf{\bibinfo{volume}{74}}, \bibinfo{pages}{195404}.

\bibitem[{\citenamefont{{Lopes dos Santos}}
  \emph{et~al.}(2007)\citenamefont{{Lopes dos Santos}, Peres, and {Castro
  Neto}}}]{LSPCN07}
\bibinfo{author}{\bibnamefont{{Lopes dos Santos}}, \bibfnamefont{J.~M.~B.}},
  \bibinfo{author}{\bibfnamefont{N.~M.~R.} \bibnamefont{Peres}}, and
  \bibinfo{author}{\bibfnamefont{A.~H.} \bibnamefont{{Castro Neto}}},
  \bibinfo{year}{2007}, \bibinfo{journal}{Phys. Rev. Lett.}
  \textbf{\bibinfo{volume}{99}}, \bibinfo{pages}{256802}.

\bibitem[{\citenamefont{Louis} \emph{et~al.}(2007)\citenamefont{Louis,
  Verg\'es, Guinea, and Chiappe}}]{LVGC07}
\bibinfo{author}{\bibnamefont{Louis}, \bibfnamefont{E.}},
  \bibinfo{author}{\bibfnamefont{J.~A.} \bibnamefont{Verg\'es}},
  \bibinfo{author}{\bibfnamefont{F.}~\bibnamefont{Guinea}}, and
  \bibinfo{author}{\bibfnamefont{G.}~\bibnamefont{Chiappe}},
  \bibinfo{year}{2007}, \bibinfo{journal}{Phys. Rev. B}
  \textbf{\bibinfo{volume}{75}}, \bibinfo{pages}{085440}.

\bibitem[{\citenamefont{Ludwig} \emph{et~al.}(1994)\citenamefont{Ludwig,
  Fisher, Shankar, and Grinstein}}]{LFSG94}
\bibinfo{author}{\bibnamefont{Ludwig}, \bibfnamefont{A.~W.~W.}},
  \bibinfo{author}{\bibfnamefont{M.~P.~A.} \bibnamefont{Fisher}},
  \bibinfo{author}{\bibfnamefont{R.}~\bibnamefont{Shankar}}, and
  \bibinfo{author}{\bibfnamefont{G.}~\bibnamefont{Grinstein}},
  \bibinfo{year}{1994}, \bibinfo{journal}{Phys. Rev. B}
  \textbf{\bibinfo{volume}{50}}, \bibinfo{pages}{7526}.

\bibitem[{\citenamefont{Lukose} \emph{et~al.}(2007)\citenamefont{Lukose,
  Shankar, and Baskaran}}]{LSB06}
\bibinfo{author}{\bibnamefont{Lukose}, \bibfnamefont{V.}},
  \bibinfo{author}{\bibfnamefont{R.}~\bibnamefont{Shankar}}, and
  \bibinfo{author}{\bibfnamefont{G.}~\bibnamefont{Baskaran}},
  \bibinfo{year}{2007}, \bibinfo{journal}{Phys. Rev. Lett.}
  \textbf{\bibinfo{volume}{98}}, \bibinfo{pages}{16802}.

\bibitem[{\citenamefont{Luk'yanchuk and Kopelevich}(2004)}]{LK04}
\bibinfo{author}{\bibnamefont{Luk'yanchuk}, \bibfnamefont{I.~A.}}, and
  \bibinfo{author}{\bibfnamefont{Y.}~\bibnamefont{Kopelevich}},
  \bibinfo{year}{2004}, \bibinfo{journal}{Phys. Rev. Lett.}
  \textbf{\bibinfo{volume}{93}}, \bibinfo{pages}{166402}.

\bibitem[{\citenamefont{{Ma\~nes}}(2007)}]{M07}
\bibinfo{author}{\bibnamefont{{Ma\~nes}}, \bibfnamefont{J.~L.}},
  \bibinfo{year}{2007}, \bibinfo{journal}{Phys. Rev. B}
  \textbf{\bibinfo{volume}{76}}, \bibinfo{pages}{045430}.

\bibitem[{\citenamefont{{Ma\~nes}} \emph{et~al.}(2007)\citenamefont{{Ma\~nes},
  Guinea, and Vozmediano}}]{MGV07}
\bibinfo{author}{\bibnamefont{{Ma\~nes}}, \bibfnamefont{J.~L.}},
  \bibinfo{author}{\bibfnamefont{F.}~\bibnamefont{Guinea}}, and
  \bibinfo{author}{\bibfnamefont{M.~A.~H.} \bibnamefont{Vozmediano}},
  \bibinfo{year}{2007}, \bibinfo{journal}{Phys. Rev. B}
  \textbf{\bibinfo{volume}{75}}, \bibinfo{pages}{155424}.

\bibitem[{\citenamefont{Maiti and Sengupta}(2007)}]{MS07}
\bibinfo{author}{\bibnamefont{Maiti}, \bibfnamefont{M.}}, and
  \bibinfo{author}{\bibfnamefont{K.}~\bibnamefont{Sengupta}},
  \bibinfo{year}{2007}, \bibinfo{journal}{Phys. Rev. B}
  \textbf{\bibinfo{volume}{76}}, \bibinfo{pages}{054513}.

\bibitem[{\citenamefont{Malard} \emph{et~al.}(2007)\citenamefont{Malard,
  Nilsson, Elias, Brant, Plentz, Alves, {Castro Neto}, and Pimenta}}]{Malard07}
\bibinfo{author}{\bibnamefont{Malard}, \bibfnamefont{L.~M.}},
  \bibinfo{author}{\bibfnamefont{J.}~\bibnamefont{Nilsson}},
  \bibinfo{author}{\bibfnamefont{D.~C.} \bibnamefont{Elias}},
  \bibinfo{author}{\bibfnamefont{J.~C.} \bibnamefont{Brant}},
  \bibinfo{author}{\bibfnamefont{F.}~\bibnamefont{Plentz}},
  \bibinfo{author}{\bibfnamefont{E.~S.} \bibnamefont{Alves}},
  \bibinfo{author}{\bibfnamefont{A.~H.} \bibnamefont{{Castro Neto}}}, and
  \bibinfo{author}{\bibfnamefont{M.~A.} \bibnamefont{Pimenta}},
  \bibinfo{year}{2007}, \bibinfo{journal}{Phys. Rev. B}
  \textbf{\bibinfo{volume}{76}}, \bibinfo{pages}{201401}.

\bibitem[{\citenamefont{Mallet} \emph{et~al.}(2007)\citenamefont{Mallet,
  Varchon, Naud, Magaud, Berger, and Veuillen}}]{Metal07}
\bibinfo{author}{\bibnamefont{Mallet}, \bibfnamefont{P.}},
  \bibinfo{author}{\bibfnamefont{F.}~\bibnamefont{Varchon}},
  \bibinfo{author}{\bibfnamefont{C.}~\bibnamefont{Naud}},
  \bibinfo{author}{\bibfnamefont{L.}~\bibnamefont{Magaud}},
  \bibinfo{author}{\bibfnamefont{C.}~\bibnamefont{Berger}}, and
  \bibinfo{author}{\bibfnamefont{J.-Y.} \bibnamefont{Veuillen}},
  \bibinfo{year}{2007}, \bibinfo{journal}{Phys. Rev. B}
  \textbf{\bibinfo{volume}{76}}, \bibinfo{pages}{041403(R)}.

\bibitem[{\citenamefont{Maple}(1998)}]{hightc}
\bibinfo{author}{\bibnamefont{Maple}, \bibfnamefont{M.~B.}},
  \bibinfo{year}{1998}, \bibinfo{journal}{Jou. Mag. Mag. Mat.}
  \textbf{\bibinfo{volume}{177}}, \bibinfo{pages}{18}.

\bibitem[{\citenamefont{Mariani} \emph{et~al.}(2007)\citenamefont{Mariani,
  Glazman, Kamenev, and {von Oppen}}}]{MGKO07}
\bibinfo{author}{\bibnamefont{Mariani}, \bibfnamefont{E.}},
  \bibinfo{author}{\bibfnamefont{L.}~\bibnamefont{Glazman}},
  \bibinfo{author}{\bibfnamefont{A.}~\bibnamefont{Kamenev}}, and
  \bibinfo{author}{\bibfnamefont{F.}~\bibnamefont{{von Oppen}}},
  \bibinfo{year}{2007}, \bibinfo{journal}{Phys. Rev. B}
  \textbf{\bibinfo{volume}{76}}, \bibinfo{pages}{165402}.

\bibitem[{\citenamefont{Mariani and {von Oppen}}(2007)}]{mVO07}
\bibinfo{author}{\bibnamefont{Mariani}, \bibfnamefont{E.}}, and
  \bibinfo{author}{\bibfnamefont{F.}~\bibnamefont{{von Oppen}}},
  \bibinfo{year}{2007}, \eprint{arXiv:cond-mat/0707.4350}.

\bibitem[{\citenamefont{Martelo} \emph{et~al.}(1997)\citenamefont{Martelo,
  Dzierzawa, Siffert, and Baeriswyl}}]{martelo97}
\bibinfo{author}{\bibnamefont{Martelo}, \bibfnamefont{L.~M.}},
  \bibinfo{author}{\bibfnamefont{M.}~\bibnamefont{Dzierzawa}},
  \bibinfo{author}{\bibfnamefont{L.}~\bibnamefont{Siffert}}, and
  \bibinfo{author}{\bibfnamefont{D.}~\bibnamefont{Baeriswyl}},
  \bibinfo{year}{1997}, \bibinfo{journal}{Z. Physik B}
  \textbf{\bibinfo{volume}{103}}, \bibinfo{pages}{335}.

\bibitem[{\citenamefont{Martin and Blanter}(2007)}]{MB07}
\bibinfo{author}{\bibnamefont{Martin}, \bibfnamefont{I.}}, and
  \bibinfo{author}{\bibfnamefont{Y.~M.} \bibnamefont{Blanter}},
  \bibinfo{year}{2007}, \eprint{cond-mat/0705.0532}.

\bibitem[{\citenamefont{Martin} \emph{et~al.}(2007)\citenamefont{Martin,
  Akerman, Ulbricht, Lohmann, Smet, {von Klitzing}, and Yacoby}}]{MAULSKY07}
\bibinfo{author}{\bibnamefont{Martin}, \bibfnamefont{J.}},
  \bibinfo{author}{\bibfnamefont{N.}~\bibnamefont{Akerman}},
  \bibinfo{author}{\bibfnamefont{G.}~\bibnamefont{Ulbricht}},
  \bibinfo{author}{\bibfnamefont{T.}~\bibnamefont{Lohmann}},
  \bibinfo{author}{\bibfnamefont{J.~H.} \bibnamefont{Smet}},
  \bibinfo{author}{\bibfnamefont{K.}~\bibnamefont{{von Klitzing}}}, and
  \bibinfo{author}{\bibfnamefont{A.}~\bibnamefont{Yacoby}},
  \bibinfo{year}{2007}, \eprint{cond-mat/0705.2180}.

\bibitem[{\citenamefont{Martins} \emph{et~al.}(2007)\citenamefont{Martins,
  Miwa, {da Silva}, and Fazzio}}]{MMSF07}
\bibinfo{author}{\bibnamefont{Martins}, \bibfnamefont{T.~B.}},
  \bibinfo{author}{\bibfnamefont{R.~H.} \bibnamefont{Miwa}},
  \bibinfo{author}{\bibfnamefont{A.~J.~R.} \bibnamefont{{da Silva}}}, and
  \bibinfo{author}{\bibfnamefont{A.}~\bibnamefont{Fazzio}},
  \bibinfo{year}{2007}, \bibinfo{journal}{Phys. Rev. Lett.}
  \textbf{\bibinfo{volume}{98}}, \bibinfo{pages}{196803}.

\bibitem[{\citenamefont{Matsui} \emph{et~al.}(2005)\citenamefont{Matsui,
  Kambara, Niimi, Tagami, Tsukada, and Fukuyama}}]{Metal05}
\bibinfo{author}{\bibnamefont{Matsui}, \bibfnamefont{T.}},
  \bibinfo{author}{\bibfnamefont{H.}~\bibnamefont{Kambara}},
  \bibinfo{author}{\bibfnamefont{Y.}~\bibnamefont{Niimi}},
  \bibinfo{author}{\bibfnamefont{K.}~\bibnamefont{Tagami}},
  \bibinfo{author}{\bibfnamefont{M.}~\bibnamefont{Tsukada}}, and
  \bibinfo{author}{\bibfnamefont{H.}~\bibnamefont{Fukuyama}},
  \bibinfo{year}{2005}, \bibinfo{journal}{Phys. Rev. Lett.}
  \textbf{\bibinfo{volume}{94}}, \bibinfo{pages}{227201}.

\bibitem[{\citenamefont{Mattausch and Pankratov}(2007)}]{MO0704}
\bibinfo{author}{\bibnamefont{Mattausch}, \bibfnamefont{A.}}, and
  \bibinfo{author}{\bibfnamefont{O.}~\bibnamefont{Pankratov}},
  \bibinfo{year}{2007}, \eprint{cond-mat/0704.0216}.

\bibitem[{\citenamefont{McCann}(2006)}]{MC06}
\bibinfo{author}{\bibnamefont{McCann}, \bibfnamefont{E.}},
  \bibinfo{year}{2006}, \bibinfo{journal}{Phys. Rev. B}
  \textbf{\bibinfo{volume}{74}}, \bibinfo{pages}{161403}.

\bibitem[{\citenamefont{McCann and Fal'ko}(2006)}]{MF06}
\bibinfo{author}{\bibnamefont{McCann}, \bibfnamefont{E.}}, and
  \bibinfo{author}{\bibfnamefont{V.~I.} \bibnamefont{Fal'ko}},
  \bibinfo{year}{2006}, \bibinfo{journal}{Phys. Rev. Lett.}
  \textbf{\bibinfo{volume}{96}}, \bibinfo{pages}{086805}.

\bibitem[{\citenamefont{McCann} \emph{et~al.}(2006)\citenamefont{McCann,
  Kechedzhi, Fal'ko, Suzuura, Ando, and Altshuler}}]{Metal06b}
\bibinfo{author}{\bibnamefont{McCann}, \bibfnamefont{E.}},
  \bibinfo{author}{\bibfnamefont{K.}~\bibnamefont{Kechedzhi}},
  \bibinfo{author}{\bibfnamefont{V.~I.} \bibnamefont{Fal'ko}},
  \bibinfo{author}{\bibfnamefont{H.}~\bibnamefont{Suzuura}},
  \bibinfo{author}{\bibfnamefont{T.}~\bibnamefont{Ando}}, and
  \bibinfo{author}{\bibfnamefont{B.~L.} \bibnamefont{Altshuler}},
  \bibinfo{year}{2006}, \bibinfo{journal}{Phys. Rev. Lett.}
  \textbf{\bibinfo{volume}{97}}, \bibinfo{pages}{146805}.

\bibitem[{\citenamefont{McClure}(1956)}]{M56}
\bibinfo{author}{\bibnamefont{McClure}, \bibfnamefont{J.~W.}},
  \bibinfo{year}{1956}, \bibinfo{journal}{Phys. Rev.}
  \textbf{\bibinfo{volume}{104}}, \bibinfo{pages}{666}.

\bibitem[{\citenamefont{McClure}(1957)}]{M57}
\bibinfo{author}{\bibnamefont{McClure}, \bibfnamefont{J.~W.}},
  \bibinfo{year}{1957}, \bibinfo{journal}{Phys. Rev.}
  \textbf{\bibinfo{volume}{108}}, \bibinfo{pages}{612}.

\bibitem[{\citenamefont{McClure}(1958)}]{M58}
\bibinfo{author}{\bibnamefont{McClure}, \bibfnamefont{J.~W.}},
  \bibinfo{year}{1958}, \bibinfo{journal}{Phys. Rev.}
  \textbf{\bibinfo{volume}{112}}, \bibinfo{pages}{715}.

\bibitem[{\citenamefont{McClure}(1960)}]{M1960}
\bibinfo{author}{\bibnamefont{McClure}, \bibfnamefont{J.~W.}},
  \bibinfo{year}{1960}, \bibinfo{journal}{Phys. Rev.}
  \textbf{\bibinfo{volume}{119}}, \bibinfo{pages}{606}.

\bibitem[{\citenamefont{McClure}(1964)}]{M64}
\bibinfo{author}{\bibnamefont{McClure}, \bibfnamefont{J.~W.}},
  \bibinfo{year}{1964}, \bibinfo{journal}{IBM J. Res. Dev.}
  \textbf{\bibinfo{volume}{8}}, \bibinfo{pages}{255}.

\bibitem[{\citenamefont{McClure}(1971)}]{mcclure_review}
\bibinfo{author}{\bibnamefont{McClure}, \bibfnamefont{J.~W.}},
  \bibinfo{year}{1971}, \emph{\bibinfo{title}{Physics of Semi-metals and
  narrow-gap semiconductors}} (\bibinfo{publisher}{D. L. Carter and R. T. Bate,
  Pergamon Press, New York}).

\bibitem[{\citenamefont{McEuen} \emph{et~al.}(1999)\citenamefont{McEuen,
  Bockrath, Cobden, Yoon, and Louie}}]{Louie99}
\bibinfo{author}{\bibnamefont{McEuen}, \bibfnamefont{P.~L.}},
  \bibinfo{author}{\bibfnamefont{M.}~\bibnamefont{Bockrath}},
  \bibinfo{author}{\bibfnamefont{D.~H.} \bibnamefont{Cobden}},
  \bibinfo{author}{\bibfnamefont{Y.-G.} \bibnamefont{Yoon}}, and
  \bibinfo{author}{\bibfnamefont{S.~G.} \bibnamefont{Louie}},
  \bibinfo{year}{1999}, \bibinfo{journal}{Phys. Rev. Lett.}
  \textbf{\bibinfo{volume}{83}}, \bibinfo{pages}{5098}.

\bibitem[{\citenamefont{Meyer}
  \emph{et~al.}(2007{\natexlab{a}})\citenamefont{Meyer, Geim, Katsnelson,
  Novoselov, Booth, and Roth}}]{meyer07}
\bibinfo{author}{\bibnamefont{Meyer}, \bibfnamefont{J.~C.}},
  \bibinfo{author}{\bibfnamefont{A.~K.} \bibnamefont{Geim}},
  \bibinfo{author}{\bibfnamefont{M.~I.} \bibnamefont{Katsnelson}},
  \bibinfo{author}{\bibfnamefont{K.~S.} \bibnamefont{Novoselov}},
  \bibinfo{author}{\bibfnamefont{T.~J.} \bibnamefont{Booth}}, and
  \bibinfo{author}{\bibfnamefont{S.}~\bibnamefont{Roth}},
  \bibinfo{year}{2007}{\natexlab{a}}, \bibinfo{journal}{Nature}
  \textbf{\bibinfo{volume}{446}}, \bibinfo{pages}{60}.

\bibitem[{\citenamefont{Meyer}
  \emph{et~al.}(2007{\natexlab{b}})\citenamefont{Meyer, Geim, Katsnelson,
  Novoselov, Obergfell, Roth, Girit, and Zettl}}]{MGKNORGZ07}
\bibinfo{author}{\bibnamefont{Meyer}, \bibfnamefont{J.~C.}},
  \bibinfo{author}{\bibfnamefont{A.~K.} \bibnamefont{Geim}},
  \bibinfo{author}{\bibfnamefont{M.~I.} \bibnamefont{Katsnelson}},
  \bibinfo{author}{\bibfnamefont{K.~S.} \bibnamefont{Novoselov}},
  \bibinfo{author}{\bibfnamefont{D.}~\bibnamefont{Obergfell}},
  \bibinfo{author}{\bibfnamefont{S.}~\bibnamefont{Roth}},
  \bibinfo{author}{\bibfnamefont{C.}~\bibnamefont{Girit}}, and
  \bibinfo{author}{\bibfnamefont{A.}~\bibnamefont{Zettl}},
  \bibinfo{year}{2007}{\natexlab{b}}, \bibinfo{journal}{Solid State Commun.}
  \textbf{\bibinfo{volume}{143}}, \bibinfo{pages}{101}.

\bibitem[{\citenamefont{Miao} \emph{et~al.}(2007)\citenamefont{Miao, Wijeratne,
  Coskun, Zhang, and Lau}}]{MWCZL07}
\bibinfo{author}{\bibnamefont{Miao}, \bibfnamefont{F.}},
  \bibinfo{author}{\bibfnamefont{S.}~\bibnamefont{Wijeratne}},
  \bibinfo{author}{\bibfnamefont{U.}~\bibnamefont{Coskun}},
  \bibinfo{author}{\bibfnamefont{Y.}~\bibnamefont{Zhang}}, and
  \bibinfo{author}{\bibfnamefont{C.~N.} \bibnamefont{Lau}},
  \bibinfo{year}{2007}, \bibinfo{journal}{Science}
  \textbf{\bibinfo{volume}{317}}, \bibinfo{pages}{1530}.

\bibitem[{\citenamefont{{Milton Pereira Junior}}
  \emph{et~al.}(2007)\citenamefont{{Milton Pereira Junior}, Vasilopoulos, and
  Peeters}}]{MPVP07}
\bibinfo{author}{\bibnamefont{{Milton Pereira Junior}}, \bibfnamefont{J.}},
  \bibinfo{author}{\bibfnamefont{P.}~\bibnamefont{Vasilopoulos}}, and
  \bibinfo{author}{\bibfnamefont{F.~M.} \bibnamefont{Peeters}},
  \bibinfo{year}{2007}, \bibinfo{journal}{Nano Letters}
  \textbf{\bibinfo{volume}{7}}, \bibinfo{pages}{946}.

\bibitem[{\citenamefont{Min} \emph{et~al.}(2006)\citenamefont{Min, Hill,
  Sinitsyn, Sahu, Kleinman, and MacDonald}}]{MHSSKM06}
\bibinfo{author}{\bibnamefont{Min}, \bibfnamefont{H.}},
  \bibinfo{author}{\bibfnamefont{J.}~\bibnamefont{Hill}},
  \bibinfo{author}{\bibfnamefont{N.}~\bibnamefont{Sinitsyn}},
  \bibinfo{author}{\bibfnamefont{B.}~\bibnamefont{Sahu}},
  \bibinfo{author}{\bibfnamefont{L.}~\bibnamefont{Kleinman}}, and
  \bibinfo{author}{\bibfnamefont{A.}~\bibnamefont{MacDonald}},
  \bibinfo{year}{2006}, \bibinfo{journal}{Phys. Rev. B}
  \textbf{\bibinfo{volume}{74}}, \bibinfo{pages}{165310}.

\bibitem[{\citenamefont{Mishchenko}(2007)}]{Mish07}
\bibinfo{author}{\bibnamefont{Mishchenko}, \bibfnamefont{E.~G.}},
  \bibinfo{year}{2007}, \bibinfo{journal}{Phys. Rev. Lett.}
  \textbf{\bibinfo{volume}{98}}, \bibinfo{pages}{216801}.

\bibitem[{\citenamefont{Morozov} \emph{et~al.}(2006)\citenamefont{Morozov,
  Novoselov, Katsnelson, Schedin, Jiang, and Geim}}]{Metal06}
\bibinfo{author}{\bibnamefont{Morozov}, \bibfnamefont{S.}},
  \bibinfo{author}{\bibfnamefont{K.}~\bibnamefont{Novoselov}},
  \bibinfo{author}{\bibfnamefont{M.}~\bibnamefont{Katsnelson}},
  \bibinfo{author}{\bibfnamefont{F.}~\bibnamefont{Schedin}},
  \bibinfo{author}{\bibfnamefont{D.}~\bibnamefont{Jiang}}, and
  \bibinfo{author}{\bibfnamefont{A.~K.} \bibnamefont{Geim}},
  \bibinfo{year}{2006}, \bibinfo{journal}{Phys. Rev. Lett.}
  \textbf{\bibinfo{volume}{97}}, \bibinfo{pages}{016801}.

\bibitem[{\citenamefont{Morozov} \emph{et~al.}(2005)\citenamefont{Morozov,
  Novoselov, Schedin, Jiang, Firsov, and Geim}}]{Metal05c}
\bibinfo{author}{\bibnamefont{Morozov}, \bibfnamefont{S.~V.}},
  \bibinfo{author}{\bibfnamefont{K.~S.} \bibnamefont{Novoselov}},
  \bibinfo{author}{\bibfnamefont{F.}~\bibnamefont{Schedin}},
  \bibinfo{author}{\bibfnamefont{D.}~\bibnamefont{Jiang}},
  \bibinfo{author}{\bibfnamefont{A.~A.} \bibnamefont{Firsov}}, and
  \bibinfo{author}{\bibfnamefont{A.~K.} \bibnamefont{Geim}},
  \bibinfo{year}{2005}, \bibinfo{journal}{Phys. Rev. B}
  \textbf{\bibinfo{volume}{72}}, \bibinfo{pages}{201401}.

\bibitem[{\citenamefont{Morpurgo and Guinea}(2006)}]{MG06}
\bibinfo{author}{\bibnamefont{Morpurgo}, \bibfnamefont{A.~F.}}, and
  \bibinfo{author}{\bibfnamefont{F.}~\bibnamefont{Guinea}},
  \bibinfo{year}{2006}, \bibinfo{journal}{Phys. Rev. Lett.}
  \textbf{\bibinfo{volume}{97}}, \bibinfo{pages}{196804}.

\bibitem[{\citenamefont{Mrozowski}(1952)}]{M1952}
\bibinfo{author}{\bibnamefont{Mrozowski}, \bibfnamefont{S.}},
  \bibinfo{year}{1952}, \bibinfo{journal}{Phys. Rev.}
  \textbf{\bibinfo{volume}{85}}, \bibinfo{pages}{609}.

\bibitem[{\citenamefont{{Mu\~noz}-Rojas}
  \emph{et~al.}(2006)\citenamefont{{Mu\~noz}-Rojas, Jacob, Fern\'andez-Rossier,
  and Palacios}}]{MJFP06}
\bibinfo{author}{\bibnamefont{{Mu\~noz}-Rojas}, \bibfnamefont{F.}},
  \bibinfo{author}{\bibfnamefont{D.}~\bibnamefont{Jacob}},
  \bibinfo{author}{\bibfnamefont{J.}~\bibnamefont{Fern\'andez-Rossier}}, and
  \bibinfo{author}{\bibfnamefont{J.~J.} \bibnamefont{Palacios}},
  \bibinfo{year}{2006}, \bibinfo{journal}{Phys. Rev. B}
  \textbf{\bibinfo{volume}{74}}, \bibinfo{pages}{195417}.

\bibitem[{\citenamefont{Nair} \emph{et~al.}(2008)\citenamefont{Nair, Bangert,
  Gass, Novoselov, Geim, and Bleloch}}]{Nairal07}
\bibinfo{author}{\bibnamefont{Nair}, \bibfnamefont{R.~R.}},
  \bibinfo{author}{\bibfnamefont{U.}~\bibnamefont{Bangert}},
  \bibinfo{author}{\bibfnamefont{M.~H.} \bibnamefont{Gass}},
  \bibinfo{author}{\bibfnamefont{K.~S.} \bibnamefont{Novoselov}},
  \bibinfo{author}{\bibfnamefont{A.~K.} \bibnamefont{Geim}}, and
  \bibinfo{author}{\bibfnamefont{A.~L.} \bibnamefont{Bleloch}},
  \bibinfo{year}{2008}, \eprint{unpublished}.

\bibitem[{\citenamefont{Nakada} \emph{et~al.}(1996)\citenamefont{Nakada,
  Fujita, Dresselhaus, and Dresselhaus}}]{Nakada96}
\bibinfo{author}{\bibnamefont{Nakada}, \bibfnamefont{K.}},
  \bibinfo{author}{\bibfnamefont{M.}~\bibnamefont{Fujita}},
  \bibinfo{author}{\bibfnamefont{G.}~\bibnamefont{Dresselhaus}}, and
  \bibinfo{author}{\bibfnamefont{M.~S.} \bibnamefont{Dresselhaus}},
  \bibinfo{year}{1996}, \bibinfo{journal}{Phys. Rev. B}
  \textbf{\bibinfo{volume}{54}}, \bibinfo{pages}{17954 }.

\bibitem[{\citenamefont{Nakamura}(2007)}]{N2007}
\bibinfo{author}{\bibnamefont{Nakamura}, \bibfnamefont{M.}},
  \bibinfo{year}{2007}, \bibinfo{journal}{Phys. Rev. B}
  \textbf{\bibinfo{volume}{76}}, \bibinfo{pages}{113301}.

\bibitem[{\citenamefont{Nakao}(1976)}]{N76}
\bibinfo{author}{\bibnamefont{Nakao}, \bibfnamefont{K.}}, \bibinfo{year}{1976},
  \bibinfo{journal}{J. Phys. Soc. Jpn.} \textbf{\bibinfo{volume}{40}},
  \bibinfo{pages}{761}.

\bibitem[{\citenamefont{Nelson} \emph{et~al.}(2004)\citenamefont{Nelson, Piran,
  and Weinberg}}]{nelson}
\bibinfo{author}{\bibnamefont{Nelson}, \bibfnamefont{D.}},
  \bibinfo{author}{\bibfnamefont{D.~R.} \bibnamefont{Piran}}, and
  \bibinfo{author}{\bibfnamefont{S.}~\bibnamefont{Weinberg}},
  \bibinfo{year}{2004}, \emph{\bibinfo{title}{Statistical mechanics of
  membranes and surfaces}} (\bibinfo{publisher}{World Scientific, Singapore}).

\bibitem[{\citenamefont{Nelson and Peliti}(1987)}]{NP87}
\bibinfo{author}{\bibnamefont{Nelson}, \bibfnamefont{D.~R.}}, and
  \bibinfo{author}{\bibfnamefont{L.}~\bibnamefont{Peliti}},
  \bibinfo{year}{1987}, \bibinfo{journal}{J. Physique}
  \textbf{\bibinfo{volume}{48}}, \bibinfo{pages}{1085}.

\bibitem[{\citenamefont{Nersesyan} \emph{et~al.}(1994)\citenamefont{Nersesyan,
  Tsvelik, and Wenger}}]{NTW94}
\bibinfo{author}{\bibnamefont{Nersesyan}, \bibfnamefont{A.~A.}},
  \bibinfo{author}{\bibfnamefont{A.~M.} \bibnamefont{Tsvelik}}, and
  \bibinfo{author}{\bibfnamefont{F.}~\bibnamefont{Wenger}},
  \bibinfo{year}{1994}, \bibinfo{journal}{Phys. Rev. Lett.}
  \textbf{\bibinfo{volume}{72}}, \bibinfo{pages}{2628}.

\bibitem[{\citenamefont{Niimi} \emph{et~al.}(2006)\citenamefont{Niimi, Kambara,
  Matsui, Yoshioka, and Fukuyama}}]{Netal06}
\bibinfo{author}{\bibnamefont{Niimi}, \bibfnamefont{Y.}},
  \bibinfo{author}{\bibfnamefont{H.}~\bibnamefont{Kambara}},
  \bibinfo{author}{\bibfnamefont{T.}~\bibnamefont{Matsui}},
  \bibinfo{author}{\bibfnamefont{D.}~\bibnamefont{Yoshioka}}, and
  \bibinfo{author}{\bibfnamefont{H.}~\bibnamefont{Fukuyama}},
  \bibinfo{year}{2006}, \bibinfo{journal}{Phys. Rev. Lett.}
  \textbf{\bibinfo{volume}{97}}, \bibinfo{pages}{236804}.

\bibitem[{\citenamefont{Nilsson and {Castro Neto}}(2007)}]{NCV06}
\bibinfo{author}{\bibnamefont{Nilsson}, \bibfnamefont{J.}}, and
  \bibinfo{author}{\bibfnamefont{A.~H.} \bibnamefont{{Castro Neto}}},
  \bibinfo{year}{2007}, \bibinfo{journal}{Phys. Rev. Lett.}
  \textbf{\bibinfo{volume}{98}}, \bibinfo{pages}{126801}.

\bibitem[{\citenamefont{Nilsson}
  \emph{et~al.}(2006{\natexlab{a}})\citenamefont{Nilsson, {Castro Neto},
  Guinea, and Peres}}]{NNGP06}
\bibinfo{author}{\bibnamefont{Nilsson}, \bibfnamefont{J.}},
  \bibinfo{author}{\bibfnamefont{A.~H.} \bibnamefont{{Castro Neto}}},
  \bibinfo{author}{\bibfnamefont{F.}~\bibnamefont{Guinea}}, and
  \bibinfo{author}{\bibfnamefont{N.~M.~R.} \bibnamefont{Peres}},
  \bibinfo{year}{2006}{\natexlab{a}}, \bibinfo{journal}{Phys. Rev. Lett.}
  \textbf{\bibinfo{volume}{97}}, \bibinfo{pages}{266801}.

\bibitem[{\citenamefont{Nilsson}
  \emph{et~al.}(2007{\natexlab{a}})\citenamefont{Nilsson, {Castro Neto},
  Guinea, and Peres}}]{NCNGP07}
\bibinfo{author}{\bibnamefont{Nilsson}, \bibfnamefont{J.}},
  \bibinfo{author}{\bibfnamefont{A.~H.} \bibnamefont{{Castro Neto}}},
  \bibinfo{author}{\bibfnamefont{F.}~\bibnamefont{Guinea}}, and
  \bibinfo{author}{\bibfnamefont{N.~M.~R.} \bibnamefont{Peres}},
  \bibinfo{year}{2007}{\natexlab{a}}, \eprint{arXiv:0712.3259}.

\bibitem[{\citenamefont{Nilsson}
  \emph{et~al.}(2007{\natexlab{b}})\citenamefont{Nilsson, {Castro Neto},
  Guinea, and Peres}}]{NCNGP06}
\bibinfo{author}{\bibnamefont{Nilsson}, \bibfnamefont{J.}},
  \bibinfo{author}{\bibfnamefont{A.~H.} \bibnamefont{{Castro Neto}}},
  \bibinfo{author}{\bibfnamefont{F.}~\bibnamefont{Guinea}}, and
  \bibinfo{author}{\bibfnamefont{N.~M.~R.} \bibnamefont{Peres}},
  \bibinfo{year}{2007}{\natexlab{b}}, \bibinfo{journal}{Phys. Rev. B}
  \textbf{\bibinfo{volume}{76}}, \bibinfo{pages}{165416}.

\bibitem[{\citenamefont{Nilsson}
  \emph{et~al.}(2006{\natexlab{b}})\citenamefont{Nilsson, {Castro Neto}, Peres,
  and Guinea}}]{NCNPG05}
\bibinfo{author}{\bibnamefont{Nilsson}, \bibfnamefont{J.}},
  \bibinfo{author}{\bibfnamefont{A.~H.} \bibnamefont{{Castro Neto}}},
  \bibinfo{author}{\bibfnamefont{N.~M.~R.} \bibnamefont{Peres}}, and
  \bibinfo{author}{\bibfnamefont{F.}~\bibnamefont{Guinea}},
  \bibinfo{year}{2006}{\natexlab{b}}, \bibinfo{journal}{Phys. Rev. B}
  \textbf{\bibinfo{volume}{73}}, \bibinfo{pages}{214418}.

\bibitem[{\citenamefont{Nilsson}
  \emph{et~al.}(2006{\natexlab{c}})\citenamefont{Nilsson, {Castro Neto}, Peres,
  and Guinea}}]{NNPG06}
\bibinfo{author}{\bibnamefont{Nilsson}, \bibfnamefont{J.}},
  \bibinfo{author}{\bibfnamefont{A.~H.} \bibnamefont{{Castro Neto}}},
  \bibinfo{author}{\bibfnamefont{N.~M.~R.} \bibnamefont{Peres}}, and
  \bibinfo{author}{\bibfnamefont{F.}~\bibnamefont{Guinea}},
  \bibinfo{year}{2006}{\natexlab{c}}, \bibinfo{journal}{Phys. Rev. B}
  \textbf{\bibinfo{volume}{73}}, \bibinfo{pages}{214418}.

\bibitem[{\citenamefont{Nomura} \emph{et~al.}(2007)\citenamefont{Nomura,
  Koshino, and Ryu}}]{NKR07}
\bibinfo{author}{\bibnamefont{Nomura}, \bibfnamefont{K.}},
  \bibinfo{author}{\bibfnamefont{M.}~\bibnamefont{Koshino}}, and
  \bibinfo{author}{\bibfnamefont{S.}~\bibnamefont{Ryu}}, \bibinfo{year}{2007},
  \bibinfo{journal}{Phys. Rev. Lett.} \textbf{\bibinfo{volume}{99}},
  \bibinfo{pages}{146806}.

\bibitem[{\citenamefont{Nomura and MacDonald}(2006)}]{NMac06}
\bibinfo{author}{\bibnamefont{Nomura}, \bibfnamefont{K.}}, and
  \bibinfo{author}{\bibfnamefont{A.~H.} \bibnamefont{MacDonald}},
  \bibinfo{year}{2006}, \bibinfo{journal}{Phys. Rev. Lett.}
  \textbf{\bibinfo{volume}{96}}, \bibinfo{pages}{256602}.

\bibitem[{\citenamefont{Nomura and MacDonald}(2007)}]{NMac07}
\bibinfo{author}{\bibnamefont{Nomura}, \bibfnamefont{K.}}, and
  \bibinfo{author}{\bibfnamefont{A.~H.} \bibnamefont{MacDonald}},
  \bibinfo{year}{2007}, \bibinfo{journal}{Phys. Rev. Lett.}
  \textbf{\bibinfo{volume}{98}}, \bibinfo{pages}{076602}.

\bibitem[{\citenamefont{Novikov}(2007{\natexlab{a}})}]{Nov07}
\bibinfo{author}{\bibnamefont{Novikov}, \bibfnamefont{D.~S.}},
  \bibinfo{year}{2007}{\natexlab{a}}, \bibinfo{journal}{Phys. Rev. B}
  \textbf{\bibinfo{volume}{76}}, \bibinfo{pages}{245435}.

\bibitem[{\citenamefont{Novikov}(2007{\natexlab{b}})}]{Novapl07}
\bibinfo{author}{\bibnamefont{Novikov}, \bibfnamefont{D.~S.}},
  \bibinfo{year}{2007}{\natexlab{b}}, \bibinfo{journal}{Appl. Phys. Lett.}
  \textbf{\bibinfo{volume}{91}}, \bibinfo{pages}{102102}.

\bibitem[{\citenamefont{Novikov}(2007{\natexlab{c}})}]{N07}
\bibinfo{author}{\bibnamefont{Novikov}, \bibfnamefont{D.~S.}},
  \bibinfo{year}{2007}{\natexlab{c}}, \bibinfo{journal}{Phys. Rev. Lett.}
  \textbf{\bibinfo{volume}{99}}, \bibinfo{pages}{056802}.

\bibitem[{\citenamefont{Novoselov}
  \emph{et~al.}(2005{\natexlab{a}})\citenamefont{Novoselov, Geim, Morozov,
  Jiang, Katsnelson, Grigorieva, Dubonos, and Firsov}}]{Netal05}
\bibinfo{author}{\bibnamefont{Novoselov}, \bibfnamefont{K.~S.}},
  \bibinfo{author}{\bibfnamefont{A.~K.} \bibnamefont{Geim}},
  \bibinfo{author}{\bibfnamefont{S.~V.} \bibnamefont{Morozov}},
  \bibinfo{author}{\bibfnamefont{D.}~\bibnamefont{Jiang}},
  \bibinfo{author}{\bibfnamefont{M.~I.} \bibnamefont{Katsnelson}},
  \bibinfo{author}{\bibfnamefont{I.~V.} \bibnamefont{Grigorieva}},
  \bibinfo{author}{\bibfnamefont{S.~V.} \bibnamefont{Dubonos}}, and
  \bibinfo{author}{\bibfnamefont{A.~A.} \bibnamefont{Firsov}},
  \bibinfo{year}{2005}{\natexlab{a}}, \bibinfo{journal}{Nature}
  \textbf{\bibinfo{volume}{438}}, \bibinfo{pages}{197}.

\bibitem[{\citenamefont{Novoselov} \emph{et~al.}(2004)\citenamefont{Novoselov,
  Geim, Morozov, Jiang, Zhang, Dubonos, Gregorieva, and Firsov}}]{Netal04}
\bibinfo{author}{\bibnamefont{Novoselov}, \bibfnamefont{K.~S.}},
  \bibinfo{author}{\bibfnamefont{A.~K.} \bibnamefont{Geim}},
  \bibinfo{author}{\bibfnamefont{S.~V.} \bibnamefont{Morozov}},
  \bibinfo{author}{\bibfnamefont{D.}~\bibnamefont{Jiang}},
  \bibinfo{author}{\bibfnamefont{Y.}~\bibnamefont{Zhang}},
  \bibinfo{author}{\bibfnamefont{S.~V.} \bibnamefont{Dubonos}},
  \bibinfo{author}{\bibfnamefont{I.~V.} \bibnamefont{Gregorieva}}, and
  \bibinfo{author}{\bibfnamefont{A.~A.} \bibnamefont{Firsov}},
  \bibinfo{year}{2004}, \bibinfo{journal}{Science}
  \textbf{\bibinfo{volume}{306}}, \bibinfo{pages}{666}.

\bibitem[{\citenamefont{Novoselov}
  \emph{et~al.}(2005{\natexlab{b}})\citenamefont{Novoselov, Jiang, Schedin,
  Booth, Khotkevich, Morozov, and Geim}}]{Netal05b}
\bibinfo{author}{\bibnamefont{Novoselov}, \bibfnamefont{K.~S.}},
  \bibinfo{author}{\bibfnamefont{D.}~\bibnamefont{Jiang}},
  \bibinfo{author}{\bibfnamefont{F.}~\bibnamefont{Schedin}},
  \bibinfo{author}{\bibfnamefont{T.~J.} \bibnamefont{Booth}},
  \bibinfo{author}{\bibfnamefont{V.~V.} \bibnamefont{Khotkevich}},
  \bibinfo{author}{\bibfnamefont{S.~V.} \bibnamefont{Morozov}}, and
  \bibinfo{author}{\bibfnamefont{A.~K.} \bibnamefont{Geim}},
  \bibinfo{year}{2005}{\natexlab{b}}, \bibinfo{journal}{Natl. Acad. Sci. USA}
  \textbf{\bibinfo{volume}{102}}, \bibinfo{pages}{10451}.

\bibitem[{\citenamefont{Novoselov} \emph{et~al.}(2007)\citenamefont{Novoselov,
  Jiang, Zhang, Morozov, Stormer, Zeitler, Maan, Boebinger, Kim, and
  Geim}}]{Netal07}
\bibinfo{author}{\bibnamefont{Novoselov}, \bibfnamefont{K.~S.}},
  \bibinfo{author}{\bibfnamefont{Z.}~\bibnamefont{Jiang}},
  \bibinfo{author}{\bibfnamefont{Y.}~\bibnamefont{Zhang}},
  \bibinfo{author}{\bibfnamefont{S.~V.} \bibnamefont{Morozov}},
  \bibinfo{author}{\bibfnamefont{H.~L.} \bibnamefont{Stormer}},
  \bibinfo{author}{\bibfnamefont{U.}~\bibnamefont{Zeitler}},
  \bibinfo{author}{\bibfnamefont{J.~C.} \bibnamefont{Maan}},
  \bibinfo{author}{\bibfnamefont{G.~S.} \bibnamefont{Boebinger}},
  \bibinfo{author}{\bibfnamefont{P.}~\bibnamefont{Kim}}, and
  \bibinfo{author}{\bibfnamefont{A.~K.} \bibnamefont{Geim}},
  \bibinfo{year}{2007}, \bibinfo{journal}{Science}
  \textbf{\bibinfo{volume}{315}}, \bibinfo{pages}{1379}.

\bibitem[{\citenamefont{Novoselov} \emph{et~al.}(2006)\citenamefont{Novoselov,
  McCann, Mozorov, Fal'ko, Katsnelson, Zeitler, Jiang, Schedin, and
  Geim}}]{Ketal06}
\bibinfo{author}{\bibnamefont{Novoselov}, \bibfnamefont{K.~S.}},
  \bibinfo{author}{\bibfnamefont{E.}~\bibnamefont{McCann}},
  \bibinfo{author}{\bibfnamefont{S.~V.} \bibnamefont{Mozorov}},
  \bibinfo{author}{\bibfnamefont{V.~I.} \bibnamefont{Fal'ko}},
  \bibinfo{author}{\bibfnamefont{M.~I.} \bibnamefont{Katsnelson}},
  \bibinfo{author}{\bibfnamefont{U.}~\bibnamefont{Zeitler}},
  \bibinfo{author}{\bibfnamefont{D.}~\bibnamefont{Jiang}},
  \bibinfo{author}{\bibfnamefont{F.}~\bibnamefont{Schedin}}, and
  \bibinfo{author}{\bibfnamefont{A.~K.} \bibnamefont{Geim}},
  \bibinfo{year}{2006}, \bibinfo{journal}{Nature Physics}
  \textbf{\bibinfo{volume}{1}}, \bibinfo{pages}{177}.

\bibitem[{\citenamefont{Nozi\`eres}(1958)}]{N58}
\bibinfo{author}{\bibnamefont{Nozi\`eres}, \bibfnamefont{P.}},
  \bibinfo{year}{1958}, \bibinfo{journal}{Phys. Rev.}
  \textbf{\bibinfo{volume}{109}}, \bibinfo{pages}{1510}.

\bibitem[{\citenamefont{Ohishi} \emph{et~al.}(2007)\citenamefont{Ohishi,
  Shiraishi, Nouchi, Nozaki, Shinjo, and Suzuki}}]{OSNNSS07}
\bibinfo{author}{\bibnamefont{Ohishi}, \bibfnamefont{M.}},
  \bibinfo{author}{\bibfnamefont{M.}~\bibnamefont{Shiraishi}},
  \bibinfo{author}{\bibfnamefont{R.}~\bibnamefont{Nouchi}},
  \bibinfo{author}{\bibfnamefont{T.}~\bibnamefont{Nozaki}},
  \bibinfo{author}{\bibfnamefont{T.}~\bibnamefont{Shinjo}}, and
  \bibinfo{author}{\bibfnamefont{Y.}~\bibnamefont{Suzuki}},
  \bibinfo{year}{2007}, \bibinfo{journal}{Jap. J. Appl. Phys.}
  \textbf{\bibinfo{volume}{46}}, \bibinfo{pages}{L605}.

\bibitem[{\citenamefont{Ohta}(1968)}]{Ohta1}
\bibinfo{author}{\bibnamefont{Ohta}, \bibfnamefont{K.}}, \bibinfo{year}{1968},
  \bibinfo{journal}{Jpn. J. Appl. Phys.} \textbf{\bibinfo{volume}{10}},
  \bibinfo{pages}{850}.

\bibitem[{\citenamefont{Ohta}(1971)}]{Ohta2}
\bibinfo{author}{\bibnamefont{Ohta}, \bibfnamefont{K.}}, \bibinfo{year}{1971},
  \bibinfo{journal}{J. Phys. Soc. Jpn.} \textbf{\bibinfo{volume}{31}},
  \bibinfo{pages}{1627}.

\bibitem[{\citenamefont{Ohta} \emph{et~al.}(2007)\citenamefont{Ohta, Bostwick,
  McChesney, Seyller, Horn, and Rotenberg}}]{Oetal07}
\bibinfo{author}{\bibnamefont{Ohta}, \bibfnamefont{T.}},
  \bibinfo{author}{\bibfnamefont{A.}~\bibnamefont{Bostwick}},
  \bibinfo{author}{\bibfnamefont{J.~L.} \bibnamefont{McChesney}},
  \bibinfo{author}{\bibfnamefont{T.}~\bibnamefont{Seyller}},
  \bibinfo{author}{\bibfnamefont{K.}~\bibnamefont{Horn}}, and
  \bibinfo{author}{\bibfnamefont{E.}~\bibnamefont{Rotenberg}},
  \bibinfo{year}{2007}, \bibinfo{journal}{Phys Rev. Lett.}
  \textbf{\bibinfo{volume}{98}}, \bibinfo{pages}{206802}.

\bibitem[{\citenamefont{Ohta} \emph{et~al.}(2006)\citenamefont{Ohta, Bostwick,
  Seyller, Horn, and Rotenberg}}]{Oetal06}
\bibinfo{author}{\bibnamefont{Ohta}, \bibfnamefont{T.}},
  \bibinfo{author}{\bibfnamefont{A.}~\bibnamefont{Bostwick}},
  \bibinfo{author}{\bibfnamefont{T.}~\bibnamefont{Seyller}},
  \bibinfo{author}{\bibfnamefont{K.}~\bibnamefont{Horn}}, and
  \bibinfo{author}{\bibfnamefont{E.}~\bibnamefont{Rotenberg}},
  \bibinfo{year}{2006}, \bibinfo{journal}{Science}
  \textbf{\bibinfo{volume}{313}}, \bibinfo{pages}{951}.

\bibitem[{\citenamefont{Oostinga} \emph{et~al.}(2007)\citenamefont{Oostinga,
  Heersche, Liu, Morpurgo, and Vandersypen}}]{OHLMV07}
\bibinfo{author}{\bibnamefont{Oostinga}, \bibfnamefont{J.~B.}},
  \bibinfo{author}{\bibfnamefont{H.~B.} \bibnamefont{Heersche}},
  \bibinfo{author}{\bibfnamefont{X.}~\bibnamefont{Liu}},
  \bibinfo{author}{\bibfnamefont{A.}~\bibnamefont{Morpurgo}}, and
  \bibinfo{author}{\bibfnamefont{L.~M.~K.} \bibnamefont{Vandersypen}},
  \bibinfo{year}{2007}, \bibinfo{journal}{Nature Materials}
  \textbf{\bibinfo{volume}{DOI: 10.1038/nmat2082}}.

\bibitem[{\citenamefont{Oshima and Nagashima}(1997)}]{ON97}
\bibinfo{author}{\bibnamefont{Oshima}, \bibfnamefont{C.}}, and
  \bibinfo{author}{\bibfnamefont{A.}~\bibnamefont{Nagashima}},
  \bibinfo{year}{1997}, \bibinfo{journal}{J. Phys.: Condens. Matter}
  \textbf{\bibinfo{volume}{9}}, \bibinfo{pages}{1}.

\bibitem[{\citenamefont{Osipov} \emph{et~al.}(2003)\citenamefont{Osipov,
  Kochetov, and Pudlak}}]{OKP03}
\bibinfo{author}{\bibnamefont{Osipov}, \bibfnamefont{V.~A.}},
  \bibinfo{author}{\bibfnamefont{E.~A.} \bibnamefont{Kochetov}}, and
  \bibinfo{author}{\bibfnamefont{M.}~\bibnamefont{Pudlak}},
  \bibinfo{year}{2003}, \bibinfo{journal}{JETP} \textbf{\bibinfo{volume}{96}},
  \bibinfo{pages}{140}.

\bibitem[{\citenamefont{Ossipov} \emph{et~al.}(2007)\citenamefont{Ossipov,
  Titov, and Beenakker}}]{OTB07}
\bibinfo{author}{\bibnamefont{Ossipov}, \bibfnamefont{A.}},
  \bibinfo{author}{\bibfnamefont{M.}~\bibnamefont{Titov}}, and
  \bibinfo{author}{\bibfnamefont{C.~W.~J.} \bibnamefont{Beenakker}},
  \bibinfo{year}{2007}, \bibinfo{journal}{Phys. Rev. B}
  \textbf{\bibinfo{volume}{75}}, \bibinfo{pages}{241401}.

\bibitem[{\citenamefont{Ostrovsky} \emph{et~al.}(2006)\citenamefont{Ostrovsky,
  Gornyi, and Mirlin}}]{OGM06}
\bibinfo{author}{\bibnamefont{Ostrovsky}, \bibfnamefont{P.~M.}},
  \bibinfo{author}{\bibfnamefont{I.~V.} \bibnamefont{Gornyi}}, and
  \bibinfo{author}{\bibfnamefont{A.~D.} \bibnamefont{Mirlin}},
  \bibinfo{year}{2006}, \bibinfo{journal}{Phys. Rev. B}
  \textbf{\bibinfo{volume}{74}}, \bibinfo{pages}{235443}.

\bibitem[{\citenamefont{Ostrovsky} \emph{et~al.}(2007)\citenamefont{Ostrovsky,
  Gornyi, and Mirlin}}]{OGM07}
\bibinfo{author}{\bibnamefont{Ostrovsky}, \bibfnamefont{P.~M.}},
  \bibinfo{author}{\bibfnamefont{I.~V.} \bibnamefont{Gornyi}}, and
  \bibinfo{author}{\bibfnamefont{A.~D.} \bibnamefont{Mirlin}},
  \bibinfo{year}{2007}, \bibinfo{journal}{Phys. Rev. Lett.}
  \textbf{\bibinfo{volume}{98}}, \bibinfo{pages}{256801}.

\bibitem[{\citenamefont{\"Ozyilmaz}
  \emph{et~al.}(2007)\citenamefont{\"Ozyilmaz, Jarillo-Herrero, Efetov, Abanin,
  Levitov, and Kim}}]{OJEALK07}
\bibinfo{author}{\bibnamefont{\"Ozyilmaz}, \bibfnamefont{B.}},
  \bibinfo{author}{\bibfnamefont{P.}~\bibnamefont{Jarillo-Herrero}},
  \bibinfo{author}{\bibfnamefont{D.}~\bibnamefont{Efetov}},
  \bibinfo{author}{\bibfnamefont{D.~A.} \bibnamefont{Abanin}},
  \bibinfo{author}{\bibfnamefont{L.~S.} \bibnamefont{Levitov}}, and
  \bibinfo{author}{\bibfnamefont{P.}~\bibnamefont{Kim}}, \bibinfo{year}{2007},
  \bibinfo{journal}{Phys. Rev. Lett.} \textbf{\bibinfo{volume}{99}},
  \bibinfo{pages}{166804}.

\bibitem[{\citenamefont{Paiva} \emph{et~al.}(2005)\citenamefont{Paiva,
  Scalettar, Zheng, Singh, and Oitmaa}}]{paiva05}
\bibinfo{author}{\bibnamefont{Paiva}, \bibfnamefont{T.}},
  \bibinfo{author}{\bibfnamefont{R.~T.} \bibnamefont{Scalettar}},
  \bibinfo{author}{\bibfnamefont{W.}~\bibnamefont{Zheng}},
  \bibinfo{author}{\bibfnamefont{R.~R.~P.} \bibnamefont{Singh}}, and
  \bibinfo{author}{\bibfnamefont{J.}~\bibnamefont{Oitmaa}},
  \bibinfo{year}{2005}, \bibinfo{journal}{Phys. Rev. B}
  \textbf{\bibinfo{volume}{72}}, \bibinfo{pages}{085123}.

\bibitem[{\citenamefont{Parr} \emph{et~al.}(1950)\citenamefont{Parr, Craig, and
  Ross}}]{Parr50}
\bibinfo{author}{\bibnamefont{Parr}, \bibfnamefont{R.~G.}},
  \bibinfo{author}{\bibfnamefont{D.~P.} \bibnamefont{Craig}}, and
  \bibinfo{author}{\bibfnamefont{I.~G.} \bibnamefont{Ross}},
  \bibinfo{year}{1950}, \bibinfo{journal}{J. Chem. Phys.}
  \textbf{\bibinfo{volume}{18}}, \bibinfo{pages}{1561}.

\bibitem[{\citenamefont{Pauling}(1972)}]{P72}
\bibinfo{author}{\bibnamefont{Pauling}, \bibfnamefont{L.}},
  \bibinfo{year}{1972}, \emph{\bibinfo{title}{The Nature of the Chemical Bond}}
  (\bibinfo{publisher}{Cornell U. P., Ithaca, NY}).

\bibitem[{\citenamefont{Peliti and Leibler}(1985)}]{PL85}
\bibinfo{author}{\bibnamefont{Peliti}, \bibfnamefont{L.}}, and
  \bibinfo{author}{\bibfnamefont{S.}~\bibnamefont{Leibler}},
  \bibinfo{year}{1985}, \bibinfo{journal}{Phys. Rev. B}
  \textbf{\bibinfo{volume}{54}}, \bibinfo{pages}{1690}.

\bibitem[{\citenamefont{Pereira} \emph{et~al.}(2006)\citenamefont{Pereira,
  Guinea, dos Santos, Peres, and {Castro Neto}}}]{Petal06}
\bibinfo{author}{\bibnamefont{Pereira}, \bibfnamefont{V.~M.}},
  \bibinfo{author}{\bibfnamefont{F.}~\bibnamefont{Guinea}},
  \bibinfo{author}{\bibfnamefont{J.~M. B.~L.} \bibnamefont{dos Santos}},
  \bibinfo{author}{\bibfnamefont{N.~M.~R.} \bibnamefont{Peres}}, and
  \bibinfo{author}{\bibfnamefont{A.~H.} \bibnamefont{{Castro Neto}}},
  \bibinfo{year}{2006}, \bibinfo{journal}{Phys. Rev. Lett.}
  \textbf{\bibinfo{volume}{96}}, \bibinfo{pages}{036801}.

\bibitem[{\citenamefont{Pereira}
  \emph{et~al.}(2007{\natexlab{a}})\citenamefont{Pereira, {Lopes dos Santos},
  and {Castro Neto}}}]{PLSCN07}
\bibinfo{author}{\bibnamefont{Pereira}, \bibfnamefont{V.~M.}},
  \bibinfo{author}{\bibfnamefont{J.~M.~B.} \bibnamefont{{Lopes dos Santos}}},
  and \bibinfo{author}{\bibfnamefont{A.~H.} \bibnamefont{{Castro Neto}}},
  \bibinfo{year}{2007}{\natexlab{a}}, \eprint{arXiv:0712.0806}.

\bibitem[{\citenamefont{Pereira}
  \emph{et~al.}(2007{\natexlab{b}})\citenamefont{Pereira, Nilsson, and {Castro
  Neto}}}]{PNCN07}
\bibinfo{author}{\bibnamefont{Pereira}, \bibfnamefont{V.~M.}},
  \bibinfo{author}{\bibfnamefont{J.}~\bibnamefont{Nilsson}}, and
  \bibinfo{author}{\bibfnamefont{A.~H.} \bibnamefont{{Castro Neto}}},
  \bibinfo{year}{2007}{\natexlab{b}}, \bibinfo{journal}{Phys. Rev. Lett.}
  \textbf{\bibinfo{volume}{99}}, \bibinfo{pages}{166802}.

\bibitem[{\citenamefont{Peres} \emph{et~al.}(2004)\citenamefont{Peres,
  Ara\'ujo, and Bozi}}]{PAD04}
\bibinfo{author}{\bibnamefont{Peres}, \bibfnamefont{N.~M.~R.}},
  \bibinfo{author}{\bibfnamefont{M.~A.~N.} \bibnamefont{Ara\'ujo}}, and
  \bibinfo{author}{\bibfnamefont{D.}~\bibnamefont{Bozi}}, \bibinfo{year}{2004},
  \bibinfo{journal}{Phys. Rev. B} \textbf{\bibinfo{volume}{70}},
  \bibinfo{pages}{195122}.

\bibitem[{\citenamefont{Peres and Castro}(2007)}]{PC07}
\bibinfo{author}{\bibnamefont{Peres}, \bibfnamefont{N.~M.~R.}}, and
  \bibinfo{author}{\bibfnamefont{E.~V.} \bibnamefont{Castro}},
  \bibinfo{year}{2007}, \bibinfo{journal}{Jou. Phys. Cond. Mat.}
  \textbf{\bibinfo{volume}{19}}, \bibinfo{pages}{406231}.

\bibitem[{\citenamefont{Peres}
  \emph{et~al.}(2006{\natexlab{a}})\citenamefont{Peres, {Castro Neto}, and
  Guinea}}]{nuno_cond}
\bibinfo{author}{\bibnamefont{Peres}, \bibfnamefont{N.~M.~R.}},
  \bibinfo{author}{\bibfnamefont{A.~H.} \bibnamefont{{Castro Neto}}}, and
  \bibinfo{author}{\bibfnamefont{F.}~\bibnamefont{Guinea}},
  \bibinfo{year}{2006}{\natexlab{a}}, \bibinfo{journal}{Phys. Rev. B}
  \textbf{\bibinfo{volume}{73}}, \bibinfo{pages}{195411}.

\bibitem[{\citenamefont{Peres}
  \emph{et~al.}(2006{\natexlab{b}})\citenamefont{Peres, {Castro Neto}, and
  Guinea}}]{nuno_confinement}
\bibinfo{author}{\bibnamefont{Peres}, \bibfnamefont{N.~M.~R.}},
  \bibinfo{author}{\bibfnamefont{A.~H.} \bibnamefont{{Castro Neto}}}, and
  \bibinfo{author}{\bibfnamefont{F.}~\bibnamefont{Guinea}},
  \bibinfo{year}{2006}{\natexlab{b}}, \bibinfo{journal}{Phys. Rev. B}
  \textbf{\bibinfo{volume}{73}}, \bibinfo{pages}{241403}.

\bibitem[{\citenamefont{Peres} \emph{et~al.}(2005)\citenamefont{Peres, Guinea,
  and {Castro Neto}}}]{PGN05b}
\bibinfo{author}{\bibnamefont{Peres}, \bibfnamefont{N.~M.~R.}},
  \bibinfo{author}{\bibfnamefont{F.}~\bibnamefont{Guinea}}, and
  \bibinfo{author}{\bibfnamefont{A.~H.} \bibnamefont{{Castro Neto}}},
  \bibinfo{year}{2005}, \bibinfo{journal}{Phys. Rev. B}
  \textbf{\bibinfo{volume}{72}}, \bibinfo{pages}{174406}.

\bibitem[{\citenamefont{Peres}
  \emph{et~al.}(2006{\natexlab{c}})\citenamefont{Peres, Guinea, and {Castro
  Neto}}}]{PGC06}
\bibinfo{author}{\bibnamefont{Peres}, \bibfnamefont{N.~M.~R.}},
  \bibinfo{author}{\bibfnamefont{F.}~\bibnamefont{Guinea}}, and
  \bibinfo{author}{\bibfnamefont{A.~H.} \bibnamefont{{Castro Neto}}},
  \bibinfo{year}{2006}{\natexlab{c}}, \bibinfo{journal}{Phys. Rev. B}
  \textbf{\bibinfo{volume}{73}}, \bibinfo{pages}{125411}.

\bibitem[{\citenamefont{Peres}
  \emph{et~al.}(2006{\natexlab{d}})\citenamefont{Peres, Guinea, and {Castro
  Neto}}}]{PGN06}
\bibinfo{author}{\bibnamefont{Peres}, \bibfnamefont{N.~M.~R.}},
  \bibinfo{author}{\bibfnamefont{F.}~\bibnamefont{Guinea}}, and
  \bibinfo{author}{\bibfnamefont{A.~H.} \bibnamefont{{Castro Neto}}},
  \bibinfo{year}{2006}{\natexlab{d}}, \bibinfo{journal}{Annals of Physics}
  \textbf{\bibinfo{volume}{321}}, \bibinfo{pages}{1559}.

\bibitem[{\citenamefont{Peres}
  \emph{et~al.}(2007{\natexlab{a}})\citenamefont{Peres, Klironomos, Tsai,
  Santos, {Lopes dos Santos}, and {Castro Neto}}}]{nit_bor}
\bibinfo{author}{\bibnamefont{Peres}, \bibfnamefont{N.~M.~R.}},
  \bibinfo{author}{\bibfnamefont{F.~D.} \bibnamefont{Klironomos}},
  \bibinfo{author}{\bibfnamefont{S.-W.} \bibnamefont{Tsai}},
  \bibinfo{author}{\bibfnamefont{J.~R.} \bibnamefont{Santos}},
  \bibinfo{author}{\bibfnamefont{J.~M.~B.} \bibnamefont{{Lopes dos Santos}}},
  and \bibinfo{author}{\bibfnamefont{A.~H.} \bibnamefont{{Castro Neto}}},
  \bibinfo{year}{2007}{\natexlab{a}}, \bibinfo{journal}{Europhys. Lett.}
  \textbf{\bibinfo{volume}{80}}, \bibinfo{pages}{67007}.

\bibitem[{\citenamefont{Peres}
  \emph{et~al.}(2007{\natexlab{b}})\citenamefont{Peres, {Lopes dos Santos}, and
  Stauber}}]{PLS07}
\bibinfo{author}{\bibnamefont{Peres}, \bibfnamefont{N.~M.~R.}},
  \bibinfo{author}{\bibfnamefont{J.~M.} \bibnamefont{{Lopes dos Santos}}}, and
  \bibinfo{author}{\bibfnamefont{T.}~\bibnamefont{Stauber}},
  \bibinfo{year}{2007}{\natexlab{b}}, \bibinfo{journal}{Phys. Rev. B}
  \textbf{\bibinfo{volume}{76}}, \bibinfo{pages}{073412}.

\bibitem[{\citenamefont{Petroski}(1989)}]{pencil}
\bibinfo{author}{\bibnamefont{Petroski}, \bibfnamefont{H.}},
  \bibinfo{year}{1989}, \emph{\bibinfo{title}{The Pencil: A History of Design
  and Circumstance}} (\bibinfo{publisher}{Alfred Knopf, New York}).

\bibitem[{\citenamefont{Phillips}(2006)}]{mottness}
\bibinfo{author}{\bibnamefont{Phillips}, \bibfnamefont{P.}},
  \bibinfo{year}{2006}, \bibinfo{journal}{Annals of Physics}
  \textbf{\bibinfo{volume}{321}}, \bibinfo{pages}{1634}.

\bibitem[{\citenamefont{Pisana} \emph{et~al.}(2007)\citenamefont{Pisana,
  Lazzeri, Casiraghi, Novoselov, Geim, Ferrari, and Mauri}}]{PLCNGFM07}
\bibinfo{author}{\bibnamefont{Pisana}, \bibfnamefont{S.}},
  \bibinfo{author}{\bibfnamefont{M.}~\bibnamefont{Lazzeri}},
  \bibinfo{author}{\bibfnamefont{C.}~\bibnamefont{Casiraghi}},
  \bibinfo{author}{\bibfnamefont{K.~S.} \bibnamefont{Novoselov}},
  \bibinfo{author}{\bibfnamefont{A.~K.} \bibnamefont{Geim}},
  \bibinfo{author}{\bibfnamefont{A.~C.} \bibnamefont{Ferrari}}, and
  \bibinfo{author}{\bibfnamefont{F.}~\bibnamefont{Mauri}},
  \bibinfo{year}{2007}, \bibinfo{journal}{Nature Materials}
  \textbf{\bibinfo{volume}{6}}, \bibinfo{pages}{198}.

\bibitem[{\citenamefont{Polini} \emph{et~al.}(2007)\citenamefont{Polini,
  Asgari, Barlas, Pereg-Barnea, and MacDonald}}]{PABPM07}
\bibinfo{author}{\bibnamefont{Polini}, \bibfnamefont{M.}},
  \bibinfo{author}{\bibfnamefont{R.}~\bibnamefont{Asgari}},
  \bibinfo{author}{\bibfnamefont{Y.}~\bibnamefont{Barlas}},
  \bibinfo{author}{\bibfnamefont{T.}~\bibnamefont{Pereg-Barnea}}, and
  \bibinfo{author}{\bibfnamefont{A.~H.} \bibnamefont{MacDonald}},
  \bibinfo{year}{2007}, \bibinfo{journal}{Solid State Commun.}
  \textbf{\bibinfo{volume}{143}}, \bibinfo{pages}{58}.

\bibitem[{\citenamefont{Polkovnikov}(2002)}]{anatoli02}
\bibinfo{author}{\bibnamefont{Polkovnikov}, \bibfnamefont{A.}},
  \bibinfo{year}{2002}, \bibinfo{journal}{Phys. Rev. B}
  \textbf{\bibinfo{volume}{65}}, \bibinfo{pages}{064503}.

\bibitem[{\citenamefont{Polkovnikov}
  \emph{et~al.}(2001)\citenamefont{Polkovnikov, Sachdev, and Vojta}}]{PSV01}
\bibinfo{author}{\bibnamefont{Polkovnikov}, \bibfnamefont{A.}},
  \bibinfo{author}{\bibfnamefont{S.}~\bibnamefont{Sachdev}}, and
  \bibinfo{author}{\bibfnamefont{M.}~\bibnamefont{Vojta}},
  \bibinfo{year}{2001}, \bibinfo{journal}{Phys. Rev. Lett.}
  \textbf{\bibinfo{volume}{86}}, \bibinfo{pages}{296}.

\bibitem[{\citenamefont{Radzihovsky and {Le Doussal}}(1992)}]{RlD92}
\bibinfo{author}{\bibnamefont{Radzihovsky}, \bibfnamefont{L.}}, and
  \bibinfo{author}{\bibfnamefont{P.}~\bibnamefont{{Le Doussal}}},
  \bibinfo{year}{1992}, \bibinfo{journal}{Phys. Rev. Lett.}
  \textbf{\bibinfo{volume}{69}}, \bibinfo{pages}{1209}.

\bibitem[{\citenamefont{Rammal}(1985)}]{rammal85}
\bibinfo{author}{\bibnamefont{Rammal}, \bibfnamefont{R.}},
  \bibinfo{year}{1985}, \bibinfo{journal}{J. Physique}
  \textbf{\bibinfo{volume}{46}}, \bibinfo{pages}{1345}.

\bibitem[{\citenamefont{Recher} \emph{et~al.}(2007)\citenamefont{Recher,
  Trauzettel, Blaner, Beenakker, and Morpurgo}}]{RTBBM07}
\bibinfo{author}{\bibnamefont{Recher}, \bibfnamefont{P.}},
  \bibinfo{author}{\bibfnamefont{B.}~\bibnamefont{Trauzettel}},
  \bibinfo{author}{\bibfnamefont{Y.~M.} \bibnamefont{Blaner}},
  \bibinfo{author}{\bibfnamefont{C.~W.~J.} \bibnamefont{Beenakker}}, and
  \bibinfo{author}{\bibfnamefont{A.~F.} \bibnamefont{Morpurgo}},
  \bibinfo{year}{2007}, \bibinfo{journal}{Phys. Rev. B}
  \textbf{\bibinfo{volume}{76}}, \bibinfo{pages}{235404}.

\bibitem[{\citenamefont{Reich} \emph{et~al.}(2002)\citenamefont{Reich,
  Maultzsch, Thomsen, and Ordej\'on}}]{Retal02}
\bibinfo{author}{\bibnamefont{Reich}, \bibfnamefont{S.}},
  \bibinfo{author}{\bibfnamefont{J.}~\bibnamefont{Maultzsch}},
  \bibinfo{author}{\bibfnamefont{C.}~\bibnamefont{Thomsen}}, and
  \bibinfo{author}{\bibfnamefont{P.}~\bibnamefont{Ordej\'on}},
  \bibinfo{year}{2002}, \bibinfo{journal}{Phys. Rev. B}
  \textbf{\bibinfo{volume}{66}}, \bibinfo{pages}{035412}.

\bibitem[{\citenamefont{Robinson and Schomerus}(2007)}]{RS06}
\bibinfo{author}{\bibnamefont{Robinson}, \bibfnamefont{J.~P.}}, and
  \bibinfo{author}{\bibfnamefont{H.}~\bibnamefont{Schomerus}},
  \bibinfo{year}{2007}, \bibinfo{journal}{Phys. Rev. B}
  \textbf{\bibinfo{volume}{76}}, \bibinfo{pages}{115430}.

\bibitem[{\citenamefont{Rollings} \emph{et~al.}(2005)\citenamefont{Rollings,
  Gweon, Zhou, Mun, McChesney, Hussain, Fedorov, First, de~Heer, and
  Lanzara}}]{Retal05}
\bibinfo{author}{\bibnamefont{Rollings}, \bibfnamefont{E.}},
  \bibinfo{author}{\bibfnamefont{G.-H.} \bibnamefont{Gweon}},
  \bibinfo{author}{\bibfnamefont{S.~Y.} \bibnamefont{Zhou}},
  \bibinfo{author}{\bibfnamefont{B.~S.} \bibnamefont{Mun}},
  \bibinfo{author}{\bibfnamefont{J.~L.} \bibnamefont{McChesney}},
  \bibinfo{author}{\bibfnamefont{B.~S.} \bibnamefont{Hussain}},
  \bibinfo{author}{\bibfnamefont{A.~V.} \bibnamefont{Fedorov}},
  \bibinfo{author}{\bibfnamefont{P.~N.} \bibnamefont{First}},
  \bibinfo{author}{\bibfnamefont{W.~A.} \bibnamefont{de~Heer}}, and
  \bibinfo{author}{\bibfnamefont{A.}~\bibnamefont{Lanzara}},
  \bibinfo{year}{2005}, \bibinfo{journal}{J. Phys. Chem. Sol.}
  \textbf{\bibinfo{volume}{67}}, \bibinfo{pages}{2172}.

\bibitem[{\citenamefont{Rong and Kuiper}(1993)}]{Rong93}
\bibinfo{author}{\bibnamefont{Rong}, \bibfnamefont{Z.~Y.}}, and
  \bibinfo{author}{\bibfnamefont{P.}~\bibnamefont{Kuiper}},
  \bibinfo{year}{1993}, \bibinfo{journal}{Phys. Rev. B}
  \textbf{\bibinfo{volume}{48}}, \bibinfo{pages}{17427}.

\bibitem[{\citenamefont{Rosenstein}
  \emph{et~al.}(1989)\citenamefont{Rosenstein, Warr, and Park}}]{RWP89}
\bibinfo{author}{\bibnamefont{Rosenstein}, \bibfnamefont{B.}},
  \bibinfo{author}{\bibfnamefont{B.~J.} \bibnamefont{Warr}}, and
  \bibinfo{author}{\bibfnamefont{S.~H.} \bibnamefont{Park}},
  \bibinfo{year}{1989}, \bibinfo{journal}{Phys. Rev. Lett.}
  \textbf{\bibinfo{volume}{62}}, \bibinfo{pages}{1433}.

\bibitem[{\citenamefont{Rosenstein}
  \emph{et~al.}(1991)\citenamefont{Rosenstein, Warr, and Park}}]{RWP91}
\bibinfo{author}{\bibnamefont{Rosenstein}, \bibfnamefont{B.}},
  \bibinfo{author}{\bibfnamefont{B.~J.} \bibnamefont{Warr}}, and
  \bibinfo{author}{\bibfnamefont{S.~H.} \bibnamefont{Park}},
  \bibinfo{year}{1991}, \bibinfo{journal}{Phys. Rep.}
  \textbf{\bibinfo{volume}{205}}, \bibinfo{pages}{59}.

\bibitem[{\citenamefont{Russo} \emph{et~al.}(2007)\citenamefont{Russo,
  Oostinga, Wehenkel, Heersche, Sobhani, Vandersypen, and
  Morpurgo}}]{ROWHSVM07}
\bibinfo{author}{\bibnamefont{Russo}, \bibfnamefont{S.}},
  \bibinfo{author}{\bibfnamefont{J.~B.} \bibnamefont{Oostinga}},
  \bibinfo{author}{\bibfnamefont{D.}~\bibnamefont{Wehenkel}},
  \bibinfo{author}{\bibfnamefont{H.~B.} \bibnamefont{Heersche}},
  \bibinfo{author}{\bibfnamefont{S.~S.} \bibnamefont{Sobhani}},
  \bibinfo{author}{\bibfnamefont{L.~M.~K.} \bibnamefont{Vandersypen}}, and
  \bibinfo{author}{\bibfnamefont{A.~F.} \bibnamefont{Morpurgo}},
  \bibinfo{year}{2007}, \eprint{arXiv:0711.1508}.

\bibitem[{\citenamefont{Rutter} \emph{et~al.}(2007)\citenamefont{Rutter, Crain,
  Guisinger, Li, First, and Stroscio}}]{RCGLFS07}
\bibinfo{author}{\bibnamefont{Rutter}, \bibfnamefont{G.~M.}},
  \bibinfo{author}{\bibfnamefont{J.~N.} \bibnamefont{Crain}},
  \bibinfo{author}{\bibfnamefont{N.~P.} \bibnamefont{Guisinger}},
  \bibinfo{author}{\bibfnamefont{T.}~\bibnamefont{Li}},
  \bibinfo{author}{\bibfnamefont{P.~N.} \bibnamefont{First}}, and
  \bibinfo{author}{\bibfnamefont{J.~A.} \bibnamefont{Stroscio}},
  \bibinfo{year}{2007}, \bibinfo{journal}{Science}
  \textbf{\bibinfo{volume}{317}}, \bibinfo{pages}{219}.

\bibitem[{\citenamefont{Rycerz} \emph{et~al.}(2007)\citenamefont{Rycerz,
  Tworzydlo, and Beenakker}}]{valleytronics}
\bibinfo{author}{\bibnamefont{Rycerz}, \bibfnamefont{A.}},
  \bibinfo{author}{\bibfnamefont{J.}~\bibnamefont{Tworzydlo}}, and
  \bibinfo{author}{\bibfnamefont{C.~W.~J.} \bibnamefont{Beenakker}},
  \bibinfo{year}{2007}, \bibinfo{journal}{Nature Physics}
  \textbf{\bibinfo{volume}{3}}, \bibinfo{pages}{172}.

\bibitem[{\citenamefont{Rydberg} \emph{et~al.}(2003)\citenamefont{Rydberg,
  Dion, Jacobson, Schr\"oder, Hyldgaard, Simak, Langreth, and
  Lundqvist}}]{langreth}
\bibinfo{author}{\bibnamefont{Rydberg}, \bibfnamefont{H.}},
  \bibinfo{author}{\bibfnamefont{M.}~\bibnamefont{Dion}},
  \bibinfo{author}{\bibfnamefont{N.}~\bibnamefont{Jacobson}},
  \bibinfo{author}{\bibfnamefont{E.}~\bibnamefont{Schr\"oder}},
  \bibinfo{author}{\bibfnamefont{P.}~\bibnamefont{Hyldgaard}},
  \bibinfo{author}{\bibfnamefont{S.~I.} \bibnamefont{Simak}},
  \bibinfo{author}{\bibfnamefont{D.~C.} \bibnamefont{Langreth}}, and
  \bibinfo{author}{\bibfnamefont{B.~I.} \bibnamefont{Lundqvist}},
  \bibinfo{year}{2003}, \bibinfo{journal}{Phys. Rev. Lett.}
  \textbf{\bibinfo{volume}{91}}, \bibinfo{pages}{126402}.

\bibitem[{\citenamefont{Ryu} \emph{et~al.}(2007)\citenamefont{Ryu, Mudry,
  Obuse, and Furusaki}}]{RMOF07}
\bibinfo{author}{\bibnamefont{Ryu}, \bibfnamefont{S.}},
  \bibinfo{author}{\bibfnamefont{C.}~\bibnamefont{Mudry}},
  \bibinfo{author}{\bibfnamefont{H.}~\bibnamefont{Obuse}}, and
  \bibinfo{author}{\bibfnamefont{A.}~\bibnamefont{Furusaki}},
  \bibinfo{year}{2007}, \bibinfo{journal}{Phys. Rev. Lett.}
  \textbf{\bibinfo{volume}{99}}, \bibinfo{pages}{116601}.

\bibitem[{\citenamefont{Sabio} \emph{et~al.}(2007)\citenamefont{Sabio, Seoanez,
  Fratini, Guinea, {Castro Neto}, and Sols}}]{SSFGCNS07}
\bibinfo{author}{\bibnamefont{Sabio}, \bibfnamefont{J.}},
  \bibinfo{author}{\bibfnamefont{C.}~\bibnamefont{Seoanez}},
  \bibinfo{author}{\bibfnamefont{S.}~\bibnamefont{Fratini}},
  \bibinfo{author}{\bibfnamefont{F.}~\bibnamefont{Guinea}},
  \bibinfo{author}{\bibfnamefont{A.~H.} \bibnamefont{{Castro Neto}}}, and
  \bibinfo{author}{\bibfnamefont{F.}~\bibnamefont{Sols}}, \bibinfo{year}{2007},
  \eprint{arXiv:0712.222}.

\bibitem[{\citenamefont{Sadowski} \emph{et~al.}(2006)\citenamefont{Sadowski,
  Martinez, Potemski, Berger, and de~Heer}}]{SMPBH06}
\bibinfo{author}{\bibnamefont{Sadowski}, \bibfnamefont{M.~L.}},
  \bibinfo{author}{\bibfnamefont{G.}~\bibnamefont{Martinez}},
  \bibinfo{author}{\bibfnamefont{M.}~\bibnamefont{Potemski}},
  \bibinfo{author}{\bibfnamefont{C.}~\bibnamefont{Berger}}, and
  \bibinfo{author}{\bibfnamefont{W.~A.} \bibnamefont{de~Heer}},
  \bibinfo{year}{2006}, \bibinfo{journal}{Phys. Rev. Lett.}
  \textbf{\bibinfo{volume}{97}}, \bibinfo{pages}{266405}.

\bibitem[{\citenamefont{Safran}(1984)}]{S1984}
\bibinfo{author}{\bibnamefont{Safran}, \bibfnamefont{S.~A.}},
  \bibinfo{year}{1984}, \bibinfo{journal}{Phys. Rev. B}
  \textbf{\bibinfo{volume}{30}}, \bibinfo{pages}{421}.

\bibitem[{\citenamefont{Safran and DiSalvo}(1979)}]{SD1979}
\bibinfo{author}{\bibnamefont{Safran}, \bibfnamefont{S.~A.}}, and
  \bibinfo{author}{\bibfnamefont{F.~J.} \bibnamefont{DiSalvo}},
  \bibinfo{year}{1979}, \bibinfo{journal}{Phys. Rev. B}
  \textbf{\bibinfo{volume}{20}}, \bibinfo{pages}{4889}.

\bibitem[{\citenamefont{Saha} \emph{et~al.}(2007)\citenamefont{Saha, Waghmare,
  Krishnamurth, and Sood}}]{SWKS07}
\bibinfo{author}{\bibnamefont{Saha}, \bibfnamefont{S.~K.}},
  \bibinfo{author}{\bibfnamefont{U.~V.} \bibnamefont{Waghmare}},
  \bibinfo{author}{\bibfnamefont{H.~R.} \bibnamefont{Krishnamurth}}, and
  \bibinfo{author}{\bibfnamefont{A.~K.} \bibnamefont{Sood}},
  \bibinfo{year}{2007}, \eprint{cond-mat/0702627}.

\bibitem[{\citenamefont{Saito} \emph{et~al.}(1998)\citenamefont{Saito,
  Dresselhaus, and Dresselhaus}}]{nanotubes}
\bibinfo{author}{\bibnamefont{Saito}, \bibfnamefont{R.}},
  \bibinfo{author}{\bibfnamefont{G.}~\bibnamefont{Dresselhaus}}, and
  \bibinfo{author}{\bibfnamefont{M.~S.} \bibnamefont{Dresselhaus}},
  \bibinfo{year}{1998}, \emph{\bibinfo{title}{Physical properties of carbon
  nanotubes}} (\bibinfo{publisher}{Imperial College Press, London}).

\bibitem[{\citenamefont{Saito}
  \emph{et~al.}(1992{\natexlab{a}})\citenamefont{Saito, Fujita, Dresselhaus,
  and Dresselhaus}}]{SFDD092b}
\bibinfo{author}{\bibnamefont{Saito}, \bibfnamefont{R.}},
  \bibinfo{author}{\bibfnamefont{M.}~\bibnamefont{Fujita}},
  \bibinfo{author}{\bibfnamefont{G.}~\bibnamefont{Dresselhaus}}, and
  \bibinfo{author}{\bibfnamefont{M.~S.} \bibnamefont{Dresselhaus}},
  \bibinfo{year}{1992}{\natexlab{a}}, \bibinfo{journal}{Appl. Phys. Lett.}
  \textbf{\bibinfo{volume}{60}}, \bibinfo{pages}{2204}.

\bibitem[{\citenamefont{Saito}
  \emph{et~al.}(1992{\natexlab{b}})\citenamefont{Saito, Fujita, Dresselhaus,
  and Dresselhaus}}]{SFDD092a}
\bibinfo{author}{\bibnamefont{Saito}, \bibfnamefont{R.}},
  \bibinfo{author}{\bibfnamefont{M.}~\bibnamefont{Fujita}},
  \bibinfo{author}{\bibfnamefont{G.}~\bibnamefont{Dresselhaus}}, and
  \bibinfo{author}{\bibfnamefont{M.~S.} \bibnamefont{Dresselhaus}},
  \bibinfo{year}{1992}{\natexlab{b}}, \bibinfo{journal}{Phys. Rev. B}
  \textbf{\bibinfo{volume}{46}}, \bibinfo{pages}{1804}.

\bibitem[{\citenamefont{San-Jose} \emph{et~al.}(2007)\citenamefont{San-Jose,
  Prada, and Golubev}}]{SPG07}
\bibinfo{author}{\bibnamefont{San-Jose}, \bibfnamefont{P.}},
  \bibinfo{author}{\bibfnamefont{E.}~\bibnamefont{Prada}}, and
  \bibinfo{author}{\bibfnamefont{D.}~\bibnamefont{Golubev}},
  \bibinfo{year}{2007}, \bibinfo{journal}{Phys. Rev. B}
  \textbf{\bibinfo{volume}{76}}, \bibinfo{pages}{195445}.

\bibitem[{\citenamefont{Saremi}(2007)}]{S07}
\bibinfo{author}{\bibnamefont{Saremi}, \bibfnamefont{S.}},
  \bibinfo{year}{2007}, \bibinfo{journal}{Phys. Rev. B}
  \textbf{\bibinfo{volume}{76}}, \bibinfo{pages}{184430}.

\bibitem[{\citenamefont{Sarma} \emph{et~al.}(2007)\citenamefont{Sarma, Hwang,
  and Tse}}]{SHT07}
\bibinfo{author}{\bibnamefont{Sarma}, \bibfnamefont{S.~D.}},
  \bibinfo{author}{\bibfnamefont{E.~H.} \bibnamefont{Hwang}}, and
  \bibinfo{author}{\bibfnamefont{W.~K.} \bibnamefont{Tse}},
  \bibinfo{year}{2007}, \bibinfo{journal}{Phys. Rev. B}
  \textbf{\bibinfo{volume}{75}}, \bibinfo{pages}{121406}.

\bibitem[{\citenamefont{Schakel}(1991)}]{Sch91}
\bibinfo{author}{\bibnamefont{Schakel}, \bibfnamefont{A.~M.~J.}},
  \bibinfo{year}{1991}, \bibinfo{journal}{Phys. Rev. D}
  \textbf{\bibinfo{volume}{43}}, \bibinfo{pages}{1428}.

\bibitem[{\citenamefont{Schedin} \emph{et~al.}(2007)\citenamefont{Schedin,
  Geim, Morozov, Jiang, Hill, Blake, and Novoselov}}]{Setal07}
\bibinfo{author}{\bibnamefont{Schedin}, \bibfnamefont{F.}},
  \bibinfo{author}{\bibfnamefont{A.~K.} \bibnamefont{Geim}},
  \bibinfo{author}{\bibfnamefont{S.~V.} \bibnamefont{Morozov}},
  \bibinfo{author}{\bibfnamefont{D.}~\bibnamefont{Jiang}},
  \bibinfo{author}{\bibfnamefont{E.~H.} \bibnamefont{Hill}},
  \bibinfo{author}{\bibfnamefont{P.}~\bibnamefont{Blake}}, and
  \bibinfo{author}{\bibfnamefont{K.~S.} \bibnamefont{Novoselov}},
  \bibinfo{year}{2007}, \bibinfo{journal}{Nature Materials}
  \textbf{\bibinfo{volume}{6}}, \bibinfo{pages}{652}.

\bibitem[{\citenamefont{Schomerus}(2007)}]{S06}
\bibinfo{author}{\bibnamefont{Schomerus}, \bibfnamefont{H.}},
  \bibinfo{year}{2007}, \bibinfo{journal}{Phys. Rev. B}
  \textbf{\bibinfo{volume}{76}}, \bibinfo{pages}{045433}.

\bibitem[{\citenamefont{Schroeder} \emph{et~al.}(1968)\citenamefont{Schroeder,
  Dresselhaus, and Javan}}]{SDJ68}
\bibinfo{author}{\bibnamefont{Schroeder}, \bibfnamefont{P.~R.}},
  \bibinfo{author}{\bibfnamefont{M.~S.} \bibnamefont{Dresselhaus}}, and
  \bibinfo{author}{\bibfnamefont{A.}~\bibnamefont{Javan}},
  \bibinfo{year}{1968}, \bibinfo{journal}{Phys. Rev. Lett.}
  \textbf{\bibinfo{volume}{20}}, \bibinfo{pages}{1292}.

\bibitem[{\citenamefont{Semenoff}(1984)}]{semenoff84}
\bibinfo{author}{\bibnamefont{Semenoff}, \bibfnamefont{G.~W.}},
  \bibinfo{year}{1984}, \bibinfo{journal}{Phys. Rev. Lett.}
  \textbf{\bibinfo{volume}{53}}, \bibinfo{pages}{2449}.

\bibitem[{\citenamefont{Sengupta and Baskaran}(2007)}]{SB07}
\bibinfo{author}{\bibnamefont{Sengupta}, \bibfnamefont{K.}}, and
  \bibinfo{author}{\bibfnamefont{G.}~\bibnamefont{Baskaran}},
  \bibinfo{year}{2007}, \bibinfo{journal}{Phys. Rev. B}
  \textbf{\bibinfo{volume}{77}}, \bibinfo{pages}{045417}.

\bibitem[{\citenamefont{Seoanez} \emph{et~al.}(2007)\citenamefont{Seoanez,
  Guinea, and {Castro Neto}}}]{SGCN07}
\bibinfo{author}{\bibnamefont{Seoanez}, \bibfnamefont{C.}},
  \bibinfo{author}{\bibfnamefont{F.}~\bibnamefont{Guinea}}, and
  \bibinfo{author}{\bibfnamefont{A.~H.} \bibnamefont{{Castro Neto}}},
  \bibinfo{year}{2007}, \bibinfo{journal}{Phys. Rev. B}
  \textbf{\bibinfo{volume}{76}}, \bibinfo{pages}{125427}.

\bibitem[{\citenamefont{Shankar}(1994)}]{S94}
\bibinfo{author}{\bibnamefont{Shankar}, \bibfnamefont{R.}},
  \bibinfo{year}{1994}, \bibinfo{journal}{Rev. Mod. Phys.}
  \textbf{\bibinfo{volume}{66}}, \bibinfo{pages}{129}.

\bibitem[{\citenamefont{Sharma} \emph{et~al.}(1974)\citenamefont{Sharma,
  Johnson, and McClure}}]{SJM1974}
\bibinfo{author}{\bibnamefont{Sharma}, \bibfnamefont{M.~P.}},
  \bibinfo{author}{\bibfnamefont{L.~G.} \bibnamefont{Johnson}}, and
  \bibinfo{author}{\bibfnamefont{J.~W.} \bibnamefont{McClure}},
  \bibinfo{year}{1974}, \bibinfo{journal}{Phys. Rev. B}
  \textbf{\bibinfo{volume}{9}}, \bibinfo{pages}{2467}.

\bibitem[{\citenamefont{Shelton} \emph{et~al.}(1974)\citenamefont{Shelton,
  Patil, and Blakely}}]{SPB74}
\bibinfo{author}{\bibnamefont{Shelton}, \bibfnamefont{J.~C.}},
  \bibinfo{author}{\bibfnamefont{H.~R.} \bibnamefont{Patil}}, and
  \bibinfo{author}{\bibfnamefont{J.~M.} \bibnamefont{Blakely}},
  \bibinfo{year}{1974}, \bibinfo{journal}{Surf. Sci.}
  \textbf{\bibinfo{volume}{43}}, \bibinfo{pages}{493}.

\bibitem[{\citenamefont{Sheng} \emph{et~al.}(2006)\citenamefont{Sheng, Sheng,
  and Wen}}]{SSW06}
\bibinfo{author}{\bibnamefont{Sheng}, \bibfnamefont{D.~N.}},
  \bibinfo{author}{\bibfnamefont{L.}~\bibnamefont{Sheng}}, and
  \bibinfo{author}{\bibfnamefont{Z.~Y.} \bibnamefont{Wen}},
  \bibinfo{year}{2006}, \bibinfo{journal}{Phys. Rev. B}
  \textbf{\bibinfo{volume}{73}}, \bibinfo{pages}{2333406}.

\bibitem[{\citenamefont{Sheng} \emph{et~al.}(2007)\citenamefont{Sheng, Sheng,
  Haldane, and Balents}}]{SSHB07}
\bibinfo{author}{\bibnamefont{Sheng}, \bibfnamefont{L.}},
  \bibinfo{author}{\bibfnamefont{D.~N.} \bibnamefont{Sheng}},
  \bibinfo{author}{\bibfnamefont{F.~D.~M.} \bibnamefont{Haldane}}, and
  \bibinfo{author}{\bibfnamefont{L.}~\bibnamefont{Balents}},
  \bibinfo{year}{2007}, \bibinfo{journal}{Phys. Rev. Lett.}
  \textbf{\bibinfo{volume}{99}}, \bibinfo{pages}{196802}.

\bibitem[{\citenamefont{Shklovskii}(2007)}]{Shk07}
\bibinfo{author}{\bibnamefont{Shklovskii}, \bibfnamefont{B.~I.}},
  \bibinfo{year}{2007}, \eprint{arXiv:0706.4425}.

\bibitem[{\citenamefont{Shon and Ando}(1998)}]{Ando7}
\bibinfo{author}{\bibnamefont{Shon}, \bibfnamefont{N.~H.}}, and
  \bibinfo{author}{\bibfnamefont{T.}~\bibnamefont{Ando}}, \bibinfo{year}{1998},
  \bibinfo{journal}{J. Phys. Soc. Jpn.} \textbf{\bibinfo{volume}{67}},
  \bibinfo{pages}{2421}.

\bibitem[{\citenamefont{Shung}(1986{\natexlab{a}})}]{S86}
\bibinfo{author}{\bibnamefont{Shung}, \bibfnamefont{K.~W.~K.}},
  \bibinfo{year}{1986}{\natexlab{a}}, \bibinfo{journal}{Phys. Rev. B}
  \textbf{\bibinfo{volume}{34}}, \bibinfo{pages}{979}.

\bibitem[{\citenamefont{Shung}(1986{\natexlab{b}})}]{S86b}
\bibinfo{author}{\bibnamefont{Shung}, \bibfnamefont{K.~W.~K.}},
  \bibinfo{year}{1986}{\natexlab{b}}, \bibinfo{journal}{Phys. Rev. B}
  \textbf{\bibinfo{volume}{34}}, \bibinfo{pages}{1264}.

\bibitem[{\citenamefont{Shytov} \emph{et~al.}(2007)\citenamefont{Shytov,
  Katsnelson, and Levitov}}]{SKL07}
\bibinfo{author}{\bibnamefont{Shytov}, \bibfnamefont{A.~V.}},
  \bibinfo{author}{\bibfnamefont{M.~I.} \bibnamefont{Katsnelson}}, and
  \bibinfo{author}{\bibfnamefont{L.~S.} \bibnamefont{Levitov}},
  \bibinfo{year}{2007}, \bibinfo{journal}{Phys. Rev. Lett.}
  \textbf{\bibinfo{volume}{99}}, \bibinfo{pages}{236801}.

\bibitem[{\citenamefont{Silvestrov and Efetov}(2007)}]{SE07}
\bibinfo{author}{\bibnamefont{Silvestrov}, \bibfnamefont{P.~G.}}, and
  \bibinfo{author}{\bibfnamefont{K.~B.} \bibnamefont{Efetov}},
  \bibinfo{year}{2007}, \bibinfo{journal}{Phys. Rev. Lett.}
  \textbf{\bibinfo{volume}{98}}, \bibinfo{pages}{016802}.

\bibitem[{\citenamefont{Sinitsyna and Yaminsky}(2006)}]{SY06}
\bibinfo{author}{\bibnamefont{Sinitsyna}, \bibfnamefont{O.~V.}}, and
  \bibinfo{author}{\bibfnamefont{I.~V.} \bibnamefont{Yaminsky}},
  \bibinfo{year}{2006}, \bibinfo{journal}{Russ. Chem. Rev.}
  \textbf{\bibinfo{volume}{75}}, \bibinfo{pages}{23}.

\bibitem[{\citenamefont{Skrypnyk and Loktev}(2006)}]{SL06}
\bibinfo{author}{\bibnamefont{Skrypnyk}, \bibfnamefont{Y.~V.}}, and
  \bibinfo{author}{\bibfnamefont{V.~M.} \bibnamefont{Loktev}},
  \bibinfo{year}{2006}, \bibinfo{journal}{Phys. Rev. B}
  \textbf{\bibinfo{volume}{73}}, \bibinfo{pages}{241402 (R)}.

\bibitem[{\citenamefont{Skrypnyk and Loktev}(2007)}]{SL07}
\bibinfo{author}{\bibnamefont{Skrypnyk}, \bibfnamefont{Y.~V.}}, and
  \bibinfo{author}{\bibfnamefont{V.~M.} \bibnamefont{Loktev}},
  \bibinfo{year}{2007}, \bibinfo{journal}{Phys. Rev. B}
  \textbf{\bibinfo{volume}{75}}, \bibinfo{pages}{245401}.

\bibitem[{\citenamefont{Slonczewski and Weiss}(1958)}]{SW58}
\bibinfo{author}{\bibnamefont{Slonczewski}, \bibfnamefont{J.~C.}}, and
  \bibinfo{author}{\bibfnamefont{P.~R.} \bibnamefont{Weiss}},
  \bibinfo{year}{1958}, \bibinfo{journal}{Phys. Rev.}
  \textbf{\bibinfo{volume}{109}}, \bibinfo{pages}{272}.

\bibitem[{\citenamefont{Snyman and Beenakker}(2007)}]{SB07b}
\bibinfo{author}{\bibnamefont{Snyman}, \bibfnamefont{I.}}, and
  \bibinfo{author}{\bibfnamefont{C.~W.~J.} \bibnamefont{Beenakker}},
  \bibinfo{year}{2007}, \bibinfo{journal}{Phys. Rev. B}
  \textbf{\bibinfo{volume}{75}}, \bibinfo{pages}{045322}.

\bibitem[{\citenamefont{Sols} \emph{et~al.}(2007)\citenamefont{Sols, Guinea,
  and {Castro Neto}}}]{SGN07b}
\bibinfo{author}{\bibnamefont{Sols}, \bibfnamefont{F.}},
  \bibinfo{author}{\bibfnamefont{F.}~\bibnamefont{Guinea}}, and
  \bibinfo{author}{\bibfnamefont{A.~H.} \bibnamefont{{Castro Neto}}},
  \bibinfo{year}{2007}, \bibinfo{journal}{Phys. Rev. Lett.}
  \textbf{\bibinfo{volume}{99}}, \bibinfo{pages}{166803}.

\bibitem[{\citenamefont{Son}(2007)}]{DTS07}
\bibinfo{author}{\bibnamefont{Son}, \bibfnamefont{D.~T.}},
  \bibinfo{year}{2007}, \bibinfo{journal}{Phys. Rev. B}
  \textbf{\bibinfo{volume}{75}}, \bibinfo{pages}{235423}.

\bibitem[{\citenamefont{Son}
  \emph{et~al.}(2006{\natexlab{a}})\citenamefont{Son, Cohen, and
  Louie}}]{SCL06b}
\bibinfo{author}{\bibnamefont{Son}, \bibfnamefont{Y.-W.}},
  \bibinfo{author}{\bibfnamefont{M.~L.} \bibnamefont{Cohen}}, and
  \bibinfo{author}{\bibfnamefont{S.~G.} \bibnamefont{Louie}},
  \bibinfo{year}{2006}{\natexlab{a}}, \bibinfo{journal}{Phys. Rev. Lett.}
  \textbf{\bibinfo{volume}{97}}, \bibinfo{pages}{216803}.

\bibitem[{\citenamefont{Son}
  \emph{et~al.}(2006{\natexlab{b}})\citenamefont{Son, Cohen, and
  Louie}}]{SCL06a}
\bibinfo{author}{\bibnamefont{Son}, \bibfnamefont{Y.-W.}},
  \bibinfo{author}{\bibfnamefont{M.~L.} \bibnamefont{Cohen}}, and
  \bibinfo{author}{\bibfnamefont{S.~G.} \bibnamefont{Louie}},
  \bibinfo{year}{2006}{\natexlab{b}}, \bibinfo{journal}{Nature}
  \textbf{\bibinfo{volume}{444}}, \bibinfo{pages}{347}.

\bibitem[{\citenamefont{Sorella and Tosatti}(1992)}]{sorella92}
\bibinfo{author}{\bibnamefont{Sorella}, \bibfnamefont{S.}}, and
  \bibinfo{author}{\bibfnamefont{E.}~\bibnamefont{Tosatti}},
  \bibinfo{year}{1992}, \bibinfo{journal}{Europhys. Lett.}
  \textbf{\bibinfo{volume}{19}}, \bibinfo{pages}{699}.

\bibitem[{\citenamefont{Soule} \emph{et~al.}(1964)\citenamefont{Soule, McClure,
  and Smith}}]{SMS64}
\bibinfo{author}{\bibnamefont{Soule}, \bibfnamefont{D.~E.}},
  \bibinfo{author}{\bibfnamefont{J.~W.} \bibnamefont{McClure}}, and
  \bibinfo{author}{\bibfnamefont{L.~B.} \bibnamefont{Smith}},
  \bibinfo{year}{1964}, \bibinfo{journal}{Phys. Rev.}
  \textbf{\bibinfo{volume}{134}}, \bibinfo{pages}{A453}.

\bibitem[{\citenamefont{Spataru} \emph{et~al.}(2001)\citenamefont{Spataru,
  Cazalilla, Rubio, Benedict, Echenique, and Louie}}]{SCABEL01}
\bibinfo{author}{\bibnamefont{Spataru}, \bibfnamefont{C.~D.}},
  \bibinfo{author}{\bibfnamefont{M.~A.} \bibnamefont{Cazalilla}},
  \bibinfo{author}{\bibfnamefont{A.}~\bibnamefont{Rubio}},
  \bibinfo{author}{\bibfnamefont{L.~X.} \bibnamefont{Benedict}},
  \bibinfo{author}{\bibfnamefont{P.~M.} \bibnamefont{Echenique}}, and
  \bibinfo{author}{\bibfnamefont{S.~G.} \bibnamefont{Louie}},
  \bibinfo{year}{2001}, \bibinfo{journal}{Phys. Rev. Lett.}
  \textbf{\bibinfo{volume}{87}}, \bibinfo{pages}{246405}.

\bibitem[{\citenamefont{Spry and Scherer}(1960)}]{SS60}
\bibinfo{author}{\bibnamefont{Spry}, \bibfnamefont{W.~J.}}, and
  \bibinfo{author}{\bibfnamefont{P.~M.} \bibnamefont{Scherer}},
  \bibinfo{year}{1960}, \bibinfo{journal}{Phys. Rev.}
  \textbf{\bibinfo{volume}{120}}, \bibinfo{pages}{826}.

\bibitem[{\citenamefont{Stauber} \emph{et~al.}(2005)\citenamefont{Stauber,
  Guinea, and Vozmediano}}]{SGV05}
\bibinfo{author}{\bibnamefont{Stauber}, \bibfnamefont{T.}},
  \bibinfo{author}{\bibfnamefont{F.}~\bibnamefont{Guinea}}, and
  \bibinfo{author}{\bibfnamefont{M.~A.~H.} \bibnamefont{Vozmediano}},
  \bibinfo{year}{2005}, \bibinfo{journal}{Phys. Rev. B}
  \textbf{\bibinfo{volume}{71}}, \bibinfo{pages}{041406}.

\bibitem[{\citenamefont{Stauber} \emph{et~al.}(2007)\citenamefont{Stauber,
  Peres, Guinea, and {Castro Neto}}}]{SPGN07}
\bibinfo{author}{\bibnamefont{Stauber}, \bibfnamefont{T.}},
  \bibinfo{author}{\bibfnamefont{N.~M.~R.} \bibnamefont{Peres}},
  \bibinfo{author}{\bibfnamefont{F.}~\bibnamefont{Guinea}}, and
  \bibinfo{author}{\bibfnamefont{A.~H.} \bibnamefont{{Castro Neto}}},
  \bibinfo{year}{2007}, \bibinfo{journal}{Phys. Rev. B}
  \textbf{\bibinfo{volume}{75}}, \bibinfo{pages}{115425}.

\bibitem[{\citenamefont{Stephan} \emph{et~al.}(1994)\citenamefont{Stephan,
  Ajayan, Colliex, Redlich, Lambert, Bernier, and Lefin}}]{nano_boron}
\bibinfo{author}{\bibnamefont{Stephan}, \bibfnamefont{O.}},
  \bibinfo{author}{\bibfnamefont{P.~M.} \bibnamefont{Ajayan}},
  \bibinfo{author}{\bibfnamefont{C.}~\bibnamefont{Colliex}},
  \bibinfo{author}{\bibfnamefont{P.}~\bibnamefont{Redlich}},
  \bibinfo{author}{\bibfnamefont{J.~M.} \bibnamefont{Lambert}},
  \bibinfo{author}{\bibfnamefont{P.}~\bibnamefont{Bernier}}, and
  \bibinfo{author}{\bibfnamefont{P.}~\bibnamefont{Lefin}},
  \bibinfo{year}{1994}, \bibinfo{journal}{Science}
  \textbf{\bibinfo{volume}{266}}, \bibinfo{pages}{1683}.

\bibitem[{\citenamefont{Stolyarova}
  \emph{et~al.}(2007)\citenamefont{Stolyarova, Rim, Ryu, Maultzsch, Kim, Brus,
  Heinz, Hybertsen, and Flynn}}]{SRRMKBHHF07}
\bibinfo{author}{\bibnamefont{Stolyarova}, \bibfnamefont{E.}},
  \bibinfo{author}{\bibfnamefont{K.~T.} \bibnamefont{Rim}},
  \bibinfo{author}{\bibfnamefont{S.}~\bibnamefont{Ryu}},
  \bibinfo{author}{\bibfnamefont{J.}~\bibnamefont{Maultzsch}},
  \bibinfo{author}{\bibfnamefont{P.}~\bibnamefont{Kim}},
  \bibinfo{author}{\bibfnamefont{L.~E.} \bibnamefont{Brus}},
  \bibinfo{author}{\bibfnamefont{T.~F.} \bibnamefont{Heinz}},
  \bibinfo{author}{\bibfnamefont{M.~S.} \bibnamefont{Hybertsen}}, and
  \bibinfo{author}{\bibfnamefont{G.~W.} \bibnamefont{Flynn}},
  \bibinfo{year}{2007}, \bibinfo{journal}{Proc. Natl. Acad. Sci. USA}
  \textbf{\bibinfo{volume}{104}}, \bibinfo{pages}{9209}.

\bibitem[{\citenamefont{Stone}(1992)}]{iqhe_si}
\bibinfo{author}{\bibnamefont{Stone}, \bibfnamefont{M.}}, \bibinfo{year}{1992},
  \emph{\bibinfo{title}{Quantum Hall Effect}} (\bibinfo{publisher}{World
  Scientific, Singapore}).

\bibitem[{\citenamefont{Su} \emph{et~al.}(1979)\citenamefont{Su, Schrieffer,
  and Heeger}}]{SSH79_PRL}
\bibinfo{author}{\bibnamefont{Su}, \bibfnamefont{W.~P.}},
  \bibinfo{author}{\bibfnamefont{J.~R.} \bibnamefont{Schrieffer}}, and
  \bibinfo{author}{\bibfnamefont{A.~J.} \bibnamefont{Heeger}},
  \bibinfo{year}{1979}, \bibinfo{journal}{Phys. Rev. Lett.}
  \textbf{\bibinfo{volume}{42}}, \bibinfo{pages}{1698}.

\bibitem[{\citenamefont{Su} \emph{et~al.}(1980)\citenamefont{Su, Schrieffer,
  and Heeger}}]{SSH79_PRB}
\bibinfo{author}{\bibnamefont{Su}, \bibfnamefont{W.~P.}},
  \bibinfo{author}{\bibfnamefont{J.~R.} \bibnamefont{Schrieffer}}, and
  \bibinfo{author}{\bibfnamefont{A.~J.} \bibnamefont{Heeger}},
  \bibinfo{year}{1980}, \bibinfo{journal}{Phys. Rev. B}
  \textbf{\bibinfo{volume}{22}}, \bibinfo{pages}{2099}.

\bibitem[{\citenamefont{Sugawara} \emph{et~al.}(2007)\citenamefont{Sugawara,
  Sato, Souma, Takahashi, and Suematsu}}]{SSSTS07}
\bibinfo{author}{\bibnamefont{Sugawara}, \bibfnamefont{K.}},
  \bibinfo{author}{\bibfnamefont{T.}~\bibnamefont{Sato}},
  \bibinfo{author}{\bibfnamefont{S.}~\bibnamefont{Souma}},
  \bibinfo{author}{\bibfnamefont{T.}~\bibnamefont{Takahashi}}, and
  \bibinfo{author}{\bibfnamefont{H.}~\bibnamefont{Suematsu}},
  \bibinfo{year}{2007}, \bibinfo{journal}{Phys. Rev. Lett.}
  \textbf{\bibinfo{volume}{98}}, \bibinfo{pages}{036801}.

\bibitem[{\citenamefont{Suzuura and Ando}(2002{\natexlab{a}})}]{SA02}
\bibinfo{author}{\bibnamefont{Suzuura}, \bibfnamefont{H.}}, and
  \bibinfo{author}{\bibfnamefont{T.}~\bibnamefont{Ando}},
  \bibinfo{year}{2002}{\natexlab{a}}, \bibinfo{journal}{Phys. Rev. Lett.}
  \textbf{\bibinfo{volume}{89}}, \bibinfo{pages}{266603}.

\bibitem[{\citenamefont{Suzuura and Ando}(2002{\natexlab{b}})}]{SA02b}
\bibinfo{author}{\bibnamefont{Suzuura}, \bibfnamefont{H.}}, and
  \bibinfo{author}{\bibfnamefont{T.}~\bibnamefont{Ando}},
  \bibinfo{year}{2002}{\natexlab{b}}, \bibinfo{journal}{Phys. Rev. B}
  \textbf{\bibinfo{volume}{65}}, \bibinfo{pages}{235412}.

\bibitem[{\citenamefont{Swain and Andelman}(1999)}]{swain99}
\bibinfo{author}{\bibnamefont{Swain}, \bibfnamefont{P.~S.}}, and
  \bibinfo{author}{\bibfnamefont{D.}~\bibnamefont{Andelman}},
  \bibinfo{year}{1999}, \bibinfo{journal}{Langmuir}
  \textbf{\bibinfo{volume}{15}}, \bibinfo{pages}{8902}.

\bibitem[{\citenamefont{Tanuma and Kamimura}(1985)}]{TK85}
\bibinfo{author}{\bibnamefont{Tanuma}, \bibfnamefont{S.}}, and
  \bibinfo{author}{\bibfnamefont{H.}~\bibnamefont{Kamimura}},
  \bibinfo{year}{1985}, \emph{\bibinfo{title}{Graphite intercalation compounds:
  Progress of Research in Japan}} (\bibinfo{publisher}{World Scientific, PA}).

\bibitem[{\citenamefont{Tersoff}(1992)}]{tersoff92}
\bibinfo{author}{\bibnamefont{Tersoff}, \bibfnamefont{J.}},
  \bibinfo{year}{1992}, \bibinfo{journal}{Phys. Rev. B}
  \textbf{\bibinfo{volume}{46}}, \bibinfo{pages}{15546}.

\bibitem[{\citenamefont{Tikhonenko}
  \emph{et~al.}(2007)\citenamefont{Tikhonenko, Horsell, Gorbachev, and
  Savchenko}}]{THGS07}
\bibinfo{author}{\bibnamefont{Tikhonenko}, \bibfnamefont{F.~V.}},
  \bibinfo{author}{\bibfnamefont{D.~W.} \bibnamefont{Horsell}},
  \bibinfo{author}{\bibfnamefont{R.~V.} \bibnamefont{Gorbachev}}, and
  \bibinfo{author}{\bibfnamefont{A.~K.} \bibnamefont{Savchenko}},
  \bibinfo{year}{2007}, \eprint{cond-mat/0707.0140}.

\bibitem[{\citenamefont{Titov}(2007)}]{T07}
\bibinfo{author}{\bibnamefont{Titov}, \bibfnamefont{M.}}, \bibinfo{year}{2007},
  \bibinfo{journal}{Europhys. Lett.} \textbf{\bibinfo{volume}{79}},
  \bibinfo{pages}{17004}.

\bibitem[{\citenamefont{Titov and Beenakker}(2006)}]{TB06}
\bibinfo{author}{\bibnamefont{Titov}, \bibfnamefont{M.}}, and
  \bibinfo{author}{\bibfnamefont{C.}~\bibnamefont{Beenakker}},
  \bibinfo{year}{2006}, \bibinfo{journal}{Phys. Rev. B}
  \textbf{\bibinfo{volume}{74}}, \bibinfo{pages}{041401}.

\bibitem[{\citenamefont{Tom\'anek} \emph{et~al.}(1987)\citenamefont{Tom\'anek,
  Louie, Mamin, Abraham, Thomson, Ganz, and Clarke}}]{Tetal87}
\bibinfo{author}{\bibnamefont{Tom\'anek}, \bibfnamefont{D.}},
  \bibinfo{author}{\bibfnamefont{S.~G.} \bibnamefont{Louie}},
  \bibinfo{author}{\bibfnamefont{H.~J.} \bibnamefont{Mamin}},
  \bibinfo{author}{\bibfnamefont{D.~W.} \bibnamefont{Abraham}},
  \bibinfo{author}{\bibfnamefont{R.~E.} \bibnamefont{Thomson}},
  \bibinfo{author}{\bibfnamefont{E.}~\bibnamefont{Ganz}}, and
  \bibinfo{author}{\bibfnamefont{J.}~\bibnamefont{Clarke}},
  \bibinfo{year}{1987}, \bibinfo{journal}{Phys. Rev. B}
  \textbf{\bibinfo{volume}{35}}, \bibinfo{pages}{7790}.

\bibitem[{\citenamefont{Tombros} \emph{et~al.}(2007)\citenamefont{Tombros,
  Jozsa, Popinciuc, Jonkman, and van Wees}}]{TJPJW07}
\bibinfo{author}{\bibnamefont{Tombros}, \bibfnamefont{N.}},
  \bibinfo{author}{\bibfnamefont{C.}~\bibnamefont{Jozsa}},
  \bibinfo{author}{\bibfnamefont{M.}~\bibnamefont{Popinciuc}},
  \bibinfo{author}{\bibfnamefont{H.~T.} \bibnamefont{Jonkman}}, and
  \bibinfo{author}{\bibfnamefont{J.}~\bibnamefont{van Wees}},
  \bibinfo{year}{2007}, \bibinfo{journal}{Nature}
  \textbf{\bibinfo{volume}{448}}, \bibinfo{pages}{571}.

\bibitem[{\citenamefont{Trushin and Schliemann}(2007)}]{TS07}
\bibinfo{author}{\bibnamefont{Trushin}, \bibfnamefont{M.}}, and
  \bibinfo{author}{\bibfnamefont{J.}~\bibnamefont{Schliemann}},
  \bibinfo{year}{2007}, \bibinfo{journal}{Phys. Rev. Lett.}
  \textbf{\bibinfo{volume}{99}}, \bibinfo{pages}{216602}.

\bibitem[{\citenamefont{Tu and Ou-Yang}(2002)}]{tu02}
\bibinfo{author}{\bibnamefont{Tu}, \bibfnamefont{Z.-C.}}, and
  \bibinfo{author}{\bibfnamefont{Z.-C.} \bibnamefont{Ou-Yang}},
  \bibinfo{year}{2002}, \bibinfo{journal}{Phys. Rev. B}
  \textbf{\bibinfo{volume}{65}}, \bibinfo{pages}{233407}.

\bibitem[{\citenamefont{Tworzydlo} \emph{et~al.}(2007)\citenamefont{Tworzydlo,
  Snyman, Akhmerov, and Beenakker}}]{TSAB07}
\bibinfo{author}{\bibnamefont{Tworzydlo}, \bibfnamefont{J.}},
  \bibinfo{author}{\bibfnamefont{I.}~\bibnamefont{Snyman}},
  \bibinfo{author}{\bibfnamefont{A.~R.} \bibnamefont{Akhmerov}}, and
  \bibinfo{author}{\bibfnamefont{C.~W.~J.} \bibnamefont{Beenakker}},
  \bibinfo{year}{2007}, \bibinfo{journal}{Phys. Rev. B}
  \textbf{\bibinfo{volume}{76}}, \bibinfo{pages}{035411}.

\bibitem[{\citenamefont{Tworzydlo} \emph{et~al.}(2006)\citenamefont{Tworzydlo,
  Trauzettel, Titov, Rycerz, and Beenakker}}]{TTTRB06}
\bibinfo{author}{\bibnamefont{Tworzydlo}, \bibfnamefont{J.}},
  \bibinfo{author}{\bibfnamefont{B.}~\bibnamefont{Trauzettel}},
  \bibinfo{author}{\bibfnamefont{M.}~\bibnamefont{Titov}},
  \bibinfo{author}{\bibfnamefont{A.}~\bibnamefont{Rycerz}}, and
  \bibinfo{author}{\bibfnamefont{C.~W.~J.} \bibnamefont{Beenakker}},
  \bibinfo{year}{2006}, \bibinfo{journal}{Phys. Rev. Lett.}
  \textbf{\bibinfo{volume}{96}}, \bibinfo{pages}{246802}.

\bibitem[{\citenamefont{Uchoa and {Castro Neto}}(2007)}]{UCN07}
\bibinfo{author}{\bibnamefont{Uchoa}, \bibfnamefont{B.}}, and
  \bibinfo{author}{\bibfnamefont{A.~H.} \bibnamefont{{Castro Neto}}},
  \bibinfo{year}{2007}, \bibinfo{journal}{Phys. Rev. Lett.}
  \textbf{\bibinfo{volume}{98}}, \bibinfo{pages}{146801}.

\bibitem[{\citenamefont{Uchoa} \emph{et~al.}(2007)\citenamefont{Uchoa, Lin, and
  {Castro Neto}}}]{ULCN07}
\bibinfo{author}{\bibnamefont{Uchoa}, \bibfnamefont{B.}},
  \bibinfo{author}{\bibfnamefont{C.-Y.} \bibnamefont{Lin}}, and
  \bibinfo{author}{\bibfnamefont{A.~H.} \bibnamefont{{Castro Neto}}},
  \bibinfo{year}{2007}, \bibinfo{journal}{Phys. Rev. B}
  \textbf{\bibinfo{volume}{77}}, \bibinfo{pages}{035420}.

\bibitem[{\citenamefont{Vafek}(2006)}]{V06}
\bibinfo{author}{\bibnamefont{Vafek}, \bibfnamefont{O.}}, \bibinfo{year}{2006},
  \bibinfo{journal}{Phys. Rev. Lett.} \textbf{\bibinfo{volume}{97}},
  \bibinfo{pages}{266406}.

\bibitem[{\citenamefont{Vafek}(2007)}]{V07}
\bibinfo{author}{\bibnamefont{Vafek}, \bibfnamefont{O.}}, \bibinfo{year}{2007},
  \bibinfo{journal}{Phys. Rev. Lett.} \textbf{\bibinfo{volume}{98}},
  \bibinfo{pages}{216401}.

\bibitem[{\citenamefont{Varchon} \emph{et~al.}(2007)\citenamefont{Varchon,
  Feng, Hass, Li, Nguyen, Naud, Mallet, Veuillen, Berger, Conrad, and
  Magaud}}]{Vetal07b}
\bibinfo{author}{\bibnamefont{Varchon}, \bibfnamefont{F.}},
  \bibinfo{author}{\bibfnamefont{R.}~\bibnamefont{Feng}},
  \bibinfo{author}{\bibfnamefont{J.}~\bibnamefont{Hass}},
  \bibinfo{author}{\bibfnamefont{X.}~\bibnamefont{Li}},
  \bibinfo{author}{\bibfnamefont{B.~N.} \bibnamefont{Nguyen}},
  \bibinfo{author}{\bibfnamefont{C.}~\bibnamefont{Naud}},
  \bibinfo{author}{\bibfnamefont{P.}~\bibnamefont{Mallet}},
  \bibinfo{author}{\bibfnamefont{J.~Y.} \bibnamefont{Veuillen}},
  \bibinfo{author}{\bibfnamefont{C.}~\bibnamefont{Berger}},
  \bibinfo{author}{\bibfnamefont{E.~H.} \bibnamefont{Conrad}}, and
  \bibinfo{author}{\bibfnamefont{L.}~\bibnamefont{Magaud}},
  \bibinfo{year}{2007}, \bibinfo{journal}{Phys. Rev. Lett.}
  \textbf{\bibinfo{volume}{99}}, \bibinfo{pages}{126805}.

\bibitem[{\citenamefont{{V\'azquez de Parga}}
  \emph{et~al.}(2007)\citenamefont{{V\'azquez de Parga}, Calleja, Borca,
  {Passeggi Jr}, Hinarejo, Guinea, and Miranda}}]{Vetal07}
\bibinfo{author}{\bibnamefont{{V\'azquez de Parga}}, \bibfnamefont{A.~L.}},
  \bibinfo{author}{\bibfnamefont{F.}~\bibnamefont{Calleja}},
  \bibinfo{author}{\bibfnamefont{B.}~\bibnamefont{Borca}},
  \bibinfo{author}{\bibfnamefont{M.~C.~G.} \bibnamefont{{Passeggi Jr}}},
  \bibinfo{author}{\bibfnamefont{J.~J.} \bibnamefont{Hinarejo}},
  \bibinfo{author}{\bibfnamefont{F.}~\bibnamefont{Guinea}}, and
  \bibinfo{author}{\bibfnamefont{R.}~\bibnamefont{Miranda}},
  \bibinfo{year}{2007}, \bibinfo{title}{Periodically rippled graphene: growth
  and spatially resolved electronic structure}, \eprint{arXiv:0709.0360}.

\bibitem[{\citenamefont{Visscher and Falicov}(1971)}]{VF71}
\bibinfo{author}{\bibnamefont{Visscher}, \bibfnamefont{P.~B.}}, and
  \bibinfo{author}{\bibfnamefont{L.~M.} \bibnamefont{Falicov}},
  \bibinfo{year}{1971}, \bibinfo{journal}{Phys. Rev. B}
  \textbf{\bibinfo{volume}{3}}, \bibinfo{pages}{2541}.

\bibitem[{\citenamefont{Vozmediano}
  \emph{et~al.}(2002)\citenamefont{Vozmediano, L\'opez-Sancho, and
  Guinea}}]{VLG02}
\bibinfo{author}{\bibnamefont{Vozmediano}, \bibfnamefont{M.~A.~H.}},
  \bibinfo{author}{\bibfnamefont{M.~P.} \bibnamefont{L\'opez-Sancho}}, and
  \bibinfo{author}{\bibfnamefont{F.}~\bibnamefont{Guinea}},
  \bibinfo{year}{2002}, \bibinfo{journal}{Phys. Rev. Lett.}
  \textbf{\bibinfo{volume}{89}}, \bibinfo{pages}{166401}.

\bibitem[{\citenamefont{Vozmediano}
  \emph{et~al.}(2003)\citenamefont{Vozmediano, L\'opez-Sancho, and
  Guinea}}]{VLG03}
\bibinfo{author}{\bibnamefont{Vozmediano}, \bibfnamefont{M.~A.~H.}},
  \bibinfo{author}{\bibfnamefont{M.~P.} \bibnamefont{L\'opez-Sancho}}, and
  \bibinfo{author}{\bibfnamefont{F.}~\bibnamefont{Guinea}},
  \bibinfo{year}{2003}, \bibinfo{journal}{Phys. Rev. B}
  \textbf{\bibinfo{volume}{68}}, \bibinfo{pages}{195122}.

\bibitem[{\citenamefont{Vozmediano}
  \emph{et~al.}(2005)\citenamefont{Vozmediano, L\'opez-Sancho, Stauber, and
  Guinea}}]{VLSG05}
\bibinfo{author}{\bibnamefont{Vozmediano}, \bibfnamefont{M.~A.~H.}},
  \bibinfo{author}{\bibfnamefont{M.~P.} \bibnamefont{L\'opez-Sancho}},
  \bibinfo{author}{\bibfnamefont{T.}~\bibnamefont{Stauber}}, and
  \bibinfo{author}{\bibfnamefont{F.}~\bibnamefont{Guinea}},
  \bibinfo{year}{2005}, \bibinfo{journal}{Phys. Rev. B}
  \textbf{\bibinfo{volume}{72}}, \bibinfo{pages}{155121}.

\bibitem[{\citenamefont{Wakabayashi}
  \emph{et~al.}(1999)\citenamefont{Wakabayashi, Fujita, Ajiki, and
  Sigrist}}]{WFA+99}
\bibinfo{author}{\bibnamefont{Wakabayashi}, \bibfnamefont{K.}},
  \bibinfo{author}{\bibfnamefont{M.}~\bibnamefont{Fujita}},
  \bibinfo{author}{\bibfnamefont{H.}~\bibnamefont{Ajiki}}, and
  \bibinfo{author}{\bibfnamefont{M.}~\bibnamefont{Sigrist}},
  \bibinfo{year}{1999}, \bibinfo{journal}{Phys. Rev. B}
  \textbf{\bibinfo{volume}{59}}, \bibinfo{pages}{8271}.

\bibitem[{\citenamefont{Wakayabashi and Sigrist}(2000)}]{WS00}
\bibinfo{author}{\bibnamefont{Wakayabashi}, \bibfnamefont{K.}}, and
  \bibinfo{author}{\bibfnamefont{M.}~\bibnamefont{Sigrist}},
  \bibinfo{year}{2000}, \bibinfo{journal}{\prl} \textbf{\bibinfo{volume}{84}},
  \bibinfo{pages}{3390}.

\bibitem[{\citenamefont{Wallace}(1947)}]{W47}
\bibinfo{author}{\bibnamefont{Wallace}, \bibfnamefont{P.~R.}},
  \bibinfo{year}{1947}, \bibinfo{journal}{Phys. Rev.}
  \textbf{\bibinfo{volume}{71}}, \bibinfo{pages}{622}.

\bibitem[{\citenamefont{Wang} \emph{et~al.}(2007)\citenamefont{Wang, Sheng,
  Sheng, and Haldane}}]{WSSH07}
\bibinfo{author}{\bibnamefont{Wang}, \bibfnamefont{H.}},
  \bibinfo{author}{\bibfnamefont{D.~N.} \bibnamefont{Sheng}},
  \bibinfo{author}{\bibfnamefont{L.}~\bibnamefont{Sheng}}, and
  \bibinfo{author}{\bibfnamefont{F.~D.~M.} \bibnamefont{Haldane}},
  \bibinfo{year}{2007}, \eprint{cond-mat/0708.0382}.

\bibitem[{\citenamefont{Wang and Chakraborty}(2007{\natexlab{a}})}]{WC07}
\bibinfo{author}{\bibnamefont{Wang}, \bibfnamefont{X.-F.}}, and
  \bibinfo{author}{\bibfnamefont{T.}~\bibnamefont{Chakraborty}},
  \bibinfo{year}{2007}{\natexlab{a}}, \bibinfo{journal}{Phys. Rev. B}
  \textbf{\bibinfo{volume}{75}}, \bibinfo{pages}{033408}.

\bibitem[{\citenamefont{Wang and Chakraborty}(2007{\natexlab{b}})}]{WC07b}
\bibinfo{author}{\bibnamefont{Wang}, \bibfnamefont{X.-F.}}, and
  \bibinfo{author}{\bibfnamefont{T.}~\bibnamefont{Chakraborty}},
  \bibinfo{year}{2007}{\natexlab{b}}, \bibinfo{journal}{Phys. Rev. B}
  \textbf{\bibinfo{volume}{75}}, \bibinfo{pages}{041404(R)}.

\bibitem[{\citenamefont{Wehling} \emph{et~al.}(2007)\citenamefont{Wehling,
  Novoselov, Morozov, Vdovin, Katsnelson, Geim, and Lichtenstein}}]{WNMVKGL07}
\bibinfo{author}{\bibnamefont{Wehling}, \bibfnamefont{T.~O.}},
  \bibinfo{author}{\bibfnamefont{K.~S.} \bibnamefont{Novoselov}},
  \bibinfo{author}{\bibfnamefont{S.~V.} \bibnamefont{Morozov}},
  \bibinfo{author}{\bibfnamefont{E.~E.} \bibnamefont{Vdovin}},
  \bibinfo{author}{\bibfnamefont{M.~I.} \bibnamefont{Katsnelson}},
  \bibinfo{author}{\bibfnamefont{A.~K.} \bibnamefont{Geim}}, and
  \bibinfo{author}{\bibfnamefont{A.~I.} \bibnamefont{Lichtenstein}},
  \bibinfo{year}{2007}, \eprint{cond-mat/0703390}.

\bibitem[{\citenamefont{Williams} \emph{et~al.}(2007)\citenamefont{Williams,
  DiCarlo, and Marcus}}]{WDM07}
\bibinfo{author}{\bibnamefont{Williams}, \bibfnamefont{J.~R.}},
  \bibinfo{author}{\bibfnamefont{L.}~\bibnamefont{DiCarlo}}, and
  \bibinfo{author}{\bibfnamefont{C.~M.} \bibnamefont{Marcus}},
  \bibinfo{year}{2007}, \bibinfo{journal}{Science}
  \textbf{\bibinfo{volume}{317}}, \bibinfo{pages}{638}.

\bibitem[{\citenamefont{Williamson}
  \emph{et~al.}(1965)\citenamefont{Williamson, Foner, and Dresselhaus}}]{WFD65}
\bibinfo{author}{\bibnamefont{Williamson}, \bibfnamefont{S.~J.}},
  \bibinfo{author}{\bibfnamefont{S.}~\bibnamefont{Foner}}, and
  \bibinfo{author}{\bibfnamefont{M.~S.} \bibnamefont{Dresselhaus}},
  \bibinfo{year}{1965}, \bibinfo{journal}{Phys. Rev.}
  \textbf{\bibinfo{volume}{140}}, \bibinfo{pages}{A1429}.

\bibitem[{\citenamefont{Wirtz and Rubio}(2004)}]{rubio}
\bibinfo{author}{\bibnamefont{Wirtz}, \bibfnamefont{L.}}, and
  \bibinfo{author}{\bibfnamefont{A.}~\bibnamefont{Rubio}},
  \bibinfo{year}{2004}, \bibinfo{journal}{Sol. Stat. Comm.}
  \textbf{\bibinfo{volume}{131}}, \bibinfo{pages}{141}.

\bibitem[{\citenamefont{Wu} \emph{et~al.}(2007)\citenamefont{Wu, Li, Song,
  Berger, and de~Heer}}]{WLSBH07}
\bibinfo{author}{\bibnamefont{Wu}, \bibfnamefont{X.}},
  \bibinfo{author}{\bibfnamefont{X.}~\bibnamefont{Li}},
  \bibinfo{author}{\bibfnamefont{Z.}~\bibnamefont{Song}},
  \bibinfo{author}{\bibfnamefont{C.}~\bibnamefont{Berger}}, and
  \bibinfo{author}{\bibfnamefont{W.~A.} \bibnamefont{de~Heer}},
  \bibinfo{year}{2007}, \bibinfo{journal}{Phys. Rev. Lett.}
  \textbf{\bibinfo{volume}{98}}, \bibinfo{pages}{136801}.

\bibitem[{\citenamefont{Wunsch} \emph{et~al.}(2006)\citenamefont{Wunsch,
  Stauber, Sols, and Guinea}}]{WSSG06}
\bibinfo{author}{\bibnamefont{Wunsch}, \bibfnamefont{B.}},
  \bibinfo{author}{\bibfnamefont{T.}~\bibnamefont{Stauber}},
  \bibinfo{author}{\bibfnamefont{F.}~\bibnamefont{Sols}}, and
  \bibinfo{author}{\bibfnamefont{F.}~\bibnamefont{Guinea}},
  \bibinfo{year}{2006}, \bibinfo{journal}{New J. Phys.}
  \textbf{\bibinfo{volume}{8}}, \bibinfo{pages}{318}.

\bibitem[{\citenamefont{Xin} \emph{et~al.}(2000)\citenamefont{Xin, Jianjun, and
  Zhong-can}}]{zhou00}
\bibinfo{author}{\bibnamefont{Xin}, \bibfnamefont{Z.}},
  \bibinfo{author}{\bibfnamefont{Z.}~\bibnamefont{Jianjun}}, and
  \bibinfo{author}{\bibfnamefont{O.-Y.} \bibnamefont{Zhong-can}},
  \bibinfo{year}{2000}, \bibinfo{journal}{Phys. Rev. B}
  \textbf{\bibinfo{volume}{62}}, \bibinfo{pages}{13692}.

\bibitem[{\citenamefont{Xu} \emph{et~al.}(1996)\citenamefont{Xu, Cao, Miller,
  Mantell, Miller, and Gao}}]{Xetal96}
\bibinfo{author}{\bibnamefont{Xu}, \bibfnamefont{S.}},
  \bibinfo{author}{\bibfnamefont{J.}~\bibnamefont{Cao}},
  \bibinfo{author}{\bibfnamefont{C.~C.} \bibnamefont{Miller}},
  \bibinfo{author}{\bibfnamefont{D.~A.} \bibnamefont{Mantell}},
  \bibinfo{author}{\bibfnamefont{R.~J.~D.} \bibnamefont{Miller}}, and
  \bibinfo{author}{\bibfnamefont{Y.}~\bibnamefont{Gao}}, \bibinfo{year}{1996},
  \bibinfo{journal}{\prl} \textbf{\bibinfo{volume}{76}}, \bibinfo{pages}{483}.

\bibitem[{\citenamefont{Yan} \emph{et~al.}(2007)\citenamefont{Yan, Zhang, Kim,
  and Pinczuk}}]{YZKP07}
\bibinfo{author}{\bibnamefont{Yan}, \bibfnamefont{J.}},
  \bibinfo{author}{\bibfnamefont{Y.}~\bibnamefont{Zhang}},
  \bibinfo{author}{\bibfnamefont{P.}~\bibnamefont{Kim}}, and
  \bibinfo{author}{\bibfnamefont{A.}~\bibnamefont{Pinczuk}},
  \bibinfo{year}{2007}, \bibinfo{journal}{Phys. Rev. Lett.}
  \textbf{\bibinfo{volume}{98}}, \bibinfo{pages}{166802}.

\bibitem[{\citenamefont{Yang}
  \emph{et~al.}(2007{\natexlab{a}})\citenamefont{Yang, Cohen, and
  Louie}}]{YCL07}
\bibinfo{author}{\bibnamefont{Yang}, \bibfnamefont{L.}},
  \bibinfo{author}{\bibfnamefont{M.~V.} \bibnamefont{Cohen}}, and
  \bibinfo{author}{\bibfnamefont{S.~G.} \bibnamefont{Louie}},
  \bibinfo{year}{2007}{\natexlab{a}}, \bibinfo{journal}{Nano Lett.}
  \textbf{\bibinfo{volume}{7}}, \bibinfo{pages}{3112}.

\bibitem[{\citenamefont{Yang}
  \emph{et~al.}(2007{\natexlab{b}})\citenamefont{Yang, Park, Son, Cohen, and
  Louie}}]{YPSCL07}
\bibinfo{author}{\bibnamefont{Yang}, \bibfnamefont{L.}},
  \bibinfo{author}{\bibfnamefont{C.-H.} \bibnamefont{Park}},
  \bibinfo{author}{\bibfnamefont{Y.-W.} \bibnamefont{Son}},
  \bibinfo{author}{\bibfnamefont{M.~L.} \bibnamefont{Cohen}}, and
  \bibinfo{author}{\bibfnamefont{S.~G.} \bibnamefont{Louie}},
  \bibinfo{year}{2007}{\natexlab{b}}, \bibinfo{journal}{Phys. Rev. Lett.}
  \textbf{\bibinfo{volume}{99}}, \bibinfo{pages}{186801}.

\bibitem[{\citenamefont{Yang and Nayak}(2002)}]{Yang02}
\bibinfo{author}{\bibnamefont{Yang}, \bibfnamefont{X.}}, and
  \bibinfo{author}{\bibfnamefont{C.}~\bibnamefont{Nayak}},
  \bibinfo{year}{2002}, \bibinfo{journal}{Phys. Rev. B}
  \textbf{\bibinfo{volume}{65}}, \bibinfo{pages}{064523}.

\bibitem[{\citenamefont{Yao} \emph{et~al.}(2007)\citenamefont{Yao, Ye, Qi,
  Zhang, and Fang}}]{YYQZF07}
\bibinfo{author}{\bibnamefont{Yao}, \bibfnamefont{Y.}},
  \bibinfo{author}{\bibfnamefont{F.}~\bibnamefont{Ye}},
  \bibinfo{author}{\bibfnamefont{X.-L.} \bibnamefont{Qi}},
  \bibinfo{author}{\bibfnamefont{S.-C.} \bibnamefont{Zhang}}, and
  \bibinfo{author}{\bibfnamefont{Z.}~\bibnamefont{Fang}}, \bibinfo{year}{2007},
  \bibinfo{journal}{Phys. Rev. B} \textbf{\bibinfo{volume}{75}},
  \bibinfo{pages}{041401}.

\bibitem[{\citenamefont{Zarea and Sandler}(2007)}]{ZS07}
\bibinfo{author}{\bibnamefont{Zarea}, \bibfnamefont{M.}}, and
  \bibinfo{author}{\bibfnamefont{N.}~\bibnamefont{Sandler}},
  \bibinfo{year}{2007}, \bibinfo{journal}{Phys. Rev. Lett.}
  \textbf{\bibinfo{volume}{99}}, \bibinfo{pages}{256804}.

\bibitem[{\citenamefont{Zhang and Joglekar}(2007)}]{ZJ0703}
\bibinfo{author}{\bibnamefont{Zhang}, \bibfnamefont{C.-H.}}, and
  \bibinfo{author}{\bibfnamefont{Y.~N.} \bibnamefont{Joglekar}},
  \bibinfo{year}{2007}, \bibinfo{journal}{Phys. Rev. B}
  \textbf{\bibinfo{volume}{75}}, \bibinfo{pages}{245414}.

\bibitem[{\citenamefont{Zhang and Fogler}(2007)}]{ZF07}
\bibinfo{author}{\bibnamefont{Zhang}, \bibfnamefont{L.~M.}}, and
  \bibinfo{author}{\bibfnamefont{M.~M.} \bibnamefont{Fogler}},
  \bibinfo{year}{2007}, \eprint{arXiv:0708.0892}.

\bibitem[{\citenamefont{Zhang} \emph{et~al.}(2006)\citenamefont{Zhang, Jiang,
  Small, Purewal, Tan, Fazlollahi, Chudow, Jaszczak, Stormer, and
  Kim}}]{Zetal06b}
\bibinfo{author}{\bibnamefont{Zhang}, \bibfnamefont{Y.}},
  \bibinfo{author}{\bibfnamefont{Z.}~\bibnamefont{Jiang}},
  \bibinfo{author}{\bibfnamefont{J.~P.} \bibnamefont{Small}},
  \bibinfo{author}{\bibfnamefont{M.~S.} \bibnamefont{Purewal}},
  \bibinfo{author}{\bibfnamefont{Y.-W.} \bibnamefont{Tan}},
  \bibinfo{author}{\bibfnamefont{M.}~\bibnamefont{Fazlollahi}},
  \bibinfo{author}{\bibfnamefont{J.~D.} \bibnamefont{Chudow}},
  \bibinfo{author}{\bibfnamefont{J.~A.} \bibnamefont{Jaszczak}},
  \bibinfo{author}{\bibfnamefont{H.~L.} \bibnamefont{Stormer}}, and
  \bibinfo{author}{\bibfnamefont{P.}~\bibnamefont{Kim}}, \bibinfo{year}{2006},
  \bibinfo{journal}{Phys. Rev. Lett.} \textbf{\bibinfo{volume}{96}},
  \bibinfo{pages}{136806}.

\bibitem[{\citenamefont{Zhang} \emph{et~al.}(2005)\citenamefont{Zhang, Tan,
  Stormer, and Kim}}]{Zetal05b}
\bibinfo{author}{\bibnamefont{Zhang}, \bibfnamefont{Y.}},
  \bibinfo{author}{\bibfnamefont{Y.-W.} \bibnamefont{Tan}},
  \bibinfo{author}{\bibfnamefont{H.~L.} \bibnamefont{Stormer}}, and
  \bibinfo{author}{\bibfnamefont{P.}~\bibnamefont{Kim}}, \bibinfo{year}{2005},
  \bibinfo{journal}{Nature} \textbf{\bibinfo{volume}{438}},
  \bibinfo{pages}{201}.

\bibitem[{\citenamefont{Zhong-can} \emph{et~al.}(1997w)\citenamefont{Zhong-can,
  Su, and Wang}}]{zhou97}
\bibinfo{author}{\bibnamefont{Zhong-can}, \bibfnamefont{O.-Y.}},
  \bibinfo{author}{\bibfnamefont{Z.-B.} \bibnamefont{Su}}, and
  \bibinfo{author}{\bibfnamefont{C.-L.} \bibnamefont{Wang}},
  \bibinfo{year}{1997w}, \bibinfo{journal}{Phys. Rev. Lett.}
  \textbf{\bibinfo{volume}{78}}, \bibinfo{pages}{4055}.

\bibitem[{\citenamefont{Zhou}
  \emph{et~al.}(2006{\natexlab{a}})\citenamefont{Zhou, Gweon, and
  Lanzara}}]{ZGL06}
\bibinfo{author}{\bibnamefont{Zhou}, \bibfnamefont{S.}},
  \bibinfo{author}{\bibfnamefont{G.-H.} \bibnamefont{Gweon}}, and
  \bibinfo{author}{\bibfnamefont{A.}~\bibnamefont{Lanzara}},
  \bibinfo{year}{2006}{\natexlab{a}}, \bibinfo{journal}{Annals of Physics}
  \textbf{\bibinfo{volume}{321}}, \bibinfo{pages}{1730}.

\bibitem[{\citenamefont{Zhou} \emph{et~al.}(2007)\citenamefont{Zhou, Gweon,
  Fedorov, First, de~Heer, Lee, Guinea, {Castro Neto}, and
  Lanzara}}]{ZGFFHLGCNL07}
\bibinfo{author}{\bibnamefont{Zhou}, \bibfnamefont{S.~Y.}},
  \bibinfo{author}{\bibfnamefont{G.-H.} \bibnamefont{Gweon}},
  \bibinfo{author}{\bibfnamefont{A.~V.} \bibnamefont{Fedorov}},
  \bibinfo{author}{\bibfnamefont{P.~N.} \bibnamefont{First}},
  \bibinfo{author}{\bibfnamefont{W.~A.} \bibnamefont{de~Heer}},
  \bibinfo{author}{\bibfnamefont{D.-H.} \bibnamefont{Lee}},
  \bibinfo{author}{\bibfnamefont{F.}~\bibnamefont{Guinea}},
  \bibinfo{author}{\bibfnamefont{A.~H.} \bibnamefont{{Castro Neto}}}, and
  \bibinfo{author}{\bibfnamefont{A.}~\bibnamefont{Lanzara}},
  \bibinfo{year}{2007}, \bibinfo{journal}{Nature Materials}
  \textbf{\bibinfo{volume}{6}}, \bibinfo{pages}{770}.

\bibitem[{\citenamefont{Zhou}
  \emph{et~al.}(2006{\natexlab{b}})\citenamefont{Zhou, Gweon, Graf, Fedorov,
  Spataru, Diehl, Kopelevich, Lee, Louie, and Lanzara}}]{lanzara06}
\bibinfo{author}{\bibnamefont{Zhou}, \bibfnamefont{S.~Y.}},
  \bibinfo{author}{\bibfnamefont{G.-H.} \bibnamefont{Gweon}},
  \bibinfo{author}{\bibfnamefont{J.}~\bibnamefont{Graf}},
  \bibinfo{author}{\bibfnamefont{A.~V.} \bibnamefont{Fedorov}},
  \bibinfo{author}{\bibfnamefont{C.~D.} \bibnamefont{Spataru}},
  \bibinfo{author}{\bibfnamefont{R.~D.} \bibnamefont{Diehl}},
  \bibinfo{author}{\bibfnamefont{Y.}~\bibnamefont{Kopelevich}},
  \bibinfo{author}{\bibfnamefont{D.-H.} \bibnamefont{Lee}},
  \bibinfo{author}{\bibfnamefont{S.~G.} \bibnamefont{Louie}}, and
  \bibinfo{author}{\bibfnamefont{A.}~\bibnamefont{Lanzara}},
  \bibinfo{year}{2006}{\natexlab{b}}, \bibinfo{journal}{Nature Physics}
  \textbf{\bibinfo{volume}{2}}, \bibinfo{pages}{595}.

\bibitem[{\citenamefont{Zhou}
  \emph{et~al.}(2006{\natexlab{c}})\citenamefont{Zhou, Gweon, Graf, Fedorov,
  Spataru, Diehl, Kopelevich, Lee, Louie, and Lanzara}}]{Zetal06}
\bibinfo{author}{\bibnamefont{Zhou}, \bibfnamefont{S.~Y.}},
  \bibinfo{author}{\bibfnamefont{G.-H.} \bibnamefont{Gweon}},
  \bibinfo{author}{\bibfnamefont{J.}~\bibnamefont{Graf}},
  \bibinfo{author}{\bibfnamefont{A.~V.} \bibnamefont{Fedorov}},
  \bibinfo{author}{\bibfnamefont{C.~D.} \bibnamefont{Spataru}},
  \bibinfo{author}{\bibfnamefont{R.~D.} \bibnamefont{Diehl}},
  \bibinfo{author}{\bibfnamefont{Y.}~\bibnamefont{Kopelevich}},
  \bibinfo{author}{\bibfnamefont{D.-H.} \bibnamefont{Lee}},
  \bibinfo{author}{\bibfnamefont{S.~G.} \bibnamefont{Louie}}, and
  \bibinfo{author}{\bibfnamefont{A.}~\bibnamefont{Lanzara}},
  \bibinfo{year}{2006}{\natexlab{c}}, \bibinfo{journal}{Nature Physics}
  \textbf{\bibinfo{volume}{2}}, \bibinfo{pages}{595}.

\bibitem[{\citenamefont{Ziegler}(1998)}]{Ziegler98}
\bibinfo{author}{\bibnamefont{Ziegler}, \bibfnamefont{K.}},
  \bibinfo{year}{1998}, \bibinfo{journal}{Phys. Rev. Lett.}
  \textbf{\bibinfo{volume}{80}}, \bibinfo{pages}{3113}.

\bibitem[{\citenamefont{Ziman}(1972)}]{Ziman}
\bibinfo{author}{\bibnamefont{Ziman}, \bibfnamefont{J.~M.}},
  \bibinfo{year}{1972}, \emph{\bibinfo{title}{Principles of the Theory of
  Solids}} (\bibinfo{publisher}{Cambridge}).

\end{thebibliography}
%------------------------------------------------------------------------------
%------------------------------------------------------------------------------

\end{document}